\UseRawInputEncoding

\documentclass[11pt,a4paper]{scrartcl}

\usepackage{CLICdp}

\usepackage{CLICdp_definitions}

\usepackage{caption}
\usepackage{subcaption}

\usepackage{placeins}

\usepackage{amsmath}
\usepackage{amsfonts}
\usepackage{amssymb}
\usepackage{adjustbox}

\usepackage{tikz}
\usetikzlibrary{arrows, decorations.markings}
\usetikzlibrary{shapes,snakes}
\usetikzlibrary{decorations.pathmorphing}	
\usetikzlibrary{decorations.markings}
\usetikzlibrary{decorations.text}

\usepackage{cleveref}

\usepackage[colorinlistoftodos]{todonotes}

\usepackage{cancel}

\definecolor{crimsonred}{RGB}{153,0,0}		
\definecolor{darkcharcoal}{RGB}{25,25,25}	
\definecolor{charcoal}{RGB}{51,51,51}		
\definecolor{ash}{RGB}{100,100,100}			
\definecolor{paleblue}{RGB}{0,102,102}		
\definecolor{turtlegreen}{RGB}{51,153,0}	
\definecolor{paleale}{RGB}{204,204,102}		
\definecolor{lager}{RGB}{140,110,10}		
\definecolor{regal}{RGB}{90,0,120}			
\definecolor{jeans}{RGB}{20,30,150}			
\definecolor{clic}{RGB}{223,102,32}			
\definecolor{clic1}{RGB}{0,161,75}			
\definecolor{clic2}{RGB}{161,152,0}			
\definecolor{clic3}{RGB}{137,0,161}			
\definecolor{clic4}{RGB}{0,83,161}			
\definecolor{clicgray}{RGB}{93,93,93}		
\definecolor{cern}{RGB}{0,83,163}     		

\tikzset{
    vector/.style={decorate, decoration={snake}, draw=clic2},
	provector/.style={decorate, decoration={snake,amplitude=2.5pt}, draw},
	antivector/.style={decorate, decoration={snake,amplitude=-2.5pt}, draw},
    fermion/.style={draw=cern, postaction={decorate},
        decoration={markings,mark=at position .55 with {\arrow[draw=cern]{>}}}},
    fermionbar/.style={draw=cern, postaction={decorate},
        decoration={markings,mark=at position .55 with {\arrow[draw=cern]{<}}}},
    fermionnoarrow/.style={draw=black},
    gluon/.style={decorate, draw=paleale,
        decoration={coil,amplitude=4pt, segment length=5pt}},
    scalar/.style={dashed,draw=clic3, postaction={decorate},
        decoration={markings,mark=at position .55 with {\arrow[draw=clic3]{>}}}},
    scalarbar/.style={dashed,draw=clic3, postaction={decorate},
        decoration={markings,mark=at position .55 with {\arrow[draw=clic3]{<}}}},
    scalarnoarrow/.style={dashed,draw=black},
    electron/.style={draw=cern, postaction={decorate},
        decoration={markings,mark=at position .55 with {\arrow[draw=black]{>}}}},
     bigvector/.style={decorate, decoration={snake,amplitude=4pt}, draw},
     line/.style={decorate, draw=black},
}

\title{Physics potential for boosted topologies in top-quark pair production at a multi-TeV Compact Linear Collider}

\clicdpnote{2020}{004}  

\date{\formatdate{12}{8}{2020}}

\addauthor{R.~Str\"om}{\institute{1}}
\addauthor{P.~Roloff\,}{\institute{1}}

\addinstitute{1}{CERN, Switzerland}

\abstract{The physics potential for boosted topologies in top-quark pair production is studied at centre-of-mass energies of 1.4\,\tev and 3\,\tev at the proposed high-luminosity linear electron-positron Compact Linear Collider (CLIC). The analyses presented in this paper focus on ``single lepton+jets'' $\ttbar$ final states and apply jet sub-structure techniques to explore the highly collimated jet environment above 1\,\tev. The charged lepton is used to determine the charge of both top quarks. We present results for the $\ttbar$ production cross section and the forward-backward asymmetry in the kinematic region $\rootsprime\geq1.2\,\tev$ ($\rootsprime\geq2.6\,\tev$) for operation at 1.4\,\tev\,(3\,\tev), where $\rootsprime$ is the effective collision energy, taking into account the CLIC luminosity spectrum and initial-state radiation. The results are based on detailed Monte Carlo simulation studies with a \geant based simulation of the \clicild detector concept and particle-flow based event reconstruction. All data samples considered include beam-induced backgrounds and other relevant background processes. The expected precision on the $\ttbar$ production cross section and the forward-backward asymmetry are 1.1\%\,(2.0\%) and 1.4\%\,(2.3\%), respectively, for operation at 1.4\,\tev\,(3\,\tev) with an integrated luminosity of $2.0\,\abinv$ ($4.0\,\abinv$) and with -80\% electron polarisation. For improved Beyond Standard Model reach, operation is also foreseen at +80\% electron polarisation, with an integrated luminosity of $0.5\,\abinv$ ($1.0\,\abinv$) at 1.4\,\tev\,(3\,\tev), where the corresponding numbers are about a factor 2.5 higher.}

\titlecomment{This work was carried out in the framework of the CLICdp Collaboration} 

\graphicspath{ {./logos/}{./figures/} }

\begin{document}

\titlepage
\clearpage
\tableofcontents
\clearpage

\section{Introduction}
\label{sec:intro}

The top quark is the heaviest known fundamental particle and the only observed fermion with a weak-scale mass, thus constituting a unique probe of the Standard Model (SM) of particle physics. Additionally it occupies an important role in many theories of physics beyond the SM (BSM). Top-quark production is precisely predicted in the SM but may receive substantial modifications from new physics effects such as extra dimensions \cite{Randall:1999ee} and compositeness \cite{Pomarol:2008bh}. The top-quark has so far only been produced in hadron collisions, at the Tevatron and Large Hadron Collider (LHC). However, top-quark production in electron-positron collisions offers both complementary and improved precision measurements.

In this paper we focus on the prospects for measurements of top-quark pair production observables in the context of the proposed Compact Linear Collider (CLIC), a high-luminosity linear electron-positron collider that reaches multi-\tev energies through a staged implementation. Precision studies of top-quark pair production and rare decays are possible already at the first energy stage of CLIC at $\roots=380\,\gev$~\cite{Abramowicz:2018rjq}. The initial stage also includes an energy scan in the top-quark pair production threshold region, which allows the top-quark mass to be extracted with a precision of around 50 MeV~\cite{Abramowicz:2018rjq}. The higher-energy stages, at $\roots=1.5\,\tev$ and $\roots=3\,\tev$, are the focus of this paper and supplement the initial energy datasets with large samples of top quarks, many of which are produced with a substantial boost. The use of dedicated strategies adapted to the more highly collimated jet environment, result in an increased sensitivity to new physics, in particular for BSM effects that grow with the centre-of-mass energy. The main results from this paper and a summary of the analysis strategy were previously summarised in \cite{Abramowicz:2018rjq} where a comprehensive view of the prospects of for the foreseen top-quark programme at CLIC was presented. This paper focus on presenting the underlying analysis in greater detail.

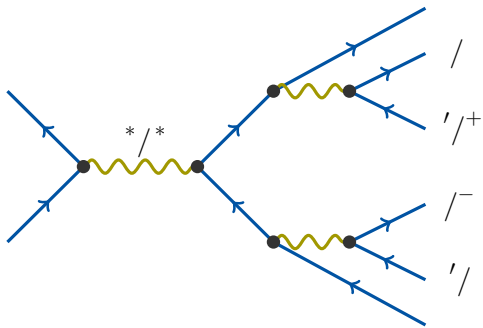
\begin{figure}[h]
\centering
\begin{tikzpicture}[line width=1.2 pt, scale=1.0]
\draw[fermionbar]  (0,1) -- (1,0) ;
\draw[fermion]  (0,-1) -- (1,0) ;
\draw[vector]  (1,0) -- (2.5,0) ;
\draw[fermion]  (2.5,0) -- (3.5,1) ;
\draw[fermionbar]  (2.5,0) -- (3.5,-1) ;
\draw (1,0) node[circle,inner sep=0pt,minimum size=5pt,fill=charcoal]{};
\draw (2.5,0) node[circle,inner sep=0pt,minimum size=5pt,fill=charcoal]{};
\draw[fermion]  (3.5,1) -- (5.5,2.1) ;
\draw[vector]  (3.5,1) -- (4.5,1) ;
\draw[fermion]  (4.5,1) -- (5.5,1.5) ;
\draw[fermionbar]  (4.5,1) -- (5.5,0.5) ;
\draw (3.5,1) node[circle,inner sep=0pt,minimum size=5pt,fill=charcoal]{};
\draw (4.5,1) node[circle,inner sep=0pt,minimum size=5pt,fill=charcoal]{};
\draw[fermionbar]  (3.5,-1) -- (5.5,-2.1) ;
\draw[vector]  (3.5,-1) -- (4.5,-1) ;
\draw[fermionbar]  (4.5,-1) -- (5.5,-1.5) ;
\draw[fermion]  (4.5,-1) -- (5.5,-0.5) ;
\draw (3.5,-1) node[circle,inner sep=0pt,minimum size=5pt,fill=charcoal]{};
\draw (4.5,-1) node[circle,inner sep=0pt,minimum size=5pt,fill=charcoal]{};
\node at (-0.3, 1) {$\Pep$};
\node at (-0.3, -1) {$\Pem$};
\node at (1.82, 0.33) {$\PZ^\ast/\,\PGg^\ast$};
\node at (3.2, 1.05) {$\PQt$};
\node at (3.2, -1.05) {$\PAQt$};
\node at (5.74, 2.1) {$\PQb$};
\node at (5.74, -2.1) {$\PAQb$};
\node at (5.92, 1.5) {$\PQq/\PGn_{\Pl}$};
\node at (6.0, 0.5) {$\PAQq'/\Pl^+$};
\node at (5.97, -1.5) {$\PAQq'/\PAGn_{\Pl}$};
\node at (5.95, -0.5) {$\PQq/\Pl^-$};
\node at (4.05, 0.65) {$\PWp$};
\node at (4.05, -0.65) {$\PWm$};
\end{tikzpicture} 
\caption{The dominant top-quark pair production process in $\epem$ interactions.\label{fig:production_diagrams:ttbar}}
\label{fig:ttbar}
\end{figure}

The top quark decays before hadronising, producing a $\PW$ boson and a bottom quark with a branching ratio close to 100\%. The dominant $\PZ^{\ast}/\gamma^{\ast}$ exchange diagram is shown in \Cref{fig:ttbar}. The analyses presented in this paper focus on ''single lepton+jets'' final states, where the reconstructed charged lepton is used to determine the charge of the hadronically decaying top quark. Note that contributions to the inclusive six-fermion final state from $\text{non-}\ttbar$ processes, such as single-top production and triple gauge boson production, cannot be fully separated due to interference. In fact, at high centre-of-mass energies, the contribution of $\text{non-}\ttbar$ events is significant~\cite{Fuster2015} and constitute an irreducible background to the analysis.

While data from different centre-of-mass energies effectively constrains new physics effects that grow with energy~\cite{AguilarSaavedra:2012vh,PerelloVosZhang}, enriching the event samples in either left-handed or right-handed top-quarks, through the use of polarised beams, allows efficient disentanglement of the photon and Z-boson contributions~\cite{Amjad:2015mma}. Measurements of the top-quark pair production cross section, $\csttbar$, and the forward-backward asymmetry, $\afb$, are presented for each of the higher energy stages of CLIC and make use of longitudinal electron spin polarisation as foreseen in the baseline accelerator design. The clean environment of lepton colliders is also well suited for the accurate measurement of observables that characterise the differential distributions of the top-quark scattering and decay kinematics. Such differential features were studied for the analyses presented here in the context of a multivariate statistical framework based on ``statistically optimal observables''~\cite{Atwood:1991ka, Davier:1992nw, Diehl:1993br}. 

The following section describes the experimental conditions at CLIC including a brief overview of the accelerator and detector concepts. The event generation, detector simulation, and particle-flow reconstruction is outlined in \Cref{sec:gensim}, while \Cref{sec:analysis_strategy} focuses on the overall analysis strategy. The subsequent sections, from \Cref{sec:lepid} to \Cref{sec:effcom}, outline the underlying analysis in detail, describing the different reconstruction methods applied, including the identification of isolated leptons, boosted hadronic top quarks, and the effective centre-of-mass energy. We also present for the first time the optimisation of the boosted top-tagger algorithm. The final event-selection step, where a system of multivariate classifier algorithm (including variables investigating the substructure of large-R jets) is applied, is described in \Cref{sec:mva}. The event selection is summarised in \Cref{sec:eventselsum}. The resulting statistical and systematical uncertainties on the considered $\ttbar$ observables are presented and discussed in \Cref{sec:results}. We conclude with a summary and outlook in \Cref{sec:summary}.
\section{Experimental environment at CLIC}
\label{sec:clic}

To optimise the physics potential, CLIC is proposed as a staged collider providing high-luminosity \epem collisions at centre-of-mass energies of 380 GeV, 1.5 TeV\footnote{The second-stage energy of 1.5\,TeV has recently been adopted and will be used for future studies. In the work presented here, the previous baseline of 1.4\,TeV is used.} and 3 TeV~\cite{staging_baseline_yellow_report} with a corresponding integrated luminosity of $1.0\,\abinv$, $2.5\,\abinv$, and $5.0\,\abinv$, respectively~\cite{Charles:2018vfv,Robson:2018zje}. The accelerator is based on an innovative two-beam scheme, in which normal-conducting high-gradient 12 GHz X-band accelerating structures are powered via a high-current drive beam~\cite{CLICCDR_vol1}. This enables a compact and cost-effective accelerator complex, with a site length ranging between 11 km and 50 km.

The baseline accelerator design foresees $\pm 80\%$ longitudinal electron spin polarisation and no positron polarisation; the polarisation state is denoted P(\Pem) in the following. Owing to the underlying chiral structure of the electroweak interaction, the $\epem\to\ttbar$ cross section is significantly enhanced for an electron polarisation of -80\%; the cross section is enhanced (reduced) by 30\% at the higher-energy stages when operating with -80\% (+80\%) beam polarisation. However, some operation at an electron polarisation of +80\% is still desired to disentangle the photon and Z-boson contributions~\cite{Amjad:2015mma}. A baseline with shared running time for -80\% and +80\% electron polarisation in the ratio 80:20 is adopted for the two higher-energy stages in the studies presented in this paper~\cite{Charles:2018vfv,Robson:2018zje}.

\begin{figure}[t!]
  \centering
  \includegraphics[scale=0.9]{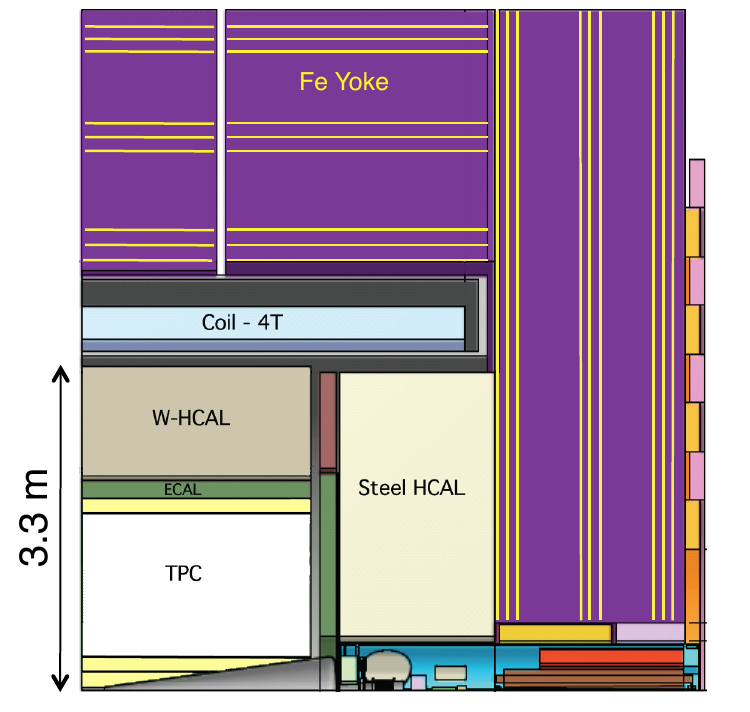}
  \caption{Longitudinal cross section of the top right quadrant of the
    \clicild detector concept \cite{CLIC_PhysDet_CDR}.}
  \label{fig:clicild}
\end{figure}

The CLIC accelerator complex is complemented by a multipurpose detector system optimised for $\epem$ physics. The \clicild detector concept, used for the study described here, is adapted from the ILD~\cite{ildloi:2009,ilctdrvol4:2013} detector concept for the International Linear Collider (ILC). Design modifications are motivated by the higher collision energy and the challenging beam conditions at CLIC.

The inner part of the \clicild detector consists of a large central gaseous time projection chamber (TPC) for tracking, enclosing an ultra-thin silicon-pixel vertex detector. Furthermore, the TPC is surrounded by a silicon strip detector envelope. Together they provide excellent track momentum resolution and high impact parameter resolution, both crucial for an accurate vertex reconstruction and flavour tagging. The detector requirements for the former is at the level of $\sigma_{\pT}/\pT^2 \lesssim 2 \cdot 10^{-5}$ $\gev^{-1}$ and the latter is defined by $a \lesssim 5\,\micron$ and $b \lesssim 15\,\micron\,\gev$ in $\sigma_{d_0}^2 = a^2 + b^2/(p^2\sin^3\theta)$. The vertex and tracking systems are surrounded by a highly-granular sampling calorimetry system, composed of an electromagnetic and a hadronic calorimeter (ECAL and HCAL), optimised for particle flow reconstruction. The resulting jet-energy resolution, for isolated central light-quark jets with energy in the range $100\,\gev$ to $1\,\tev$, is $\sigma_E/E \lesssim 3.5\,\%$~\cite{Marshall2013153}. A strong solenoidal magnet located outside the HCAL provides a 4\,T magnetic field. The magnetic flux return is contained in a large iron yoke instrumented with detectors for muon identification. Dedicated calorimeters in the very forward region are used for luminosity measurements and extended coverage for electromagnetic particles. The detector layout is shown in \Cref{fig:clicild} and is discussed in more detail in~\cite{CLIC_PhysDet_CDR}.\footnote{The detector is described using a right-handed coordinate system with the $z$-axis along the electron beam direction. Here, $\theta$ denotes the polar angle from the $z$-axis.} 

The beam and bunch structure of CLIC is rather distinct, with a bunch train repetition rate of 50\,Hz. Each bunch train consists of 312 bunches with 0.5\,ns separation. High luminosity is reached by a very small beam emittance that is maintained through the accelerator chain. The resulting highly-focused and intense electron and positron beams at the interaction point give rise to significant beamstrahlung from interactions between colliding bunches~\cite{CLICCDR_vol1}. This constitutes a large experimental background of $\epem$ pairs and $\gghadrons$ processes. Note that the low bunch train repetition rate allows for a trigger-less readout of the full detector between bunch trains. The energy deposits from hard physics events and those from beam-induced backgrounds in other bunch-crossings can be sufficiently distinguished through the sub-ns time resolution achieved for reconstructed particle flow objects. The beamstrahlung background is reduced to a manageable level by applying detector-system dependent and $\pT$-dependent timing cuts. The cuts also depend on the reconstructed particle type. In these studies we consider the so-called \texttt{default} and \texttt{tight} cuts, the latter applying a more stringent selection. The former is applied for operation at $1.4\,\tev$ and the latter at $3\,\tev$, where the beam backgrounds are more significant. In addition, the use of hadron-collider-like jet-clustering algorithms with beam-jets further reduces the impact of these backgrounds on physics measurements~\cite{Boronat:2016tgd}.

\begin{figure}[t!]
  \centering
  \includegraphics[scale=0.42]{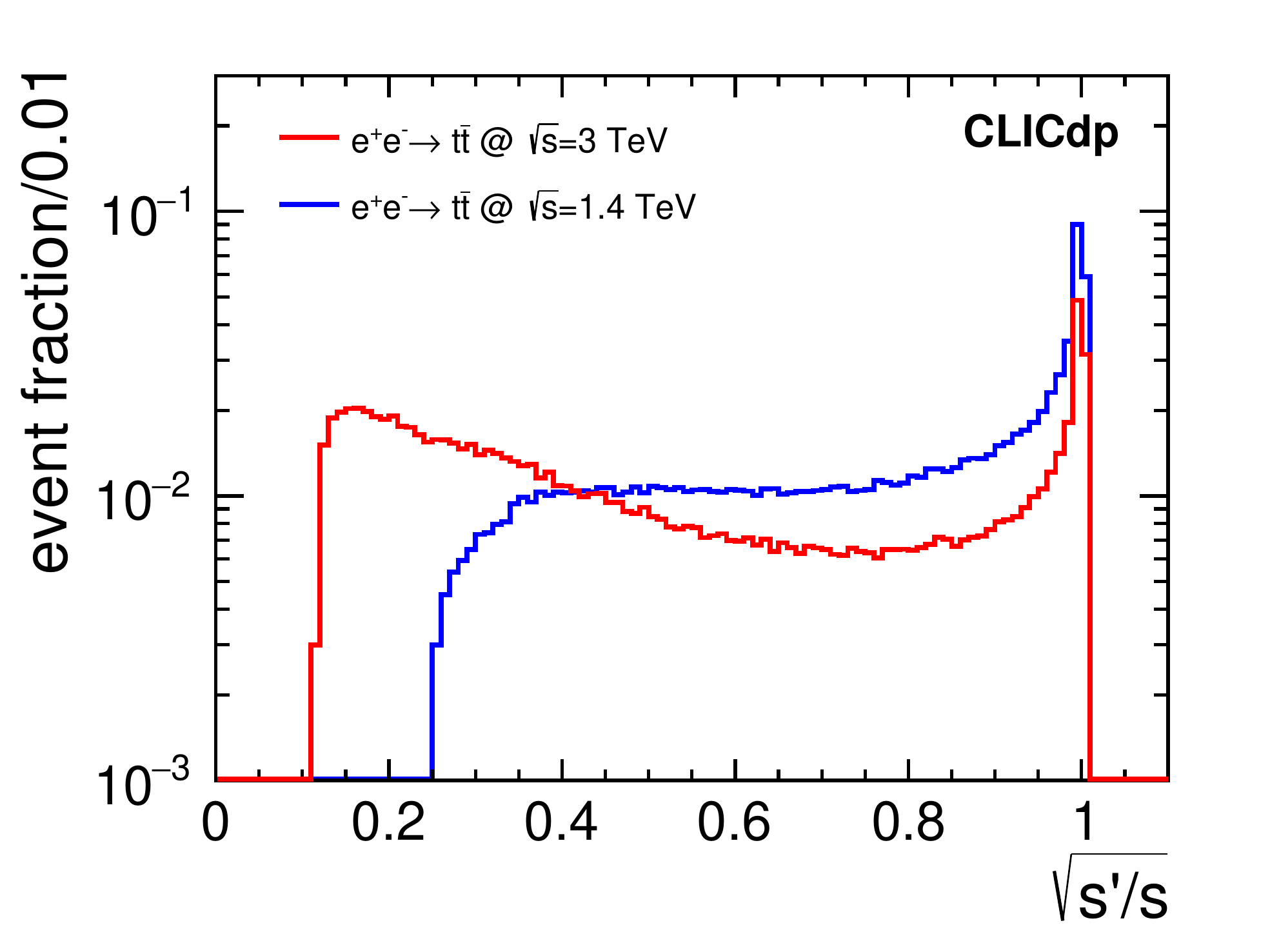}
  \caption{Fraction of the nominal collision energy that is carried by all final-state products in $\ttbar$ production at $\roots=1.4\,\tev$ (blue) and $\roots=3\,\tev$ (red), including ISR, electroweak corrections, and the CLIC luminosity spectrum. Both distributions are normalised to unity.}
  \label{fig:sprimefrac}
\end{figure}

As a result of beamstrahlung, the luminosity spectrum features a long lower-energy tail. Initial-state radiation (ISR) has a similar effect. The convolution of these effects with the top-quark pair production cross-section, that decreases with centre-of-mass energy, is shown in \Cref{fig:sprimefrac} for operation at $\roots=1.4\,\tev$ (blue) and $3\,\tev$ (red)~\cite{CLIC_PhysDet_CDR}. The measurements presented in this paper are studied in the region close to the nominal collision energy $\roots$ for each collider stage. For $\roots=1.4\,\tev$ about 36\% of the events collide with an energy above $1.2\,\tev$ ($\sim$85\% of $\roots$). For the highest energy stage at $\roots=3\,\tev$ the corresponding  number is about 18\% for events that collide with energy above $2.6\,\tev$ ($\sim$85\% of $\roots$). The reconstruction of the effective centre-of-mass energy $\rootsprime$ is presented in detail in \Cref{sec:effcom}. So-called radiative top-quark pair production events, produced below the nominal collision energy, have been studied in detail elsewhere~\cite{Abramowicz:2018rjq}.

\section{Event generation, detector simulation, and reconstruction}
\label{sec:gensim}

The studies reported here are based on detailed Monte Carlo (MC) simulation studies with the \clicild detector concept. Events are generated with \whizard 1.95~\cite{Kilian:2007gr}, while the detector response is simulated with the detector simulation toolkit \mokka~\cite{Mokka} based on \geant~\cite{Agostinelli2003, Allison2006}. Fragmentation and hadronisation is simulated using \pythia 6.4 \cite{Sjostrand2006} tuned to OPAL $\epem$ data recorded at LEP \cite{Alexander:1995bk}. The impact of other \pythia tunes in top-quark pair production events is illustrated in~\cite{Chekanov:2289960}. The decays of $\PGt$ leptons are simulated using \tauola~\cite{tauola}. ISR is characterised by the leading logarithmic approximation structure function~\cite{Skrzypek:1990qs}, which includes hard collinear photons up to third order. The top-quark mass and width are set to $m_{\PQt}=174.0\,\gev$ and $\Gamma_{top}=1.523\,\gev$, respectively.

The beam-induced background from \gghadrons processes was simulated separately using \mbox{\pythia}, with photon spectra from \guineapig \cite{guineapig}. The resulting background particles were overlaid on the physics events\footnote{Considering the bunch train structure of CLIC as well as the expected timing resolution and performance of the detector read-out electronics, backgrounds corresponding to 60 bunch-crossings were overlaid on top of each physics event.}, corresponding to about 1.3 (3.2) \gghadrons interactions per bunch crossing at \roots=1.4\,\tev (3\,\tev).

Track reconstruction is performed using the \marlin software package, and \pandora~\cite{thomson:pandora, Marshall2013153, Marshall:2015rfa} is used for calorimeter clustering and particle flow reconstruction, creating a collection of so-called Particle-Flow Objects (PFOs). The \textsc{LcfiPlus} package \cite{Suehara:2015ura} is used for vertex reconstruction and for tagging jets for charm and beauty probabilities. These are based on variables such as secondary vertex decay lengths, multiplicities and masses, as well as track impact parameters. 

Dedicated sections in this document provide details on the reconstruction of isolated leptons as well as the tagging of hadronically decaying tops. The event simulation and reconstruction is performed using the \ilcdirac grid production tools \cite{Grefe:2014sca, Tsaregorodtsev:2008zz}.

\subsection{Simulation samples}

\begin{figure}
  \centering
  \includegraphics[scale=0.33]{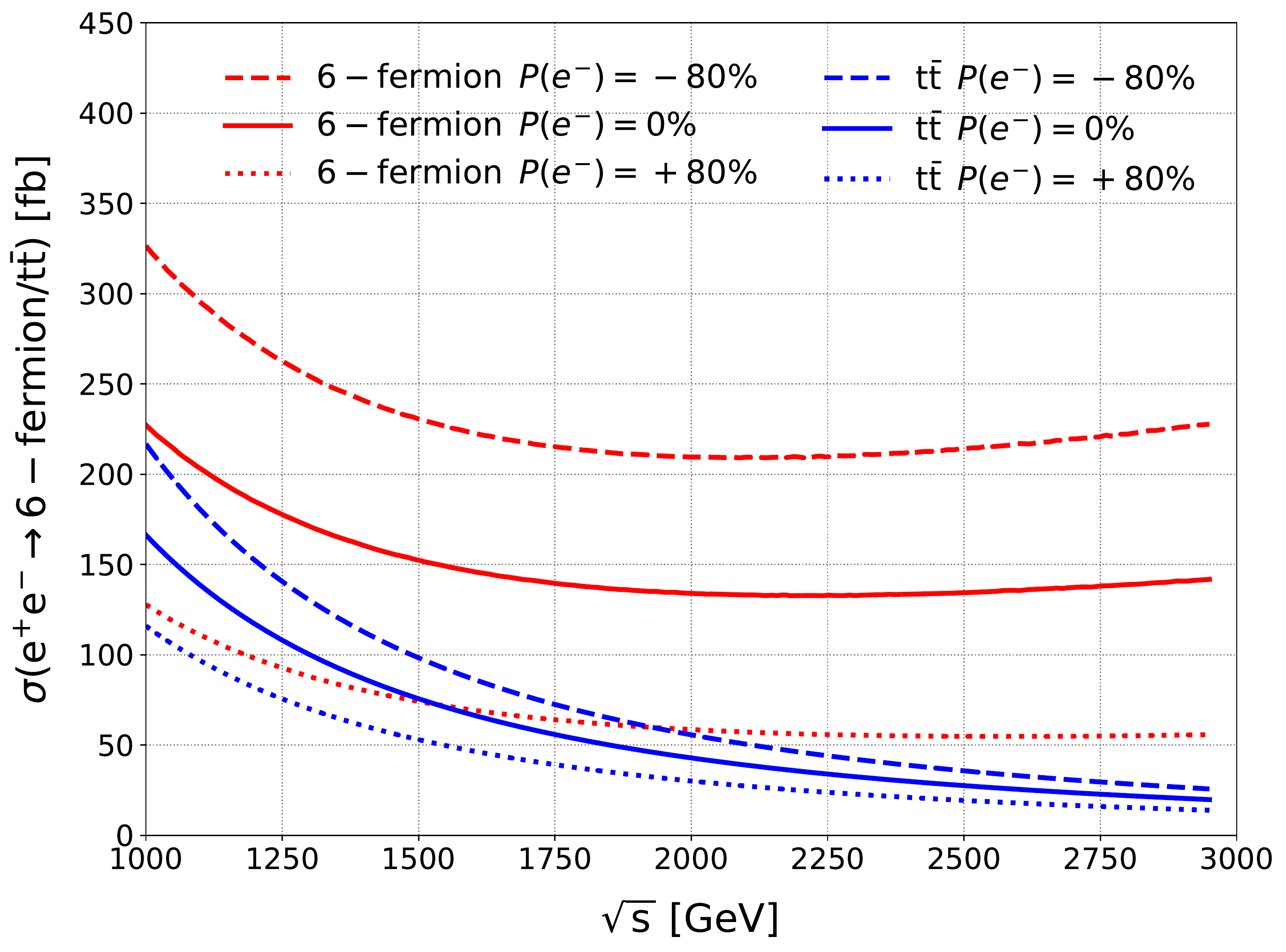}
  \caption{Inclusive \ttbar (blue) and 6-fermion (red) cross sections for different electron polarisations, P(e). The values shown do not include the effect of ISR.}
  \label{fig:xsec}
\end{figure}

As discussed in \Cref{sec:intro}, final states with six fermions are generally dominated by the $\ttbar$ production process, but have a growing contribution from non-$\ttbar$ processes such as single-top production and triple gauge boson production for higher collision energies. This is illustrated in \Cref{fig:xsec} that shows the \whizard cross section of top-quark pair production (blue) and inclusive six-fermion production (red) for different electron polarisations.

The different contributions to the six-fermion final state cannot be separated fully due to interference. Therefore the top-quark pair production signal samples used in the analyses were simulated as part of an inclusive six-fermion sample. The signal sample was extracted using a parton-level categorisation requiring two on-shell top-quark candidates. Each candidate consists of three of the six final state particles and should have a mass within $\sim$7.6\,GeV of the generated top-quark mass; a value that corresponds to five times the generated top-quark width. The $\text{non-}\ttbar$ contributions, denoted $\not\to\ttbar$, are treated as backgrounds in the following. While no algorithm can separate them from the signal completely, attempts are made in the event selection to reduce the fraction of $\text{non-}\ttbar$ events using some of the characteristic features of the $\ttbar$ process. 

The analyses presented in this paper consider a large range of additional relevant background processes, including di-quark final states and final states resulting from $\PW\PW$- and $\PZ\PZ$-fusion events. Note that for events with a centre-of-mass energy close to the nominal collision energy, as studied here, we expect a negligible contribution from hard background processes such as $\PGg\PGg$ and $\Pe\PGg$; these are therefore not studied further. Summaries of the signal and background samples considered are presented in \Cref{tab:samples:1400} and \Cref{tab:samples:3000}.

\begin{table}[t!]
\begin{minipage}{\columnwidth}
\centering
\begin{tabular}{lcccc}
\vspace{1.0mm}
{} & \multicolumn{2}{c}{$\sigma\,[\fb]$} & \multicolumn{2}{c}{N} \rule{0pt}{3ex} \\
\vspace{1.0mm}
P(\Pem) & -80\% & +80\% & -80\% & +80\% \\
\midrule
$\epem(\to\ttbar)\to\PQq\PQq\PQq\PQq\Pl\PGn\,(\Pl=\Pe,\PGm)$\footnote{Kinematic region defined as $\rootsprime\geq1.2\,\tev$} & 18.4 & 9.83 & 36,800 & 4,915 \\
\midrule
$\epem(\to\ttbar)\to\PQq\PQq\PQq\PQq\Pl\PGn\,(\Pl=\Pe,\PGm)$\footnote{$\rootsprime<1.2\,\tev$} & 28.5 & 14.9 & 57,000 & 7,450 \\
$\epem(\to\ttbar)\to\PQq\PQq\PQq\PQq\Pl\PGn\,(\Pl=\PGt)$ &23.2 & 12.3 & 46,400 & 6,150 \\
$\epem(\not\to\ttbar)\to\PQq\PQq\PQq\PQq\Pl\PGn$ & 72.2 & 16.5 & 144,400 & 8,250 \\
$\epem\to\PQq\PQq\PQq\PQq\PQq\PQq$ & 116 & 44.9 & 232,000 & 22,450 \\
$\epem\to\PQq\PQq\Pl\PGn\Pl\PGn$ & 44.1 & 15.3 & 88,200 & 7,650 \\
$\epem\to\PQq\PQq\PQq\PQq$ & 2,300 & 347 & 4,600,000 & 173,500 \\
$\epem\to\PQq\PQq\Pl\PGn$ & 6,980 & 1,640 & 13,960,000 & 820,000 \\
$\epem\to\PQq\PQq\Pl\Pl$ & 2,680 & 2,530 & 5,360,000 & 1,265,000 \\
$\epem\to\PQq\PQq$ & 4,840 & 3,170 & 9,680,000 & 1,585,000 \\
\bottomrule
\end{tabular}
\end{minipage}
\caption{Cross sections and number of events of the simulated samples used in the analysis of $\ttbar$ events at $\roots=1.4\,\tev$, assuming $2.0\,\abinv$ and $0.5\,\abinv$ for $P(\Pem)=\text{-}80\%$ and $P(\Pem)=\text{+}80\%$, respectively. The cross section quoted for the signal sample in the uppermost row is defined in the kinematic region $\rootsprime\geq1.2\,\tev$ \label{tab:samples:1400}}
\end{table}
 
\begin{table}[t!]
\begin{minipage}{\columnwidth}
\centering
\begin{tabular}{lcccc}
\vspace{1.0mm}
{} & \multicolumn{2}{c}{$\sigma\,[\fb]$} & \multicolumn{2}{c}{N} \rule{0pt}{3ex} \\
\vspace{1.0mm}
P(\Pem) & -80\% & +80\% & -80\% & +80\% \\
\midrule
$\epem(\to\ttbar)\to\PQq\PQq\PQq\PQq\Pl\PGn\,(\Pl=\Pe,\PGm)$\footnote{Kinematic region defined as $\rootsprime\geq2.6\,\tev$} & 3.48 & 1.89 & 13,920 & 1,890 \\
\midrule
$\epem(\to\ttbar)\to\PQq\PQq\PQq\PQq\Pl\PGn\,(\Pl=\Pe,\PGm)$\footnote{$\rootsprime<2.6\,\tev$} & 13.7 & 7.26 & 54,800 & 7,260 \\
$\epem(\to\ttbar)\to\PQq\PQq\PQq\PQq\Pl\PGn\,(\Pl=\PGt)$ & 8.45 & 4.51 & 33,800 & 4,510 \\
$\epem(\not\to\ttbar)\to\PQq\PQq\PQq\PQq\Pl\PGn$ & 99.6 & 22.6 & 398,400 & 22,600 \\
$\epem\to\PQq\PQq\PQq\PQq\PQq\PQq$ & 54.0 & 18.0 & 216,000 & 18,000 \\
$\epem\to\PQq\PQq\Pl\PGn\Pl\PGn$ & 59.7 & 14.9 & 238,800 & 14,900 \\
$\epem\to\PQq\PQq\PQq\PQq$ & 963 & 130 & 3,852,000 & 130,000 \\
$\epem\to\PQq\PQq\Pl\PGn$ & 8,810 & 2,310 & 35,240,000 & 2,310,000 \\
$\epem\to\PQq\PQq\Pl\Pl$ & 3,230 & 3,060 & 12,920,000 & 3,060,000 \\
$\epem\to\PQq\PQq$ & 3,510 & 2,390 & 14,040,000 & 2,390,000 \\
\bottomrule
\end{tabular}
\end{minipage}
\caption{Cross sections and number of events of the simulated samples used in the analysis of $\ttbar$ events at $\roots=3\,\tev$, assuming $4.0\,\abinv$ and $1.0\,\abinv$ for $P(\Pem)=\text{-}80\%$ and $P(\Pem)=\text{+}80\%$, respectively. The cross section quoted for the signal sample in the uppermost row is defined in the kinematic region $\rootsprime\geq2.6\,\tev$ \label{tab:samples:3000}}
\end{table}
\section{Analysis strategy}
\label{sec:analysis_strategy}

Precision studies of observables such as the $\ttbar$ production cross section, $\csttbar$, and the top-quark forward-backward asymmetry, $\afb$, are powerful tools for discovery.

\begin{figure}
  \centering
    \includegraphics[width=0.68\columnwidth]{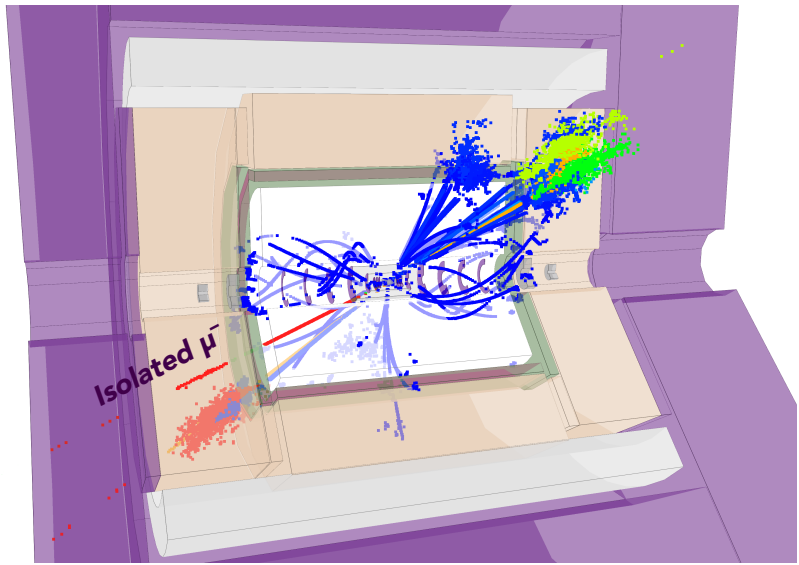}
  \vspace{0.7cm}
  \caption{Example display of an $\ttbar\to\PQq\PQq\PQq\PQq\PGm\PGnGm$ events in \clicild at $\roots=3\,\tev$. The event includes overlay of beam-induced $\gghadrons$ background as described in \Cref{sec:gensim}. An isolated lepton is clearly seen along with four isotropically distributed jets. The colour scale represents the energy of the individual particles shown, where red (blue) indicates the highest (lowest) energy in the event collection. \label{fig:ttbar:boosted:eventdisplay}}
\end{figure}

The analyses presented in this paper focus on ``single lepton+jets'' final states, $\ttbar\to\PQq\PQq\PQq\PQq\Pl\PGn$. The branching ratio of this process is about 30\%. The charged lepton is used to determine the charge of each of the top quarks, thus enabling the measurement of $\afb$. Identification of isolated leptons thus constitutes an important part of the analyses and is covered in more detail in \Cref{sec:lepid}. In addition, for operation above $\sim1\,\tev$ a large fraction of the top quarks will be produced with significant boosts, leading to a significantly different event topology compared to production close to the top-quark pair production threshold. This is illustrated by the event display in  \Cref{fig:ttbar:boosted:eventdisplay} showing a boosted semi-leptonic $\ttbar$ event at $\roots=3\,\tev$ featuring a clear separation between the decay products of the top- and anti-top quark, respectively.

The event selection proceeds through the identification of one isolated charged lepton in association with one large-R top-quark jet, the latter being identified using the dedicated top-quark tagger algorithm whose details and performance are described in \Cref{sec:jetreco}. In addition we require that no isolated high-energy photons are present, as these might be indicative of large energy losses due to beamstrahlung or ISR. See more details on the reconstruction of isolated photons in \Cref{sec:photonreco}. The remaining events are analysed using multivariate algorithms as described in detail in \Cref{sec:mva}. 

The signal events are restricted to the kinematic region defined as $\rootsprime\geq1.2\,\tev$ and $\rootsprime\geq2.6\,\tev$, for $\roots=1.4\,\tev$ and $\roots=3\,\tev$, respectively. A corresponding cut is applied to the reconstructed collision energy, $\rootsprimereco$, described in \Cref{sec:effcom}. The event selection is based on PFOs with \texttt{default} timing cuts at 1.4\tev and \texttt{tight} timing cuts at 3\tev; see the discussion in \Cref{sec:clic}.

The differential $\ttbar$ cross section, as a function of the polar angle of the top quark in the $\ttbar$ centre-of-mass system (defined with respect to the $\Pem$ beam)\footnote{In the \clicild coordinate system, both beams are tilted by half the crossing angle (10 mrad) with respect to the z-axis. Thus the impact of the crossing angle on the reconstructed forward-backward asymmetry cancels out.}, is described by
\begin{equation}\label{eq:AFBFit}
   \frac{d\sigma}{d(\cos(\theta^*))} = \sigma_1(1+\cos(\theta^*))^2 + \sigma_2(1-\cos(\theta^*))^2 + \sigma_3(1-\cos^2(\theta^*)).
\end{equation}
At tree level, the three terms can be related to the top-quark pair production cross sections for different helicity combinations in the final state, $\sigma_{1,2,3}$. The forward and backward cross sections, $\sigma_{\mathrm{F}}$ and $\sigma_{\mathrm{B}}$, can be obtained by integrating the differential cross section over the top-quark polar angle ranges, $0<\theta^*<\pi/2$ and $\pi/2<\theta^*<\pi$, respectively. The total production cross section, $\csttbar$, can be expressed as 
\begin{equation} \label{eq:totcs}
\csttbar = \sigma_{\mathrm{F}} + \sigma_{\mathrm{B}} = (4/3)(2\,\sigma_1+2\,\sigma_2+\sigma_3),
\end{equation}
while the top-quark forward-backward asymmetry is defined as
\begin{equation} \label{eq:afb}
\afb \equiv \frac {\sigma_{\mathrm{F}} - \sigma_{\mathrm{B}}} {\sigma_{\mathrm{F}} + \sigma_{\mathrm{B}}} = \frac{1}{\csttbar}\,2\,(\sigma_1-\sigma_2).
\end{equation}

The $\csttbar$ and $\afb$ observables are extracted for each analysis by fitting \Cref{eq:AFBFit} to the reconstructed polar-angle distribution of the hadronically decaying top quarks (or anti-top quarks). Note that the sign of $\cos(\theta^{*})$ is inverted for events with hadronically decaying anti-top quarks. The fit is performed after background subtraction and correction for finite selection efficiencies, see \Cref{sec:results} for details. Note that the measured cross sections represent a convolution of $\csttbar$ with the CLIC luminosity spectrum.
\section{Isolated lepton identification}
\label{sec:lepid}

The classification of candidate top-quark events as either fully-hadronic, where both the $\PW$ bosons decay hadronically, or semi-leptonic, where one of the $\PW$ bosons decays leptonically (to a lepton and a neutrino) and the other hadronically, relies on efficient identification of high-energy charged leptons. The lepton tagging applied is based on the Isolated Lepton Processor in \marlin~\cite{MarlinLCCD}, and is optimised to identify the $\Pepm$ and $\PGmpm$ from the final state of semi-leptonic $\ttbar$ events at $\roots=1.4\,\tev$. These leptons are typically well-isolated from other activity in the event and of higher energy than leptons from hadronic decays inside jets. Note that the same settings are used for the events at $\roots=3\,\tev$, leaving room for further improvements.

The tagging algorithm considers all PFOs in an event and identifies isolated charged leptons by studying the energy depositions in the ECAL and HCAL, impact parameters, and isolation in a cone around each PFO. This section discusses the observables considered, parameter optimisation, special considerations taken for the boosted environment, and presents the resulting tagging efficiencies. Fully-leptonic events, where both $\PW$ bosons decay leptonically, have not been studied so far. In addition, we do not consider semi-leptonic $\ttbar$ events with a tau lepton as part of the signal sample since these are more difficult to reconstruct due to the additional missing energy. Note that the lepton charge is determined by the curvature of the helix from a standard Kalman-filter-based track reconstruction of the associated hits in the tracking system.

\subsection{Observables}

The identification and optimisation of observables for isolated charged lepton tagging is studied at reconstruction level, where we use simulator-level information to match the reconstructed PFOs to parton-level information. In the following we consider so-called ``truth-matched'' electrons (muons), namely reconstructed PFOs matched to the final state charged electron (muon). Other reconstructed particles, i.e. non truth-matched PFOs, are denoted ``other'' in the figures. 

The plots in this section are produced for PFOs with a reconstructed charge different from zero ($\mathrm{q}\neq0$). In addition, a fiducial region cut is applied for both of the parton-level top-quarks (for $\ttbar$ events) requiring them to fulfil $|\cos(\theta^{\mathrm{MC}})|\leq0.80$, where $\theta^{\mathrm{MC}}$ denotes the top-quark polar angle at parton-level in the laboratory frame and after ISR\footnote{The detector coverage goes down to about $8^\circ$. Excluding a larger area in the forward direction for the optimisation reduces the effect of losing energy down the beam pipe and adds some margin for the finite size of the jets.}. Note that the electrons are reconstructed without photon recovery.

\begin{figure}
  \centering
  \includegraphics[width=0.48\columnwidth]{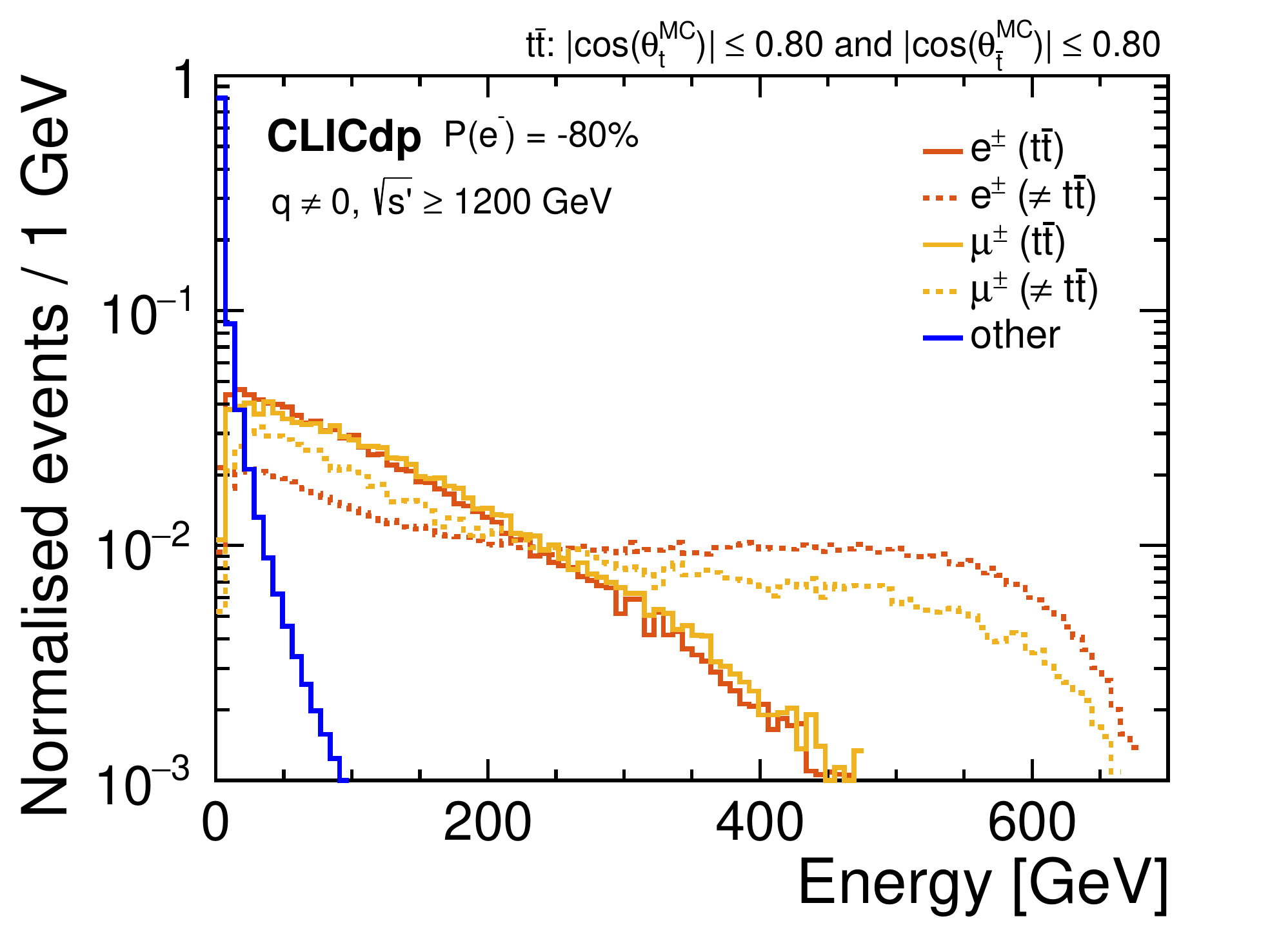}
  ~~~
  \includegraphics[width=0.48\columnwidth]{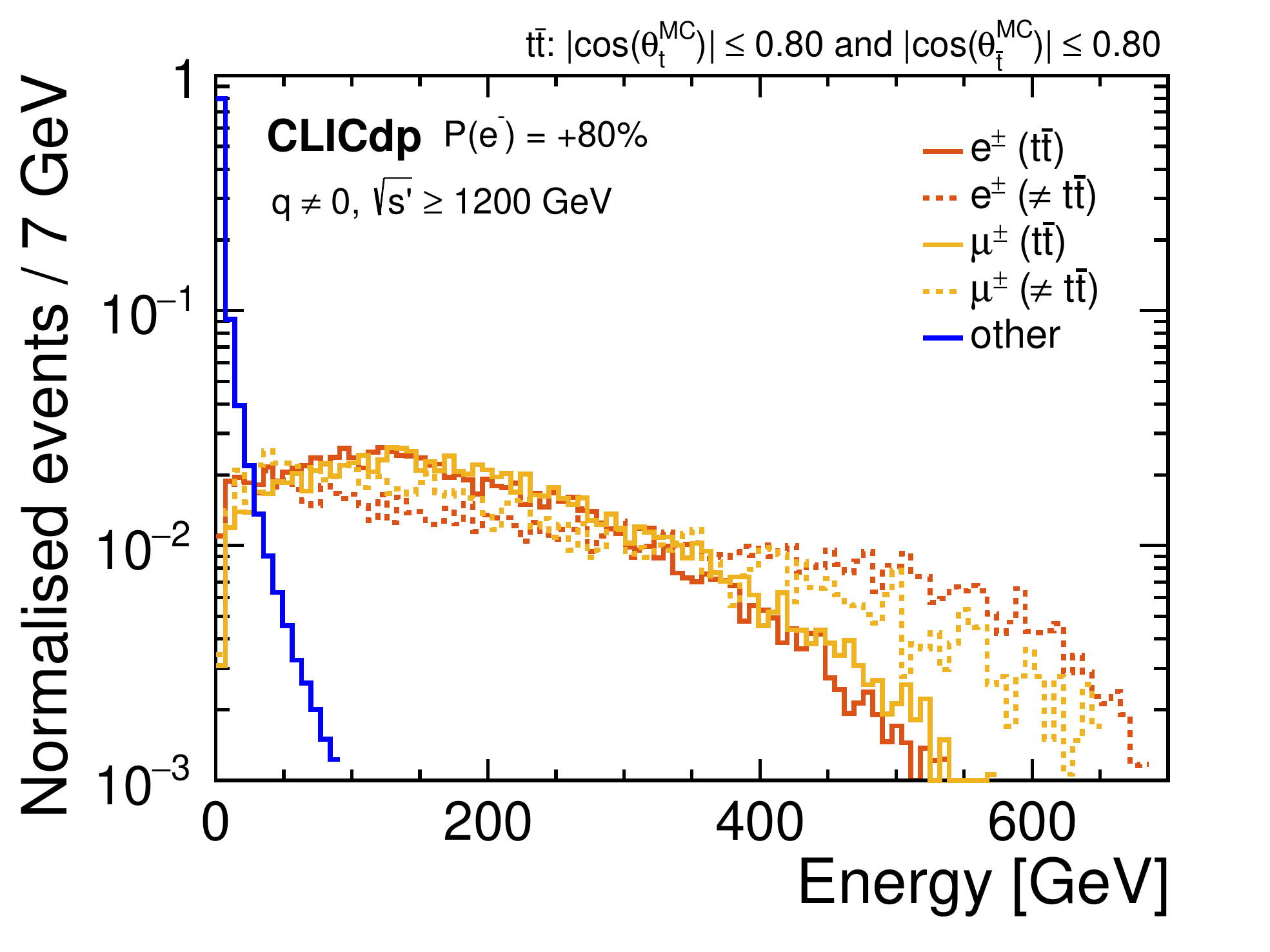}
  \caption{Energy distribution for PFOs in boosted $\ttbar$ events for operation at $P(\Pem)=-80\%$ (left) and $P(\Pem)=+80\%$ (right). The red and yellow lines represent truth-matched final state charge leptons while the blue line represents other charged PFOs. \label{fig:analysis:leptonid:trackE}}
\end{figure}

\begin{figure}
  \centering
  \includegraphics[width=0.48\columnwidth]{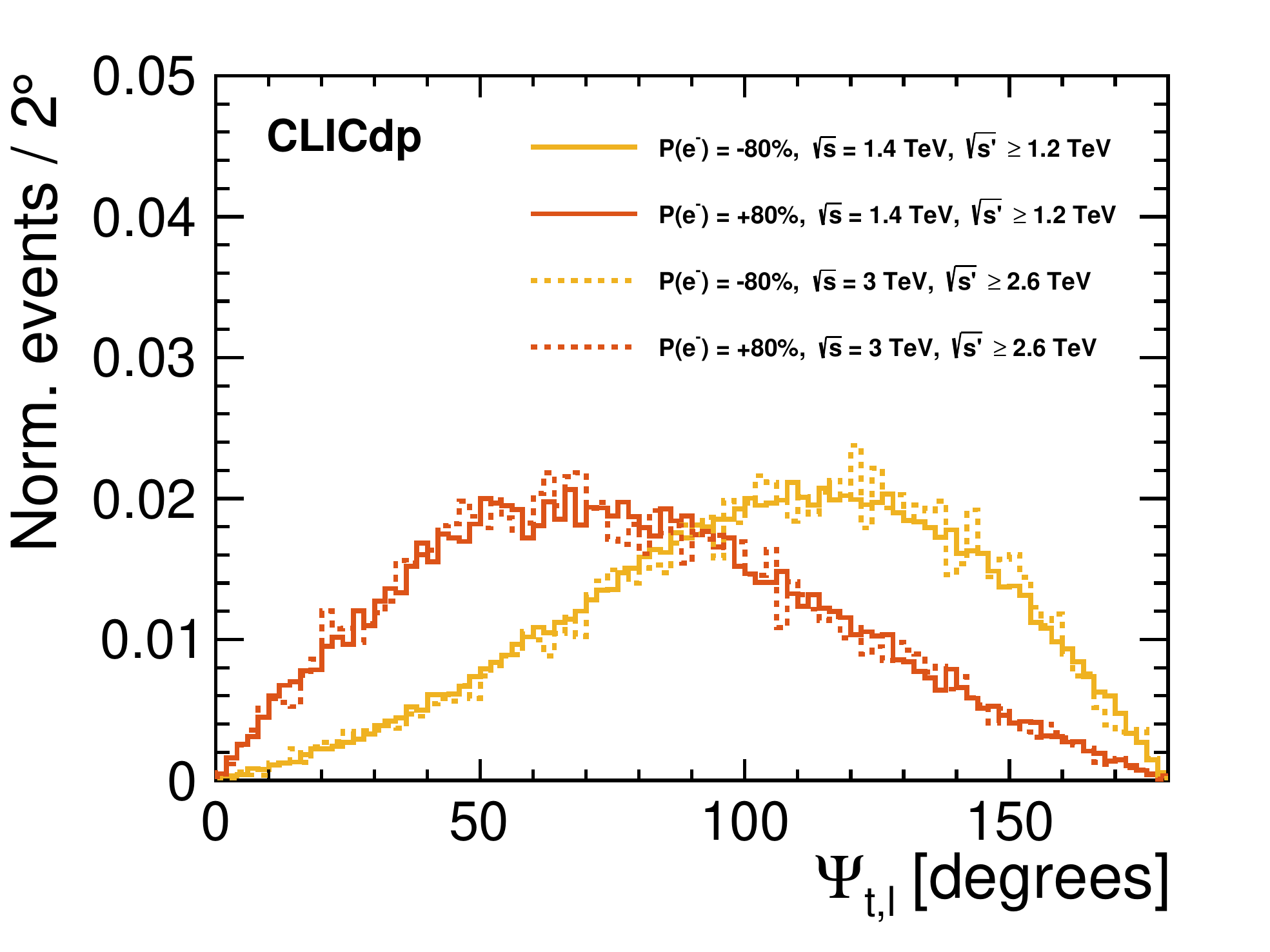}  
  ~~~
  \includegraphics[width=0.48\columnwidth]{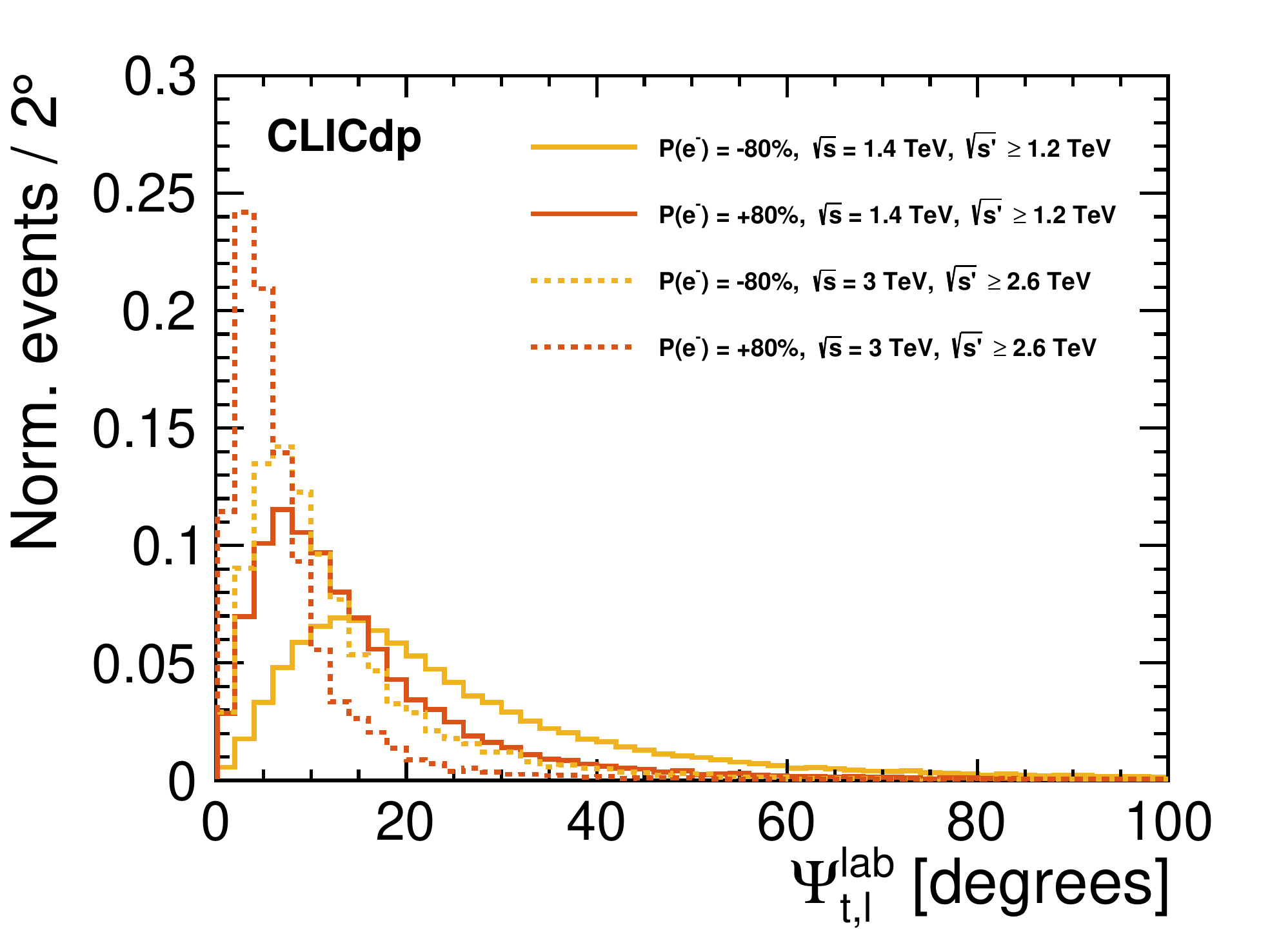}
  \caption{Angle between the final state lepton and the associated top-quark, in semi-leptonic $\ttbar$ events, in the rest frame of the latter (left) and in the laboratory frame (right). \label{fig:analysis:leptonid:psi}}
\end{figure}

\paragraph{Particle energy:}\hspace{-3mm}
\Cref{fig:analysis:leptonid:trackE} shows the PFO energy of both truth-matched (yellow/orange) and ``other'' (blue) particles. The energy of the truth-matched electrons and muons are generally higher than for typical particles in jets. For boosted $\ttbar$ events at $P(\Pem)=-80\%$, shown in the figure to the left, the peak is at a significantly lower energy than the corresponding peak for $P(\Pem)=+80\%$ shown in the figure to the right. This is due to the fact that the $\PW$ boson is emitted with a larger angle to the top-quark flight-direction for top quarks with left-handed helicity (enriched in the $P(\Pem)=-80\%$ sample), as illustrated in the left panel of \Cref{fig:analysis:leptonid:psi}, showing the angle between the final state lepton and the associated top-quark, in the rest frame of the latter. The right panel of \Cref{fig:analysis:leptonid:psi} shows the the same angle in the laboratory frame. Considering the energy distribution for the $P(\Pem)=-80\%$ case, a soft cut at 10\,\gev is chosen. 

\begin{figure}
  \centering
  \includegraphics[width=0.65\columnwidth]{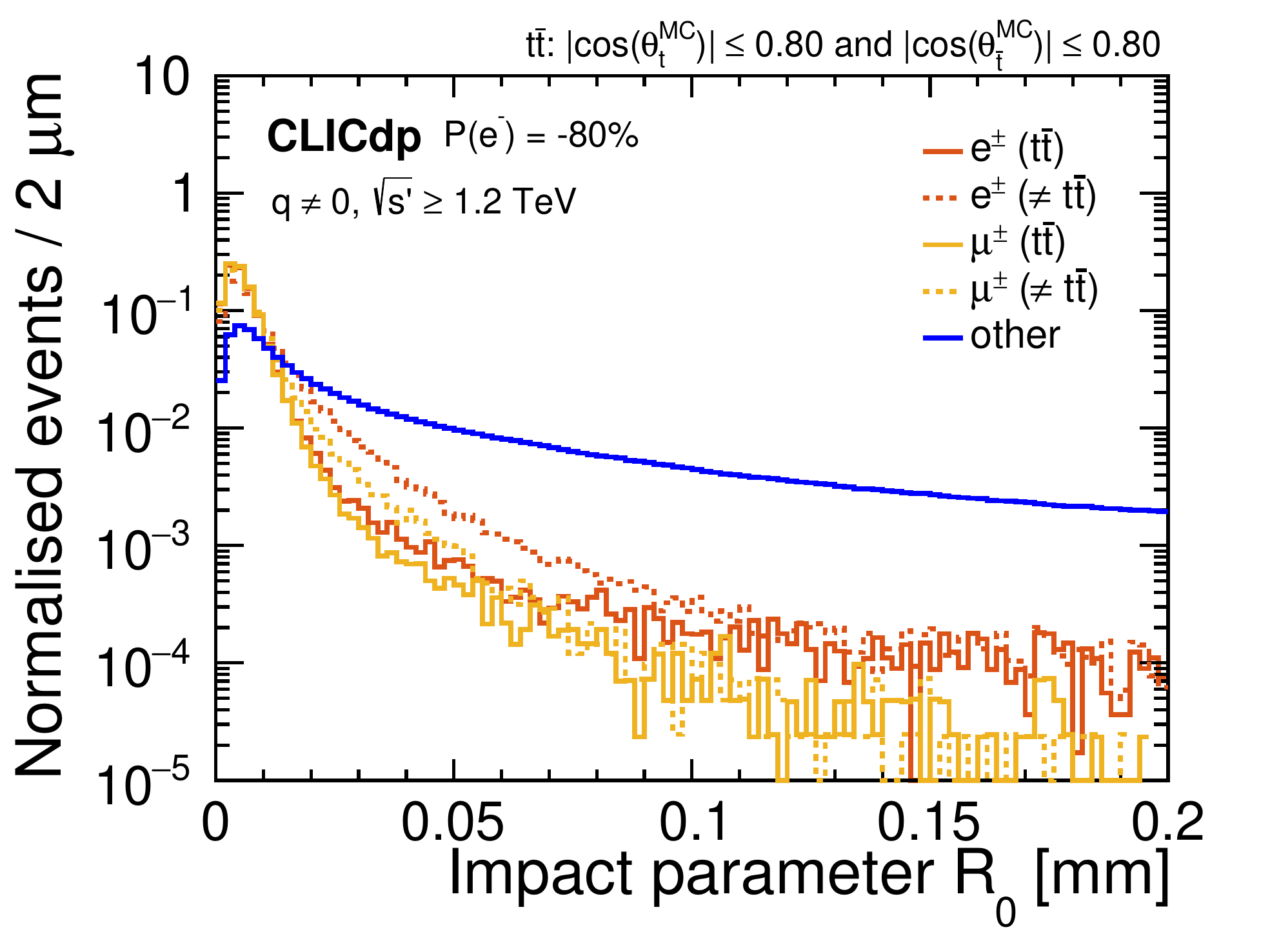}
   \caption{Distributions of the impact parameter $\mathrm{R}_0$ for truth-matched electrons (orange) and muons (yellow), as well as for non truth-matched PFOs (blue). Solid (dashed) lines represent the distribution for the $\ttbar$ (non-$\ttbar$) final states in the six-fermion samples under study.\label{fig:analyis:leptonid:trackR0}}
\end{figure}

\paragraph{Impact parameter:}\hspace{-3mm}
Further, since the top-quarks have a short lifetime, the final state electrons and muons from the $\PW$ decay typically originate from the primary event vertex. Conversely, b-quarks and tau leptons both have longer lifetimes and may thus originate from displaced vertices. The longitudinal ($\mathrm{Z}_0$) and radial ($\mathrm{d}_0$) components of the track impact parameter, characterising the perpendicular distance between the track and the primary vertex at the point of closest approach, are combined into the parameter $\mathrm{R}_0$ defined as
\begin{equation}
  \mathrm{R}_0 = \sqrt{\mathrm{Z}_0^2 + \mathrm{d}_0^2}.
\end{equation}
\Cref{fig:analyis:leptonid:trackR0} shows $\mathrm{R}_0$ for truth-matched and ``other'' PFOs. A soft cut is placed selecting PFOs with an impact parameter $\mathrm{R}_0<0.1$ mm.

\begin{figure}
  \centering
  \includegraphics[width=0.48\columnwidth]{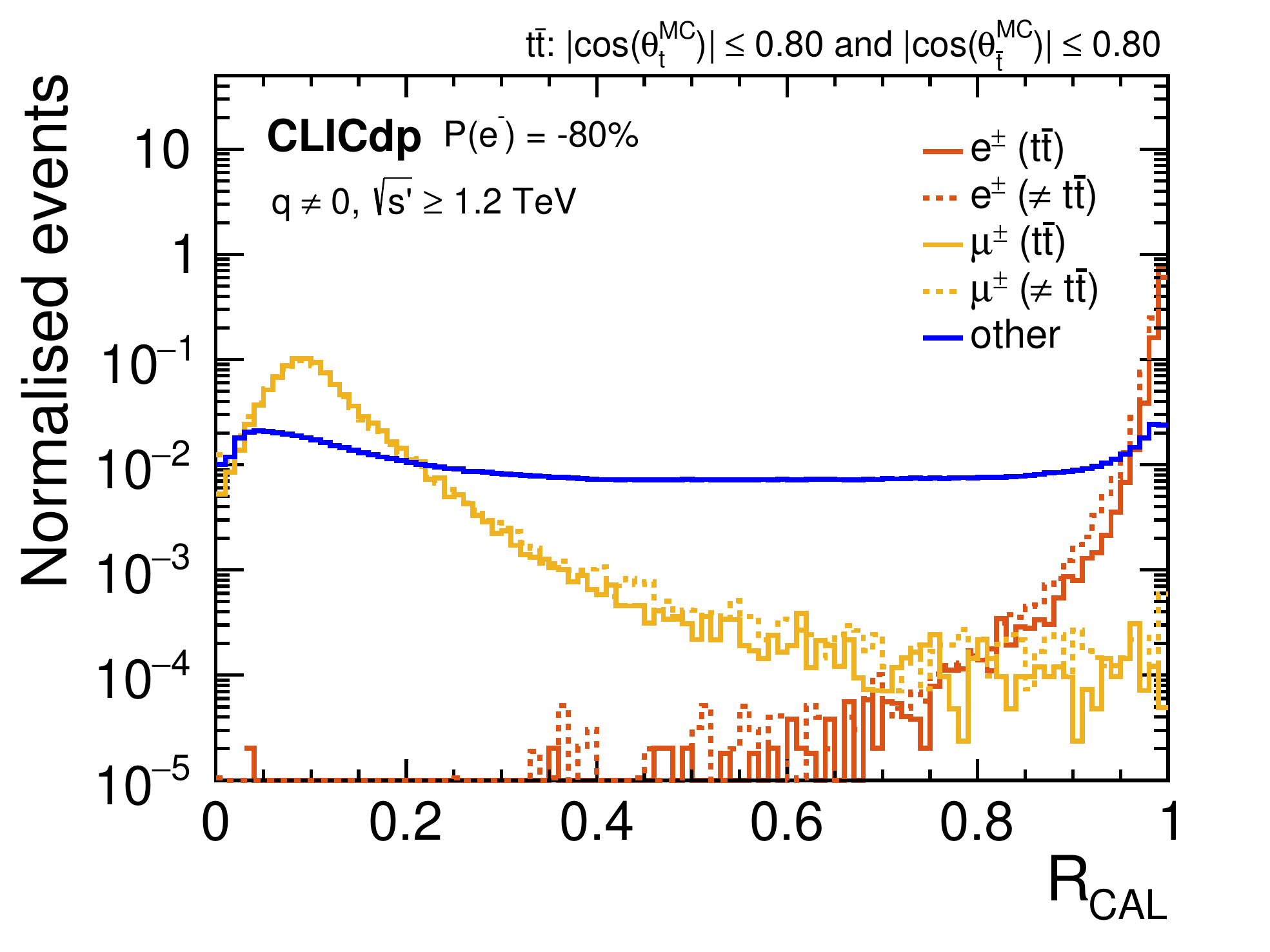}
  ~~
  \includegraphics[width=0.48\columnwidth]{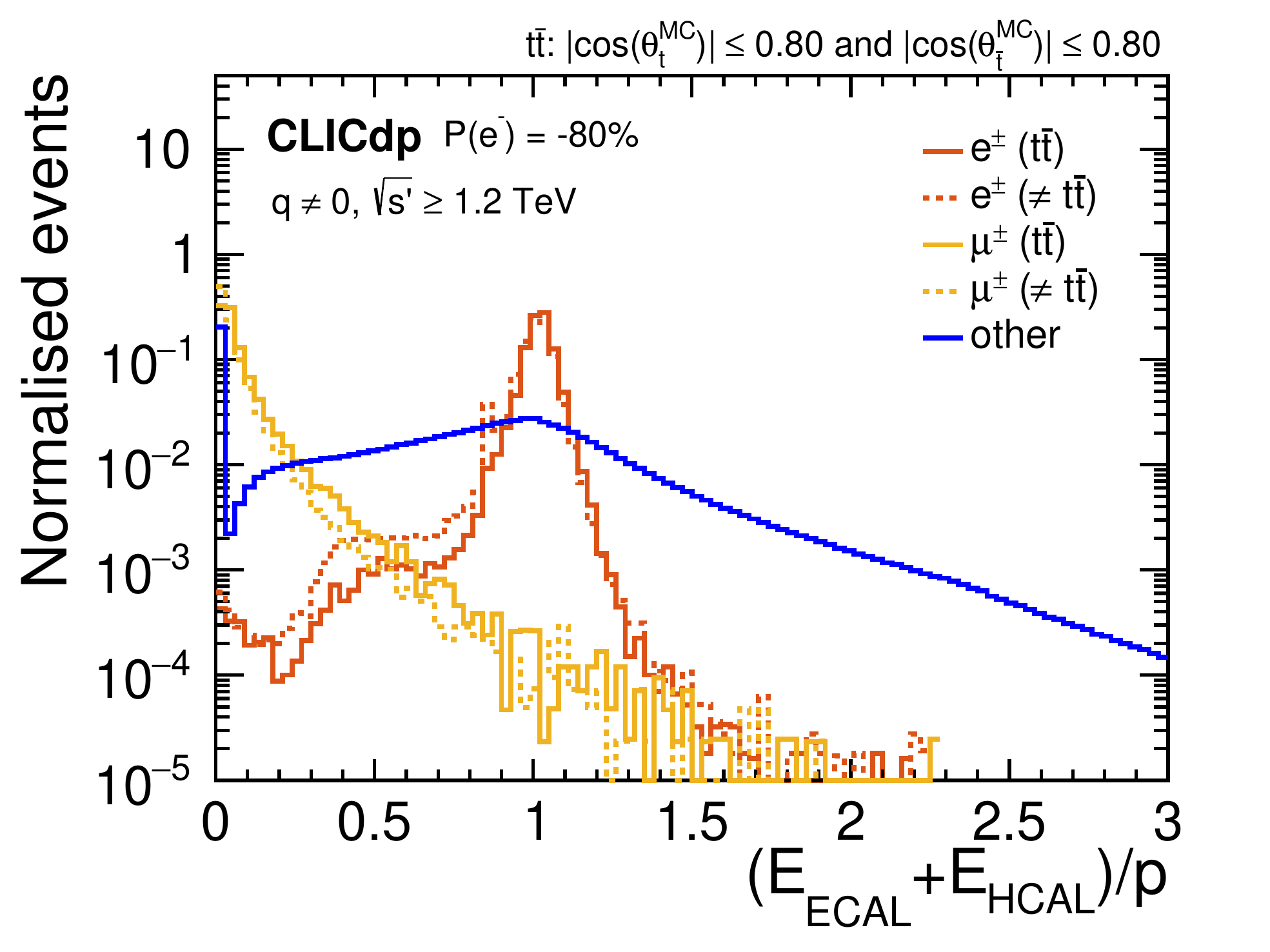}
  \caption{Distribution of the energy depositions in the calorimeter system: energy deposited in ECAL w.r.t the total deposited energy (left), total deposited energy w.r.t. momentum (right).\label{fig:analyis:leptonid:calo}}
\end{figure}

\begin{figure}
  \centering
  \begin{subfigure}[b]{0.48\columnwidth}
  \includegraphics[width=\textwidth]{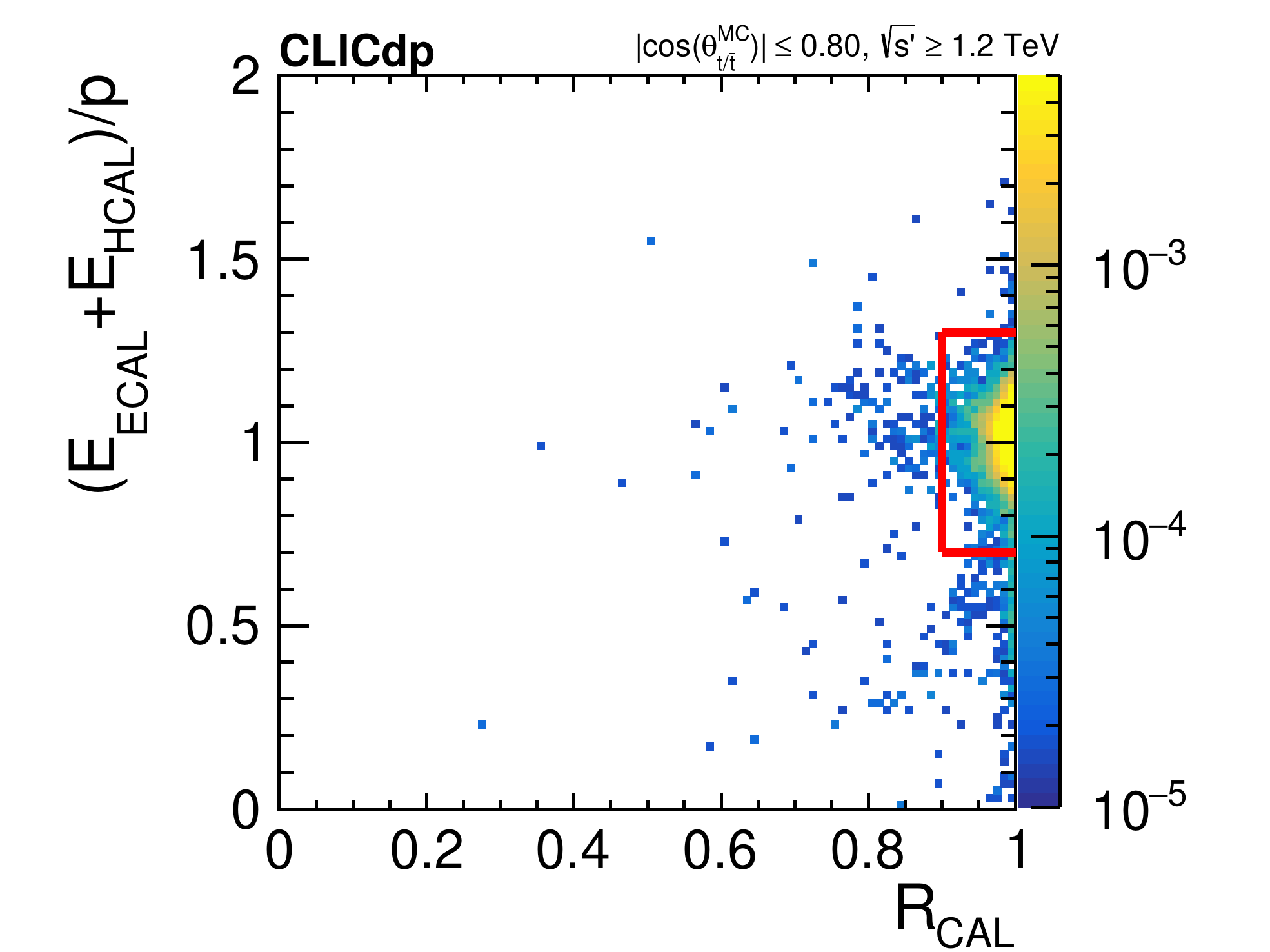}
  \caption{Truth-matched electrons\label{fig:analyis:leptonid:calo2d:e}}
  \end{subfigure}
  \begin{subfigure}[b]{0.48\columnwidth}
  \includegraphics[width=\textwidth]{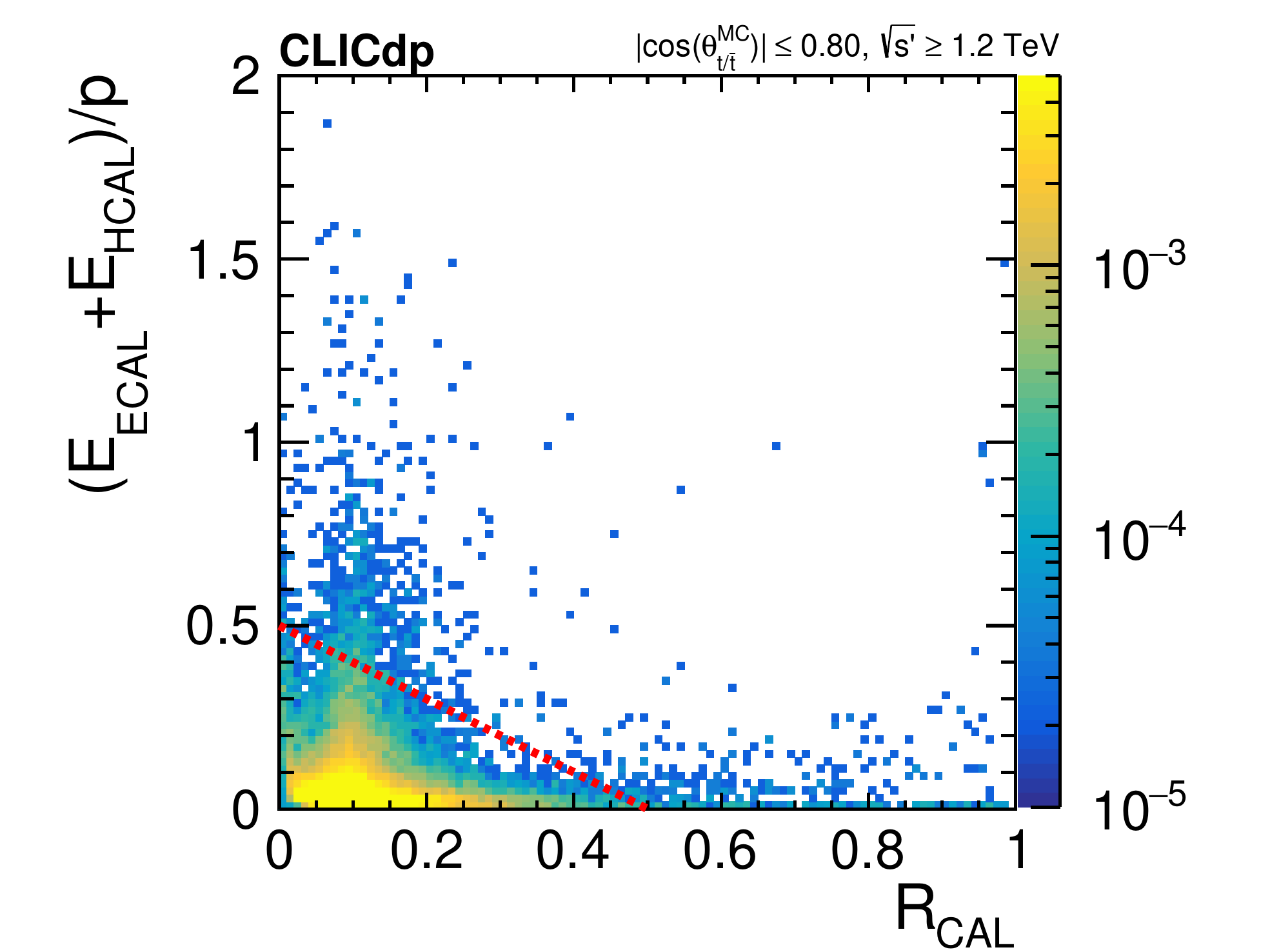}
  \caption{Truth-matched muons\label{fig:analyis:leptonid:calo2d:mu}}
  \end{subfigure}\\
  \vspace{4mm}
  \begin{subfigure}[b]{0.48\columnwidth}
  \includegraphics[width=\textwidth]{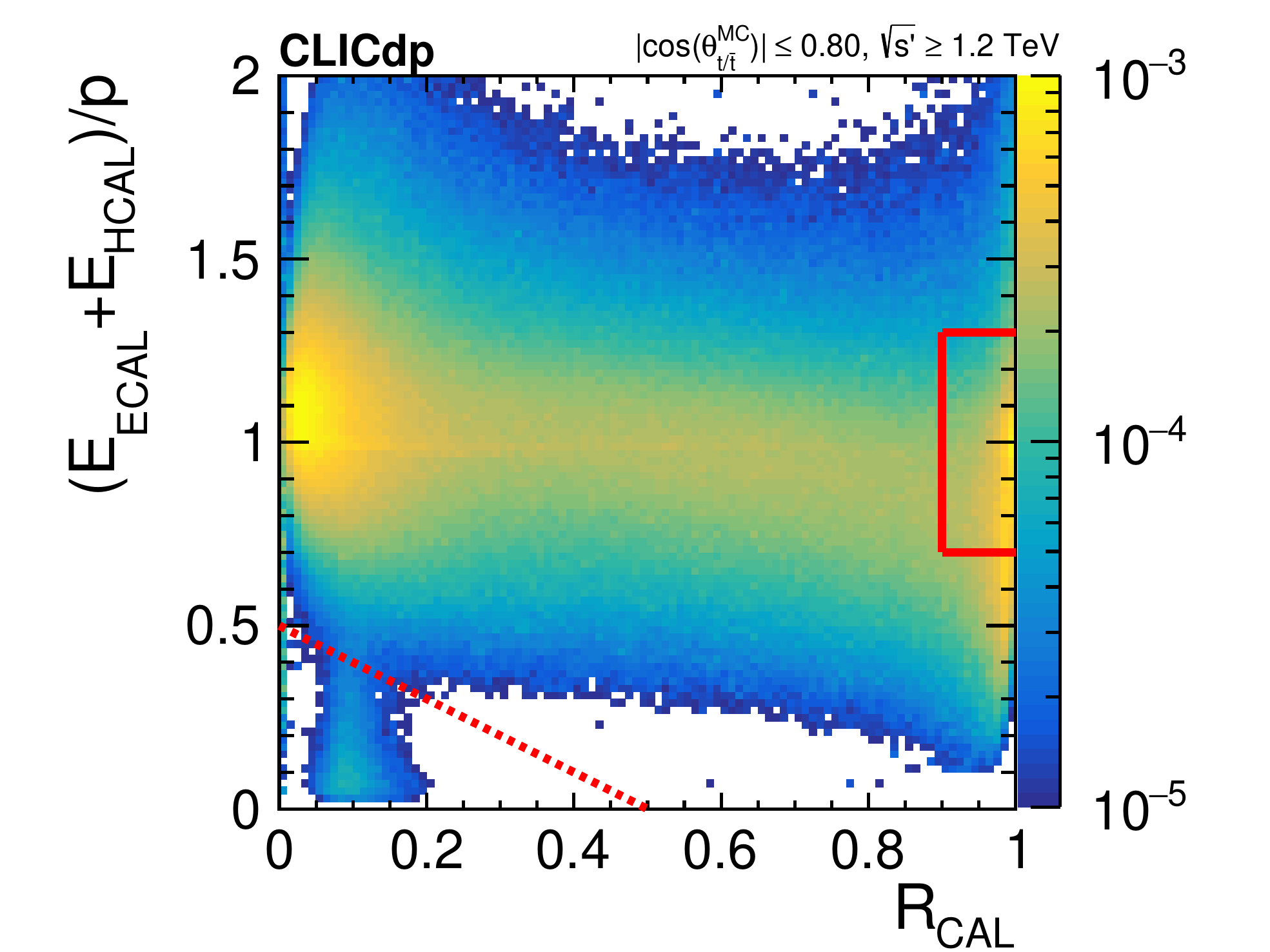}
  \caption{Non truth-matched PFOs (``other'')\label{fig:analyis:leptonid:calo2d:other}}
  \end{subfigure}
  \caption{Correlation between the two observables studied for the depositions in the calorimeter system, shown individually for truth-matched electron and muon PFOs and non truth-matched particles. The red box and dashed line indicate the corresponding cuts applied.\label{fig:analyis:leptonid:calo2d}}
\end{figure}

\paragraph{Energy fraction in the ECAL:}\hspace{-3mm}
The energy depositions in the calorimeter system are studied to distinguish between electrons and muons, and to further distinguish these from depositions from non truth-matched particles such as hadrons. The ratio between the energy deposited in the ECAL system w.r.t. the depositions in the full calorimeter system, ECAL and HCAL, is defined as
\begin{equation}
\mathrm{R}_{\mathrm{CAL}} = \frac{\mathrm{E}_{\mathrm{ECAL}}}{\mathrm{E}_{\mathrm{ECAL}} + \mathrm{E}_{\mathrm{HCAL}}}.
\end{equation}
In addition, we study the ratio between $\mathrm{E}_{\mathrm{ECAL}} + \mathrm{E}_{\mathrm{HCAL}}$ and the PFO momentum. \Cref{fig:analyis:leptonid:calo} shows the distribution of these observables for three categories: truth-matched electron PFOs (yellow), truth-matched muon PFOs (orange), and non truth-matched particles (blue). While electrons are mainly contained within the ECAL, muons only deposit a minimum amount of ionisation energy throughout the calorimeters system. The correlation between the two observables is studied in \Cref{fig:analyis:leptonid:calo2d}. Electrons and muons are selected within the red box and line indicated in \Cref{fig:analyis:leptonid:calo2d:e} and \Cref{fig:analyis:leptonid:calo2d:mu}, respectively. The corresponding regions are also indicated in \Cref{fig:analyis:leptonid:calo2d:other} showing the distribution for non truth-matched PFOs (``other''). These regions correspond to
\begin{equation}
\begin{aligned}
  &0.7\leq(\mathrm{E}_{\mathrm{ECAL}} + \mathrm{E}_{\mathrm{HCAL}})/\mathrm{p}\,\leq1.3,\,\,\,\,\,0.9\leq \mathrm{R}_{\mathrm{CAL}} \leq 1.0,           &&\mathrm{for\,\,electrons,}\\
  &(\mathrm{E}_{\mathrm{ECAL}} + \mathrm{E}_{\mathrm{HCAL}})/\mathrm{p}\,\,\leq(0.5 - \mathrm{R}_{\mathrm{CAL}}),       &&\mathrm{for\,\,muons.}\\
\end{aligned}
\end{equation}

\begin{figure}
  \centering
  \begin{subfigure}[b]{0.48\columnwidth}
  \includegraphics[width=\textwidth]{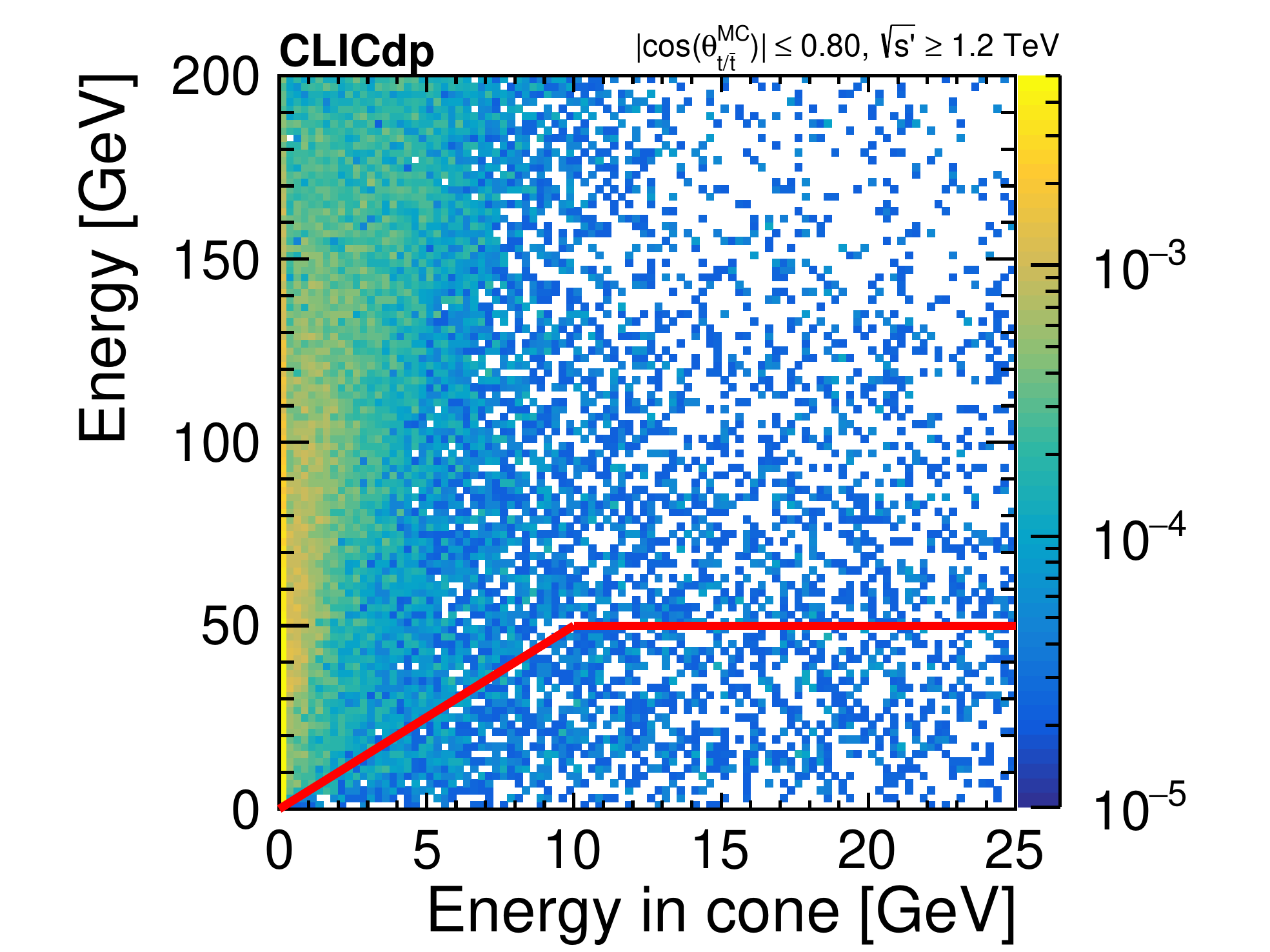}
  \caption{Truth-matched electrons\label{fig:analyis:leptonid:isolation:e}}
  \end{subfigure}
  \begin{subfigure}[b]{0.48\columnwidth}
  \includegraphics[width=\textwidth]{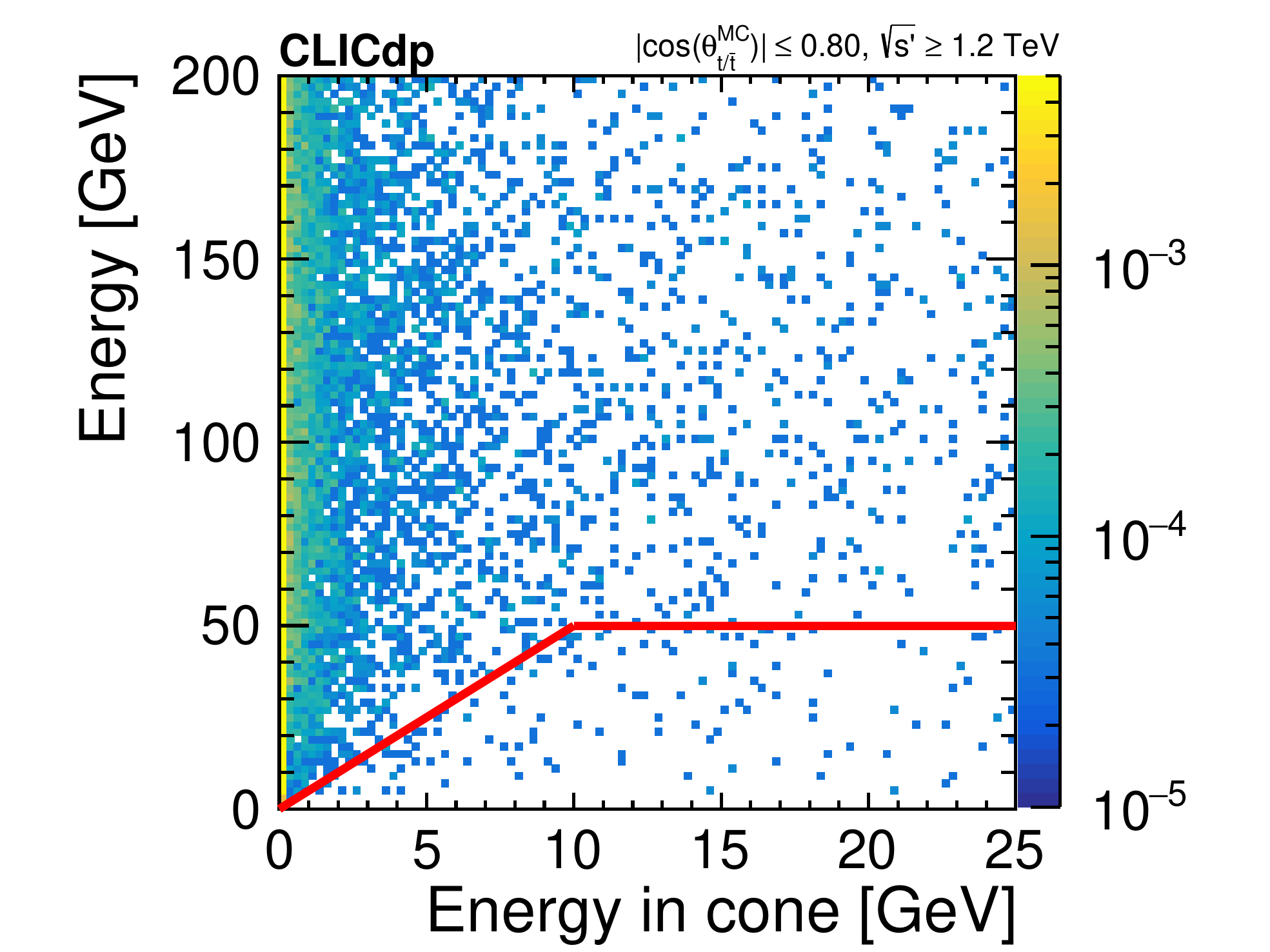}
  \caption{Truth-matched muons\label{fig:analyis:leptonid:isolation:mu}}
  \end{subfigure}\\
  \vspace{4mm}
  \begin{subfigure}[b]{0.48\columnwidth}
  \includegraphics[width=\textwidth]{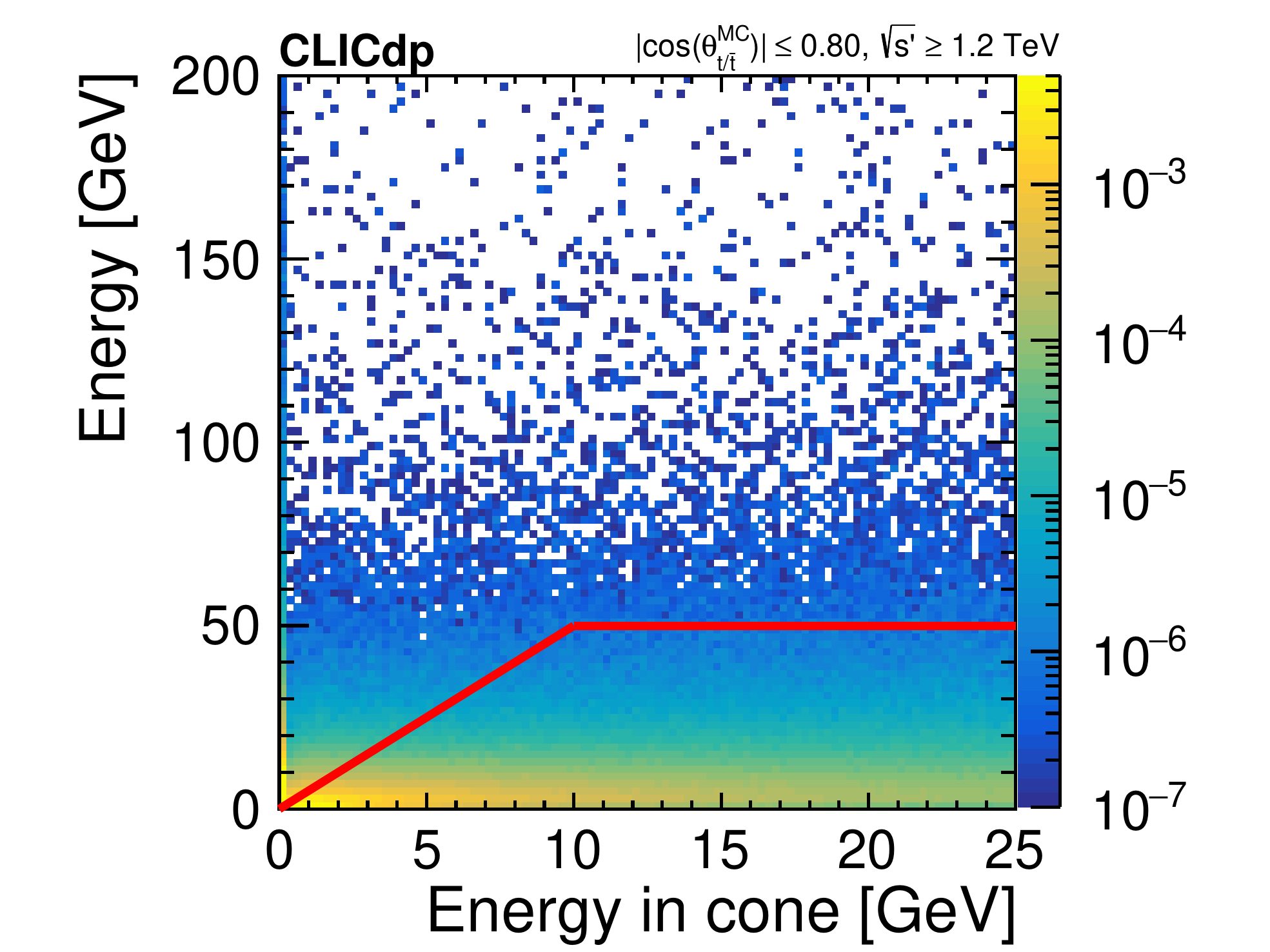}
  \caption{Non truth-matched PFOs (``other'')\label{fig:analyis:leptonid:isolation:other}}
  \end{subfigure}
  \caption{Particle energy as a function of cone energy shown individually for truth-matched electron and muon PFOs and non truth-matched particles. The red lines indicate the isolation cut applied. \label{fig:analyis:leptonid:isolation}}
\end{figure}

\paragraph{Isolation:}\hspace{-3mm}
The isolation of the candidate PFOs is studied by looking at the energy in a cone around the particle as a function of its energy. While the truth-matched charged leptons are isolated from the rest of the activity in the events, particles that originate from showers within jets are instead reconstructed in regions with high occupancy. This is clearly illustrated in \Cref{fig:analyis:leptonid:isolation} that shows the PFO energy as function of the cone energy, the latter defined as the sum of energies from PFOs located inside a cone of $\cos(\theta)=0.999$ w.r.t. the particle. Muons, as expected, are observed to radiate less than electrons. In addition, a clear difference is seen for the truth-matched particles w.r.t. non truth-matched particles such as hadrons. The red lines indicate the isolation cut applied, defined as
\[
  \mathrm{Particle\,\,energy} > 
  \begin{cases}
  \,50\,\,\gev,\hspace{3.15cm}\mathrm{if\,\,cone\,\,energy\,\,}>\,10\,\gev.\\
  \,5\,\times\,\mathrm{cone\,\,energy\,\,}[\gev],\hspace{0.8cm}\mathrm{if\,\,cone\,\,energy\,\,}\leq\,10\,\gev.
   \end{cases}
\]
These cuts are constructed to remove the majority of the non-truth matched PFOs, while at the same time retaining isolated muons and electrons including those at high energy which are more likely to radiate a photon.

\begin{figure}[t!]
  \centering
  \includegraphics[width=0.48\columnwidth]{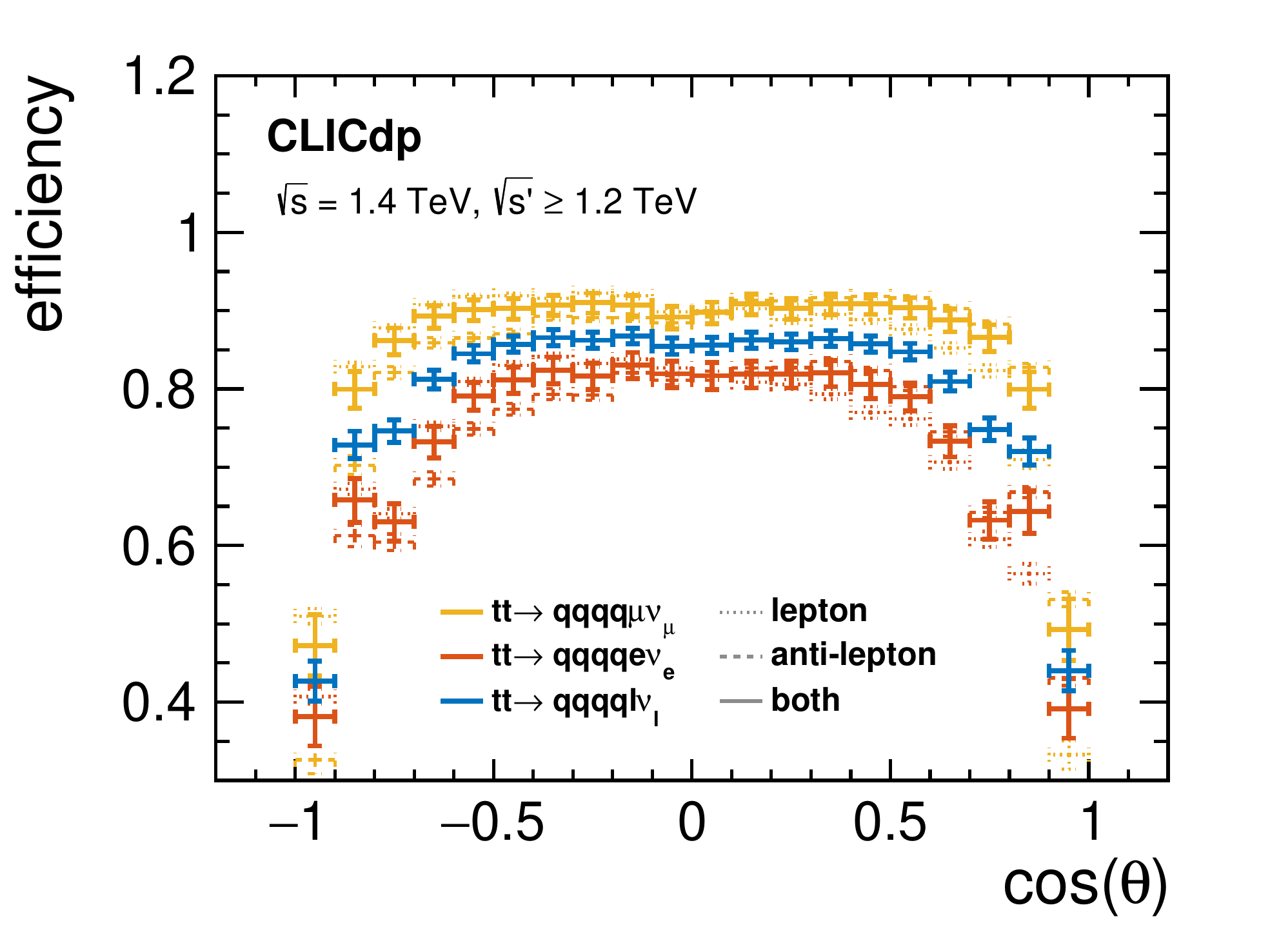}
  ~~
  \includegraphics[width=0.48\columnwidth]{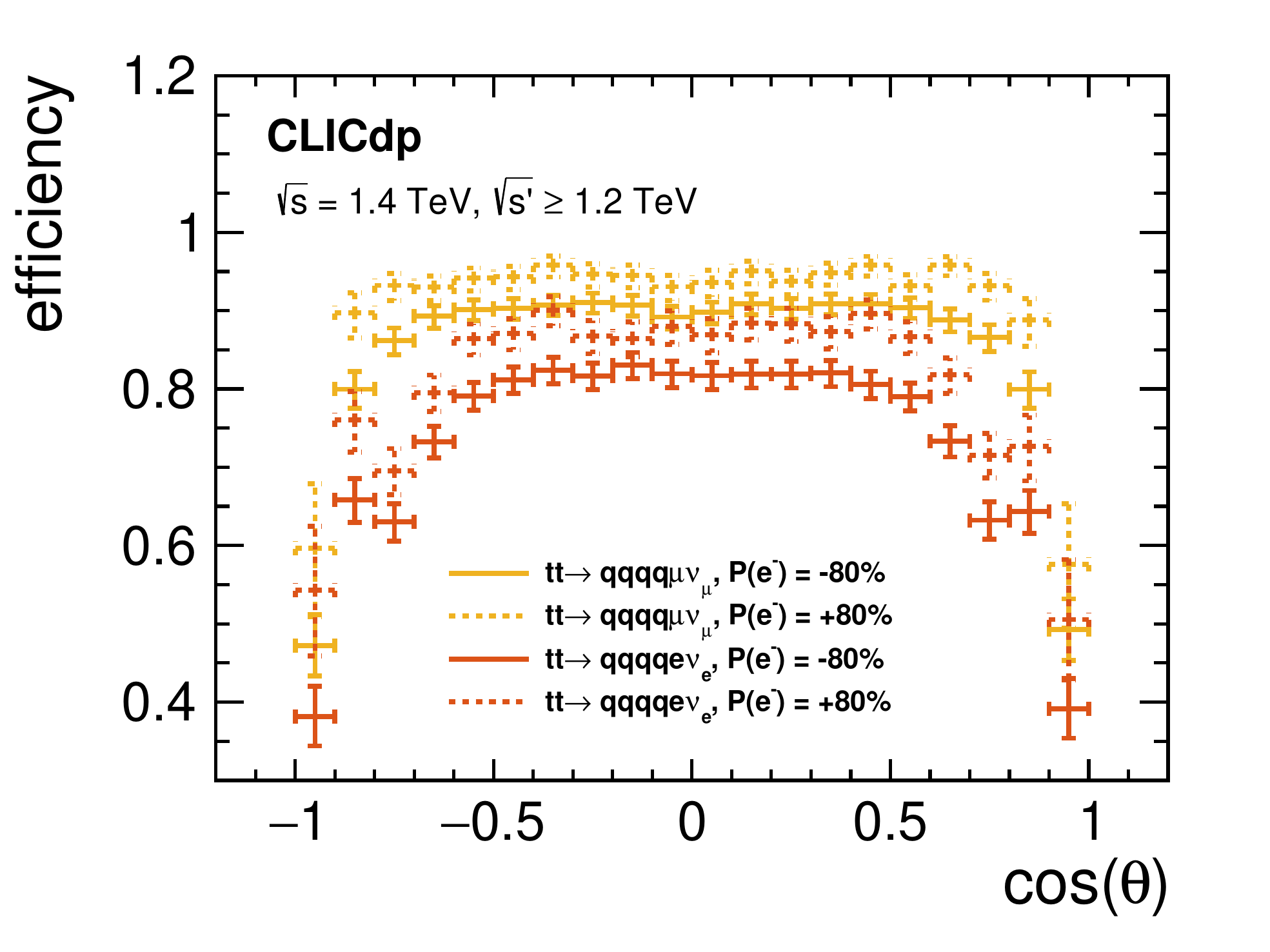}
  \caption{Charged lepton tagging efficiency as a function of polar angle of the $\ttbar$ final state lepton for $P(\mathrm{e}^{-})=-80\%$ (left) and a comparison of different electron beam polarisations (right). \label{fig:analysis:leptonid:efftheta}}
\end{figure}

\subsection{Selection efficiency}

In addition to the cuts described above we require that the $\pT$ of the isolated lepton candidate is larger than $10\,\gev$. In cases where several candidates exist, the candidate with the highest $\pT$ is selected. The application of the isolated lepton finder thus results in the identification of either zero or one charged lepton per event. 

\begin{figure}
  \centering
  \includegraphics[width=0.62\columnwidth]{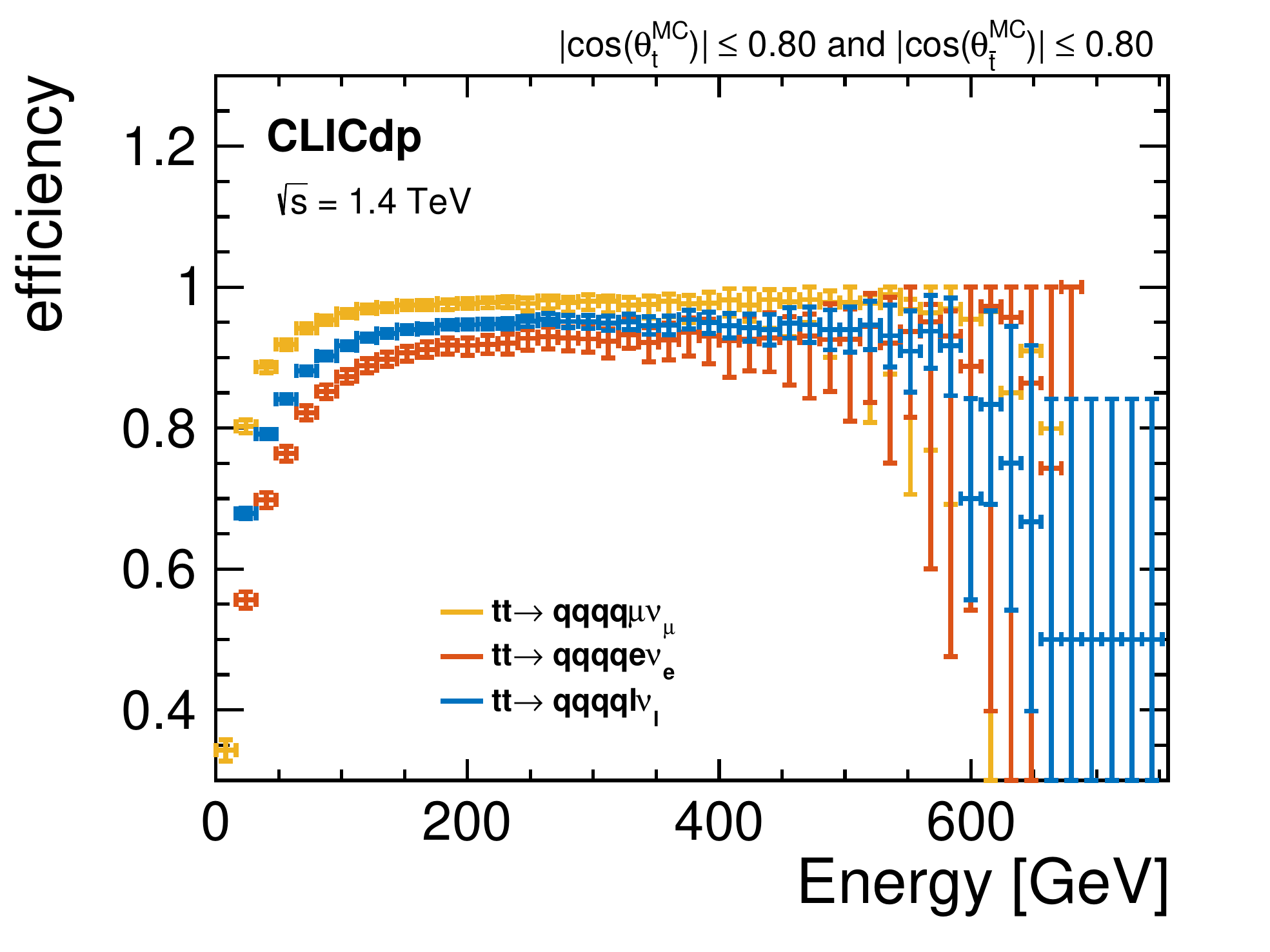}
  \caption{Charged lepton tagging efficiency as a function of energy of the $\ttbar$ final state lepton for $P(\mathrm{e}^{-})=-80\%$. \label{fig:analysis:leptonid:effpt}}
\end{figure}

The charged lepton tagging efficiency of semi-leptonic $\ttbar$ events is defined as the ratio of the number of events with a candidate isolated lepton truth-matched to the generated final state lepton and reconstructed within a cone of $1^\circ$ around it w.r.t. the number of generated events. A fiducial region where both final state top-quarks have a polar angle $\cos(\theta)\leq0.80$ is applied. The resulting efficiency of muons (solid yellow) and electrons (solid red) in the final state of boosted $\ttbar$ events is about 90\%, respectively 80\%, and is shown as a function of the generated lepton polar angle $\theta$ in \Cref{fig:analysis:leptonid:efftheta}; the left panel shows the distribution for $P(\mathrm{e}^{-})=-80\%$ while the right panel shows a comparison of the distributions for the different electron beam polarisations considered. The overall shape is well understood in terms of the detector design and acceptance; a small dip is expected in the central region due to the mechanical division of the two detector halves, and the drop around $\cos(\theta)=0.7$, observed for electrons, coincides with the challenging transition region between the ECAL barrel and endcap. The efficiency is presented as a function of energy in \Cref{fig:analysis:leptonid:effpt}, showing a near constant level above $\sim100\,\gev$ with a sharp decrease towards lower energies. As expected, the tagging efficiency for electrons is observed to be worse than for muons, an effect caused by an increased level of Bremsstrahlung for the significantly lighter electrons. This could partially be recovered with a refined algorithm in the future.

A somewhat larger tagging efficiency is observed for the configuration $P(\mathrm{e}^{-})=+80\%$, for which the leptons are generally of higher energy, as observed in \Cref{fig:analysis:leptonid:trackE}. Further, in \Cref{fig:analysis:leptonid:efftheta} where the distributions for leptons (dotted) and anti-leptons (dashed) are shown separately, a clear asymmetry is visible. This is again due to the helicity-dependence of the $\ttbar$ final state.

\begin{figure}
  \centering
  \includegraphics[width=0.48\columnwidth]{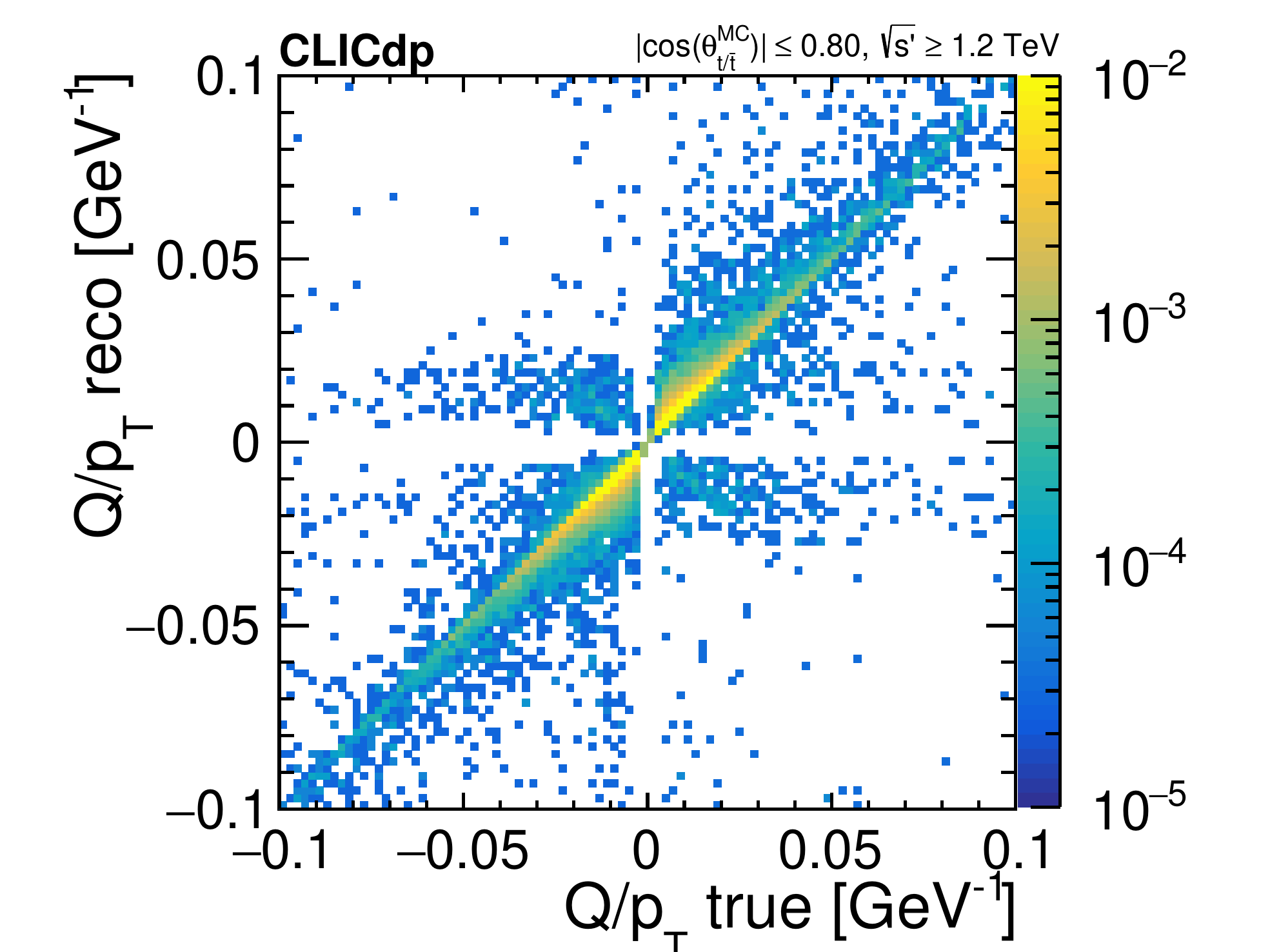}
  ~~
  \includegraphics[width=0.48\columnwidth]{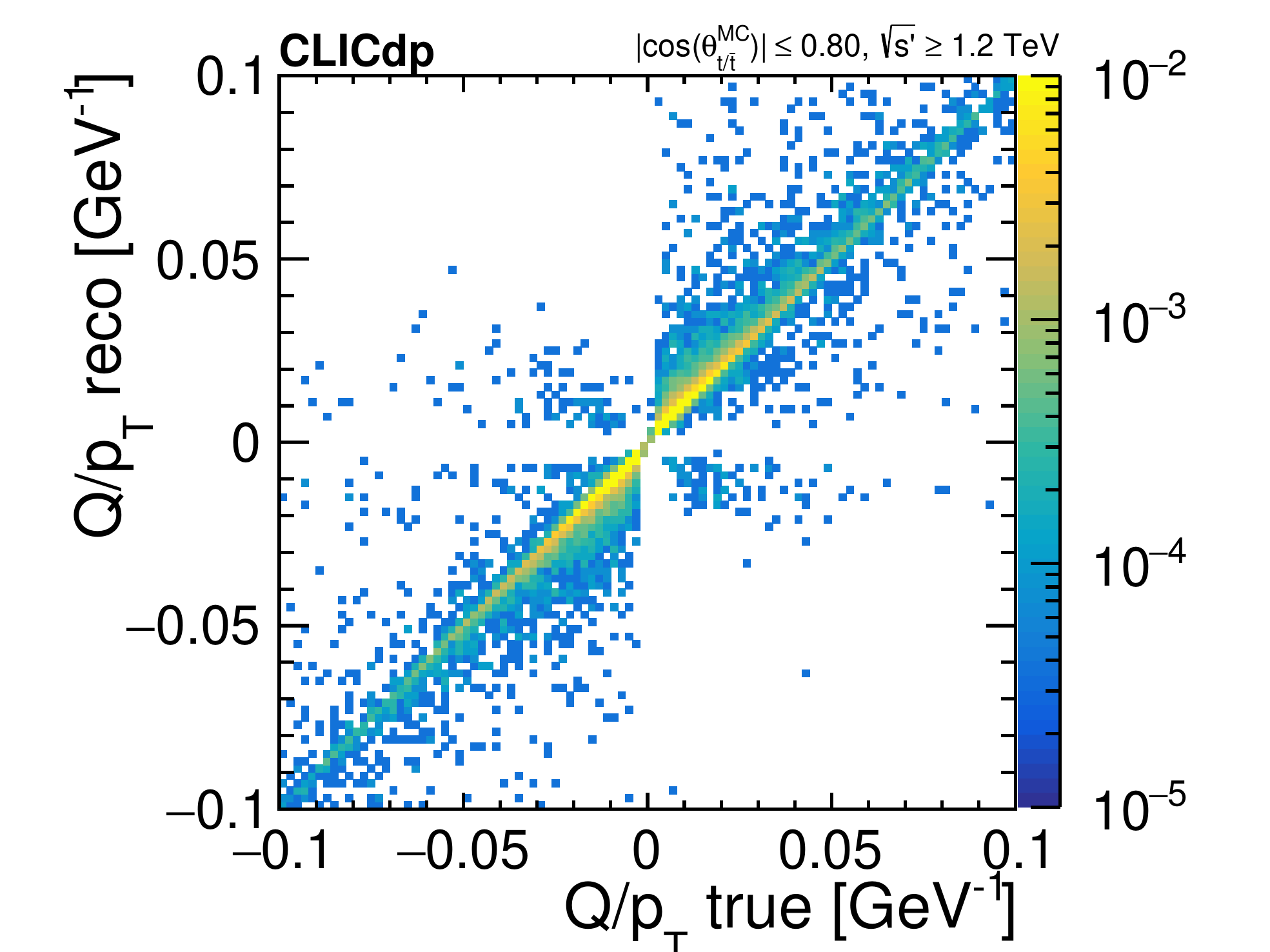}
  \caption{$\pT$-weighted charged distribution for truth-matched electrons (left) and muons (right) for boosted events ($\rootsprime\geq1.2\,\tev$) at a nominal collision energy $\roots=1.4\,\tev$. \label{fig:analyis:leptonid:effcharge}}
\end{figure}

\begin{figure}
  \centering
  \includegraphics[width=0.48\columnwidth]{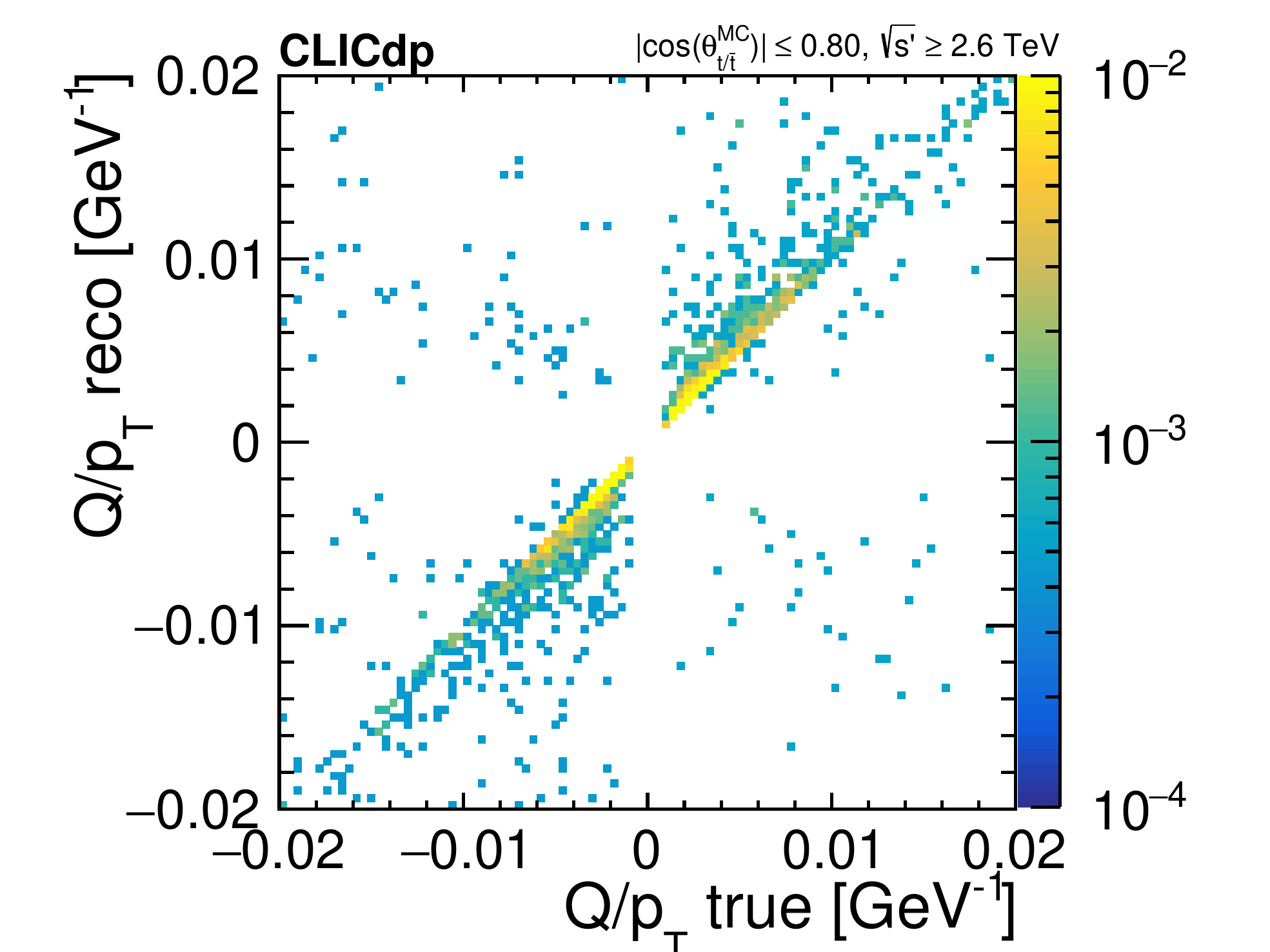}
  ~~
  \includegraphics[width=0.48\columnwidth]{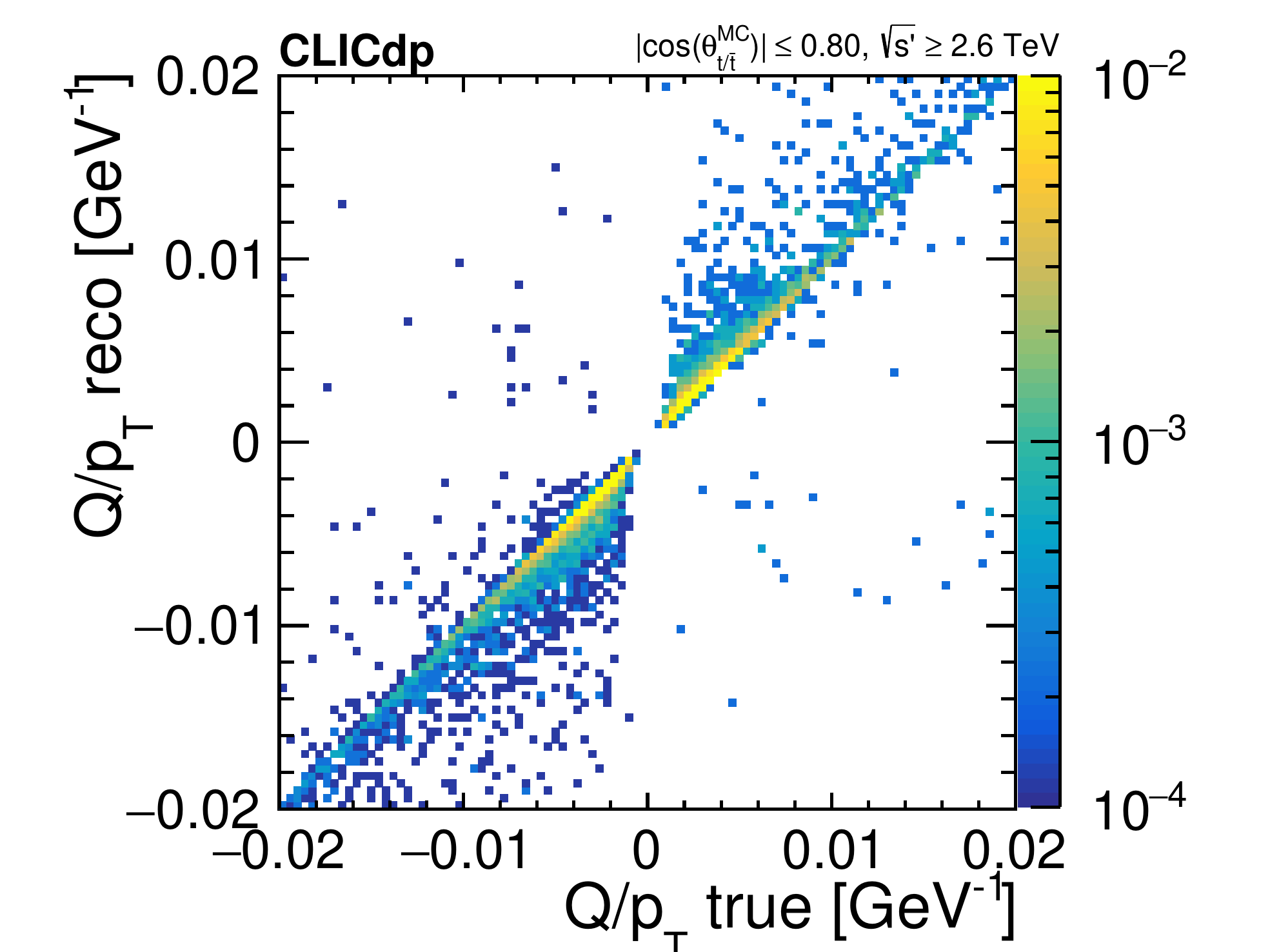}
  \caption{$\pT$-weighted charged distribution for truth-matched electrons (left) and muons (right) for boosted events ($\rootsprime\geq2.6\,\tev$) at a nominal collision energy $\roots=3\,\tev$.  \label{fig:analyis:leptonid:effcharge3tev}}
\end{figure}

Out of the identified muons (electrons), 99\% (98\%) are reconstructed with the correct charge. The charge tagging efficiency is illustrated in \Cref{fig:analyis:leptonid:effcharge} for $\roots=1.4\,\tev$ and in \Cref{fig:analyis:leptonid:effcharge3tev} for $\roots=3\,\tev$, showing the $\pT$-weighted charged distribution. Here, most events are populating the diagonal representing the correctly identified sign of the final state lepton, going from the bottom-left to the top-right corner. The efficiencies quoted above are calculated as the fraction of entries in the bottom-left and top-right quadrants to to the total number of entries.

\begin{figure}
  \centering
  \hspace{-2cm}
  \includegraphics[width=0.7\columnwidth]{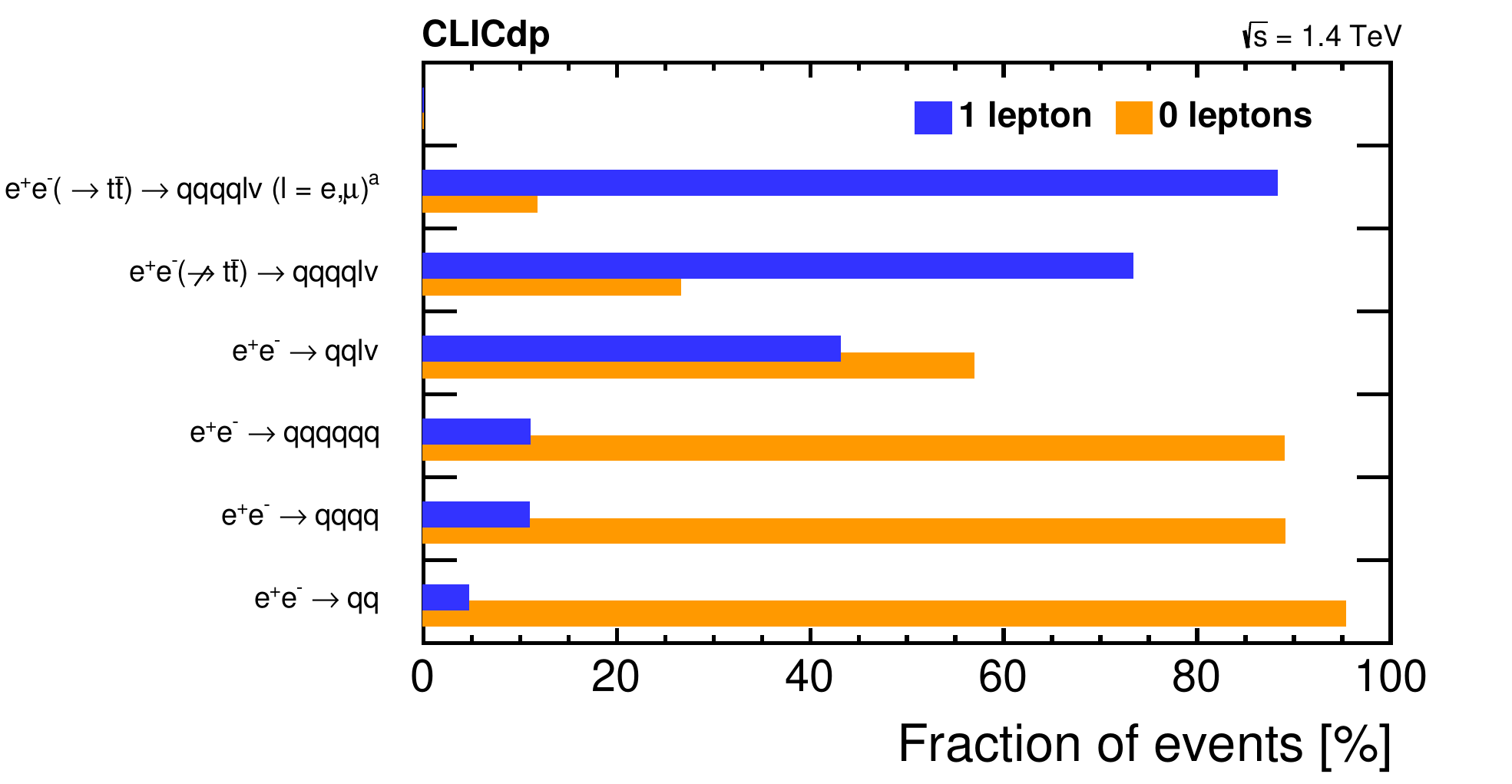}
  ~~~~\\
  \vspace{0.7cm}
  \hspace{-2.4cm}
  \includegraphics[width=0.7\columnwidth]{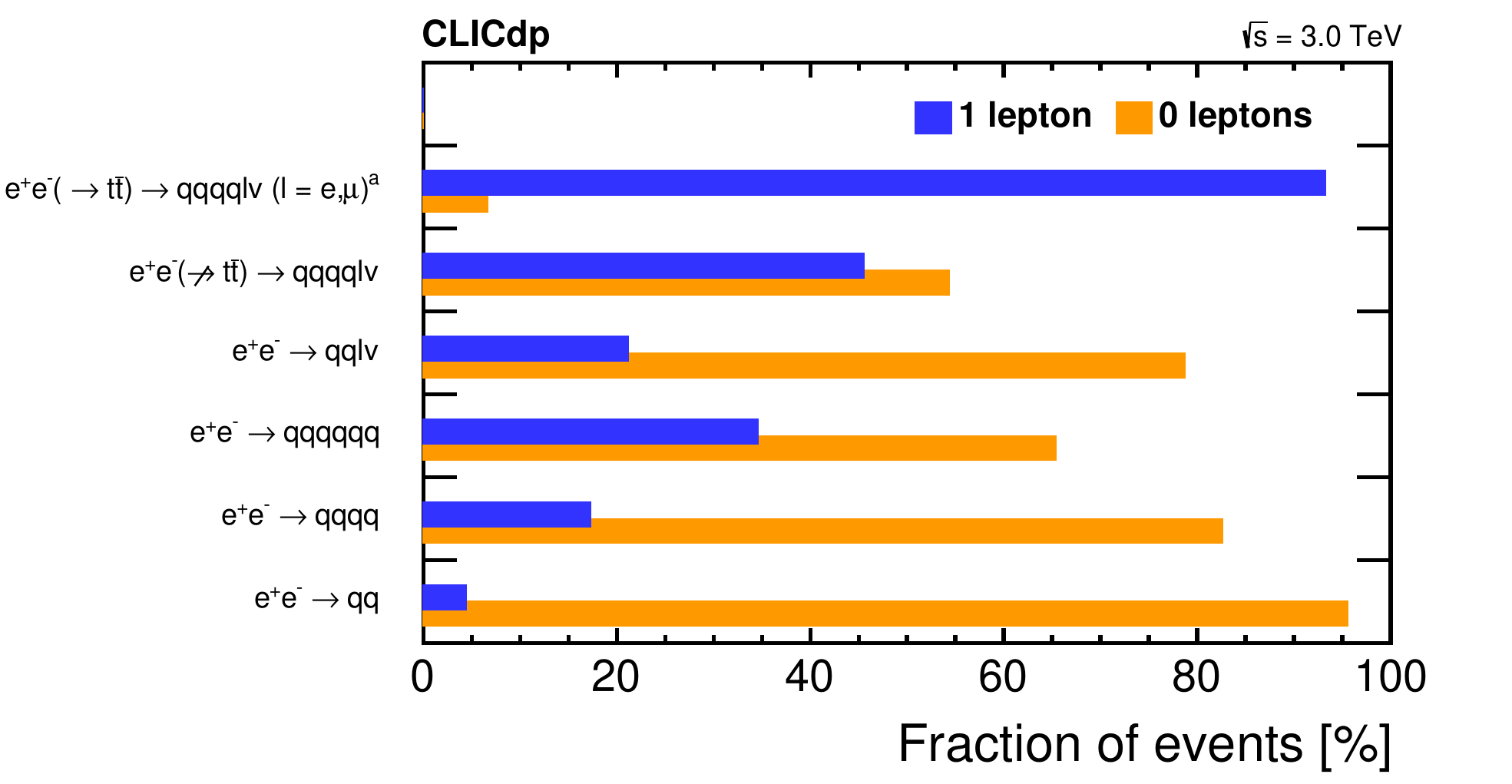}
  \caption{Fraction of events with 0 or 1 reconstructed leptons at $\roots=1.4\,\tev$ (top) and $\roots=3\,\tev$ (bottom) for $P(\Pem)=-80\%$ and for the considered semi-leptonic and fully-hadronic processes. The superscript `a' refers to the kinematic region $\rootsprime\geq1.2\,\tev$. \label{fig:analysis:leptonid:effevent}}
\end{figure}

Events without an identified lepton are discarded. The fraction of events with either zero or one identified charged leptons is illustrated in \Cref{fig:analysis:leptonid:effevent} for the semi-leptonic and fully-hadronic processes considered. 

\clearpage
\section{Top-tagging of large-R jets}
\label{sec:jetreco}

\begin{figure}
  \centering
  \includegraphics[width=0.65\columnwidth]{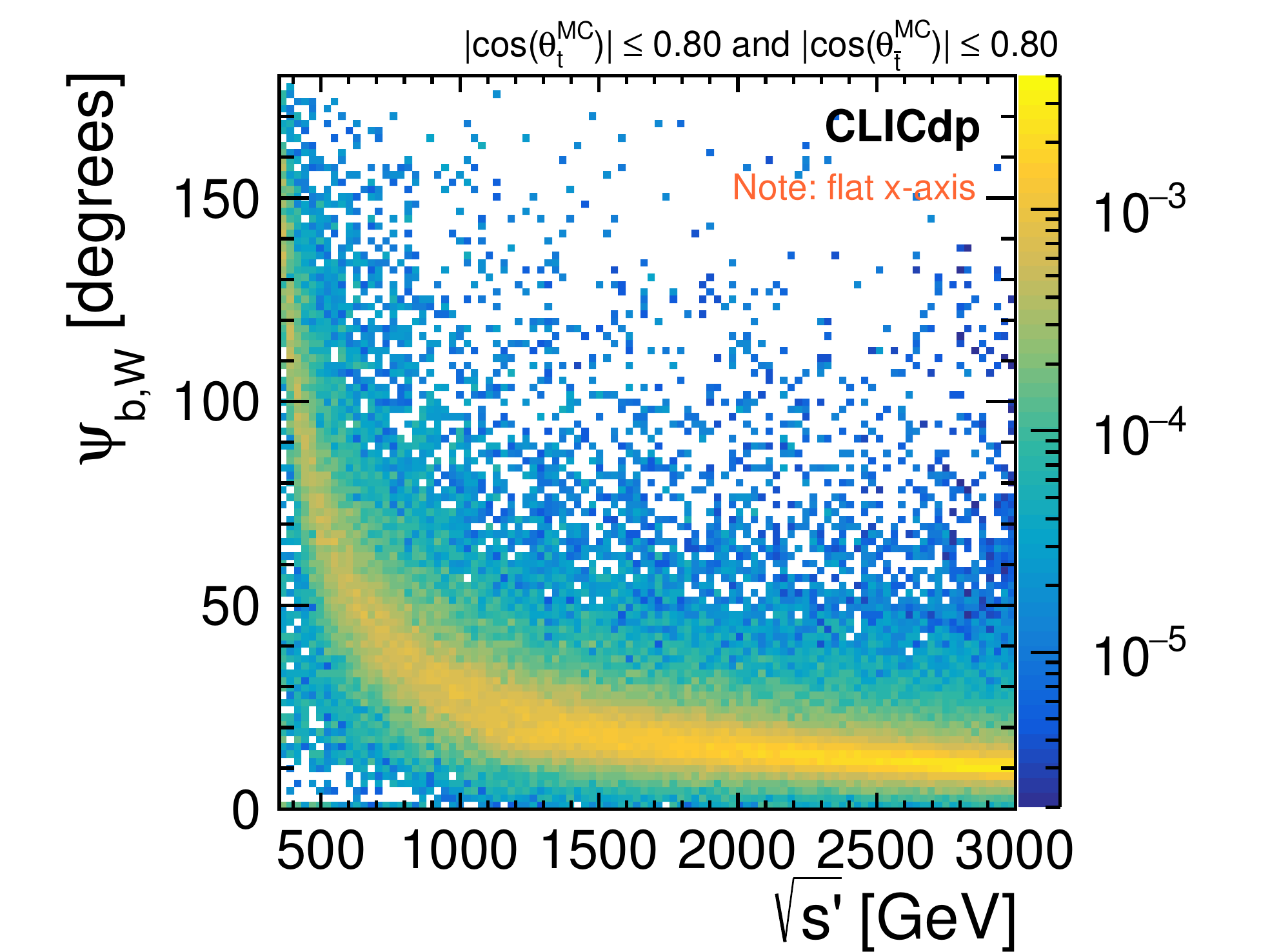}
  \caption{Angle between the W boson and the associated $\PQb$-quark from top-quark decays at parton-level, shown for top-quark decays at a nominal collision energy of 3\,\tev, as a function of the effective centre-of-mass energy. To disentangle the effect of the shape of the effective centre-of-mass energy spectrum, the horizontal axis was weighted so that each vertical column contains the same number of entries, leading to a flat $\rootsprime$ distribution. \label{fig:analysis:jetreco:anglebW}}
\end{figure}

At the higher energy stages of CLIC, a large proportion of the top quarks in $\epem\to\ttbar$ events is produced with significant boosts, resulting in a more collimated jet environment. While this often leads to a clear separation between the decay products of the top- and anti-top quark respectively, the separation between the individual top-quark decay products is generally very small. This is illustrated in \Cref{fig:analysis:jetreco:anglebW} showing the angle between the top-quark decay products for a large range of effective collision energies at $\roots=3\,\tev$. The more dense jet environment constitutes a challenge in particular for traditional jet reconstruction methods where the top-quark is reconstructed by combining individually reconstructed objects to find a suitable candidate. In addition, for highly-boosted top-quark events, a significant fraction of the resulting B-hadrons decay outside the vertex detector, motivating the development of reconstructions methods that do not rely heavily on traditional methods of flavour tagging. For example, about 20\% of the $\mathrm{B}^{\pm}$ hadrons in central $\PQb$-quark jets of 500 GeV decay after the vertex detector~\cite{bhadron}.

The reconstruction of boosted top quarks is studied in full simulation using the \clicild detector model and including $\gghadrons$ background. In the detailed studies of the reconstruction performance presented in this paper, the approach of using large-$R$ jets was found to outperform methods based on the combination of individual jets to form top-quark candidates.

The isolated charged lepton is identified using the isolated lepton finding procedure described in \Cref{sec:lepid}. The remaining PFOs are clustered in two subsequent steps, resulting in two exclusive large-$R$ jets that are used as input to a top-quark tagging algorithm. The latter constitutes the basis for identification of the hadronically decaying top quark in the analyses. In addition, the identified inner jet structure, the so-called ``subjet'' constituents, provide further handles for discrimination against background as illustrated in \Cref{sec:mva}.

\subsection{Jet clustering algorithm}
\label{ssec:jetalgo}

The clustering algorithms used here are based on sequential recombination, where the pair of input particles that are closest in some distance measure are recombined, a process that is repeated until some stopping criterion is reached, for example when a predefined number of so-called ``exclusive jets'' have been identified. Below we simultaneously consider two different distance measures: $d_{ij}$, between particle four-vectors, and $d_{iB}$, between particle and beam four-vectors. While the former, if closest, would combine input $i$ and $j$ into a new four-vector considered for further clustering, the latter, if closest, would declare $i$ as part of a so-called ``beam'' jet or a final ``inclusive'' jet, depending on the clustering mode.

The background from \gghadrons events, discussed in \Cref{sec:gensim}, yields a diffuse background superposed on the signal events, and becomes increasingly challenging for jet reconstructions at the high energy operation of CLIC. The longitudinally invariant algorithms developed for hadron colliders are found to be more robust against this background than classic jet reconstruction algorithms developed for $\Pep\Pem$ colliders~\cite{Boronat:2016tgd}. Even better background resilience can be achieved with the VLC algorithm~\cite{Boronat:2016tgd}, that is based on a classical $\Pep\Pem$ inter-particle distance criterion, but with a beam distance criterion that has a reduced solid angle in the forward region of the detector. Similar to the algorithms used at hadron colliders, when operated in ``exclusive'' clustering mode, particles that are found to be closer to the beam axis than to other particles become part of the ``beam jet''. These are assumed to have originated in beam-beam interactions and are therefore removed from the event.

The VLC algorithm uses the particle energies $E$, and angular separation $\theta$, to compute a clustering distance parameter
\begin{equation}
d_{ij} = 2\min({E_i^{2\beta},E_j^{2\beta}})(1-\cos{\theta_{ij}})/R^2,
\end{equation}
where $R$ is the radius parameter that determines the maximum area of the jet and $\beta$ regulates the clustering order. The default choice is $\beta=1.0$ unless otherwise specified. The distance to the beam axis is measured by
\begin{equation}
d_{iB} = E_i^{2\beta}{(p_{\mathrm{T}\,i}/E_i)}^{2\gamma},
\end{equation}
where the $\PGg$ parameter controls the rate of shrinking in jet size in the forward region\footnote{We apply the beam distance measure as implemented in the ValenciaPlugin of \fastjet `contrib' versions up to 1.039. Note that this differs slightly from the one quoted in \cite{Boronat:2016tgd}.}; the default choice is $\PGg=1.0$ unless otherwise specified.

The PFOs in each event are clustered in two subsequent steps following the approach described in~\cite{Nachman:2014kla}. A pre-clustering is performed in an inclusive mode using the Generalised-\kT algorithm for $\epem$ collisions (``gen-\kT algorithm'')~\cite{Fastjet} with a minimum $\pT$ threshold. The clustered PFOs are re-clustered into two exclusive jets using the VLC algorithm described above. The effect of this two-stage clustering is similar to that of grooming/trimming in that it reduces the effective area of the jet and removes soft contributions to not obscure the underlying jet substructure. The jet clustering steps are performed by the \fastjet package \cite{Fastjet}.

\subsection{Jet clustering optimisation}
\label{ssec:jetopt}

The optimisation of the jet clustering parameters was studied using fully-hadronic $\ttbar$ events with an effective centre-of-mass energy close to the nominal collision energy. The parameters were optimised to fully enclose the decay products of hadronically decaying boosted top-quarks while reducing the effect from including extra background particles. More specifically, the parameters, for both clustering stages, were selected as the best trade-off between achieving a narrow top-quark mass peak close to the generated parton-level top-quark mass, and minimising the contributions to the mass peak at $m_{\PW}$. 

\begin{figure}
\centering
\includegraphics[width=0.48\columnwidth]{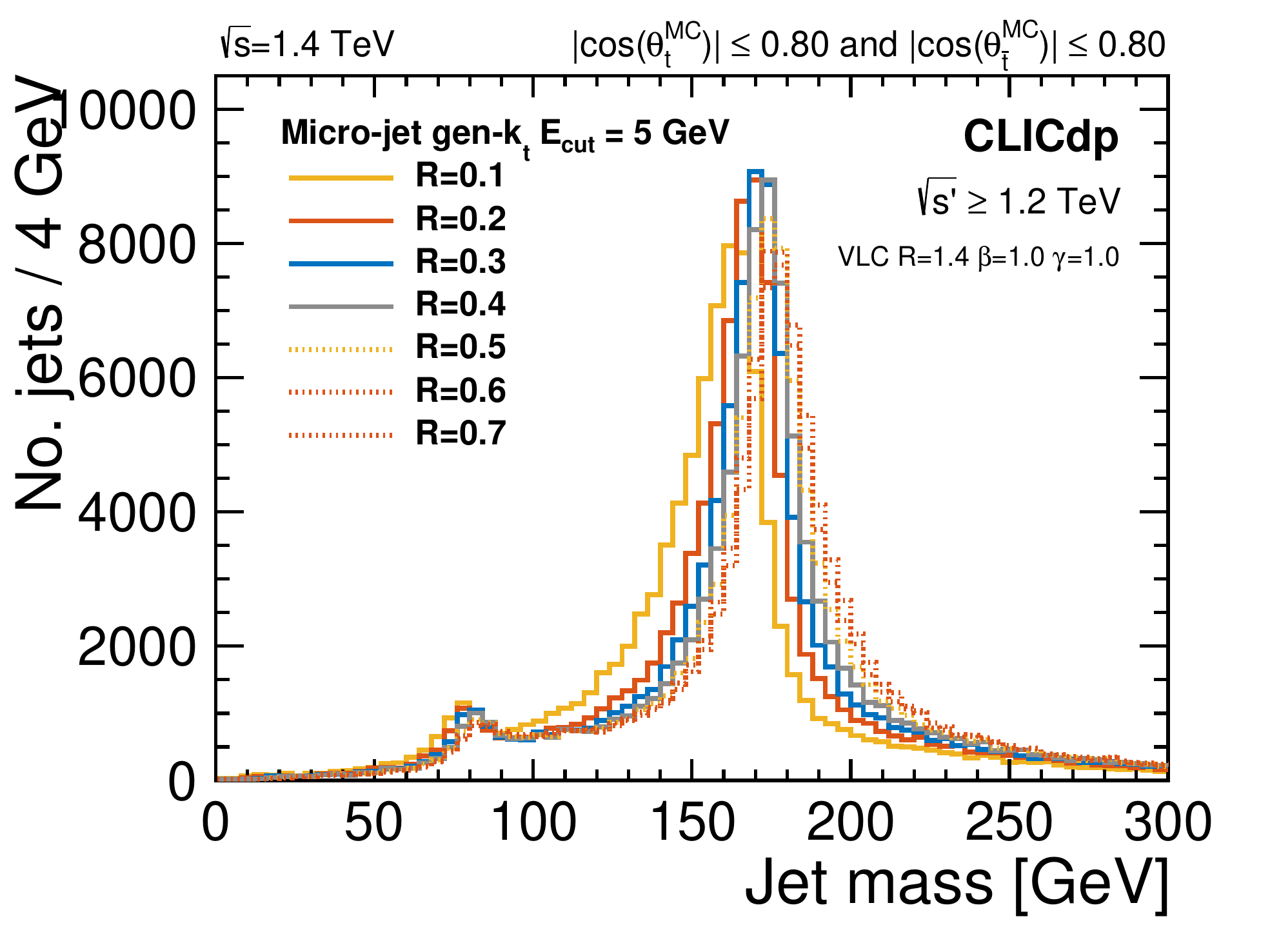}
~~~
\includegraphics[width=0.48\columnwidth]{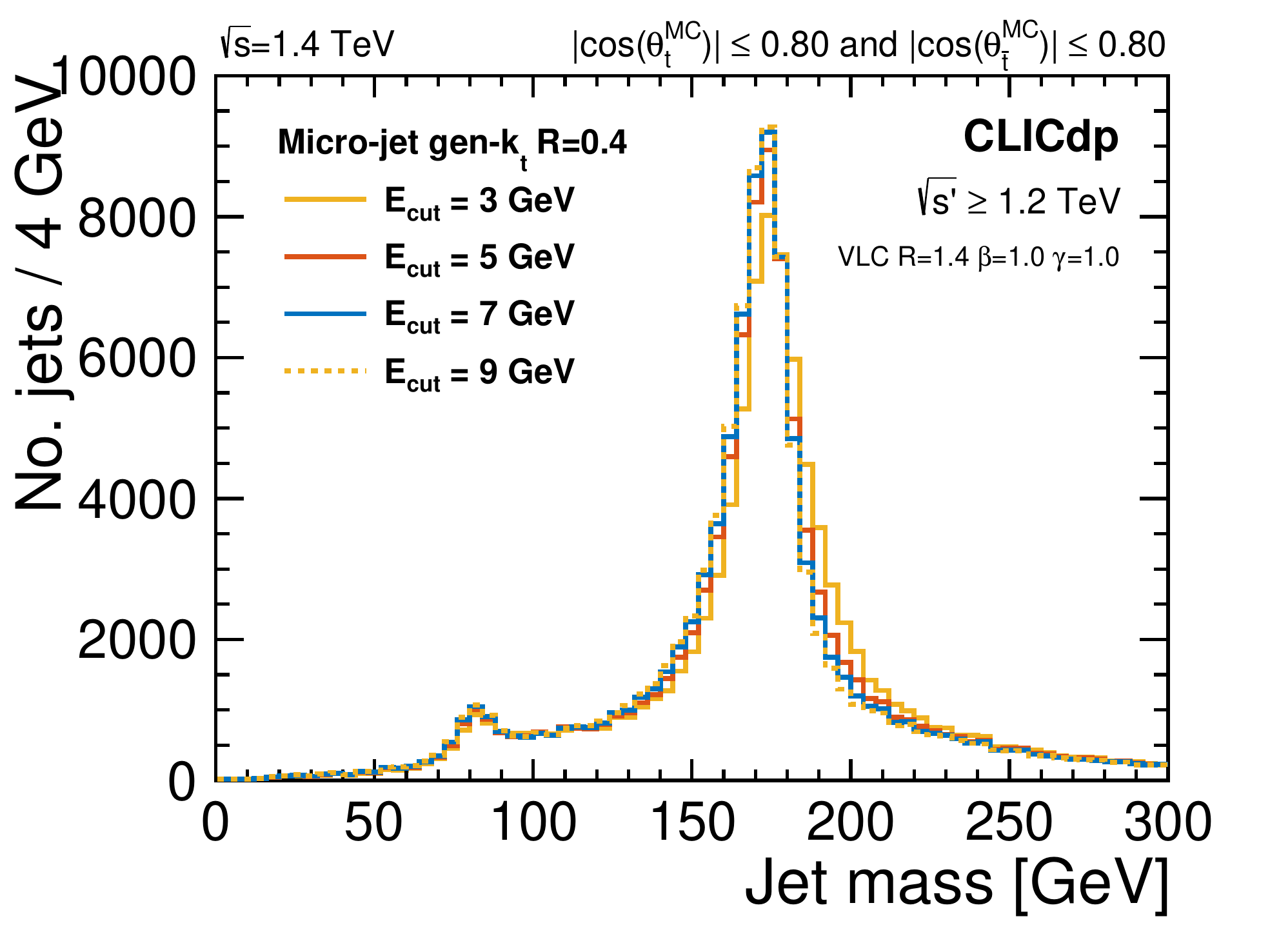}
\caption{Resulting large-R jet mass distributions for different micro-jet radii (left) and $\pT$ threshold in the pre-clustering step (right), shown for events at $\roots=1.4\,\tev$. A large-R jet radius of 1.4 and default parameters for $\beta$ and $\gamma$ were assumed.\label{fig:analysis:jetreco:preclusteringradius}}
\end{figure}

\Cref{fig:analysis:jetreco:preclusteringradius} shows the large-R jet mass distribution for different radii and $\pT$ cuts used in the pre-clustering step. For the optimisation of the radius ($\pT$), the $\pT$ cut (radius) was fixed to $5\,\mathrm{GeV}$ (0.4). For both examples shown, a large-R jet radius of 1.4 and default parameters for $\beta$ and $\gamma$ were assumed. We found that a jet radius of $R=0.4$ and a minimum $\pT$ threshold of $5\,\mathrm{GeV}$ were optimal in the pre-clustering step. In particular we find only a small dependence on the latter in the tested region. Note that the same conclusions hold when varying the assumed large-R jet clustering parameters.

\begin{figure}
\centering
\includegraphics[width=0.48\columnwidth]{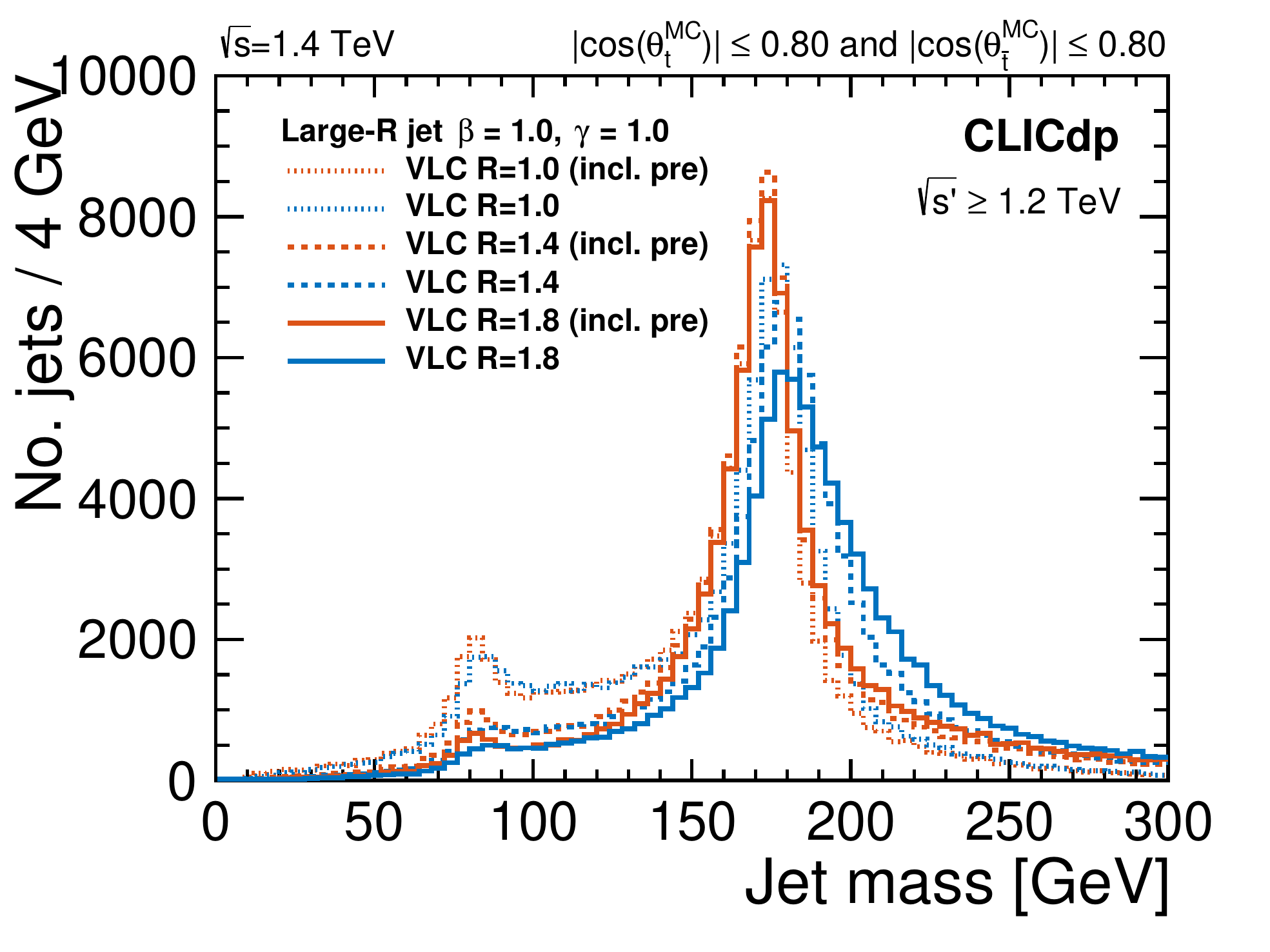}
~~~
\includegraphics[width=0.48\columnwidth]{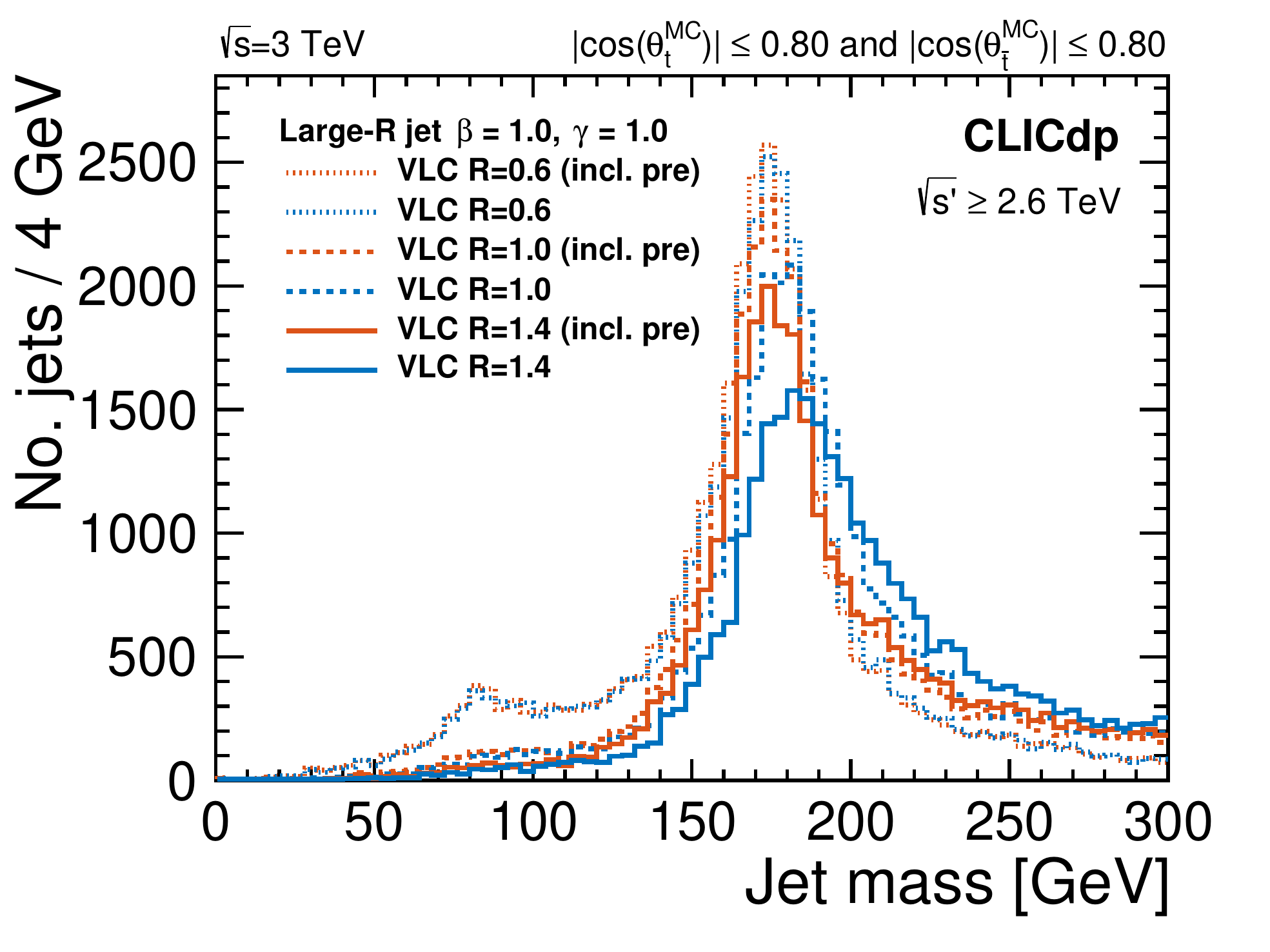}
\caption{Reconstructed large-$R$ jet mass for fully-hadronic $\ttbar$ events in CLIC at $\roots=1.4\,\tev$ (left) and $\roots=3\,\tev$ (right), illustrating different choices of jet clustering radius $R$ including the effect of applying a pre-clustering step. Right-hand side figure taken from \cite{Abramowicz:2018rjq}. \label{fig:analysis:jetreco:largeRradius}}
\end{figure}

\Cref{fig:analysis:jetreco:largeRradius} shows the reconstructed large-$R$ jet mass distribution for different choices of jet clustering radius $R$, for $\roots=1.4\,\tev$ (left) and $\roots=3\,\tev$ (right). The figures also illustrate the effect of applying the pre-clustering step prior to the large-$R$ jet clustering, as described above. It is clear from the figure that too small a jet radius does not enclose the entire top-quark decay products, leading to a significant peak close to the mass of the $\PW$ boson. In contrast, a larger jet radius includes a growing contribution from background processes leading to a long tail in the distribution towards higher masses. We found that a large-$R$ jet radius of $R=1.4$ and $R=1.0$, were optimal for operation at $\roots=\,1.4\,\tev$ and $\roots=\,3\,\tev$, respectively. No significant difference was observed for varying the $\beta$ and $\gamma$ parameters, therefore the default value of 1.0 is chosen for both.

\Cref{fig:analysis:jetreco:largeRradius:boosted} and \Cref{fig:analysis:jetreco:largeRradius:nonboosted} (in Appendix A) show the detailed jet mass distributions, including pre-clustering, at $\roots=1.4\,\tev$ (left) and $\roots=3\,\tev$ (right), for various radii. While the distributions in the former only show events with an effective centre-of-mass energy close to the nominal collision energy, the latter display distributions for events above $\rootsprime=400\,\gev$. Similarly, \Cref{fig:analysis:jetreco:largeRradius:semilep:boosted} and \Cref{fig:analysis:jetreco:largeRradius:semilep:nonboosted} (in Appendix A) show the distributions for the highest energy jet in reconstructed semi-leptonic $\ttbar$ events.

\begin{figure}
\centering
\includegraphics[width=0.48\columnwidth]{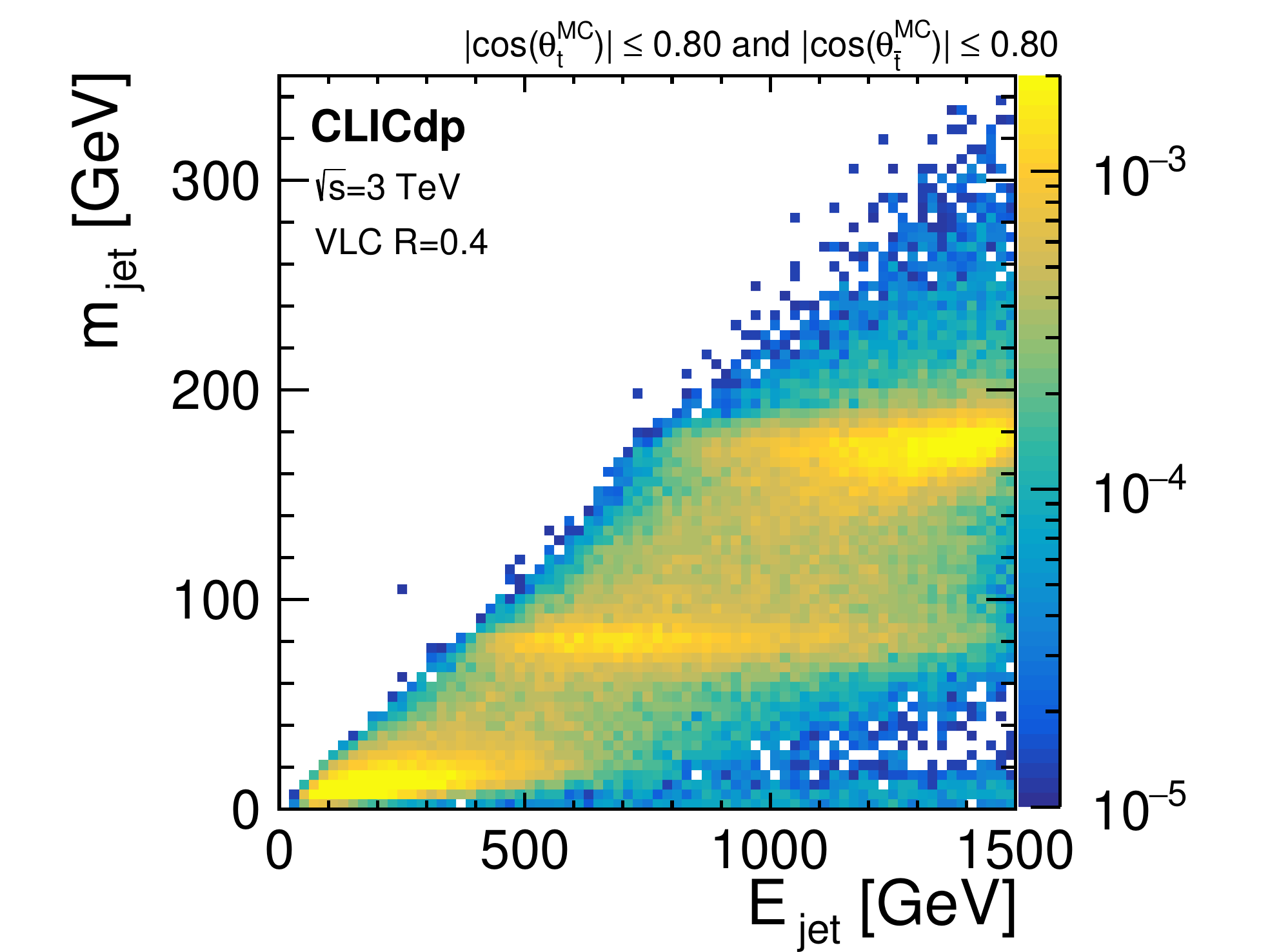}
~~~
\includegraphics[width=0.48\columnwidth]{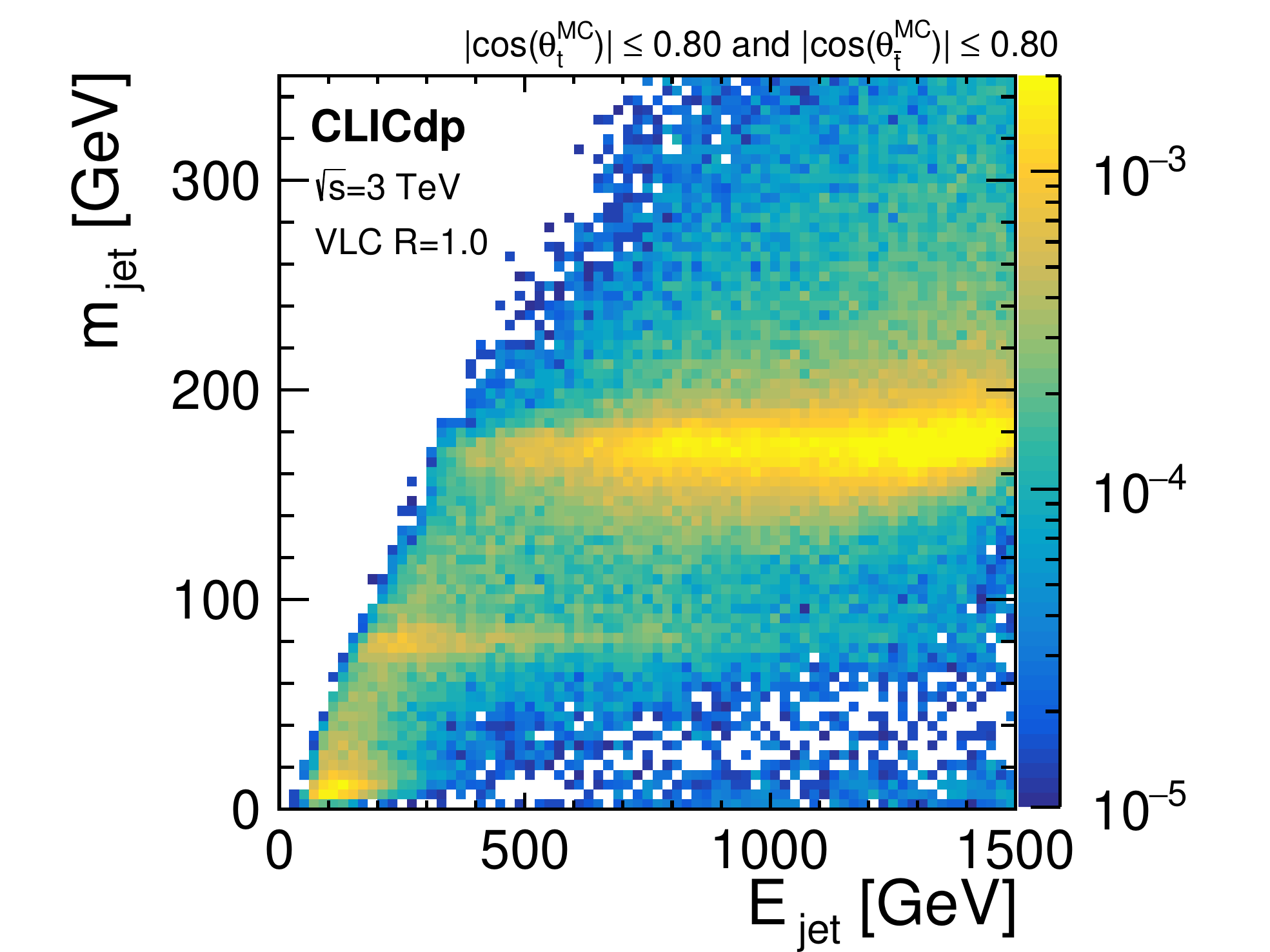}
\caption{Normalised distribution of the number of events as a function of reconstructed jet mass and jet energy for fully-hadronic $\ttbar$ events at $\roots=3\,\tev$. The particles are reconstructed in two exclusive jets using the VLC algorithm and a jet clustering radius $R=0.4$ (left) and $R=1.0$ (right). The jet mass represents the distribution including a pre-clustering step with a jet radius of $R=0.4$ and a minimum $\pT$ threshold of $5\,\mathrm{GeV}$. Right-hand side figure taken from \cite{Abramowicz:2018rjq}.\label{fig:analysis:jetreco:MassVSE}}
\end{figure}

\begin{figure}
\centering
\includegraphics[width=0.48\columnwidth]{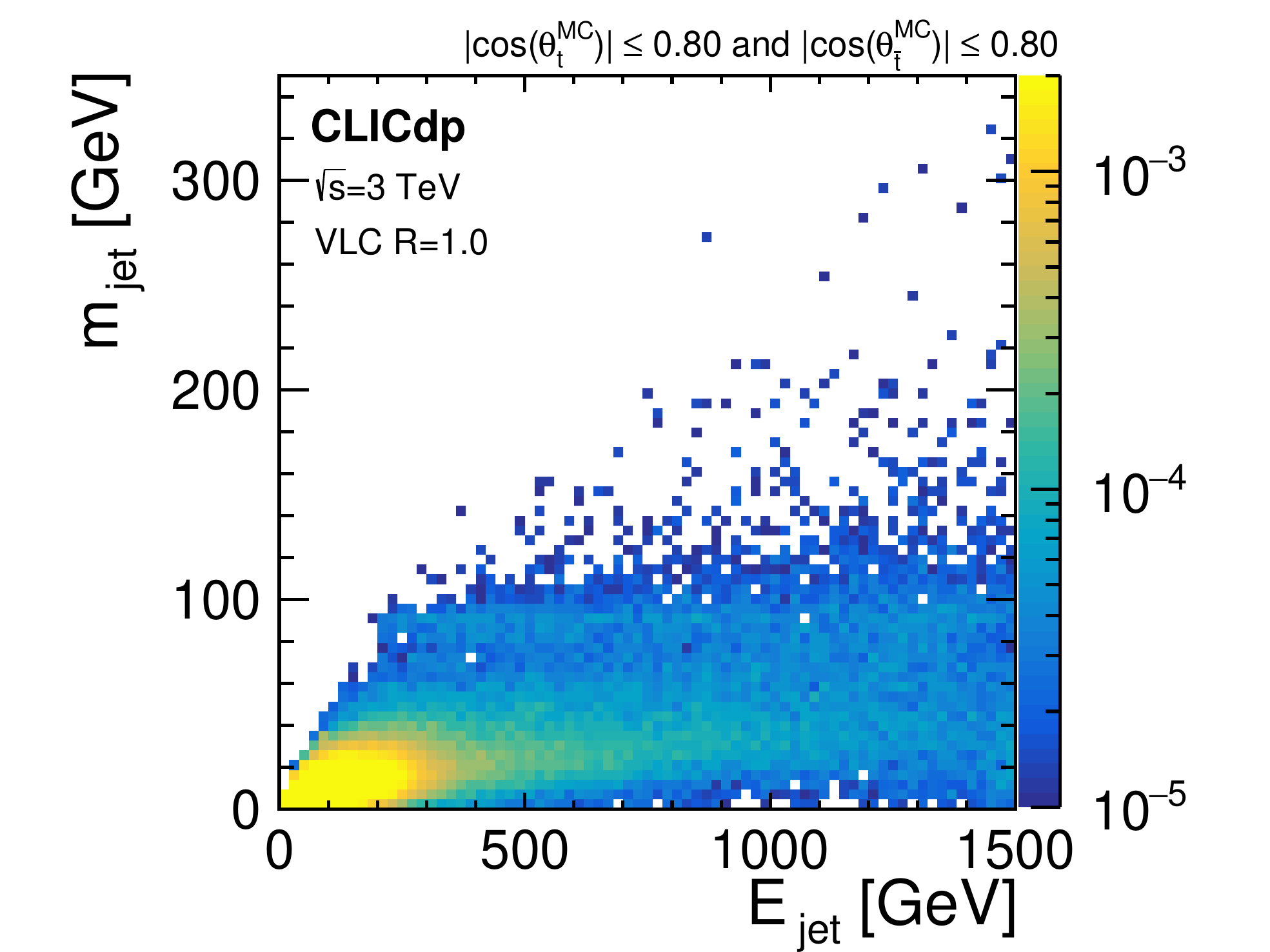}
\includegraphics[width=0.48\columnwidth]{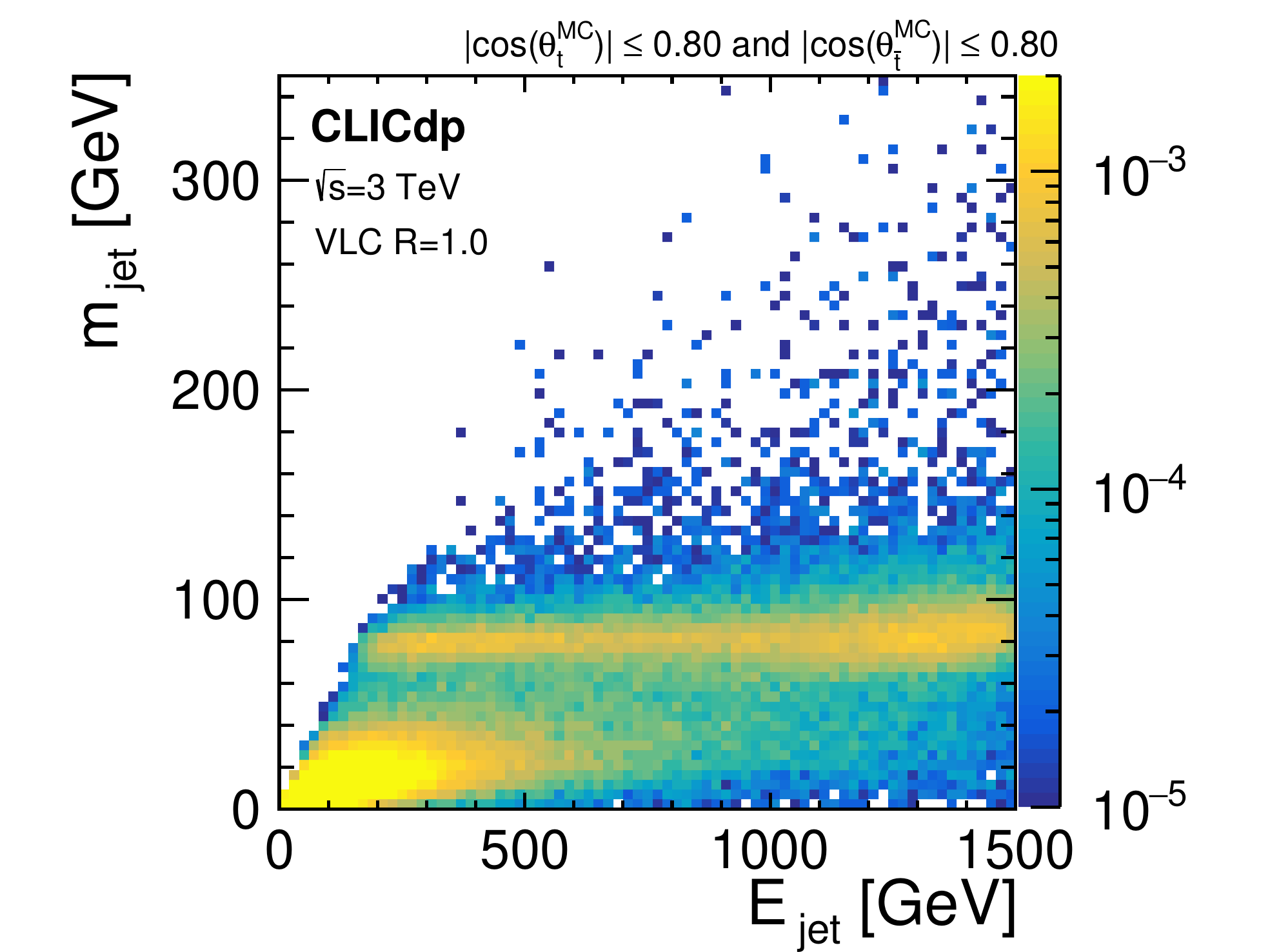}
\caption{Normalised distribution of the number of events as a function of reconstructed jet mass and jet energy for fully-hadronic di-jet (left) and four-quark (right) events at $\roots=3\,\tev$. The particles are reconstructed in two exclusive jets using the VLC algorithm and a large-R jet clustering radius of $R=1.0$, and includes a pre-clustering step with a jet radius of $R=0.4$ and a minimum $\pT$ threshold of $5\,\mathrm{GeV}$.\label{fig:analysis:jetreco:MassVSEBkg}}
\end{figure}

\Cref{fig:analysis:jetreco:MassVSE} shows the reconstructed jet mass as a function of the jet energy for fully-hadronic $\ttbar$ events at $\roots=3\,\tev$, for two different values of the radius parameter $R$. The uppermost of the three visible yellow bands indicates top quarks that are fully captured within the large-$R$ jet, while the lower two bands represent partially captured top quarks close to the mass of $m_{\PW}$ and $m_{\PQb}$, respectively. As expected, the large-$R$ jet approach performs well for jets at higher energy, while the ability to capture the full top-quark jet is significantly reduced in the non-boosted regime, below $\sim500\,\gev$. The equivalent result for two of the largest background processes, four-jet and di-jet events, are shown in \Cref{fig:analysis:jetreco:MassVSEBkg}.

\begin{figure}
\centering
\includegraphics[width=0.65\columnwidth]{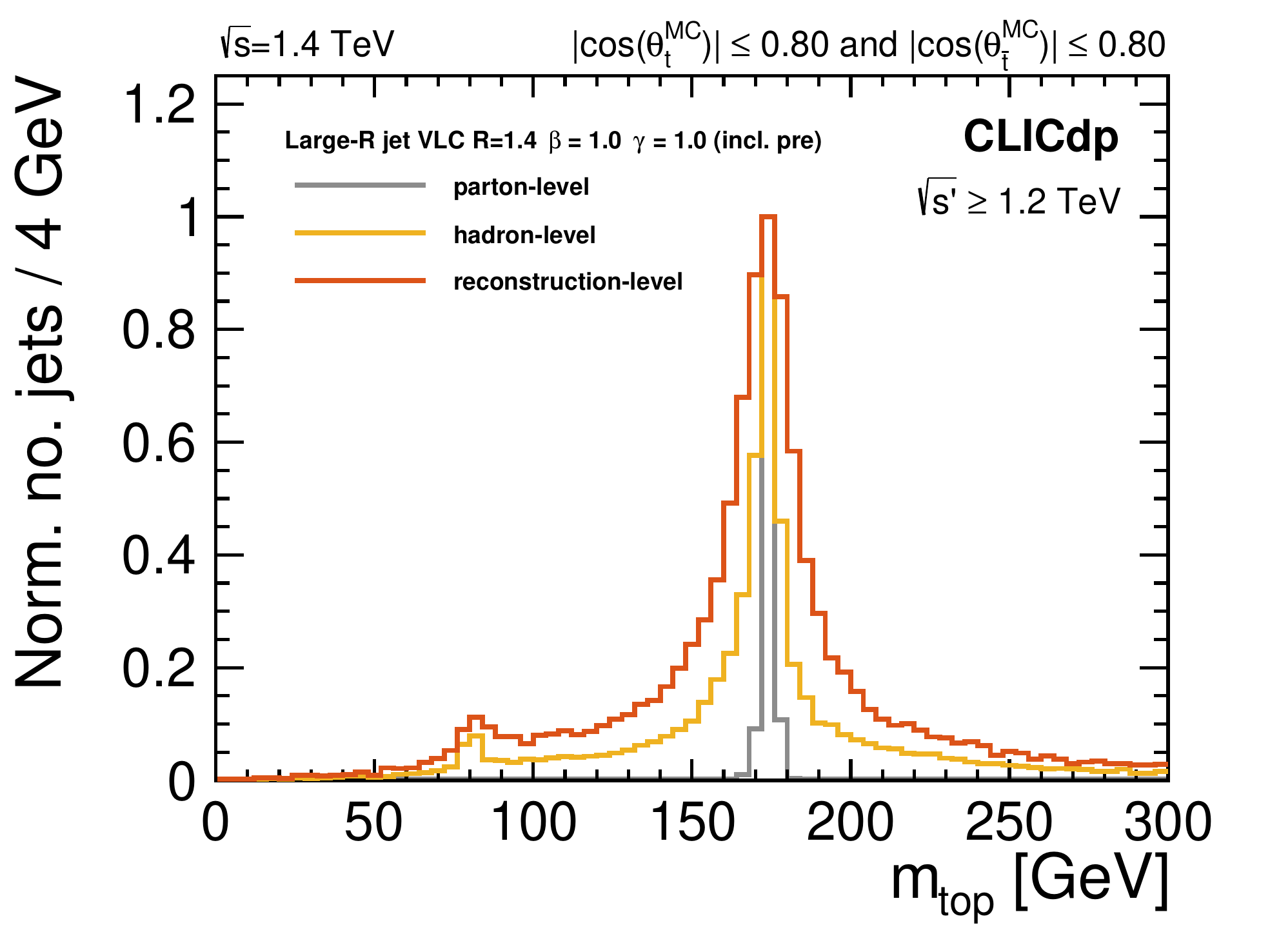}
\caption{Reconstructed large-$R$ jet mass for fully-hadronic $\ttbar$ events in CLIC at $\roots=1.4\,\tev$ (red). The MC hadron-level distribution for the same jet clustering parameters is shown in yellow. The grey curve shows the parton-level distribution. Note that for easier comparison of the shape, all curves are normalised, with a maximum at 1. \label{fig:analysis:jetreco:massdistatdifflevels}}
\end{figure}

\Cref{fig:analysis:jetreco:massdistatdifflevels} shows the large-R jet mass distribution for optimal jet clustering parameters, at $\roots=1.4\,\tev$, in red. Similarly, the yellow curve indicates the hadron-level distributions for the same parameters. The grey curve shows the parton-level distribution. 

\subsection{Top-tagging algorithm}
\label{ssec:toptaggeralgo}

\textit{This section follows closely the description of the top-tagger algorithm in \cite{Abramowicz:2018rjq}.}

Each of the resulting large-$R$ jets, from the two-step clustering described above, serve as input to a top tagger algorithm, based on the Johns Hopkins top tagger~\cite{Kaplan:2008ie} as implemented in \fastjet~\cite{Fastjet, Fastjet:2006}. The algorithm is designed to identify top quarks by reversing the final steps of the jet clustering, looking for up to three or four hard subjets consistent with a top-quark decay. This de-clustering procedure provides strong discrimination-power for hadronically decaying top quarks against QCD-induced quark jets. In the following sections we outline the tagging algorithm and characterise its performance using fully-hadronic $\ttbar$ events. Later it is applied to semi-leptonic $\ttbar$ events where it aims to reconstruct the hadronically decaying top quark.

\begin{figure}
\centering
\includegraphics[width=0.65\columnwidth]{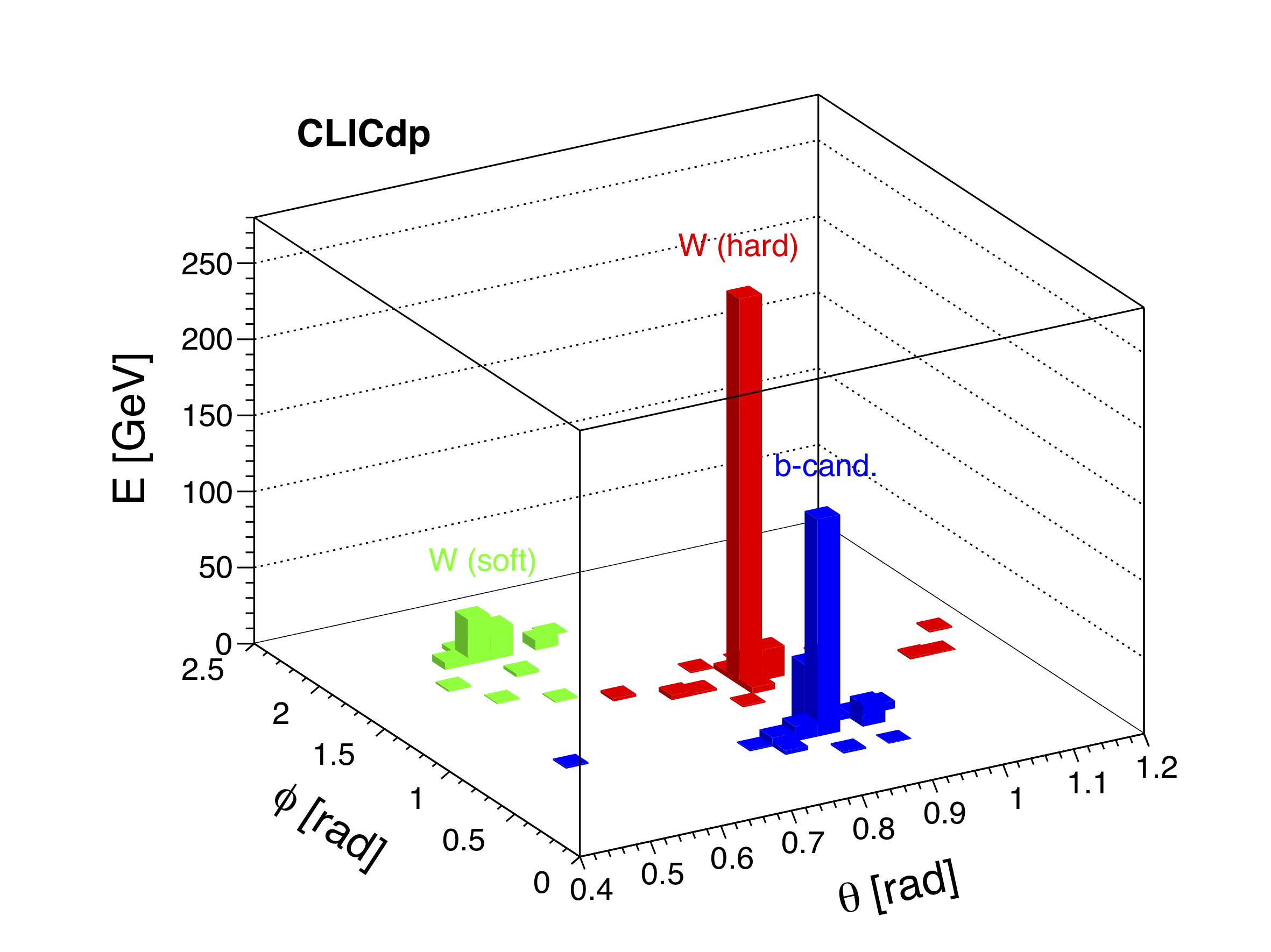}
\caption{Energy depositions of a hadronically decaying top-quark as function of $\theta$ and $\phi$. The energy depositions are shown individually for the soft (green) and hard (red) components of the $\PW$ decay, as well as for the $\PQb$-quark (blue). \label{fig:analysis:jetreco:toptaggerdisplay}}
\end{figure}

\Cref{fig:analysis:jetreco:toptaggerdisplay} shows the energy depositions of a hadronically decaying top-quark, as function of the detector coordinates $\theta$ and $\phi$: the soft (green) and hard (red) component of the $\PW$ decay, and the b-quark (blue).

The top-tagging algorithm is governed by two parameters: $\delta_r$, the subjet distance; and $\delta_p$, the fraction of subjet $\pT$ relative to the $\pT$ of the large-R input jet. These parameters control whether to accept the objects, resulting from the split, as subjets for further de-clustering or whether, for example, the de-clustering should continue only on the harder of the two objects. An object is rejected if its $\pT$ fraction is lower than $\delta_p$ or if its distance to another object is smaller than $\delta_r$. The de-clustering loop is terminated when two successive splittings have been accepted resulting in two, three, or four subjets of the input jet. The case with two final subjets is rejected and the other cases are further analysed. The jet is considered a top quark if the total invariant mass of the subjets is within $\pm55$\,GeV of $m_{\PQt}$ and one subjet pair has an invariant mass within $\pm30$\,GeV of $m_{\PW}$. Further, the reconstructed helicity angle $\theta_{\PW}$, measured in the rest frame of the reconstructed W boson and defined as the opening angle of the top quark to the softer of the two \PW boson decay subjets, is studied and gives additional separation power. Too shallow an angle would be an indication of a false splitting, where one of the pairs of subjets produces a small mass compatible with QCD-like emission. Note however, that in the analysis presented in this paper, we do not included this variable in the definition of the top-tagger. Instead, it is used as one of the inputs to the multivariate analysis introduced in \Cref{sec:mva}.

\subsection{Top-tagging optimisation}

The optimisation of the top tagging algorithm was studied using fully-hadronic $\ttbar$ events, four-jet events $\qqbar\qqbar\,(\mathrm{\PQu,\PQd,\PQs,\PQc,\PQb})$, and dijet events $\qqbar\,(\mathrm{\PQu,\PQd,\PQs,\PQc,\PQb})$, with an effective centre-of-mass energy close to the nominal collision energy. In this study, the final-state partons (parton-level) and the reconstructed input large-R jets each fulfil $|\cos(\theta)|\leq0.80$.

\begin{figure}[p]
	\centering
	\begin{subfigure}{0.48\columnwidth}
	\includegraphics[width=\textwidth]{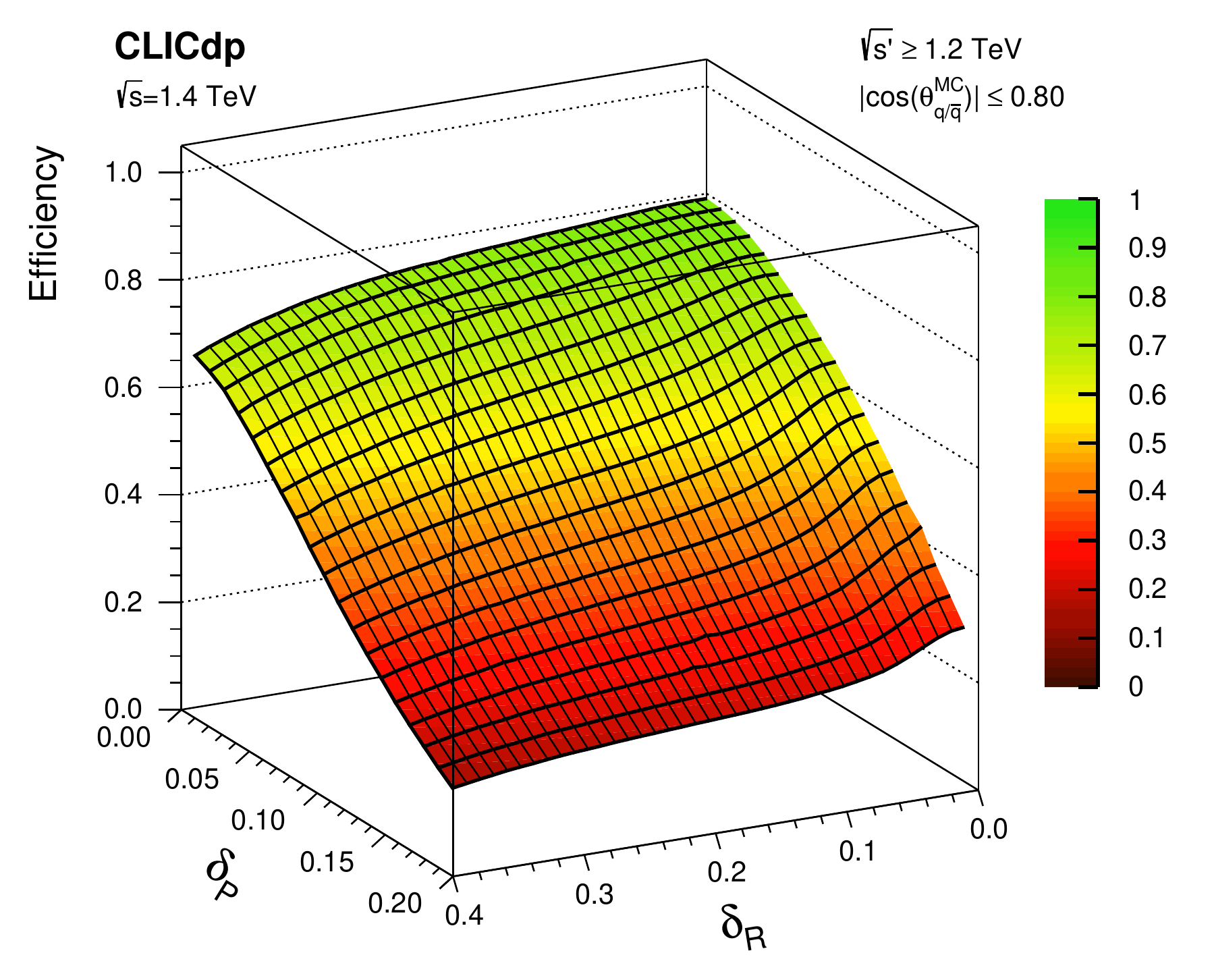}
	\caption{Fully-hadronic $\ttbar$ at $\roots=1.4\,\tev$}
	\end{subfigure}
	~~~
	\begin{subfigure}{0.48\columnwidth}
	\includegraphics[width=\textwidth]{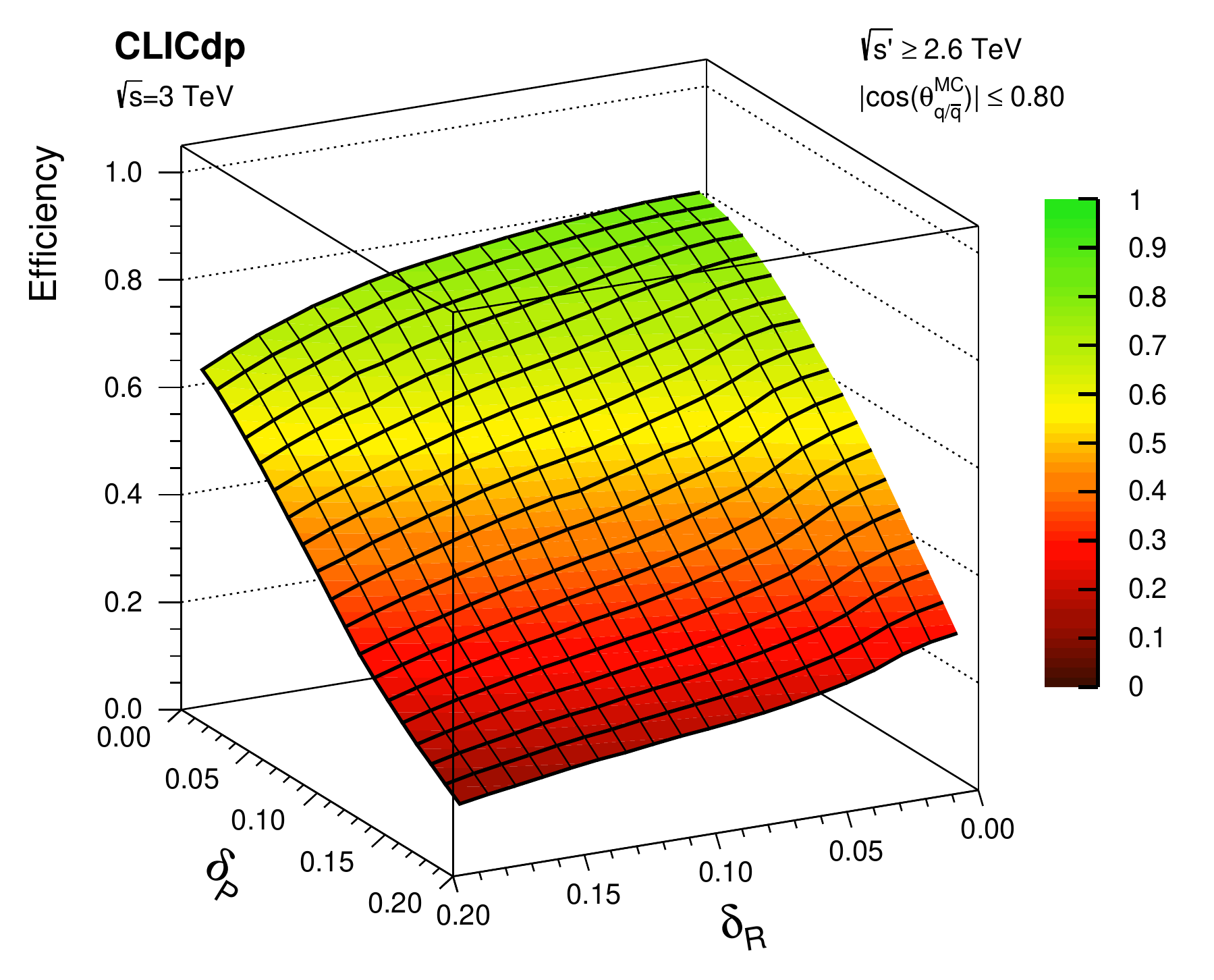}
	\caption{Fully-hadronic $\ttbar$ at $\roots=3\,\tev$}
	\end{subfigure}\\
	\vspace{5mm}
	\begin{subfigure}{0.48\columnwidth}
	\includegraphics[width=\textwidth]{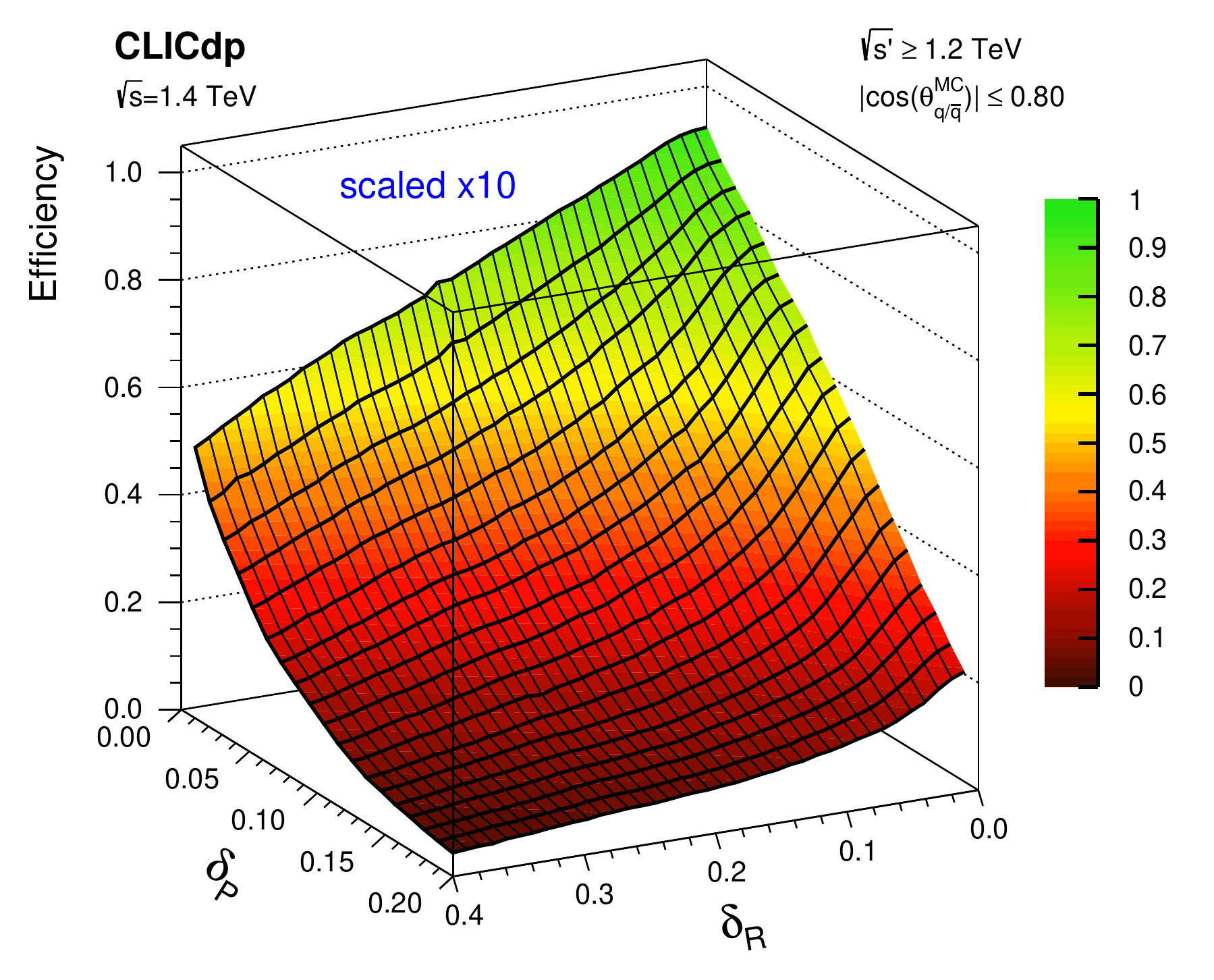}
	\caption{Four-jet background at $\roots=1.4\,\tev$\label{sfig:toptagger:qqqq:1.4}}
	\end{subfigure}
	~~~
	\begin{subfigure}{0.48\columnwidth}
	\includegraphics[width=\textwidth]{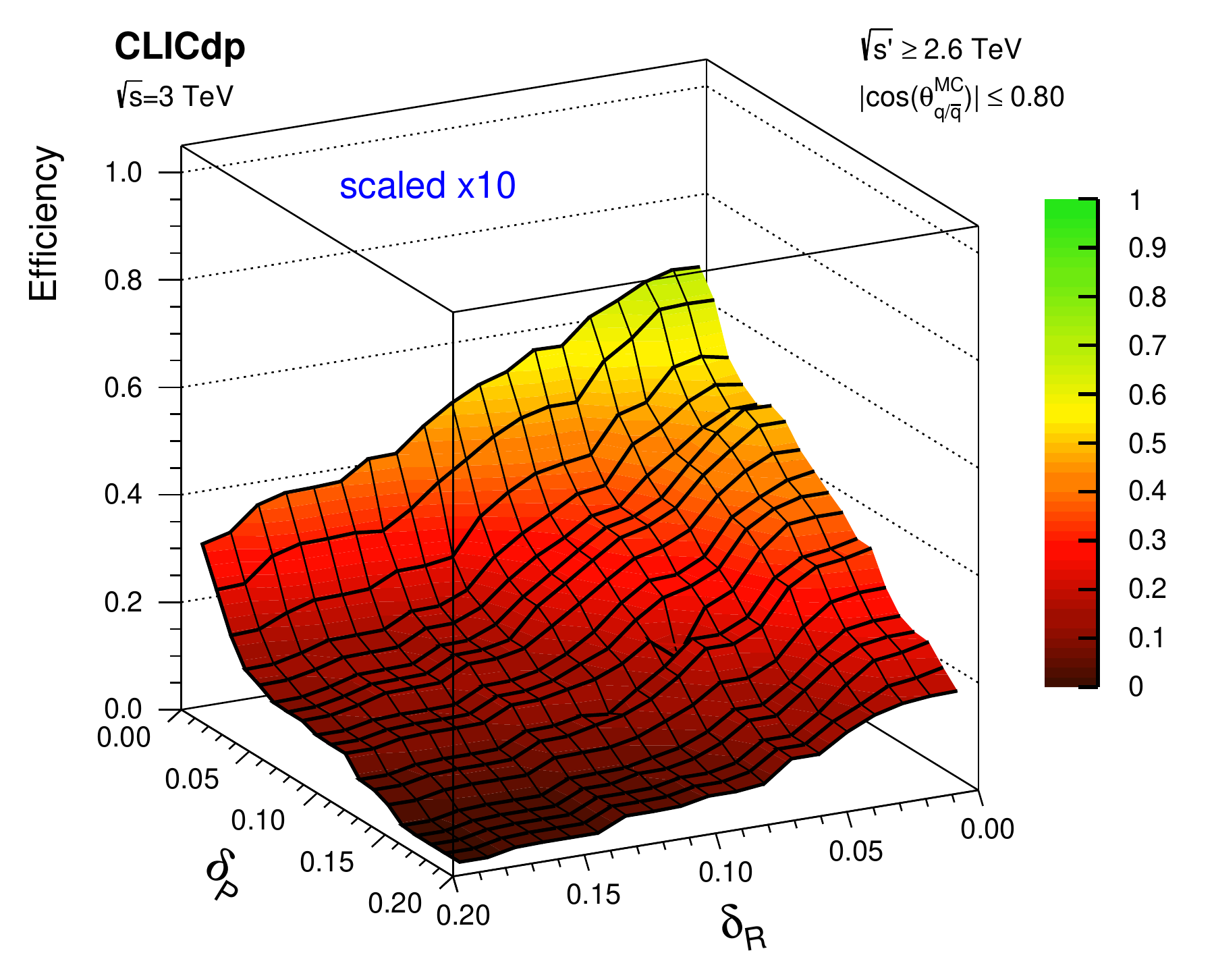}
	\caption{Four-jet background at $\roots=3\,\tev$\label{sfig:toptagger:qqqq:3}}
	\end{subfigure}\\
	\vspace{5mm}
	\begin{subfigure}{0.48\columnwidth}
	\includegraphics[width=\textwidth]{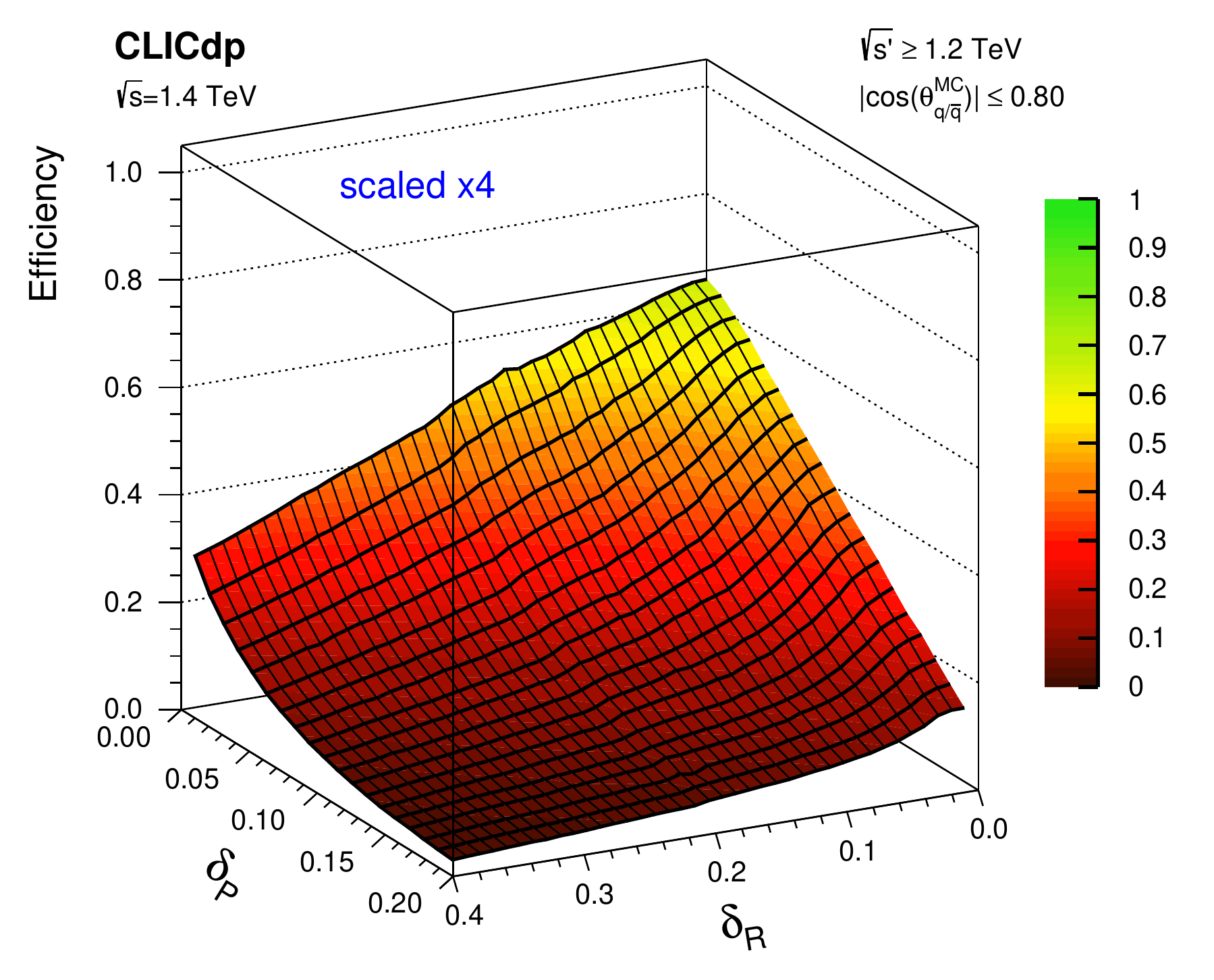}
	\caption{Dijet background at $\roots=1.4\,\tev$\label{sfig:toptagger:qq:1.4}}
	\end{subfigure}
	~~~
	\begin{subfigure}{0.48\columnwidth}
	\includegraphics[width=\textwidth]{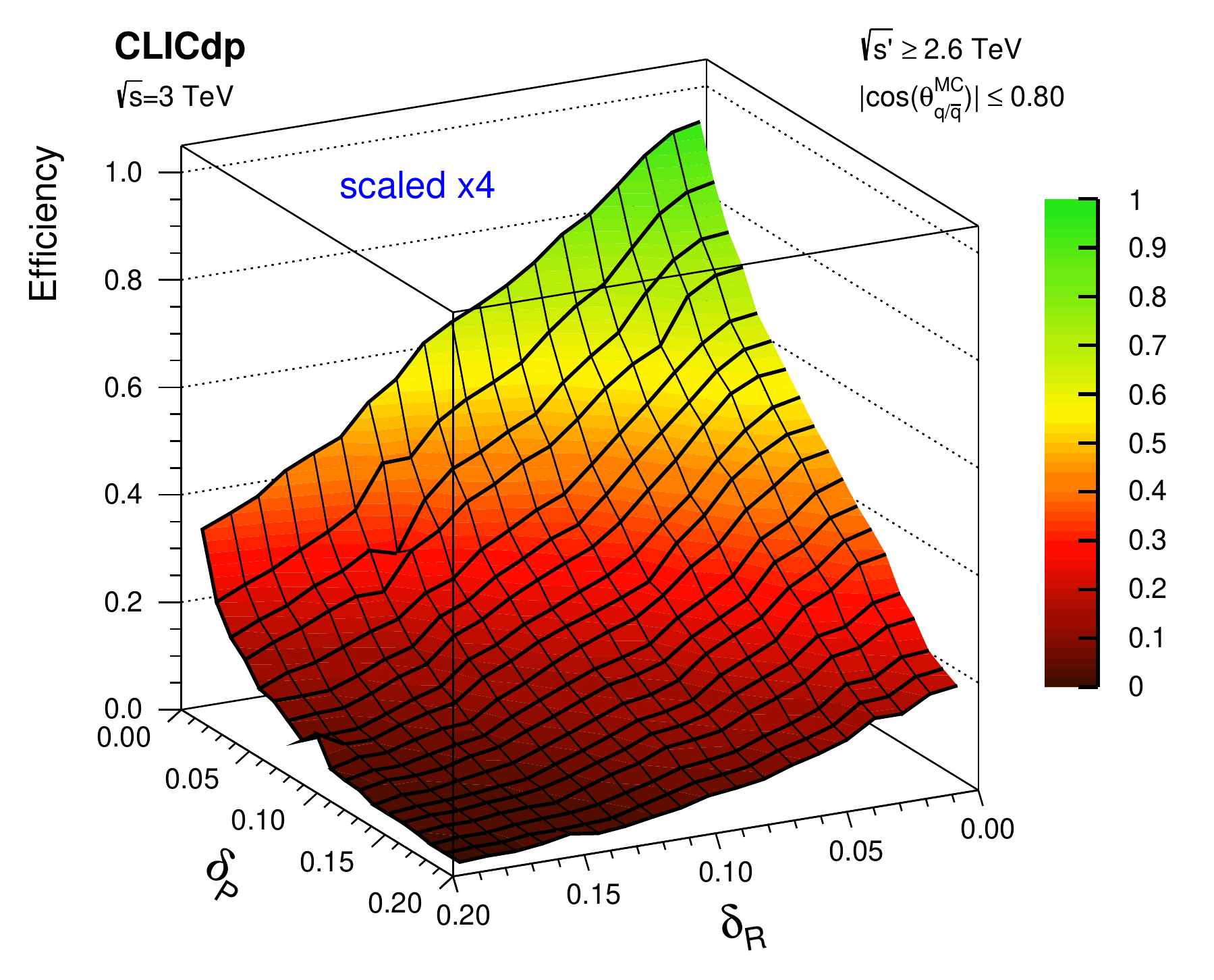}
	\caption{Dijet background at $\roots=3\,\tev$\label{sfig:toptagger:qq:3}}
	\end{subfigure}
\caption{Top-quark candidate tagging efficiency as a function of two parameters in the Johns-Hopkins top tagger algortithm: the subjet distance, $\delta_r$, and the fraction of subjet $\pT$ relative to the $\pT$ of the large-R input jet, $\delta_p$. Note that the efficiencies in  \Cref{sfig:toptagger:qqqq:1.4}, \Cref{sfig:toptagger:qqqq:3}, \Cref{sfig:toptagger:qq:1.4}, and \Cref{sfig:toptagger:qqqq:3} are scaled. The scaling factor is shown in blue at the top of each figure. \label{fig:analysis:jetreco:efficiencyvsdeltas}}
\end{figure}

\Cref{fig:analysis:jetreco:efficiencyvsdeltas} shows the top-tagging efficiency including mass cuts as a function of the subjet distance, $\delta_r$, and the fraction of subjet $\pT$, $\delta_p$. The distributions on the left show the efficiency for operation at $\roots=1.4\,\tev$, while on the right the equivalent distributions for operation at $\roots=3\,\tev$ are shown. Note that the efficiencies for $\qqbar\qqbar$ ($\qqbar$), shown in the second (third) row, are scaled with a factor 10 (4).
             
\begin{figure}
\centering
\includegraphics[width=0.48\columnwidth,clip]{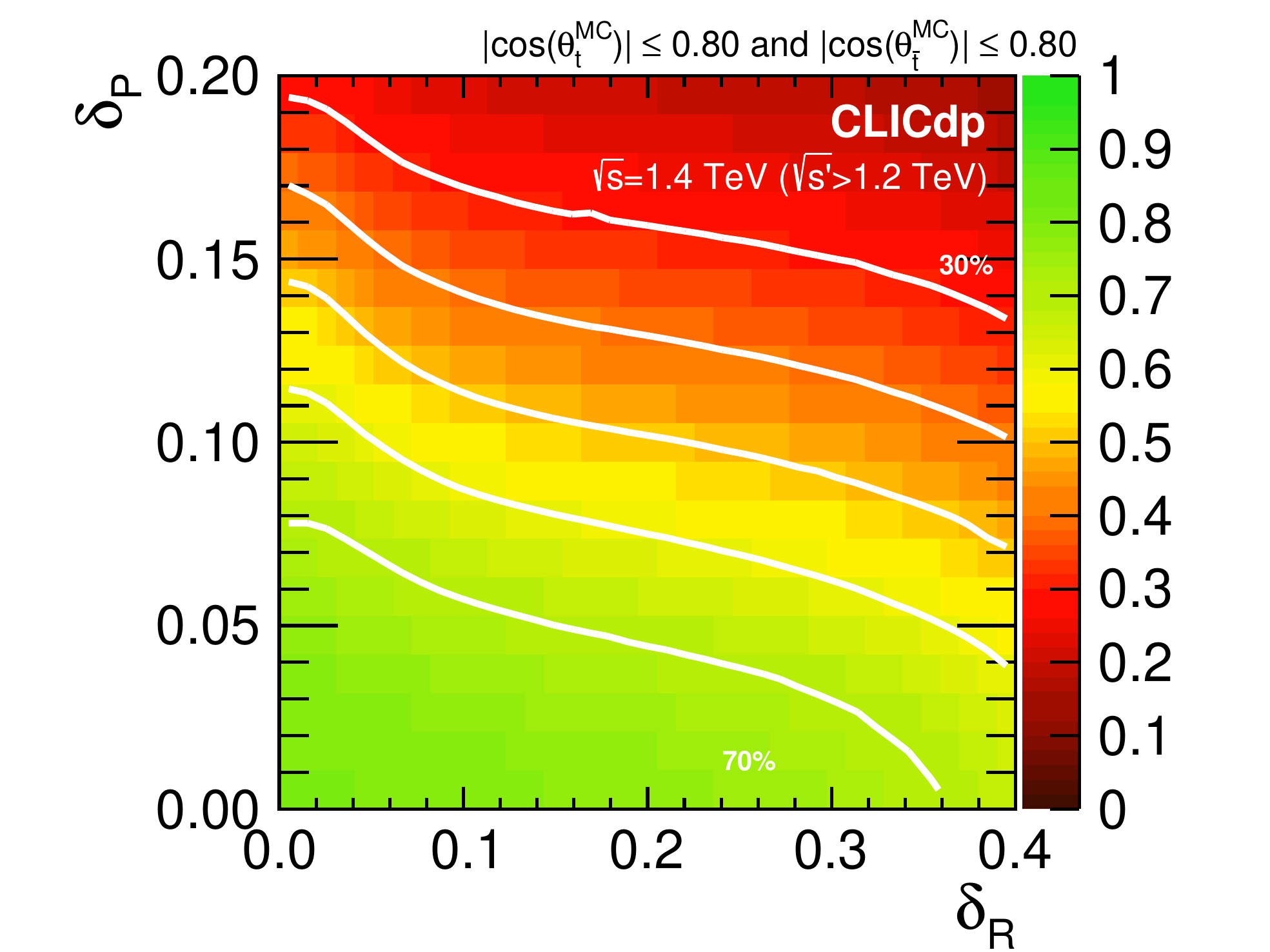}
~~~
\includegraphics[width=0.48\columnwidth,clip]{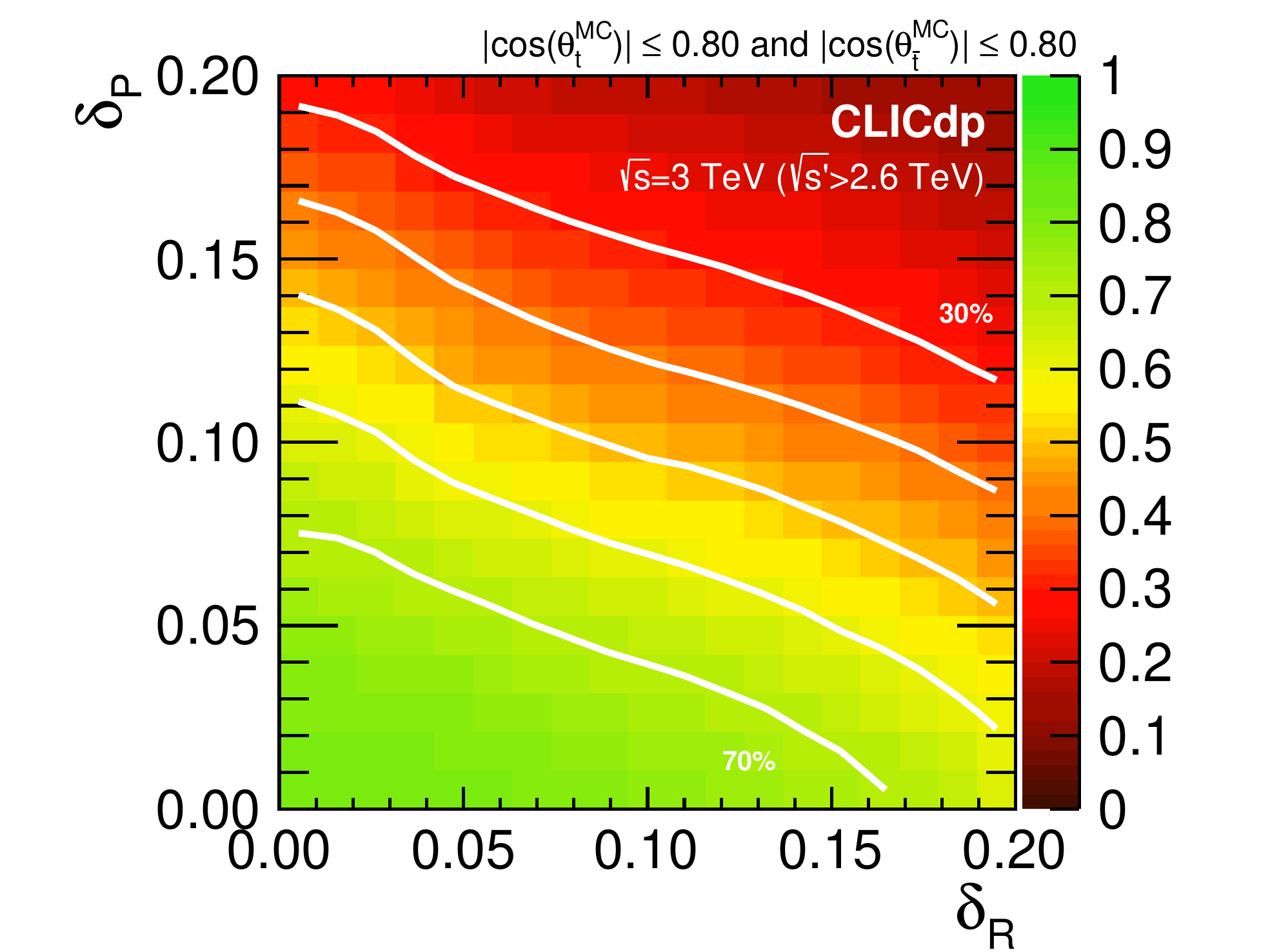}
\caption{Top-tagger efficiency for the fully-hadronic $\ttbar$ signal sample at $\roots=1.4\,\tev$ (left) and $\roots=3\,\tev$ (right). The white contours represent a fixed signal efficiency ranging from 30\% (top) to 70\% (bottom).}
\label{fig:analysis:jetreco:efficiencycontours}
\end{figure}

Since the amount of background at a lepton collider is substantially lower than at a hadron collider, a somewhat higher rate of wrongly tagged quark-jets $(\mathrm{\PQu,\PQd,\PQs,\PQc,\PQb})$ is acceptable for a given top-quark jet tagging efficiency; the optimisation of the algorithm is tuned to a high-efficiency operating point for the fully-hadronic $\ttbar$ sample. The corresponding top tagger parameters are chosen by minimising the rate of wrongly tagged light-quark jets from the four-jet sample. The white contours shown in \Cref{fig:analysis:jetreco:efficiencycontours}, each represents a fixed signal efficiency ranging from 30\% (top) to 70\% (bottom) and indicates the values of $\delta_r$ and $\delta_p$ studied when minimising the background efficiency. \Cref{tab:analysis:jetreco:taggereff} and \Cref{tab:analysis:jetreco:taggereff3tev} show the benchmark efficiencies considered along with the minimal background efficiency and associated top-tagger settings. For the studies presented in the following, we use a benchmark efficiency of 70\%. The corresponding top tagger parameters are thus $\delta_r=0.25\,(0.11)$ and $\delta_p=0.03\,(0.03)$, for the samples at $\roots=1.4\,(3)\,\tev$, respectively. Note that similar optimal points were found when studying the significance, defined as $S/\sqrt{S+B}$, over the whole considered range of $\delta_r$ and $\delta_p$.

\begin{table}
\centering
\begin{tabular}{cccc}
\vspace{1.0mm}
Benchmark efficiency [\%] & Bkg. efficiency [\%] & $\delta_r$ & $\delta_p$ \rule{0pt}{3ex} \\
\midrule
70\% & 4.8\% (7.9\%) & 0.25 (0.26) & 0.03 (0.03) \\
50\% & 2.1\% (2.7\%) & 0.38 (0.39) & 0.07 (0.07) \\
30\% & 0.9\% (1.2\%) & 0.39 (0.39) & 0.13 (0.13) \\
\bottomrule
\end{tabular}
\caption{Benchmark efficiencies for the fully-hadronic $\ttbar$ sample and the corresponding top-tagger settings for operation at $\roots=1.4\,\tev$. The efficiencies for the background four-jet sample are shown for each working point. The corresponding background efficiencies for when the $\ttbar$ sample is instead optimised against the di-jet sample are presented in brackets. \label{tab:analysis:jetreco:taggereff}}
\end{table}

\begin{table}
\centering
\begin{tabular}{cccc}
\vspace{1.0mm}
Benchmark efficiency [\%] & Bkg. efficiency [\%] & $\delta_r$ & $\delta_p$ \rule{0pt}{3ex} \\
\midrule
70\% & 2.8\% (8.6\%) & 0.11 (0.13) & 0.03 (0.02) \\
50\% & 1.3\% (3.3\%) & 0.19 (0.17) & 0.06 (0.07) \\
30\% & 0.7\% (1.8\%) & 0.19 (0.19) & 0.12 (0.12) \\
\bottomrule
\end{tabular}
\caption{Benchmark efficiencies for the fully-hadronic $\ttbar$ sample and the corresponding top-tagger settings for operation at $\roots=3\,\tev$. The efficiencies for the background four-jet sample are shown for each working point. The corresponding background efficiencies for when the $\ttbar$ sample is instead optimised against the di-jet sample are presented in brackets. \label{tab:analysis:jetreco:taggereff3tev}}
\end{table}

\subsection{Top-tagging efficiency}

\begin{figure}
\centering
\includegraphics[width=0.9\columnwidth,clip]{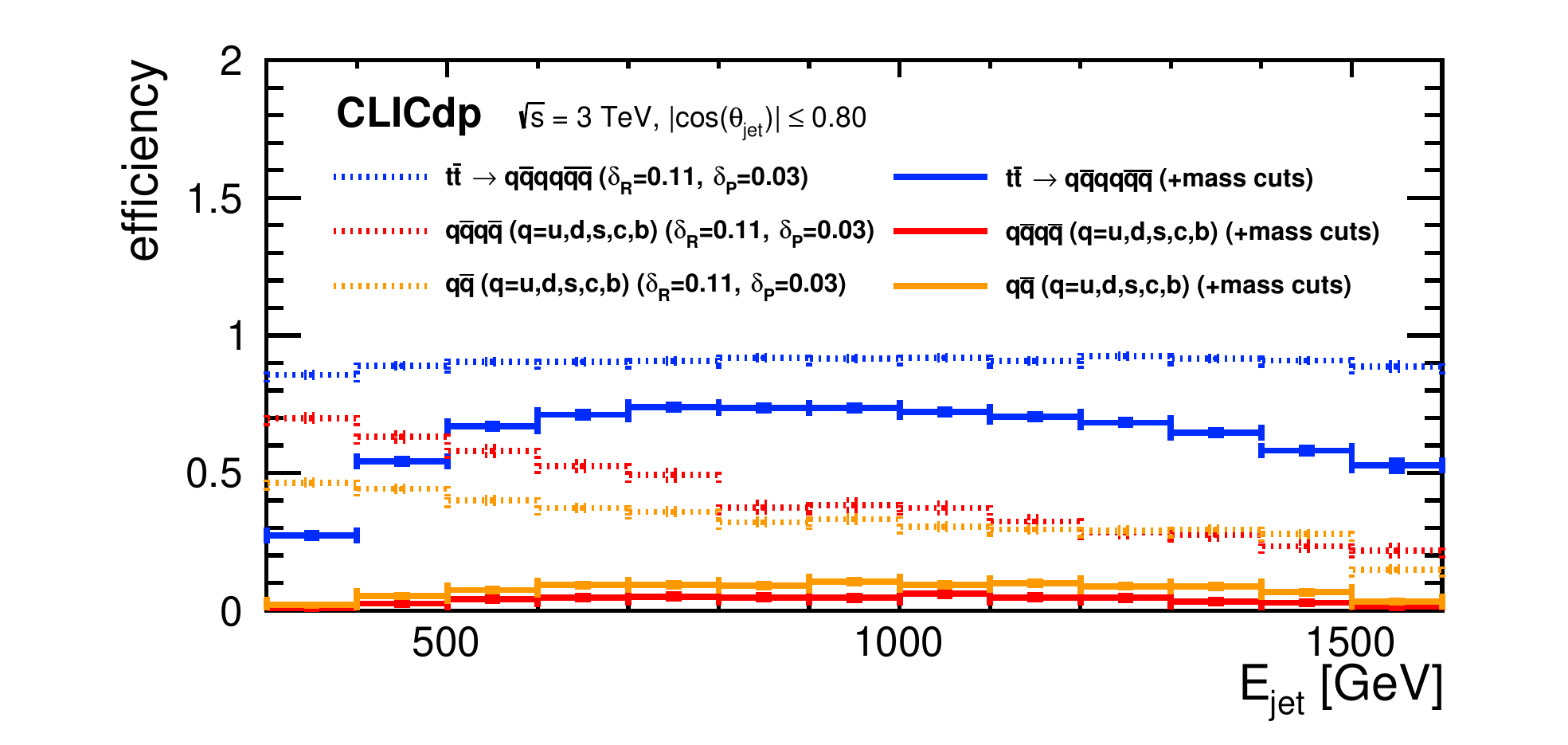}
\caption{Top tagger efficiency for fully-hadronic $\ttbar$ events (blue), four-jet events (red), and dijet events (orange) as function of jet energy. The dashed lines show the effect of applying the de-clustering procedure outlined in the text, while the solid lines show the efficiency including also the cuts on the reconstructed invariant mass of the top-quark and $\PW$ candidates. Figure taken from \cite{Abramowicz:2018rjq}.}
\label{fig:analysis:jetreco:toptaggerefficiencyenergy}
\end{figure}

\begin{figure}
\centering
\includegraphics[width=0.9\columnwidth,clip]{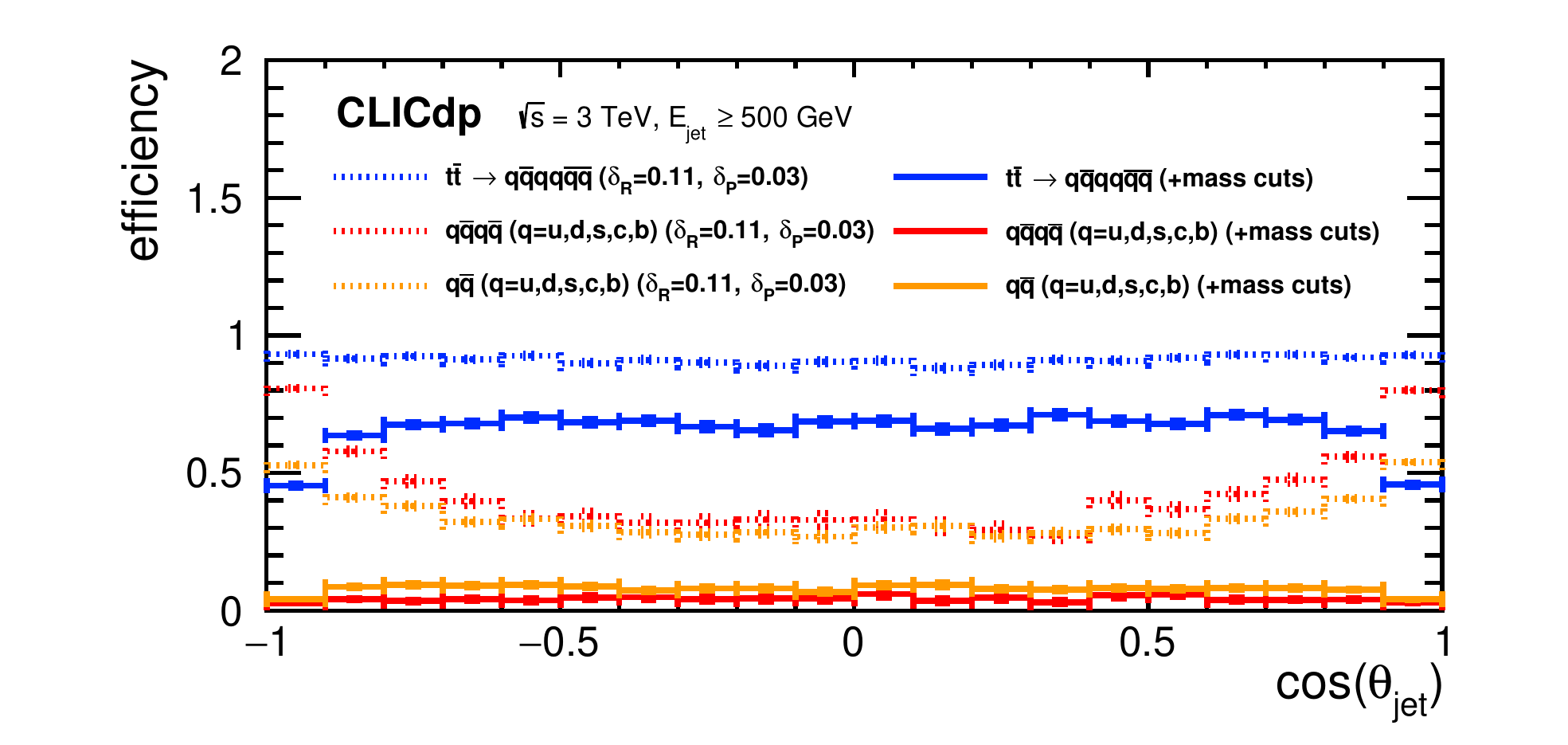}
\caption{Top tagger efficiency for fully-hadronic $\ttbar$ events (blue), four-jet events (red), and dijet events (orange) as function of jet polar angle $\theta$. See \Cref{fig:analysis:jetreco:toptaggerefficiencyenergy} for more details. Figure taken from \cite{Abramowicz:2018rjq}.}
\label{fig:analysis:jetreco:toptaggerefficiencytheta}
\end{figure}

\Cref{fig:analysis:jetreco:toptaggerefficiencyenergy} and \Cref{fig:analysis:jetreco:toptaggerefficiencytheta} show the top-quark tagging efficiency for operation at $\roots=3\,\tev$, as a function of the large-$R$ jet energy and polar angle $\theta$. The solid lines represent the tagging efficiency as described in \Cref{ssec:toptaggeralgo}, while the dashed lines show the same distributions after the initial de-clustering step (excluding mass cuts). Note that the de-clustering step has a limited effect in the forward region. This is caused by the larger beam induced background, that effectively mimics a prongy topology. In addition, as expected, the overall efficiency, including the mass cuts, drops at energies below $500\,\gev$ where the jets are no longer sufficiently boosted to be contained within one large-$R$ jet. The slightly lower efficiency for large jet energies is also anticipated and is mainly due to a more challenging environment for the \pandora algorithm and the subjet de-clustering. Furthermore, the limited detector acceptance in the forward direction reduces the overall efficiency in the corresponding region significantly.

The resulting tagging efficiency for top-quark jets from the $\roots=3\,\tev$ $\ttbar$ dataset is 69\% in the central region of the detector (defined as $|\cos\theta|\leq0.8$) and for energies in the range from 500\,\gev to 1500\,\gev. The corresponding efficiency for wrongly tagged light-quark jets is substantially lower: 4.4\% and 8.8\% for the four-jet and dijet background samples, respectively.\footnote{Alternatively, adopting a tighter operating point at $\roots=3\,\tev$ results in a top-quark jet efficiency of 54\% and an efficiency for wrongly tagged light-quark jets of 2.7\% (3.7\%)} The resulting efficiency for top-quark jets from the $\roots=1.4\,\tev$ dataset is 71\% in the central region of the detector (defined as $|\cos\theta|\leq0.8$) and for energies in the range from 400\,\gev to 700\,\gev. The corresponding efficiency for wrongly tagged light-quark jets is 5.7\% (6.9\%) for jets from the four-jet (dijet) background sample. Note that these values  differs slightly from the ones presented in \Cref{tab:analysis:jetreco:taggereff} and \Cref{tab:analysis:jetreco:taggereff3tev}. The tabulated values represent a convolution of the efficiencies as function of energy and polar angle with the spectra observed for  each dataset.

\begin{figure}
\centering
\includegraphics[width=0.65\columnwidth,clip]{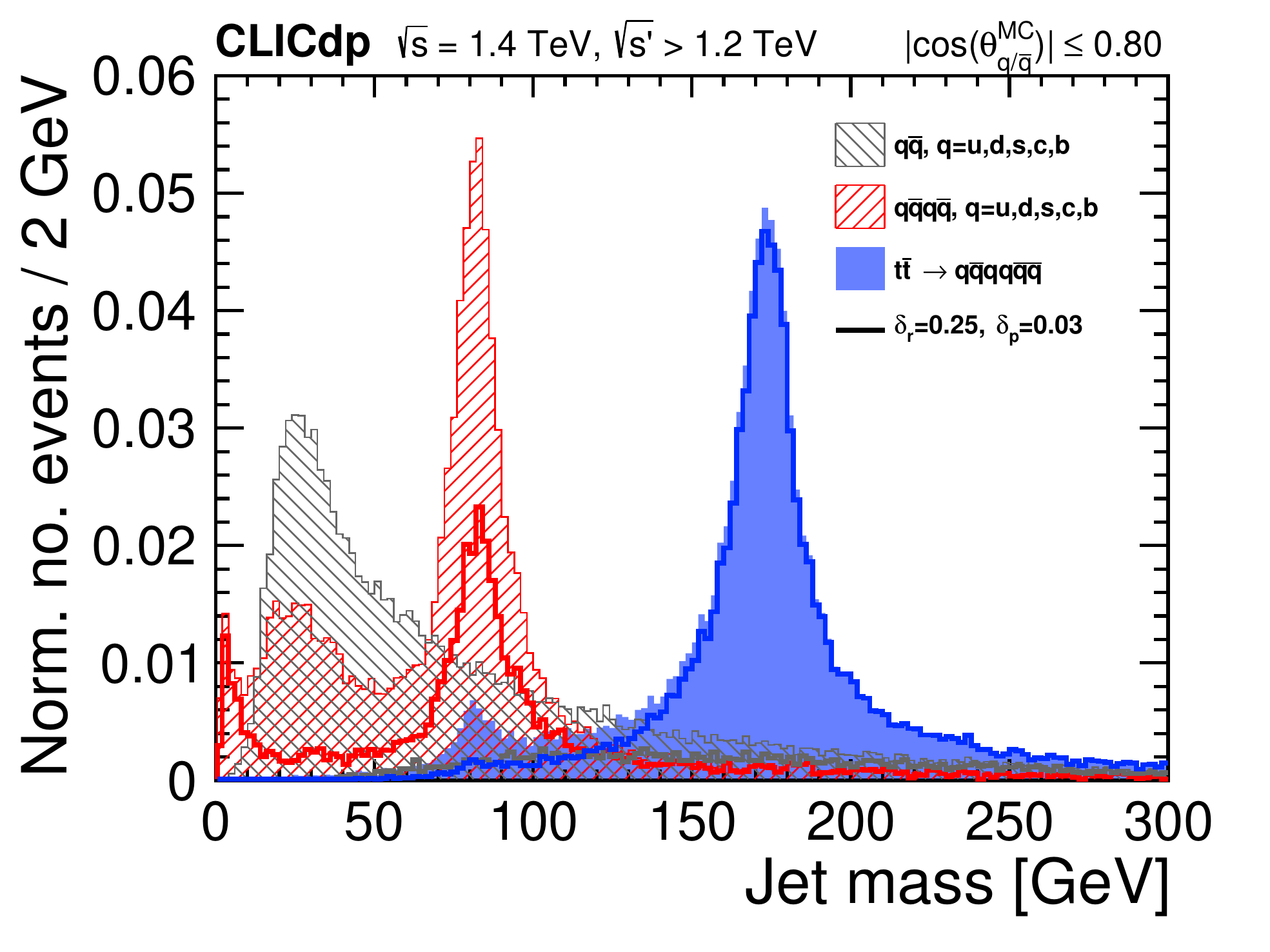}
\caption{Reconstructed top-quark candidate mass distributions at $\roots=1.4\,\tev$ for events with $\rootsprime\geq1.2\tev$. The filled distributions represent the jet mass before application of the de-clustering step of the top-tagger and are normalised to unity. The solid lines show the effect of applying the de-clustering. Fully-hadronic $\ttbar$ events are shown in blue, four-jet events in red, and dijet events in gray.}
\label{fig:analysis:jetreco:declusteringmass}
\end{figure}

\begin{figure}
\centering
\includegraphics[width=0.65\columnwidth,clip]{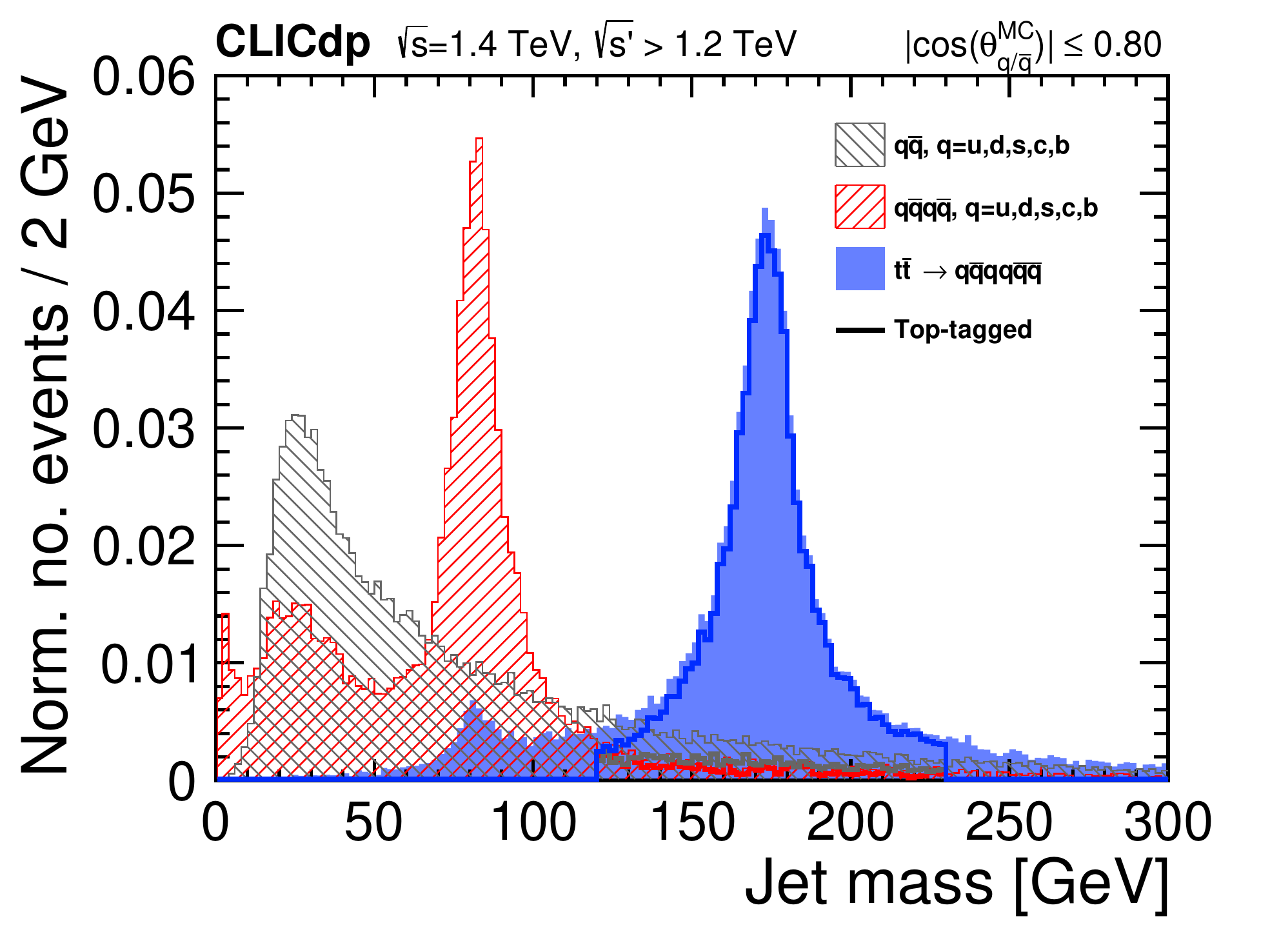}
\caption{Reconstructed top-quark candidate mass distributions at $\roots=1.4\,\tev$ for events with $\rootsprime\geq1.2\tev$. The filled distributions represent the jet mass before application of the top-tagger and are normalised to unity. The solid lines show the effect of applying the top-tagger. Fully-hadronic $\ttbar$ events are shown in blue, four-jet events in red, and dijet events in gray.}
\label{fig:analysis:jetreco:toptaggermass}
\end{figure}

\Cref{fig:analysis:jetreco:declusteringmass} shows the reconstructed jet mass before and after application of the top tagger de-clustering step, for operation at $\roots=1.4\,\tev$. \Cref{fig:analysis:jetreco:toptaggermass} shows the reconstructed jet mass before and after the application of the full top-tagger, including cuts on both the top-quark and $\PW$ mass. By comparing the solid lines with the corresponding filled distributions, the improvement of the tagger compared to a simple cut on the jet mass is clearly visible. The top tagger algorithm increases the significance, estimated as $S/\sqrt{B}$ where $S$ represents the number of top-quark jets from the fully-hadronic $\ttbar$ sample and $B$ the number of wrongly tagged light-quark jets from either the four-jet or dijet sample, by between 18-26\% (depending on the background process and collision energy considered), compared to a simple cut on the reconstructed large-$R$ jet mass in the corresponding range (within $\pm55$\,GeV of $m_{\PQt}$). In addition, the de-clustering procedure provides additional handles on the jet substructure such as the kinematic variables of the $\PW$ boson candidate including the helicity angle $\theta_{\PW}$ that examine whether the identified subjets are consistent with a top-quark decay. These handles are useful to discriminate against the remaining background events and are studied in more detail in \Cref{sec:mva}.

\begin{figure}
  \centering
  \includegraphics[width=0.65\columnwidth]{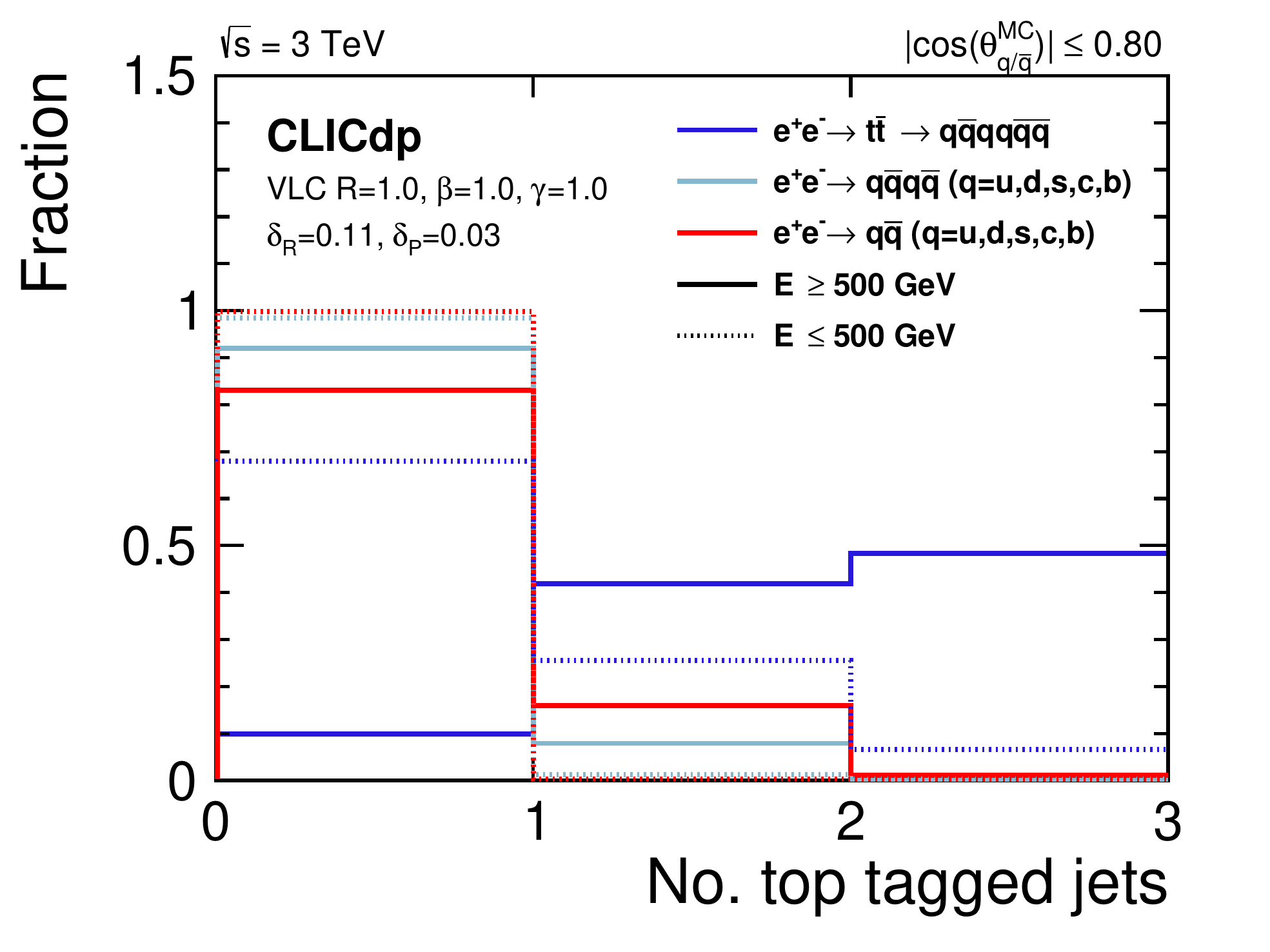}
  \caption{Fraction of top-tagged large-$R$ jets for various fully-hadronic samples. Solid  (dashed) lines represent the distribution considering only jets with an energy above (below) 500\,\gev. \label{fig:analysis:jetreco:toptaggernumber}}
\end{figure}

The final number of top-tagged jets, for the samples considered in this section, is presented in \Cref{fig:analysis:jetreco:toptaggernumber}, where solid lines represent jets with an energy above 500\,\gev and dashed lines jets with an energy below 500\,\gev. In agreement with a benchmark top-tagging efficiency of 70\% as defined above, we find that about 50\% of the fully-hadronic $\ttbar$ events are correctly reconstructed with two top-tagged jets, while in about 40\% of the events only one of the jets is successfully tagged. Meanwhile, only 10\% of the $\ttbar$ events studied are reconstructed without a top-tagged jet.

\section{Sub-structure of large-R jets}
\label{sec:substructure}

The substructure of the boosted large-R jets is further analysed using variables that describe the multi-body kinematics of the jets, such as N-subjettiness~\cite{Thaler:2010tr} and energy correlation functions~\cite{Larkoski:2013eya}. These variables were implemented in \marlin and are part of the \fastjet-contrib library~\cite{fastjetcontrib}. The performance of these variables are studied both before and after the application of the top-tagger procedure outlined in \Cref{sec:jetreco}. The variables presented in this section are later used in the multivariate classifier discussed in detail in \Cref{sec:mva}.

\paragraph{N-subjettiness}
N-subjettiness, $\tau_{N}$, is defined in \Cref{eq:tau} and generally describes to what degree the substructure of a jet can be regarded as composed of $N$ or fewer subjets. The quantity is evaluated for all particles in a jet along $N$ candidate subjet axes, here defined as the final steps of the large-R VLC clustering.
\begin{equation} \label{eq:tau}
\tau_{N} = \frac{1}{d_{0}} \sum_k \ensuremath{p_{\mathrm{T},k}}\xspace \min \{ \Delta R_{1,k}, \Delta R_{2,k}, \ldots, \Delta R_{N,k} \},
\end{equation}
where $k$ runs over the constituent particles of the jet, each with transverse momentum $\ensuremath{p_{\mathrm{T},k}}$. The distance in the pseudorapidity-azimuth plane, between each candidate subjet $J$ and constituent particle $k$, is denoted 
\begin{equation}
\Delta R_{J,k}^2={\Delta\eta^2+\Delta\phi^2}
\end{equation}
and 
\begin{equation}
d_0 = R_0 \cdot \sum_k \ensuremath{p_{\mathrm{T},k}},
\end{equation}
where $R_0$ is the jet radius used in the large-$R$ jet clustering described in \Cref{ssec:jetalgo} and \Cref{ssec:jetopt}.

While a large N-subjettiness value would indicate that the jets have a large fraction of their energy distributed away from the candidate subjet directions, a low value would rather point towards agreement with the subjet hypothesis. In practice, we study ratios of N-subjettiness variables, $\tau_{N+1}/\tau_N$, that have shown to be particularly powerful in the discrimination of multi-body structures against QCD background jets. 

\begin{figure}[p]
	\centering
	\begin{subfigure}{0.48\columnwidth}
	\includegraphics[width=\textwidth]{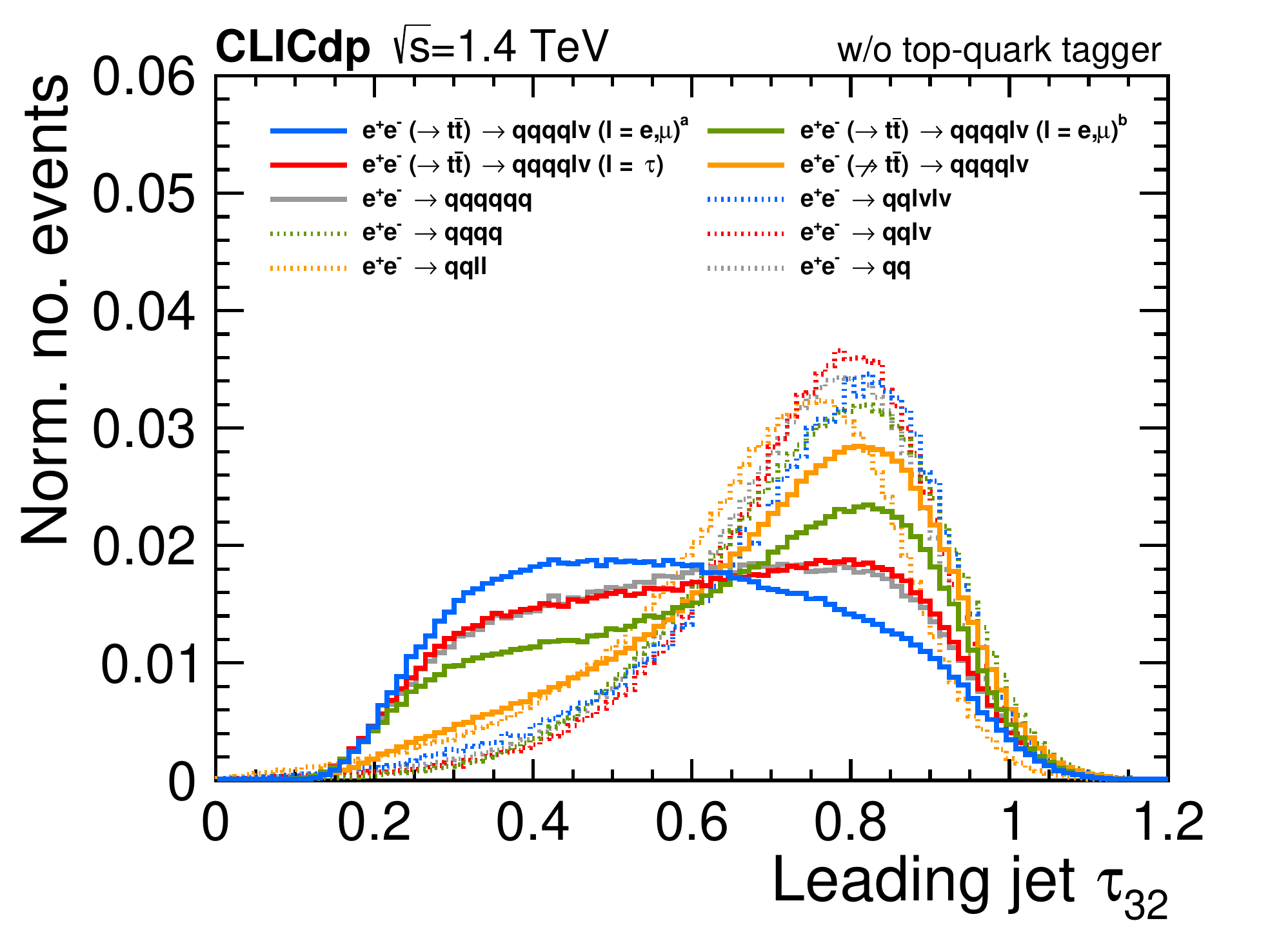}
	\caption{$\tau_{32}$ without applying the top-quark tagger.}
	\end{subfigure}
	~~~
	\begin{subfigure}{0.48\columnwidth}
	\includegraphics[width=\textwidth]{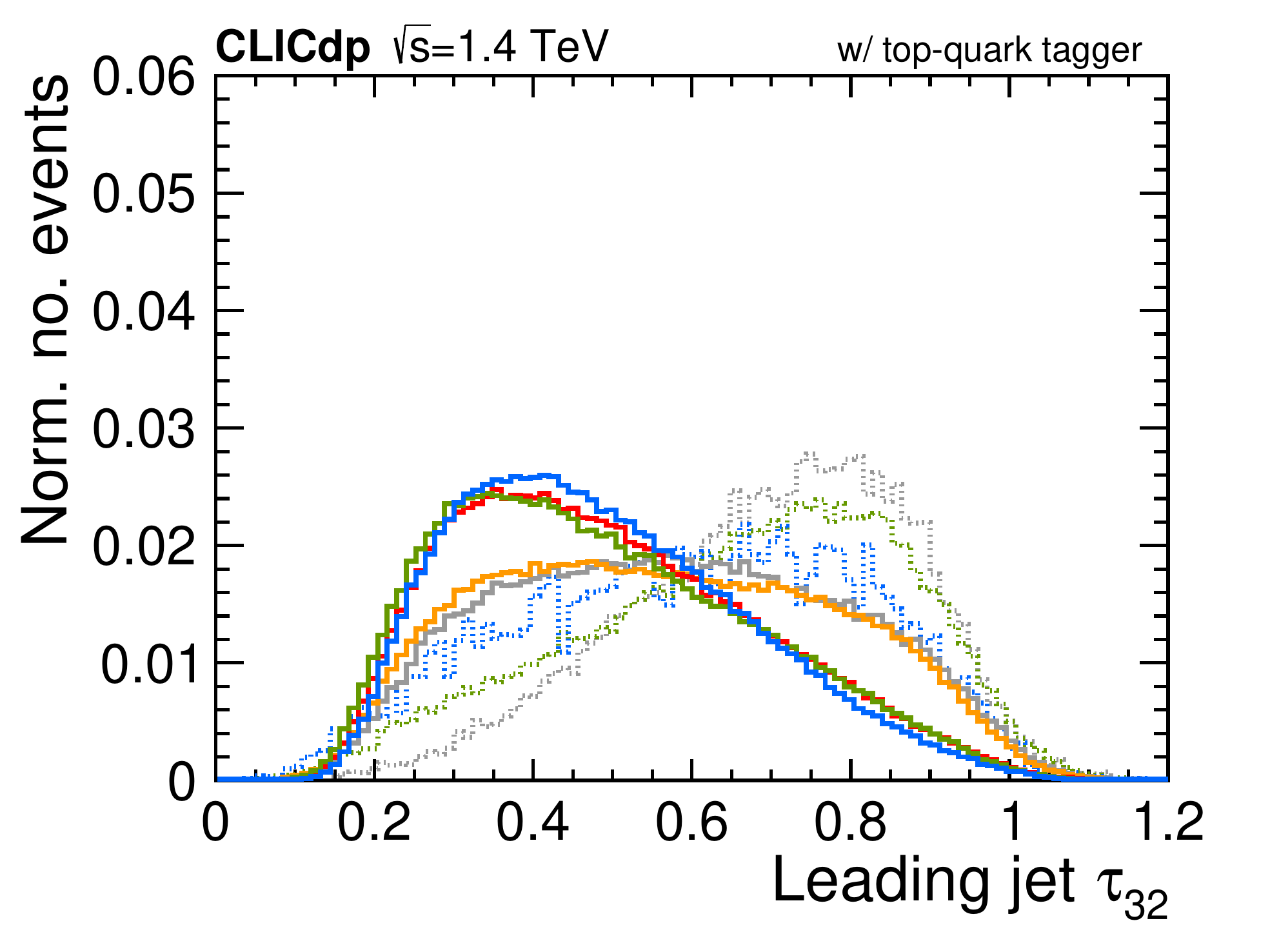}
	\caption{$\tau_{32}$ after applying the top-quark tagger.}
	\end{subfigure}\\
	\vspace{5mm}
	\begin{subfigure}{0.48\columnwidth}
	\includegraphics[width=\textwidth]{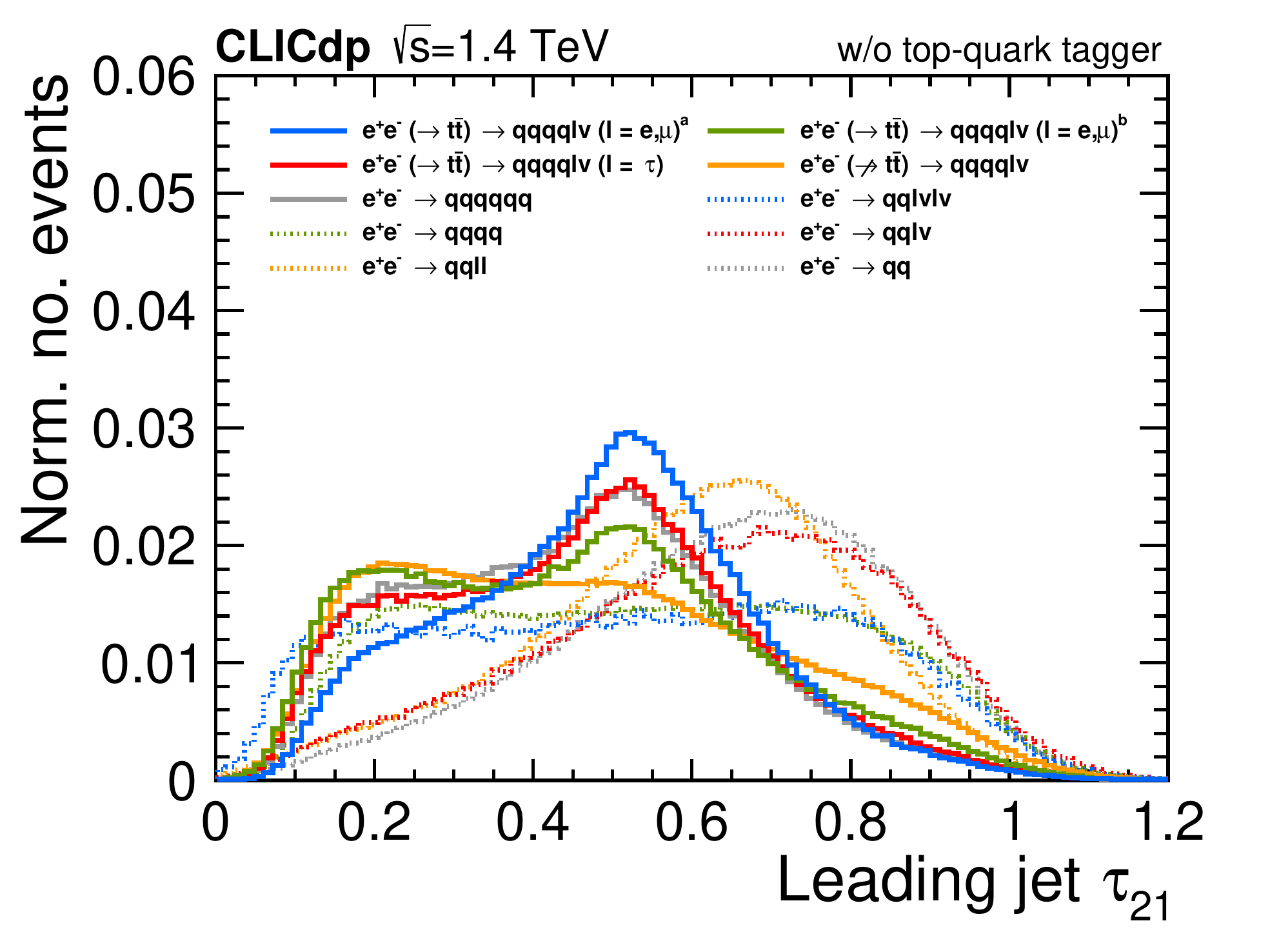}
	\caption{$\tau_{21}$ without applying the top-quark tagger.}
	\end{subfigure}
	~~~
	\begin{subfigure}{0.48\columnwidth}
	\includegraphics[width=\textwidth]{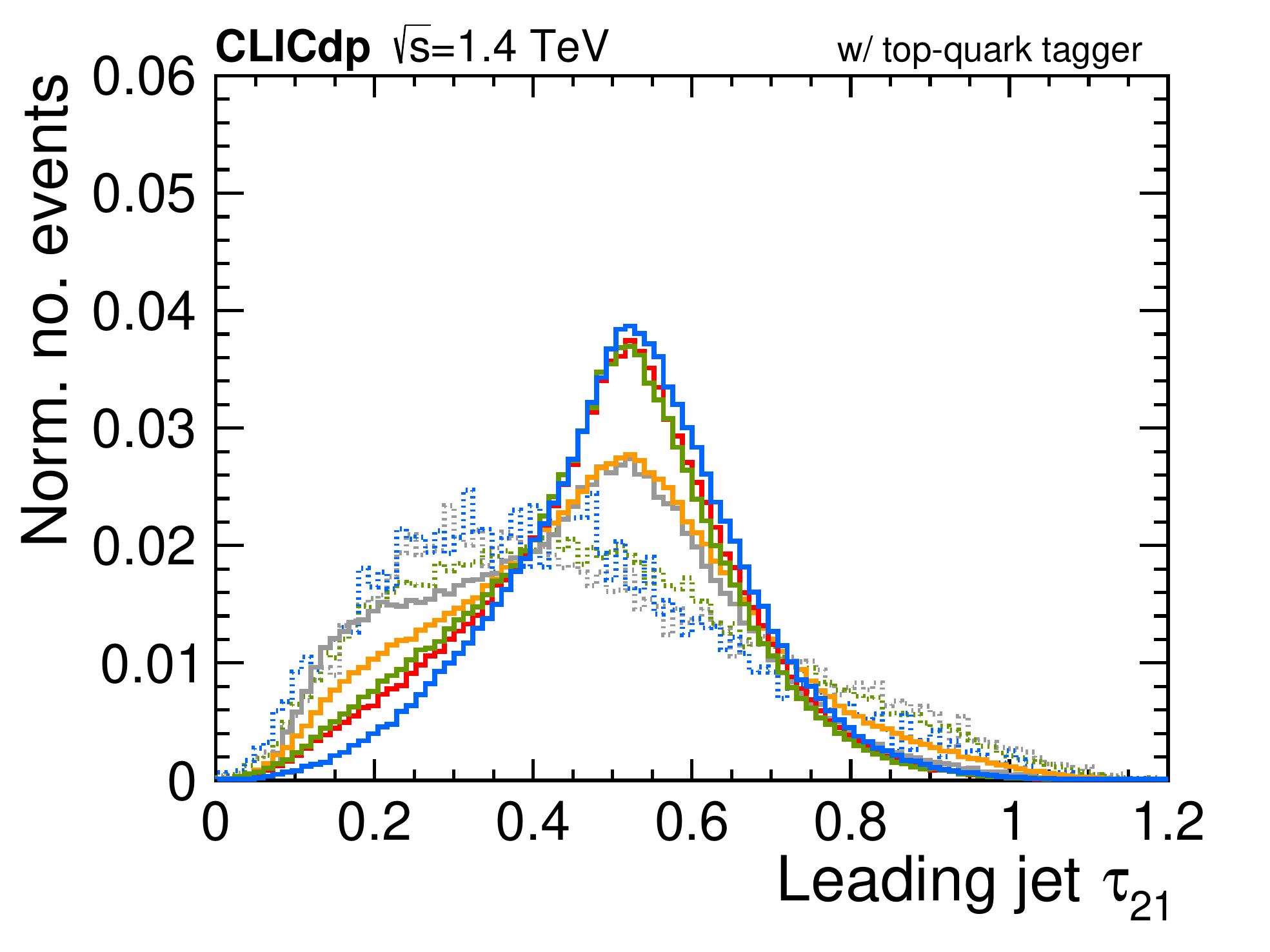}
	\caption{$\tau_{21}$ after applying the top-quark tagger.}
	\end{subfigure}\\
	\vspace{5mm}
	\begin{subfigure}{0.48\columnwidth}
	\includegraphics[width=\textwidth]{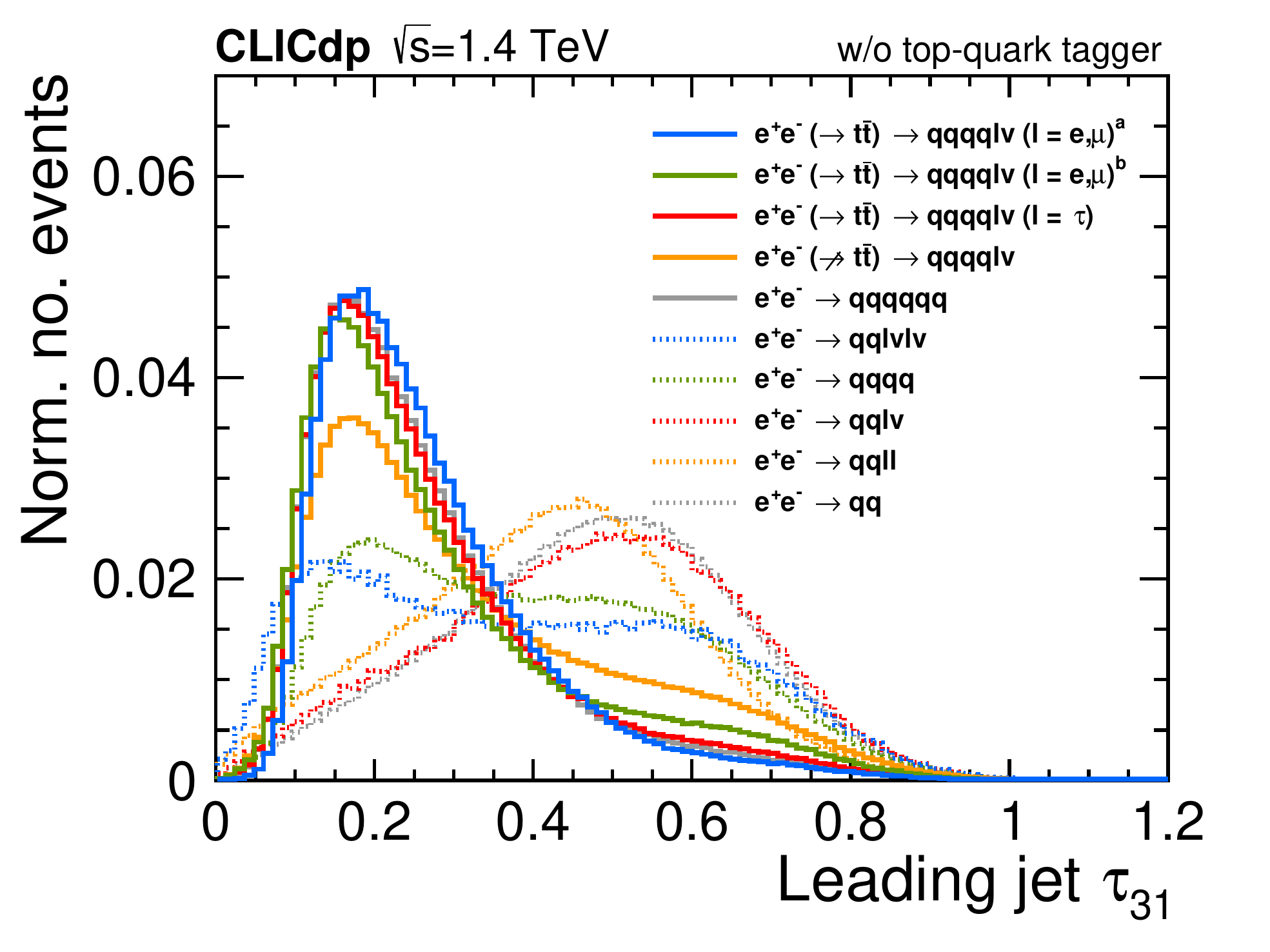}
	\caption{$\tau_{31}$ without applying the top-quark tagger.}
	\end{subfigure}
	~~~
	\begin{subfigure}{0.48\columnwidth}
	\includegraphics[width=\textwidth]{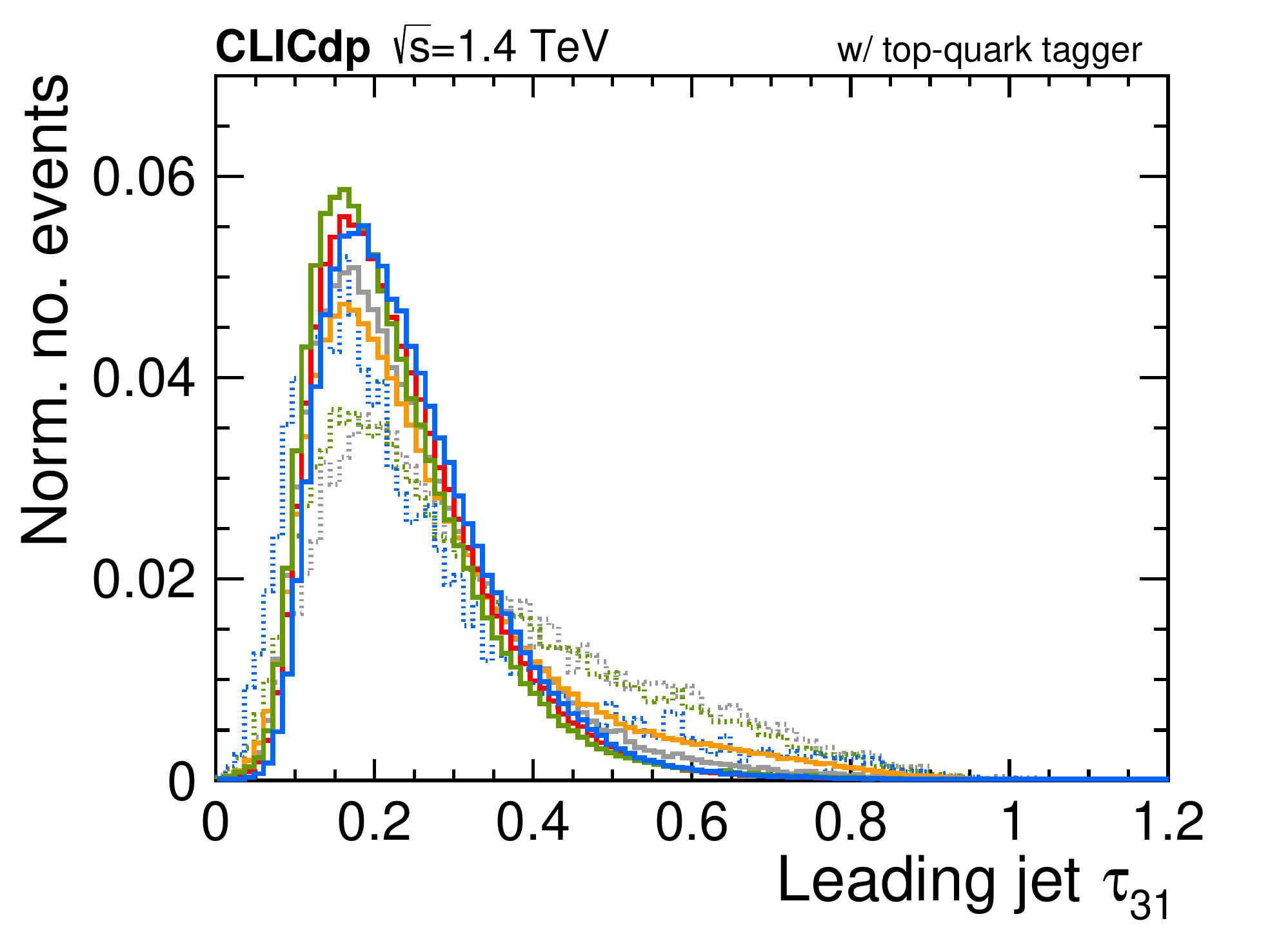}
	\caption{$\tau_{31}$ after applying the top-quark tagger.}
	\end{subfigure}
\caption{N-subjettiness ratios for the highest energy ''leading`` large-R jet. The superscript `a' (`b') refers to the kinematic region $\rootsprime\geq1.2\,\tev$ ($\rootsprime<1.2\,\tev$). Note that the qqlv and qqll backgrounds have been omitted in the figures in the right column due to low available statistics. The retention of these backgrounds after the full event selection is negligible. \label{fig:analysis:mva:variables:NsubjettinessJ1}}
\end{figure}

\begin{figure}[p]
	\centering
	\begin{subfigure}{0.48\columnwidth}
	\includegraphics[width=\textwidth]{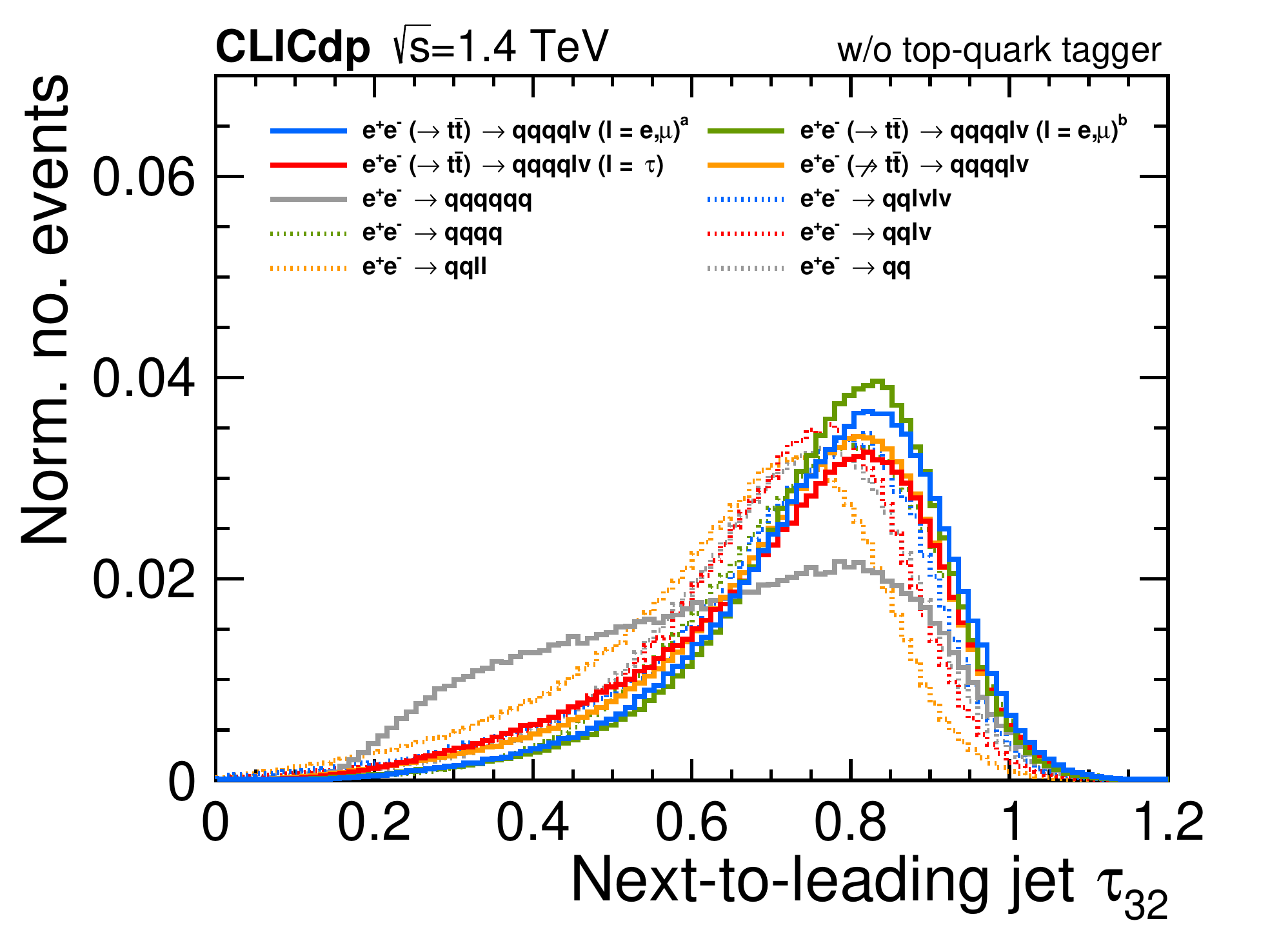}
	\caption{$\tau_{32}$ without applying the top-quark tagger.}
	\end{subfigure}
	~~~
	\begin{subfigure}{0.48\columnwidth}
	\includegraphics[width=\textwidth]{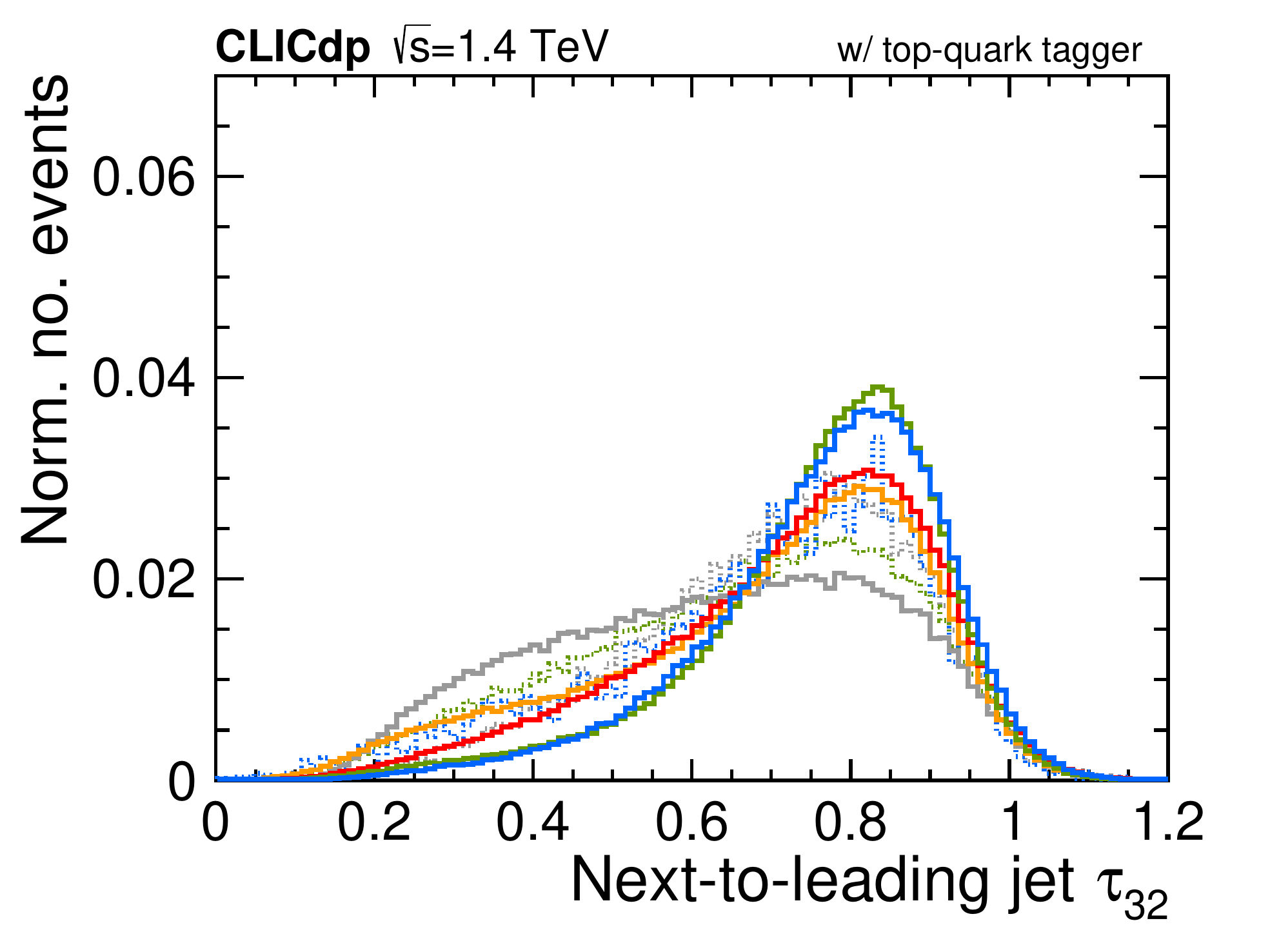}
	\caption{$\tau_{32}$ after applying the top-quark tagger.}
	\end{subfigure}\\
	\vspace{5mm}
	\begin{subfigure}{0.48\columnwidth}
	\includegraphics[width=\textwidth]{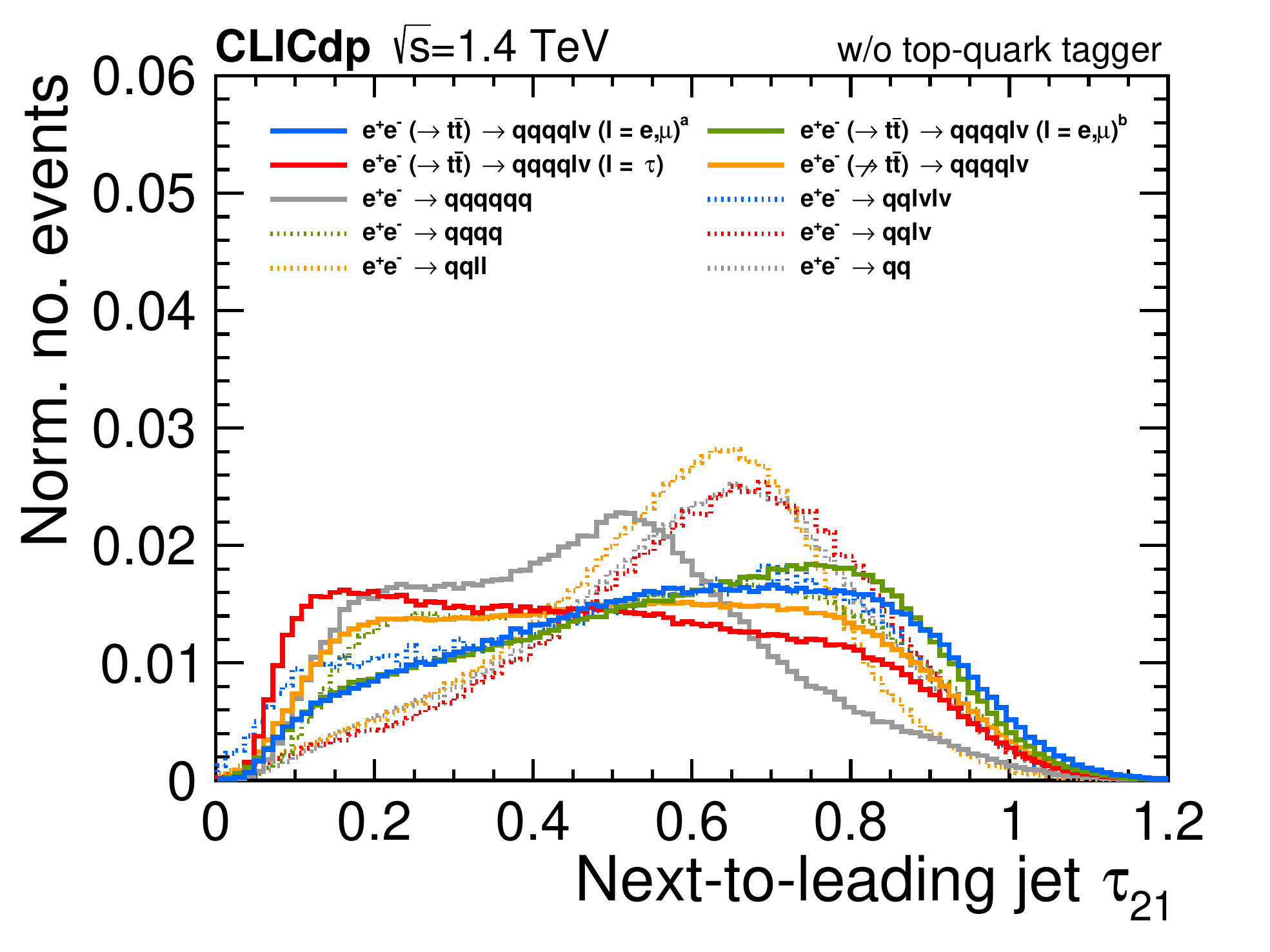}
	\caption{$\tau_{21}$ without applying the top-quark tagger.}
	\end{subfigure}
	~~~
	\begin{subfigure}{0.48\columnwidth}
	\includegraphics[width=\textwidth]{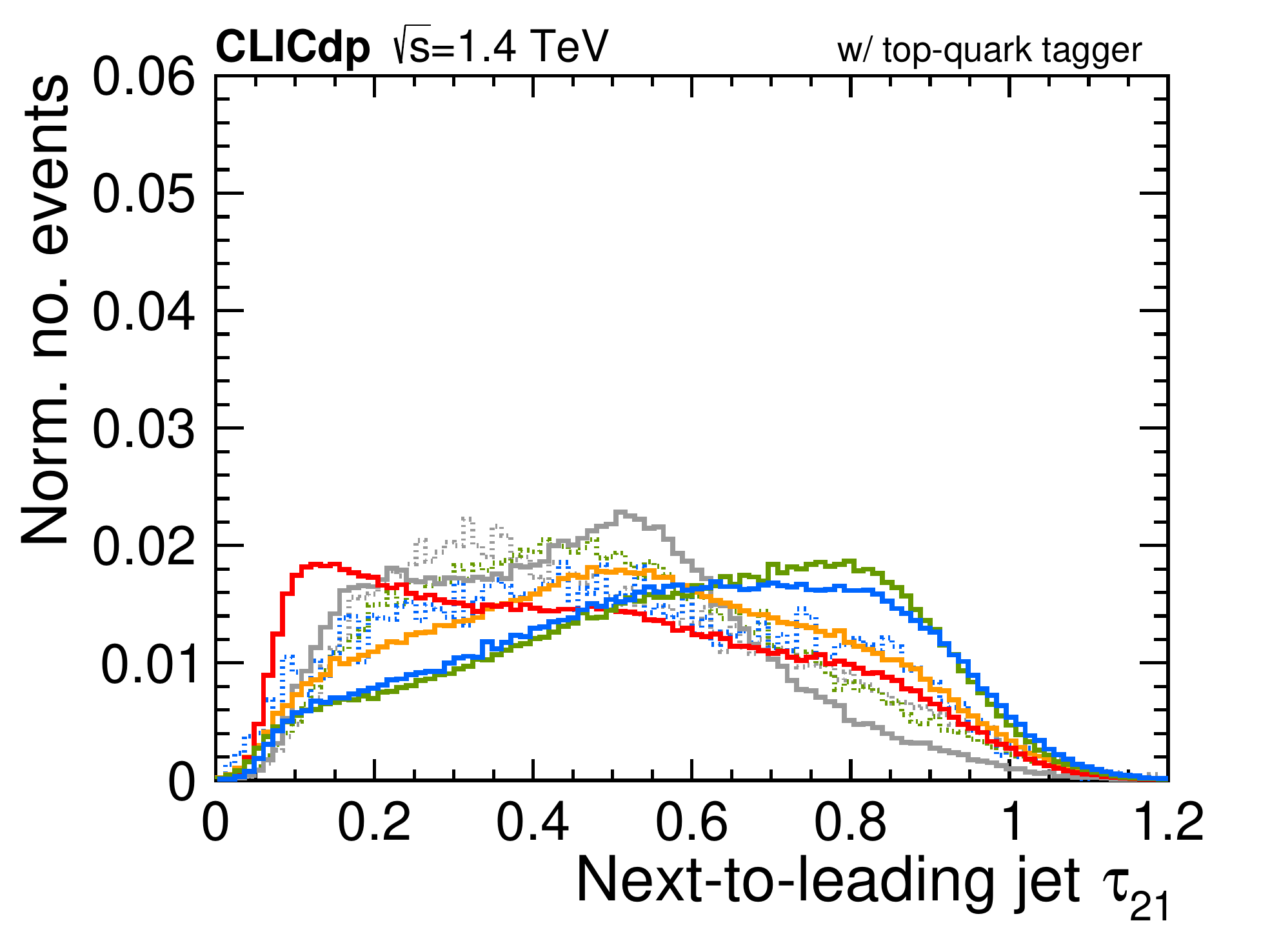}
	\caption{$\tau_{21}$ after applying the top-quark tagger.}
	\end{subfigure}\\
	\vspace{5mm}
	\begin{subfigure}{0.48\columnwidth}
	\includegraphics[width=\textwidth]{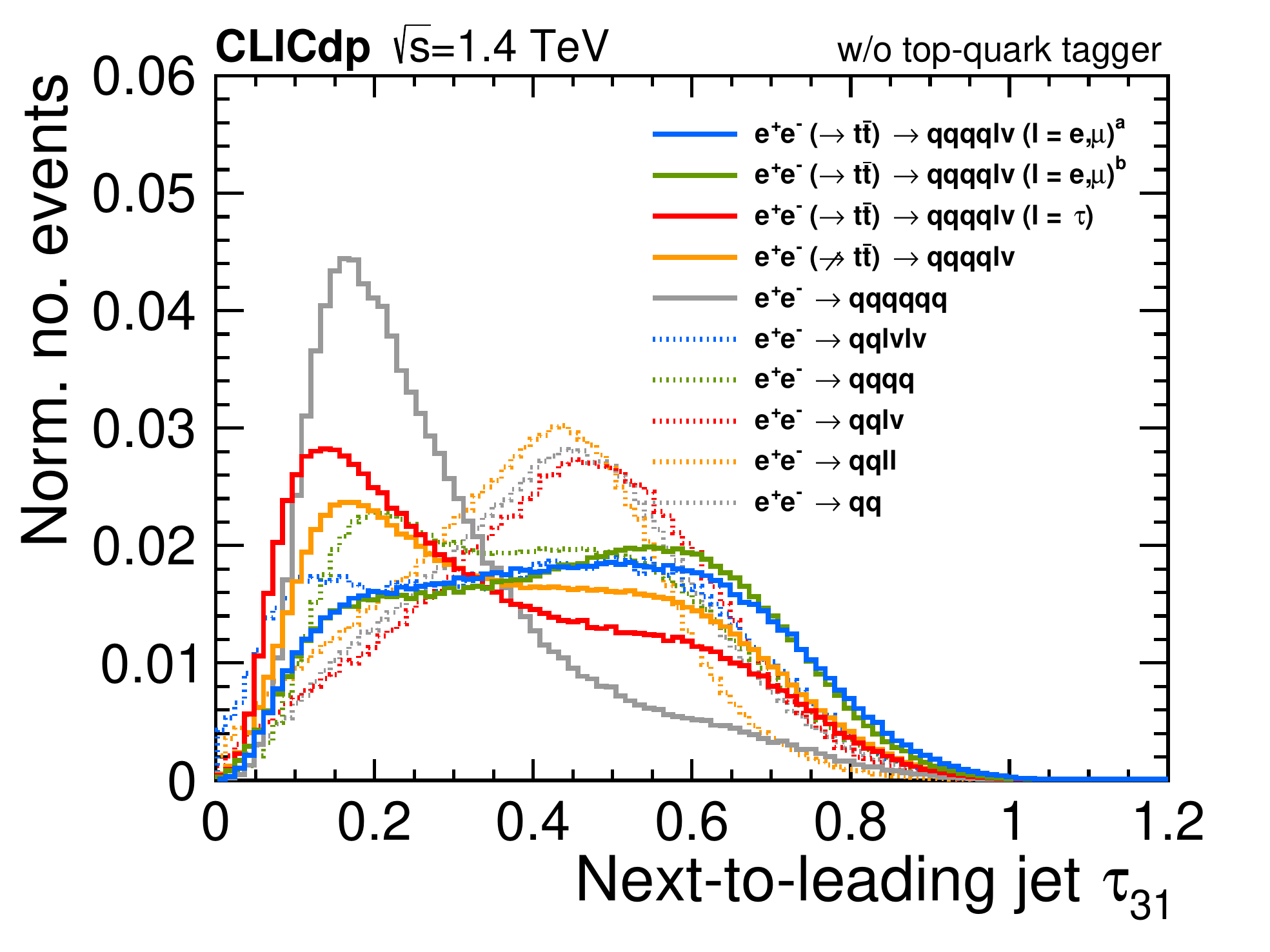}
	\caption{$\tau_{31}$ without applying the top-quark tagger.}
	\end{subfigure}
	~~~
	\begin{subfigure}{0.48\columnwidth}
	\includegraphics[width=\textwidth]{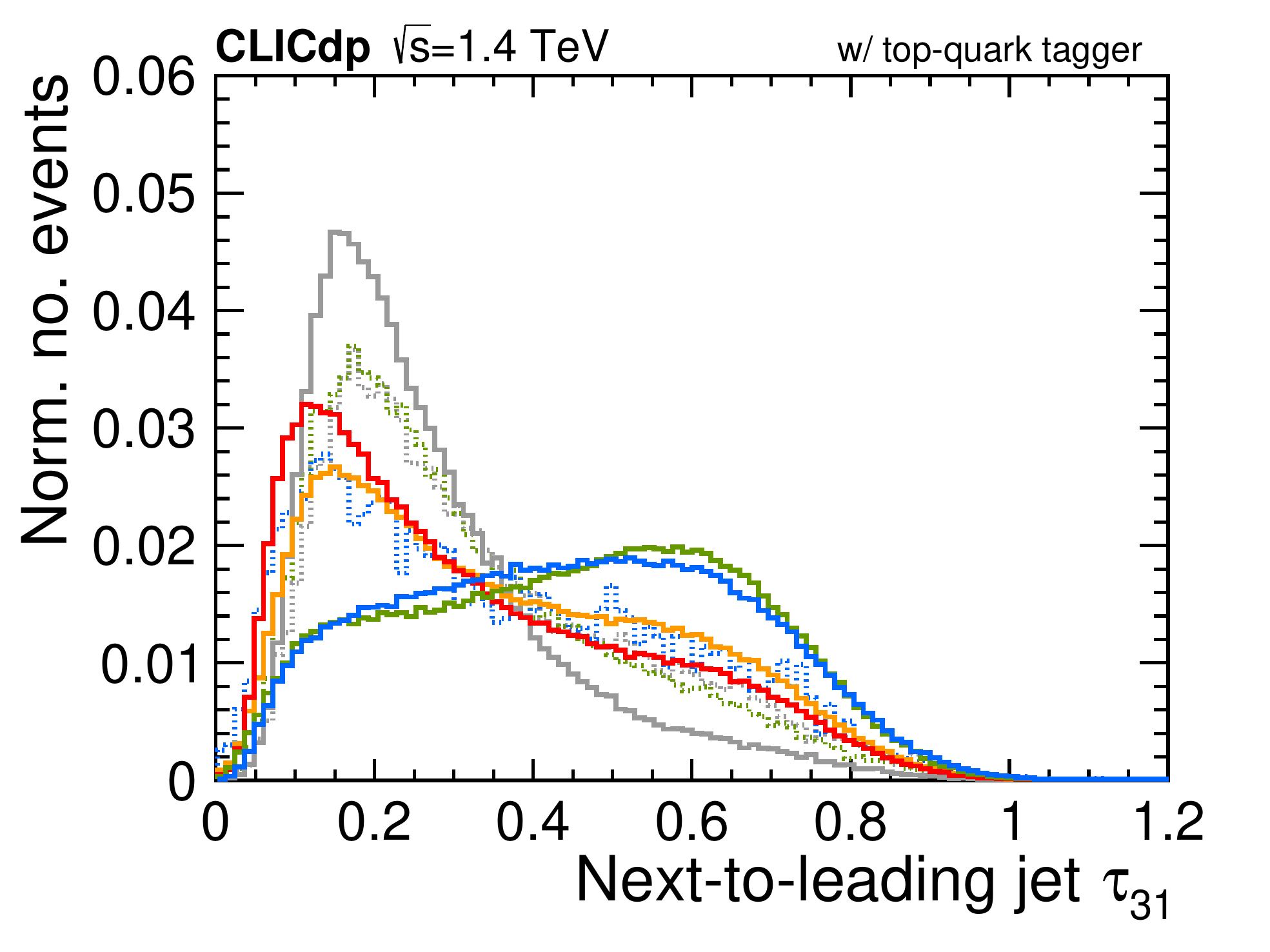}
	\caption{$\tau_{31}$ after applying the top-quark tagger.}
	\end{subfigure}
\caption{N-subjettiness ratios for the lowest energy ''next-to-leading`` large-R jet. The superscript `a' (`b') refers to the kinematic region $\rootsprime\geq1.2\,\tev$ ($\rootsprime<1.2\,\tev$). Note that the qqlv and qqll backgrounds have been omitted in the figures in the right column due to low available statistics. The retention of these backgrounds after the full event selection is negligible. \label{fig:analysis:mva:variables:NsubjettinessJ2}}
\end{figure}

\Cref{fig:analysis:mva:variables:NsubjettinessJ1} displays the distributions for the highest-energy (denoted ''leading``) large-R jet before (left) and after (right) the top-tagger, while \Cref{fig:analysis:mva:variables:NsubjettinessJ2} shows the corresponding distributions for the ''next-to-leading jet``. Note that with this classification, the leading jet represents the fully-hadronically decaying top-quark jet for semi-leptonic $\ttbar$ event. Conversely, the next-to-leading jet represents the $\PQb$-quark of the leptonic top-quark decay. Each figure shows the distributions after pre-selection for a large number of processes considered in the analysis and follows the nomenclature introduced in \Cref{sec:analysis_strategy}. As clearly demonstrated, the leading-jet variables are powerful in discrimination against single-top, four-jet and di-jet events, to some extend even after applying the top-tagger. As expected, the next-to-leading jet variables are instead powerful in discriminating against fully-hadronic $\ttbar$ events: in particular $\tau_{31}$.

\paragraph{Energy correlation functions}
The energy correlation functions are defined as:
\begin{equation}
\mathrm{ECF}(N) = \sum_{i_1<i_2<\ldots< i_N\in J} \bigg( \prod_{a=1}^{N}E_{i_a} \bigg) \xspace \bigg( \prod_{b=1}^{N-1} \prod_{c=b+1}^{N} \theta_{i_b i_c} \bigg),
\end{equation}
where the sum runs over the constituent particles of the jet $J$ and each term consists of $N$ energies multiplied with $\big( \binom{N}{2} \big)$ pairwise angles~\cite{Larkoski:2013eya}.

In this case we study the double ratios $C_{2,3}$~\cite{Larkoski:2013eya} and $D_{2,3}$~\cite{Larkoski:2014gra,Larkoski:2014zma} defined as:
\begin{equation}
C_2 = \mathrm{ECF}(3)\cdot \frac{\mathrm{ECF}(1)}{(\mathrm{ECF}(2))^2}\,\,, \,\,\,\,\,\,\,\,\,\,\,
C_3 = \mathrm{ECF}(4)\cdot \frac{\mathrm{ECF}(2)}{(\mathrm{ECF}(3))^2}\,\,,
\end{equation}
and 
\begin{equation}
D_2 = \mathrm{ECF}(3)\cdot \frac{(\mathrm{ECF}(1))^3}{(\mathrm{ECF}(2))^3}\,\,, \,\,\,\,\,\,\,\,\,\,\,
D_3 = \mathrm{ECF}(4)\cdot \frac{(\mathrm{ECF}(2))^3}{(\mathrm{ECF}(3))^3}\,\,.
\end{equation}

\begin{figure}
	\centering
	\begin{subfigure}{0.48\columnwidth}
	\includegraphics[width=\textwidth]{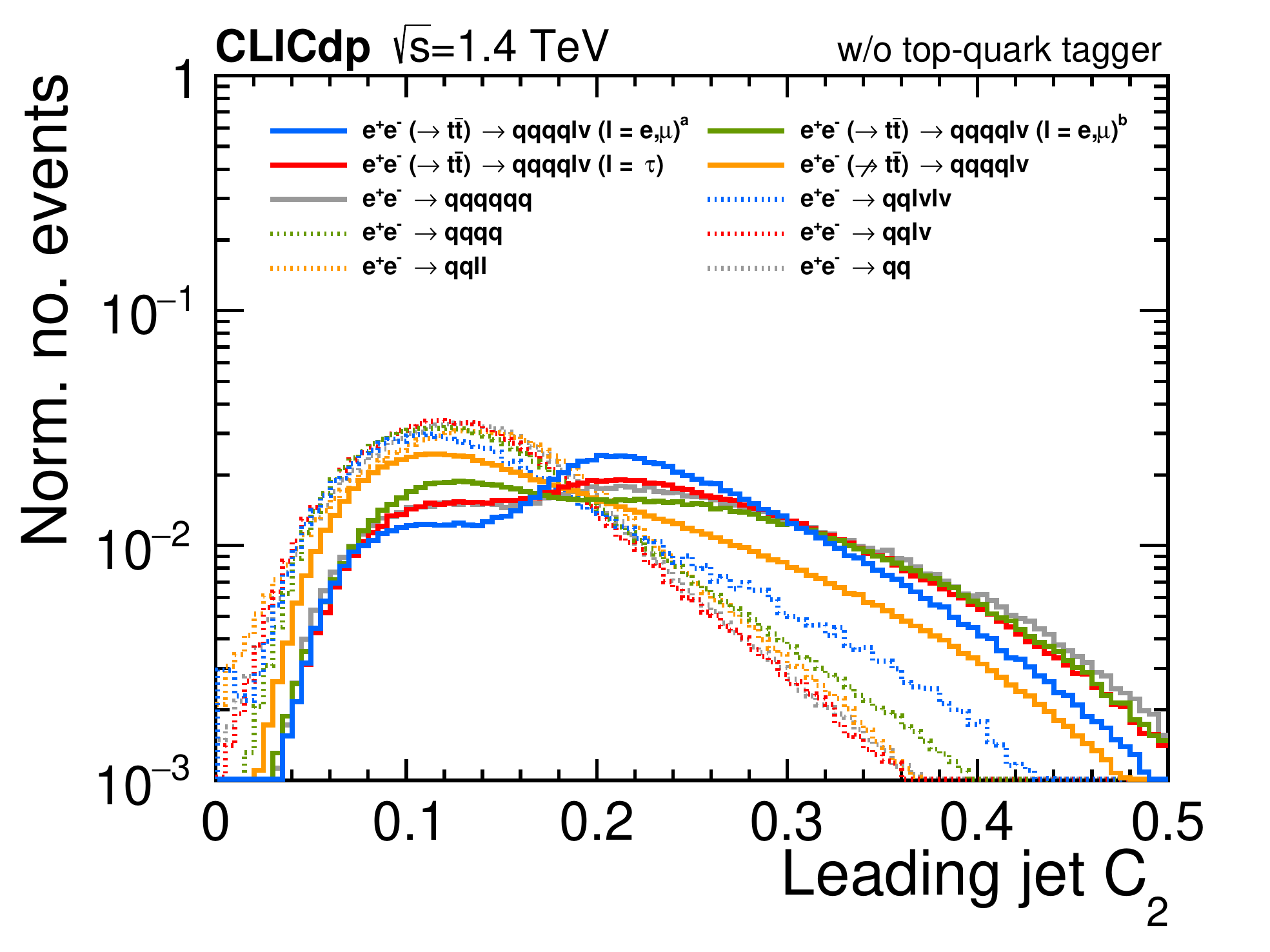}
	\caption{$C_{2}$ without applying the top-quark tagger.}
	\end{subfigure}
	~~~
	\begin{subfigure}{0.48\columnwidth}
	\includegraphics[width=\textwidth]{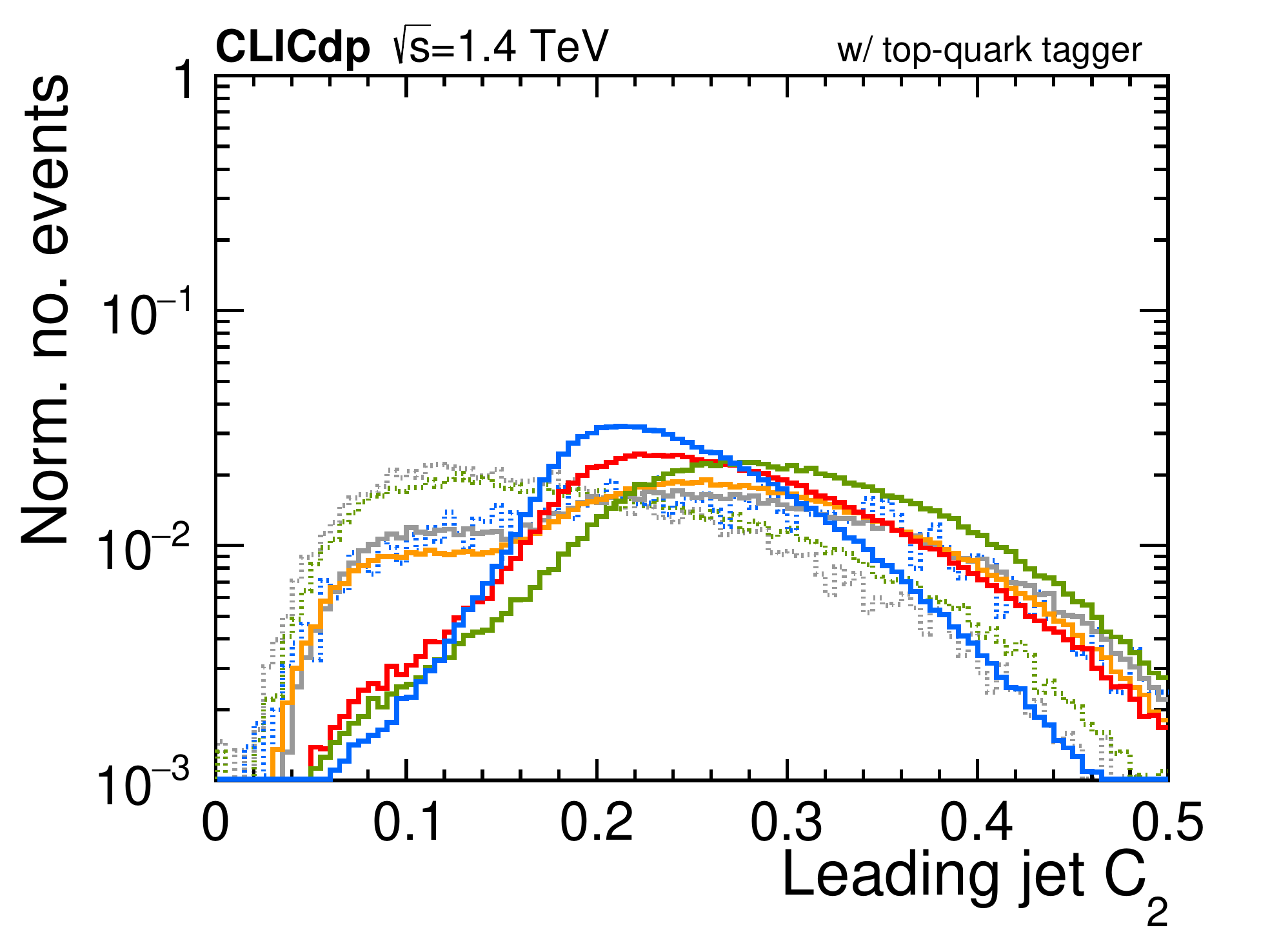}
	\caption{$C_{2}$ after applying the top-quark tagger.}
	\end{subfigure}\\
	\vspace{5mm}
	\begin{subfigure}{0.48\columnwidth}
	\includegraphics[width=\textwidth]{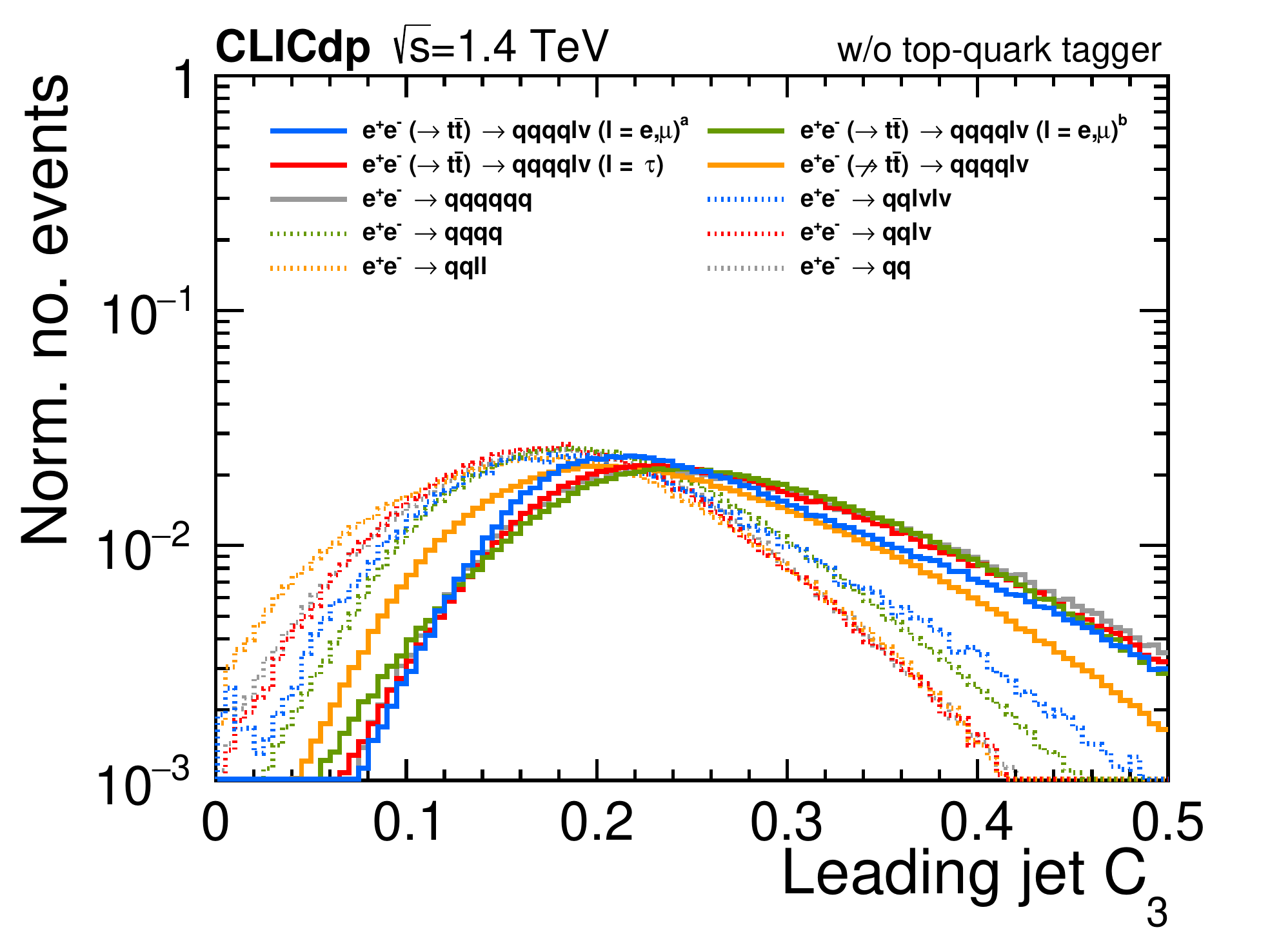}
	\caption{$C_{3}$ without applying the top-quark tagger.}
	\end{subfigure}
	~~~
	\begin{subfigure}{0.48\columnwidth}
	\includegraphics[width=\textwidth]{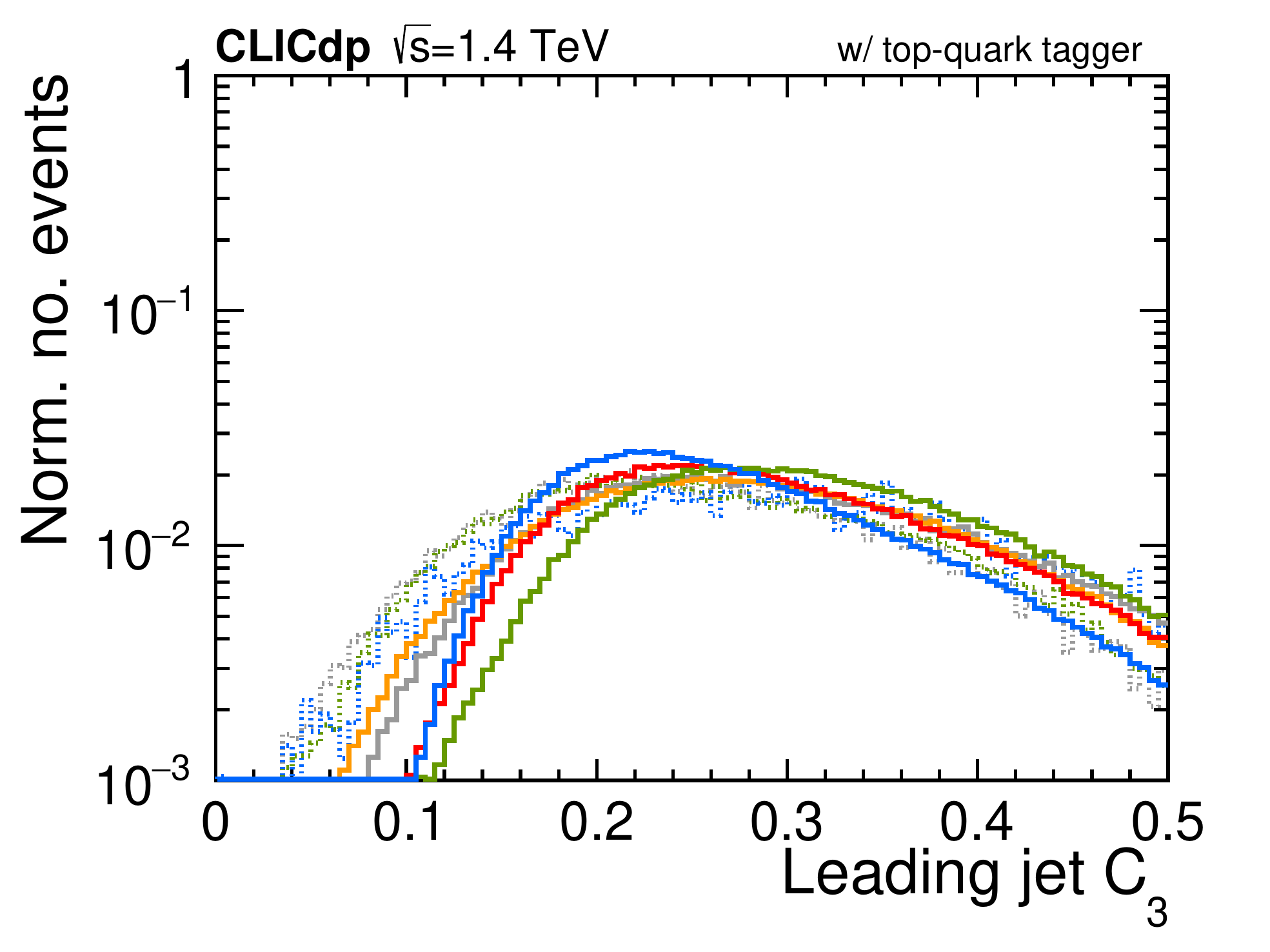}
	\caption{$C_{3}$ after applying the top-quark tagger.}
	\end{subfigure}\\
\caption{Energy correlation functions $C_2$ and $C_3$ for the highest energy ''leading`` large-R jet. The superscript `a' (`b') refers to the kinematic region $\rootsprime\geq1.2\,\tev$ ($\rootsprime<1.2\,\tev$). Note that the qqlv and qqll backgrounds have been omitted in the figures in the right column due to low available statistics. The retention of these backgrounds after the full event selection is negligible.
\label{fig:analysis:mva:variables:energycorrCJ1}}
\end{figure}

\begin{figure}[p!]
	\centering
	\begin{subfigure}{0.48\columnwidth}
	\includegraphics[width=\textwidth]{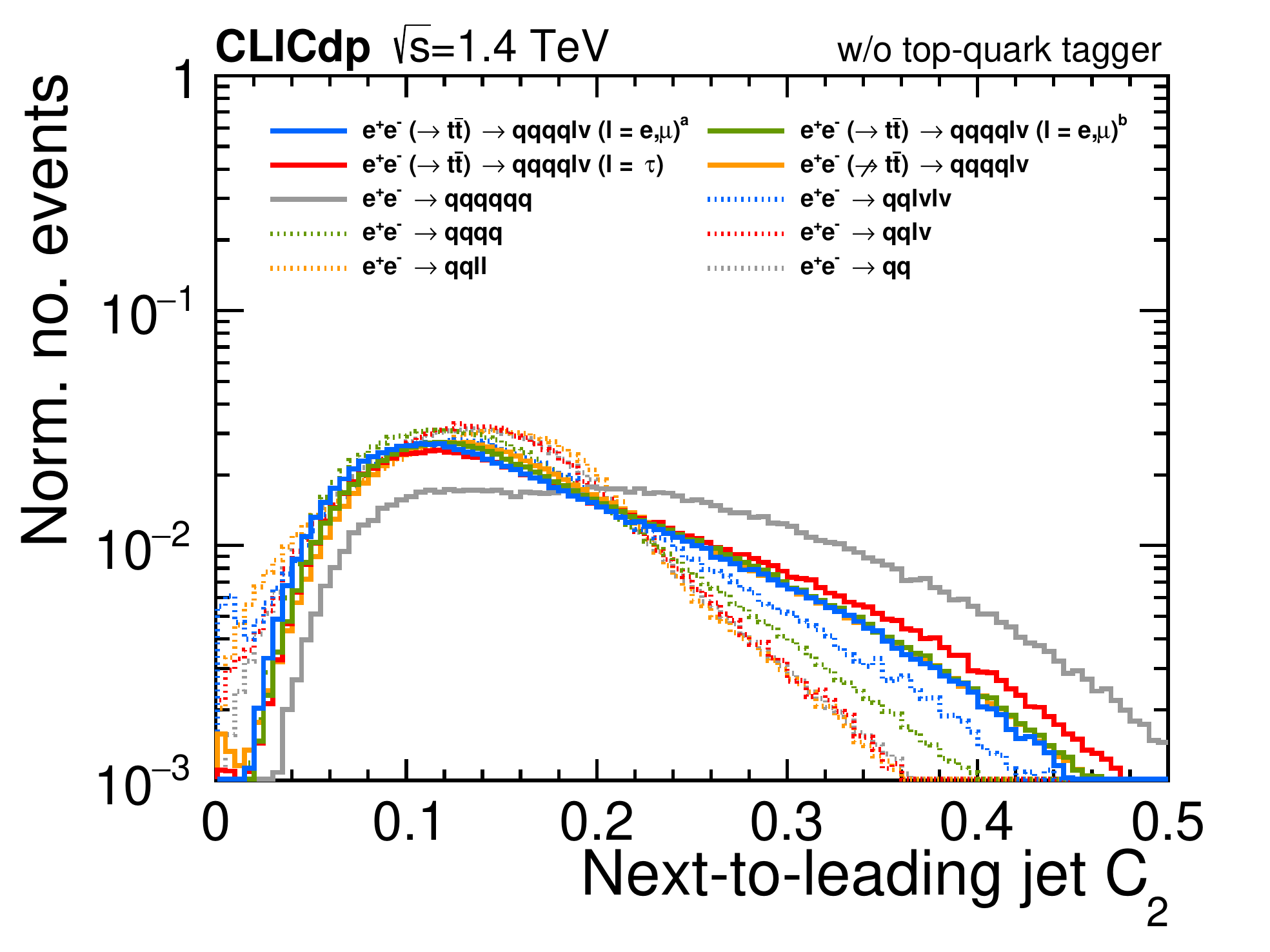}
	\caption{$C_{2}$ without applying the top-quark tagger.}
	\end{subfigure}
	~~~
	\begin{subfigure}{0.48\columnwidth}
	\includegraphics[width=\textwidth]{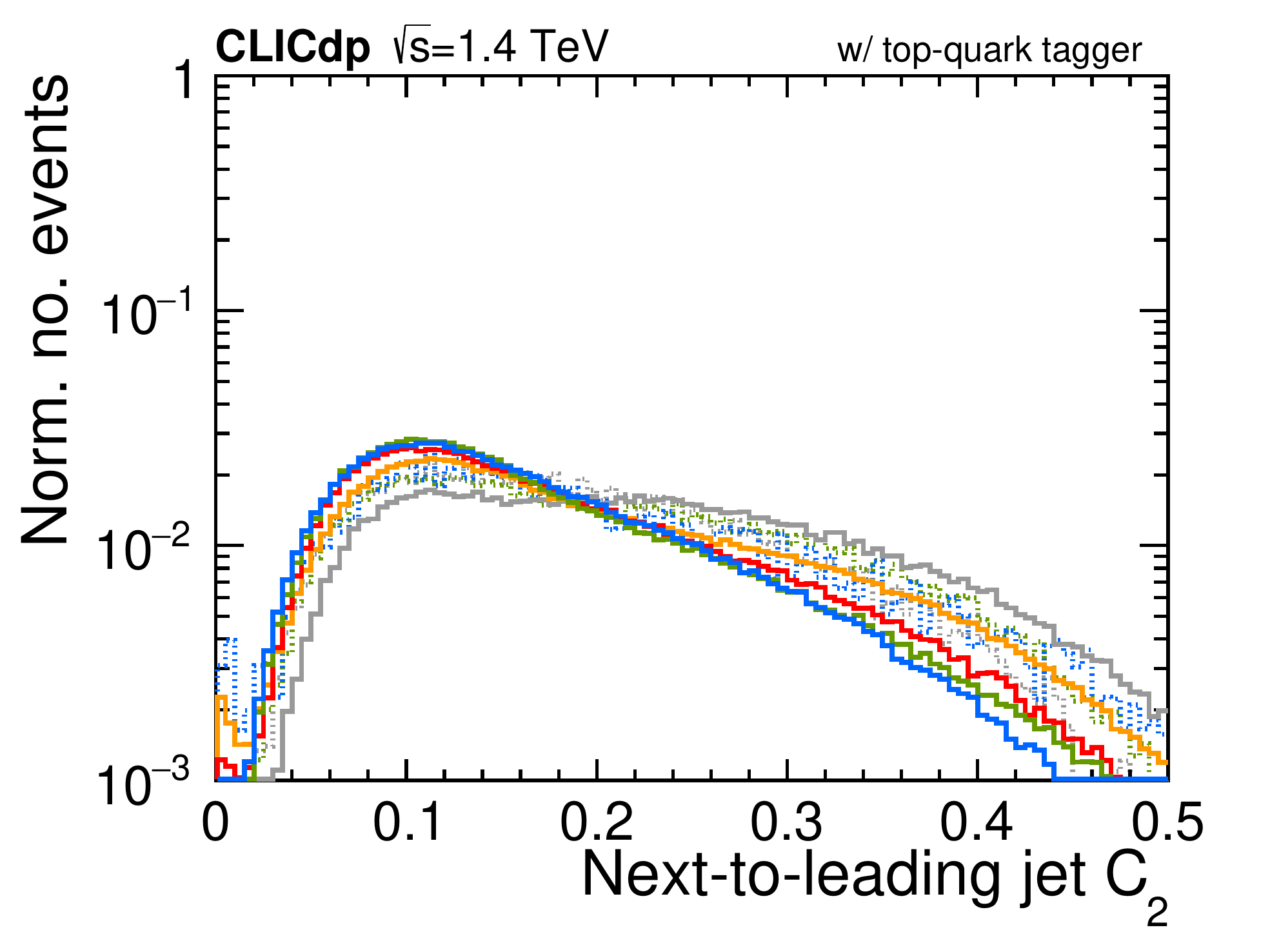}
	\caption{$C_{2}$ after applying the top-quark tagger.}
	\end{subfigure}\\
	\vspace{5mm}
	\begin{subfigure}{0.48\columnwidth}
	\includegraphics[width=\textwidth]{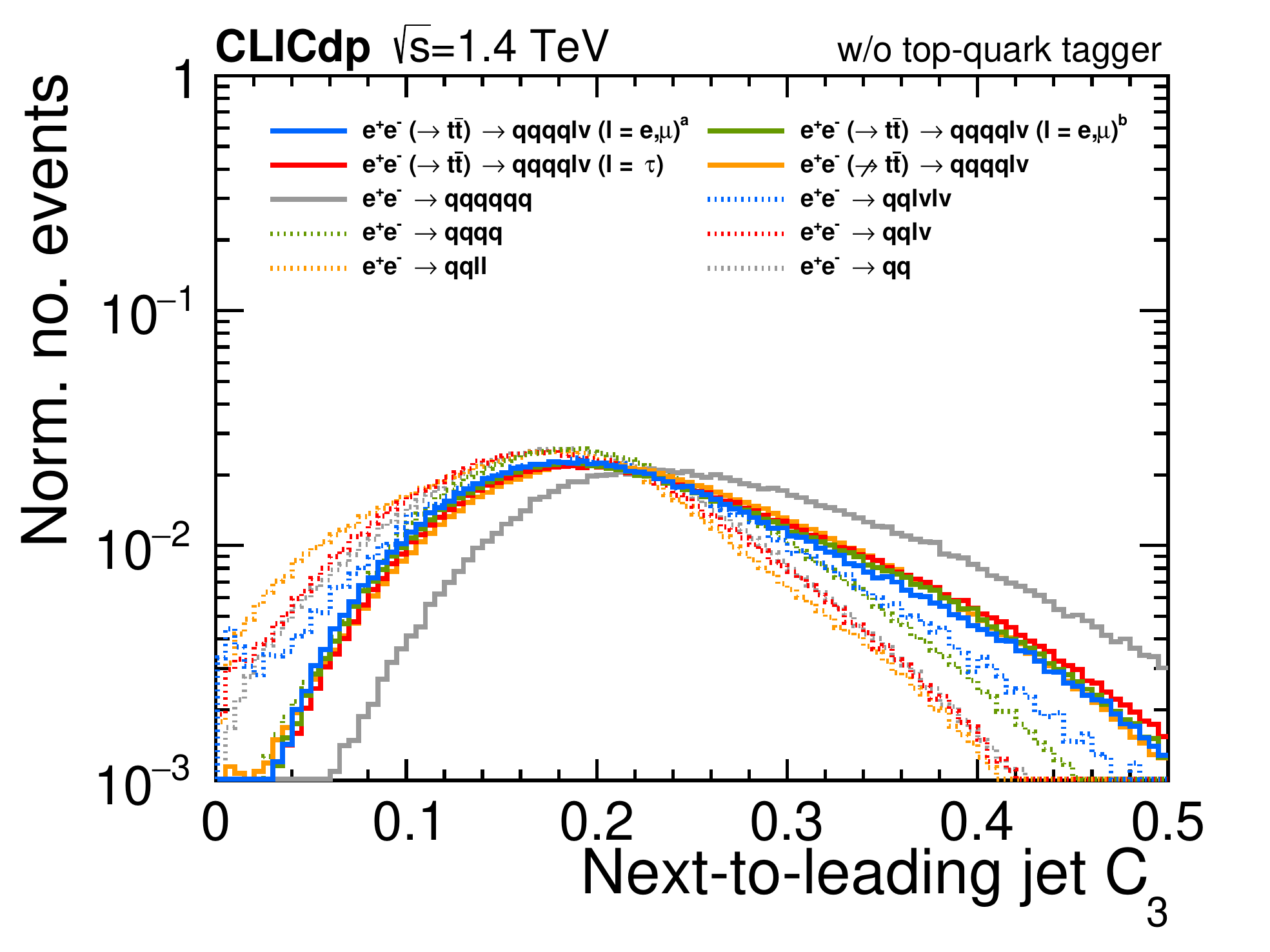}
	\caption{$C_{3}$ without applying the top-quark tagger.}
	\end{subfigure}
	~~~
	\begin{subfigure}{0.48\columnwidth}
	\includegraphics[width=\textwidth]{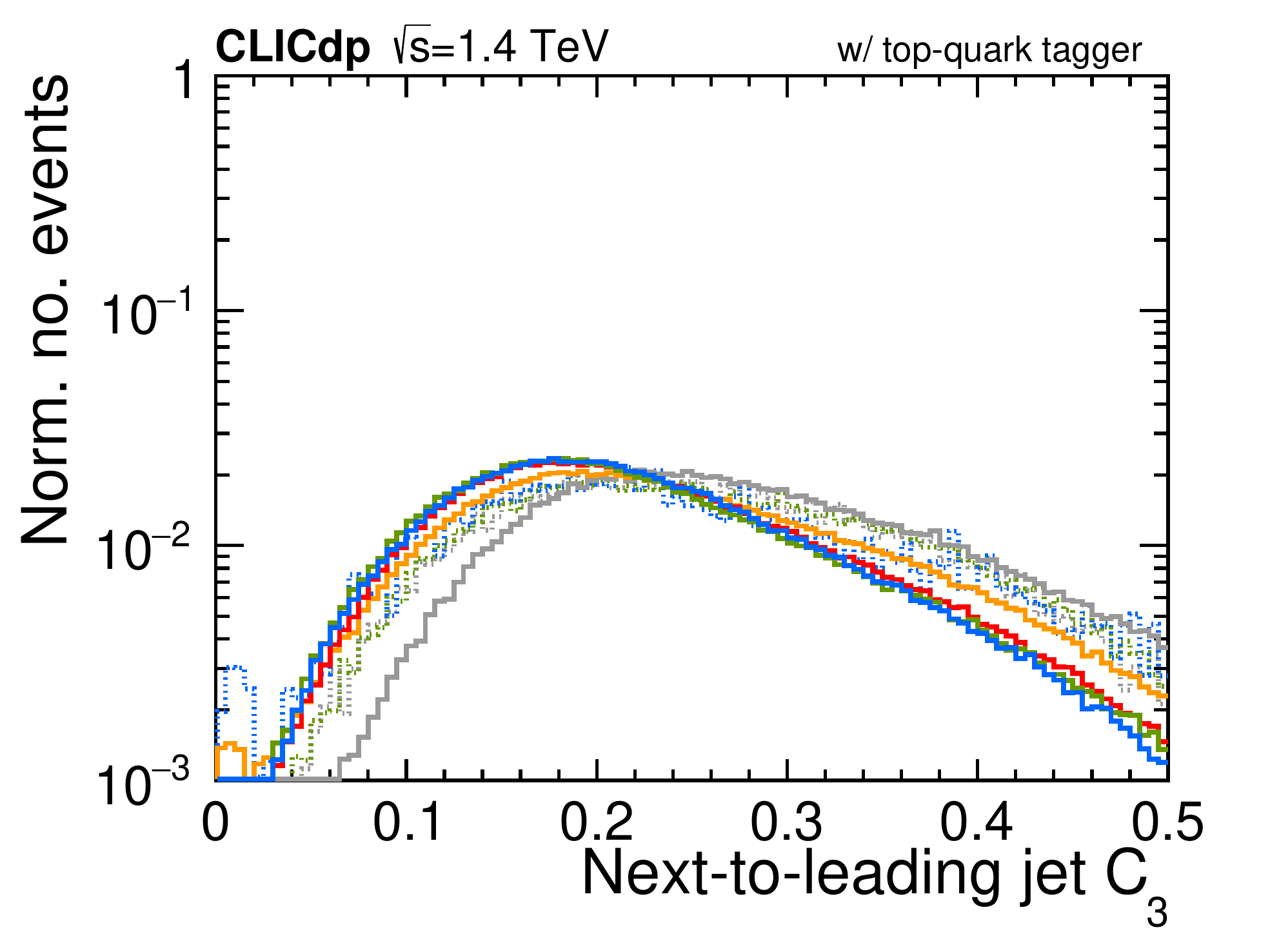}
	\caption{$C_{3}$ after applying the top-quark tagger.}
	\end{subfigure}\\
\caption{Energy correlation functions $C_2$ and $C_3$ for the lowest energy ''next-to-leading`` large-R jet. The superscript `a' (`b') refers to the kinematic region $\rootsprime\geq1.2\,\tev$ ($\rootsprime<1.2\,\tev$). Note that the qqlv and qqll backgrounds have been omitted in the figures in the right column due to low available statistics. The retention of these backgrounds after the full event selection is negligible.
\label{fig:analysis:mva:variables:energycorrCJ2}}
\end{figure}

\begin{figure}[p!]
	\centering
	\begin{subfigure}{0.48\columnwidth}
	\includegraphics[width=\textwidth]{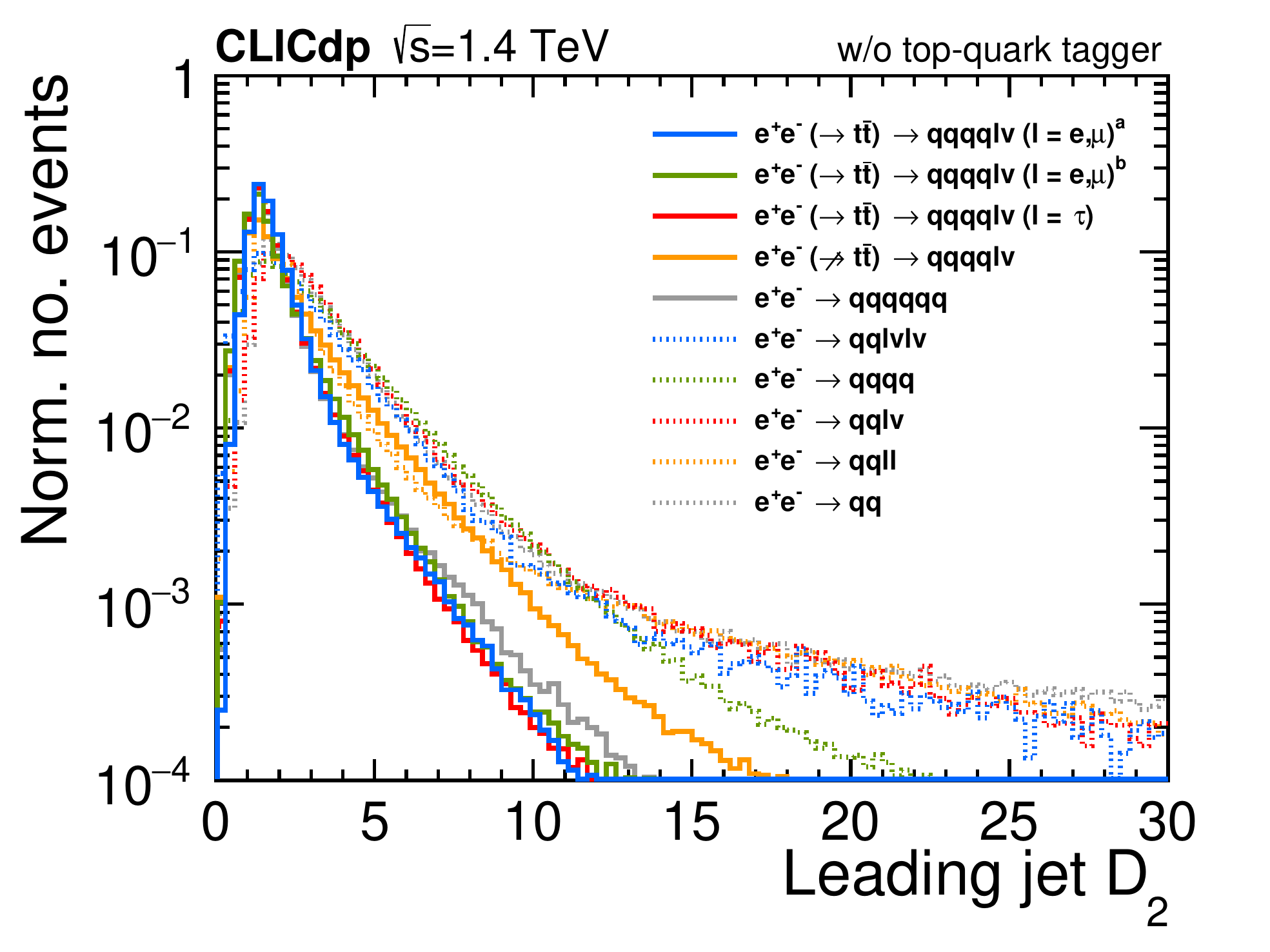}
	\caption{$D_{2}$ without applying the top-quark tagger.}
	\end{subfigure}
	~~~
	\begin{subfigure}{0.48\columnwidth}
	\includegraphics[width=\textwidth]{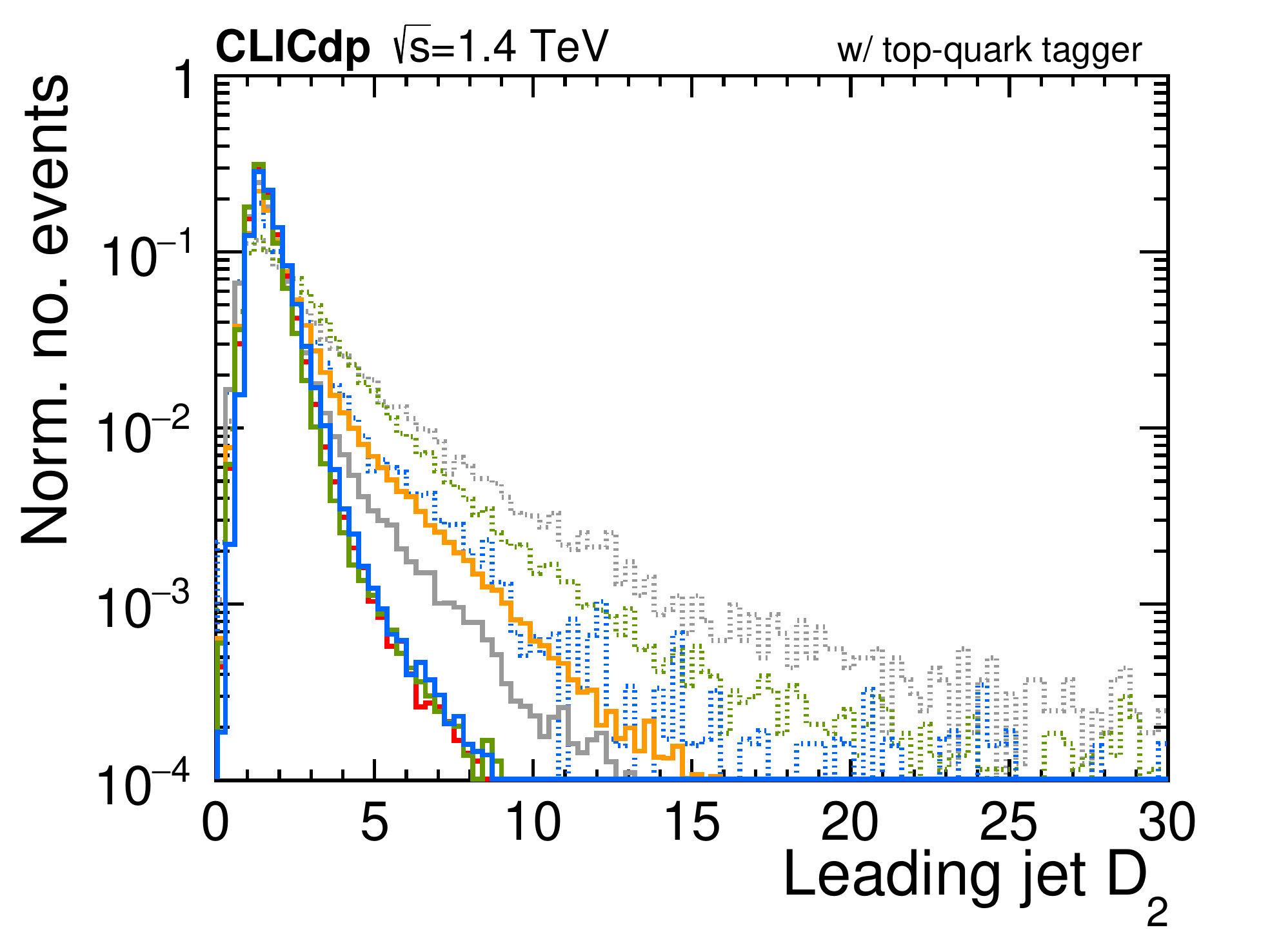}
	\caption{$D_{2}$ after applying the top-quark tagger.}
	\end{subfigure}\\
	\vspace{5mm}
	\begin{subfigure}{0.48\columnwidth}
	\includegraphics[width=\textwidth]{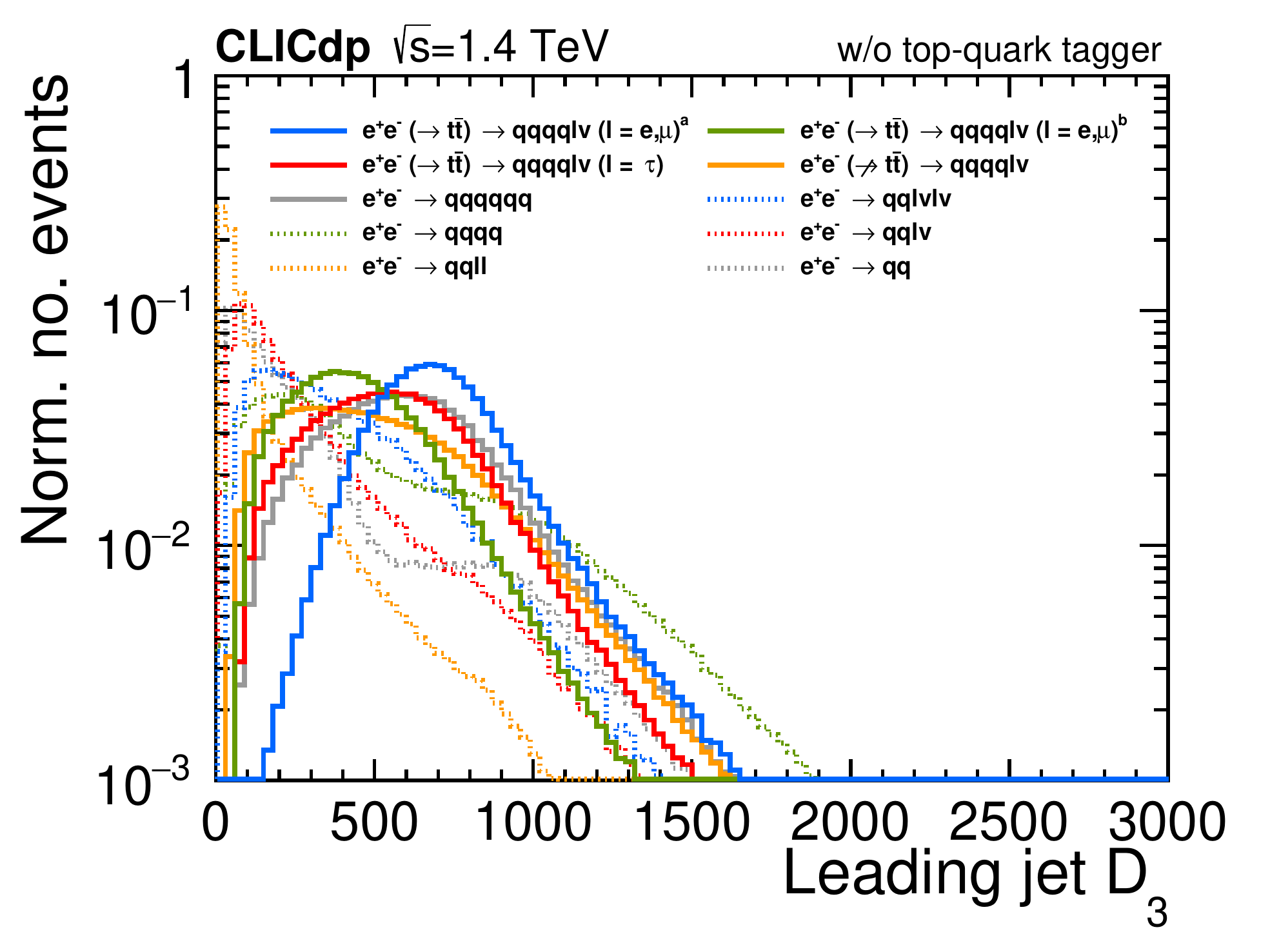}
	\caption{$D_{3}$ without applying the top-quark tagger.}
	\end{subfigure}
	~~~
	\begin{subfigure}{0.48\columnwidth}
	\includegraphics[width=\textwidth]{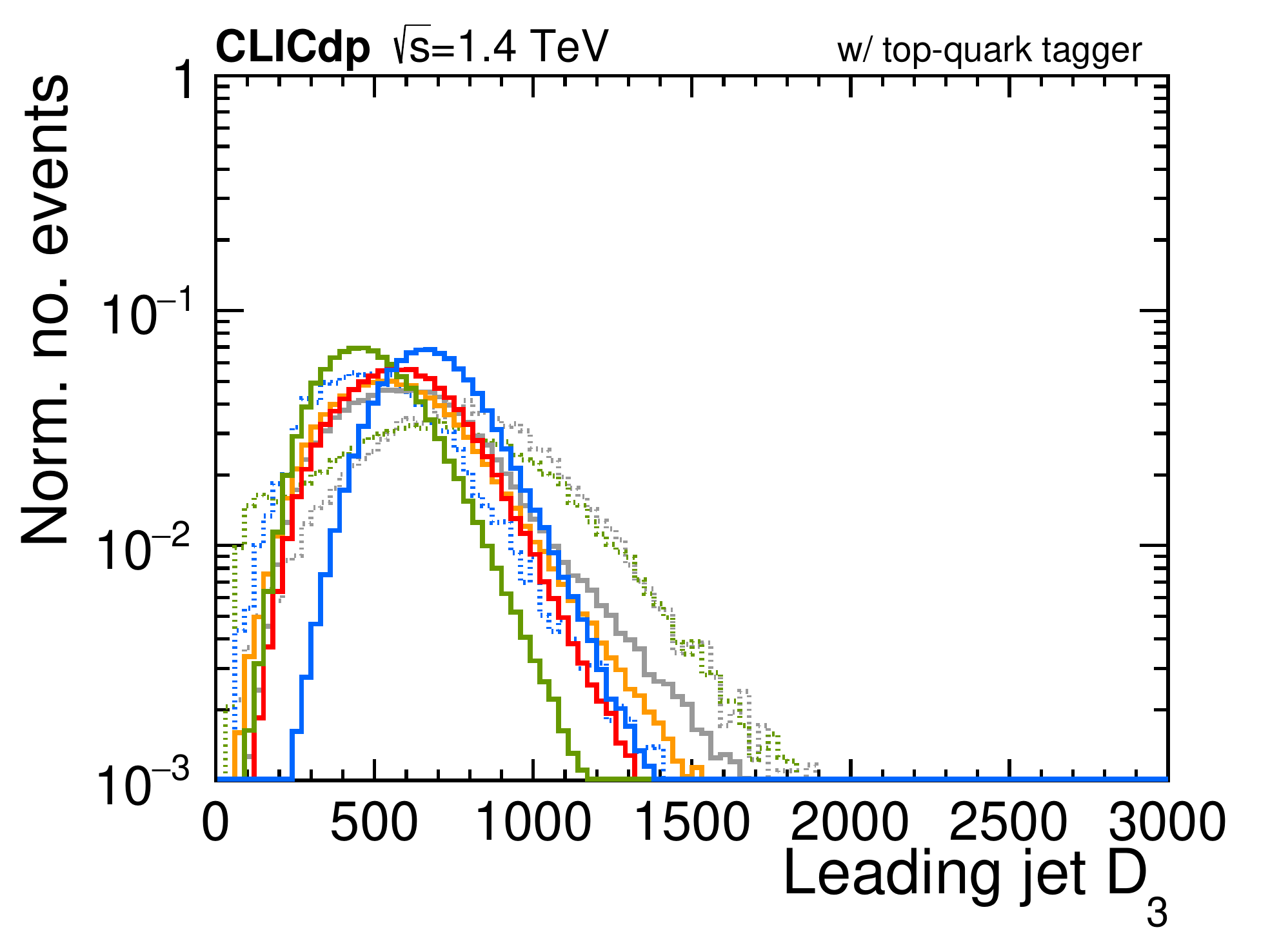}
	\caption{$D_{3}$ after applying the top-quark tagger.}
	\end{subfigure}\\
\caption{Energy correlation functions $D_2$ and $D_3$ for the highest energy ''leading`` large-R jet. The superscript `a' (`b') refers to the kinematic region $\rootsprime\geq1.2\,\tev$ ($\rootsprime<1.2\,\tev$). Note that the qqlv and qqll backgrounds have been omitted in the figures in the right column due to low available statistics. The retention of these backgrounds after the full event selection is negligible. \label{fig:analysis:mva:variables:energycorrDJ1}}
\end{figure}

\begin{figure}[p!]
	\centering
	\begin{subfigure}{0.48\columnwidth}
	\includegraphics[width=\textwidth]{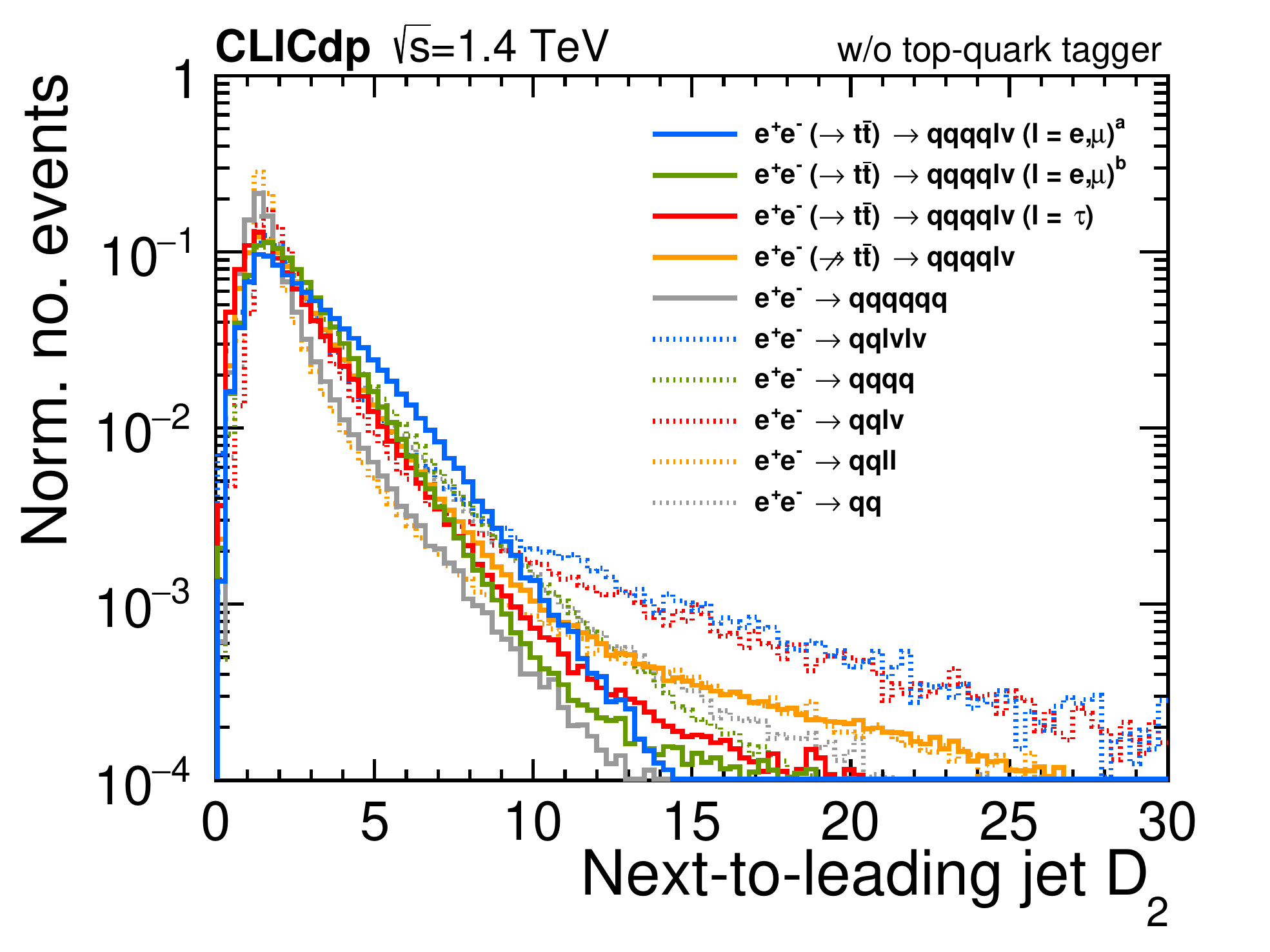}
	\caption{$D_{2}$ without applying the top-quark tagger.}
	\end{subfigure}
	~~~
	\begin{subfigure}{0.48\columnwidth}
	\includegraphics[width=\textwidth]{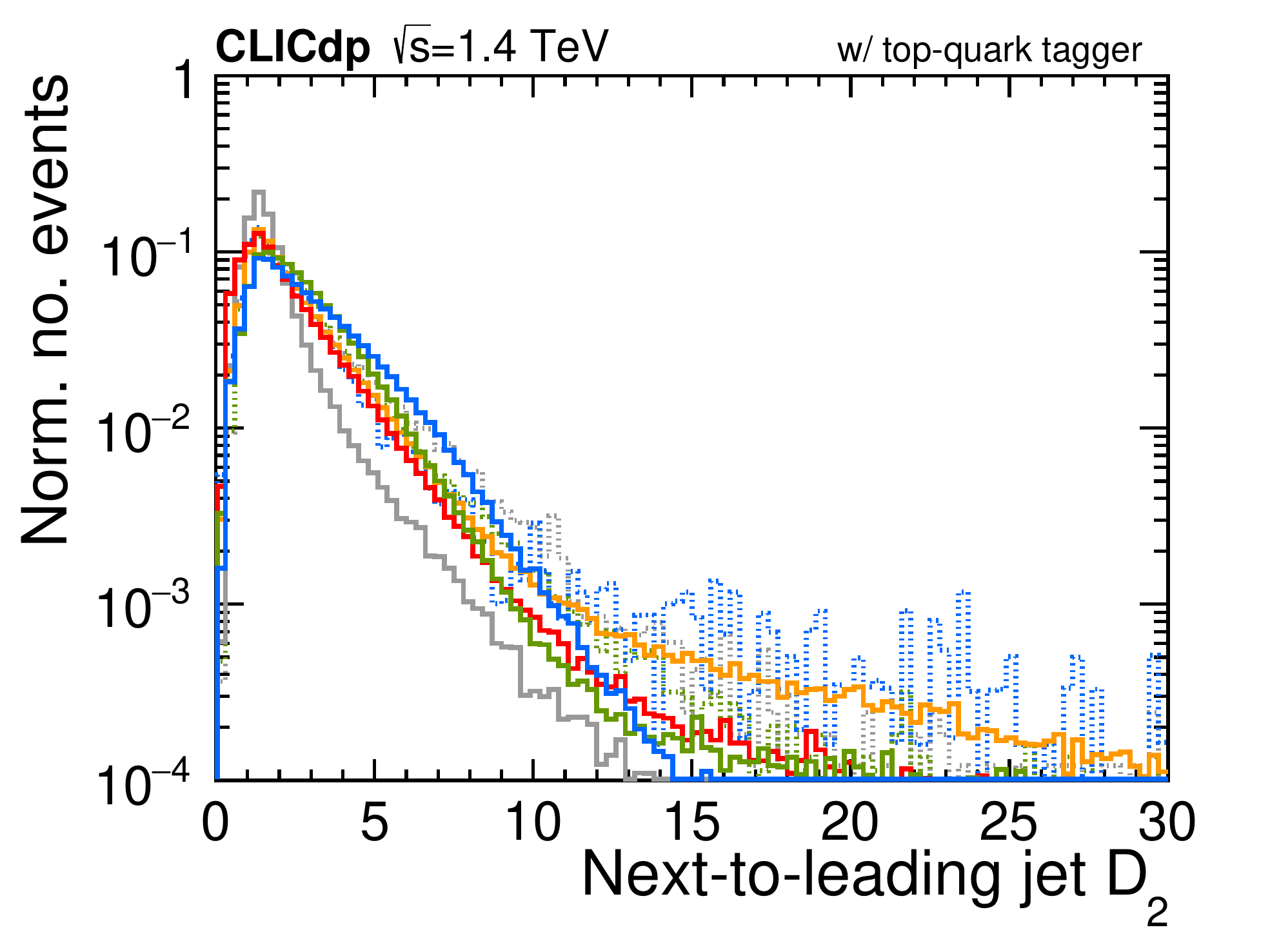}
	\caption{$D_{2}$ after applying the top-quark tagger.}
	\end{subfigure}\\
	\vspace{5mm}
	\begin{subfigure}{0.48\columnwidth}
	\includegraphics[width=\textwidth]{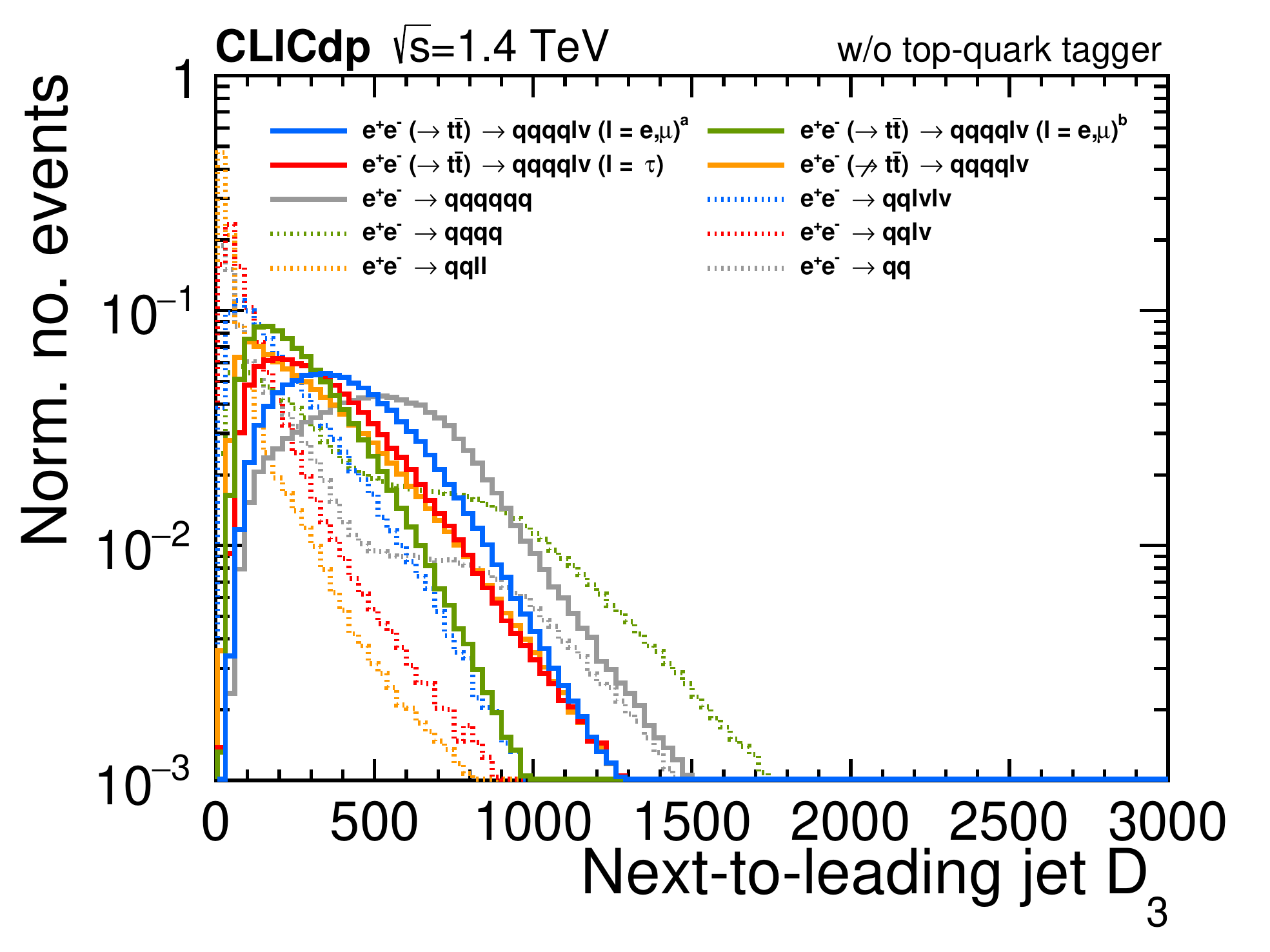}
	\caption{$D_{3}$ without applying the top-quark tagger.}
	\end{subfigure}
	~~~
	\begin{subfigure}{0.48\columnwidth}
	\includegraphics[width=\textwidth]{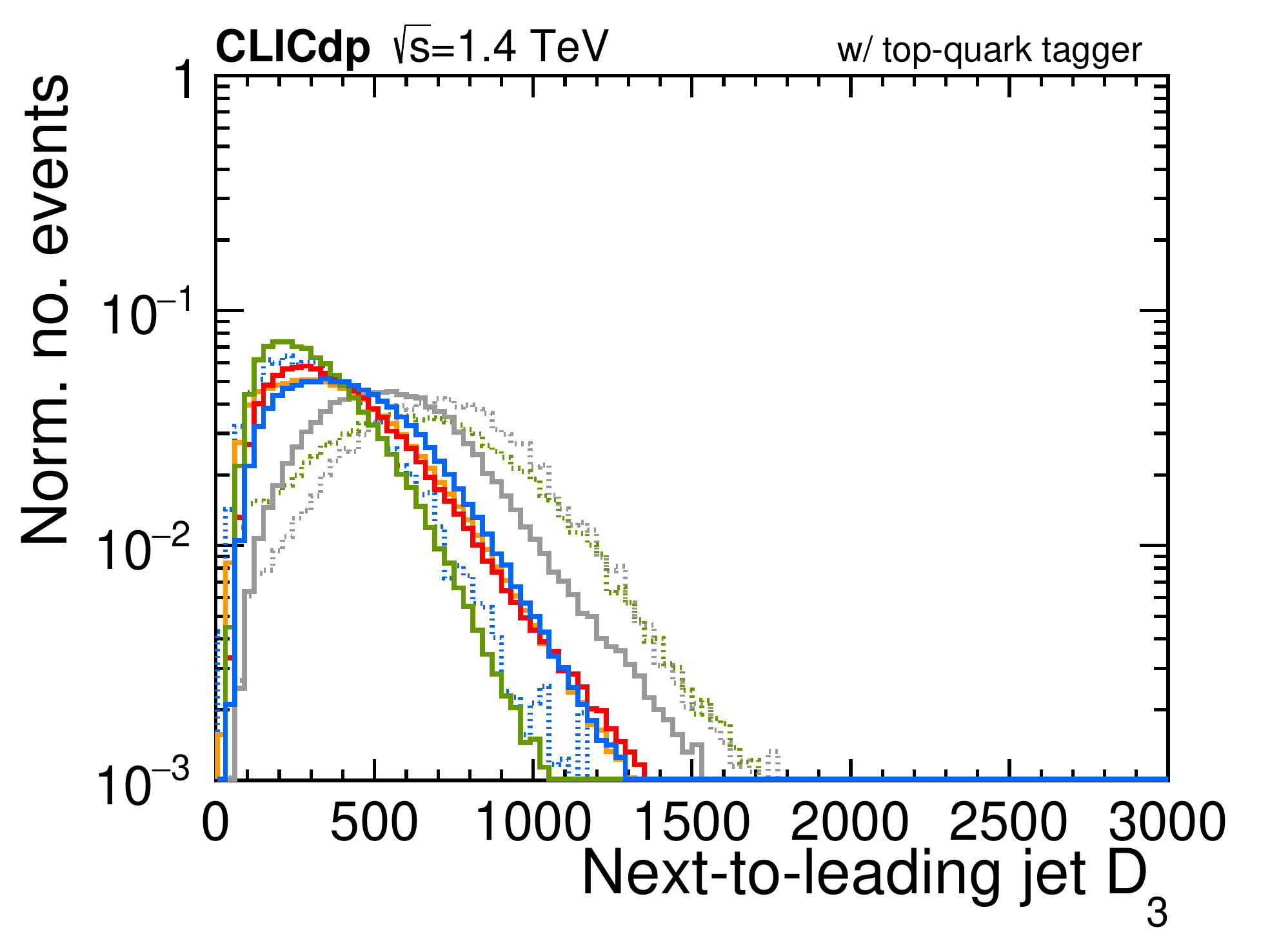}
	\caption{$D_{3}$ after applying the top-quark tagger.}
	\end{subfigure}\\
\caption{Energy correlation functions $D_2$ and $D_3$ for the lowest energy ''next-to-leading`` large-R jet. The superscript `a' (`b') refers to the kinematic region $\rootsprime\geq1.2\,\tev$ ($\rootsprime<1.2\,\tev$). Note that the qqlv and qqll backgrounds have been omitted in the figures in the right column due to low available statistics. The retention of these backgrounds after the full event selection is negligible.
\label{fig:analysis:mva:variables:energycorrDJ2}}
\end{figure}

\Cref{fig:analysis:mva:variables:energycorrCJ1} ($C_2$ and $C_3$) and \Cref{fig:analysis:mva:variables:energycorrDJ1} ($D_2$ and $D_3$) display the distributions for the highest-energy (denoted ''leading``) large-R jet before (left) and after (right) the top-tagger, while \Cref{fig:analysis:mva:variables:energycorrCJ2} ($C_2$ and $C_3$) and \Cref{fig:analysis:mva:variables:energycorrDJ2} ($D_2$ and $D_3$) show the corresponding distributions for the ''next-to-leading jet``. 

In conclusion, these sub-structure variables, in similarity to the N-subjettiness observables, display a powerful separation between signal events and single-top, four-jet and di-jet events for the leading jet, and likewise against fully-hadronic $\ttbar$ events for the next-to-leading jet. Again, this ability is somewhat retained after applying the top-tagger.  

\clearpage
\section{High-energy photon veto}
\label{sec:photonreco}

ISR effectively lowers the centre-of-mass energy of an event w.r.t. the nominal collision energy $\roots$. Some of these radiative events where the ISR photon is not colinear with the beam direction can be found by searching for isolated high-energy photons within the detector acceptance. These are identified as PFOs tagged as photons, with a high $\pT$, and low activity in a surrounding cone. Since the measurements presented in this paper aim to study observables close to the nominal collision energy, events with one or more isolated photons are vetoed.

\begin{figure}[htpb]
\centering
\includegraphics[width=0.48\columnwidth]{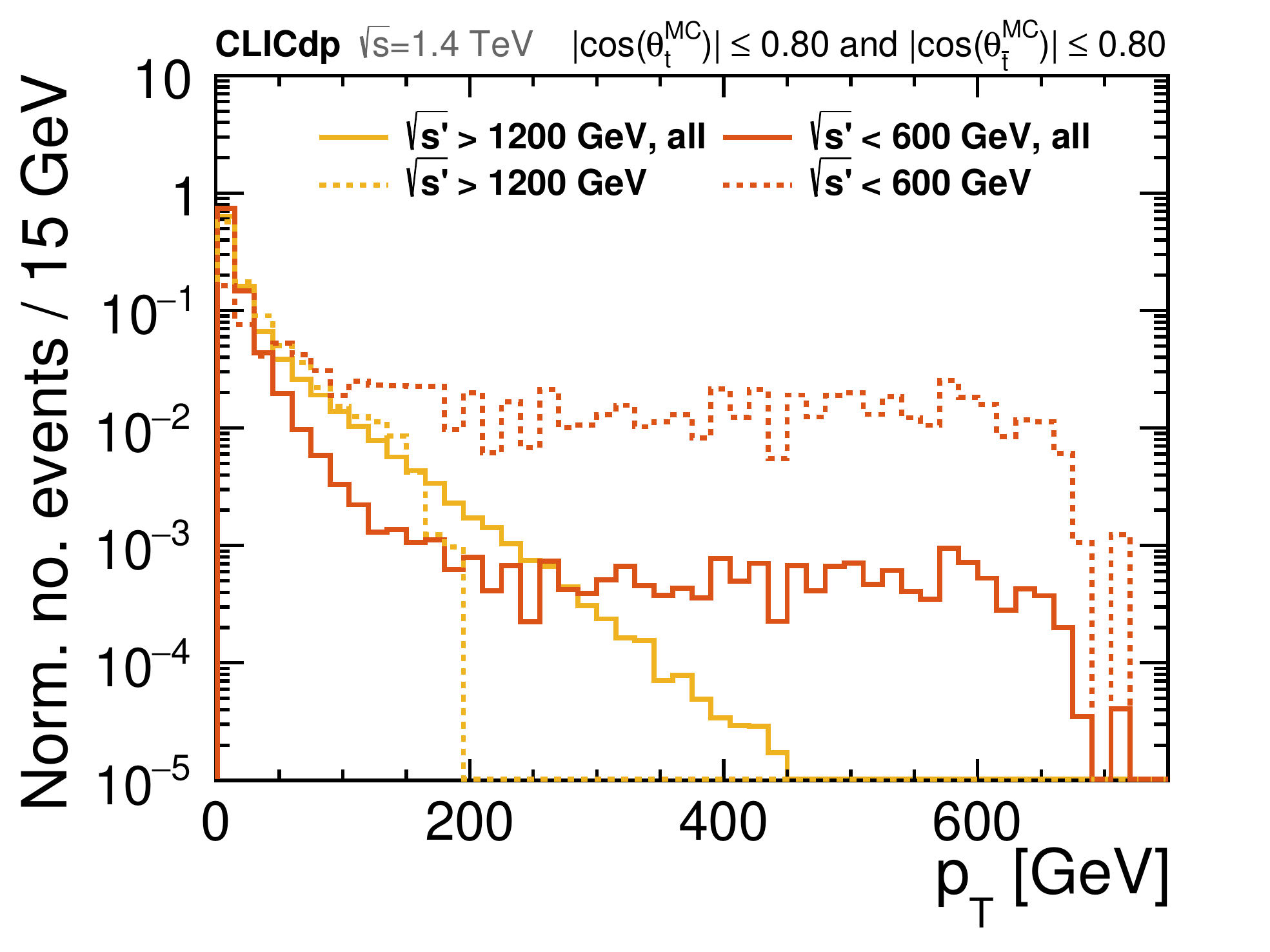}
~~~
\includegraphics[width=0.48\columnwidth]{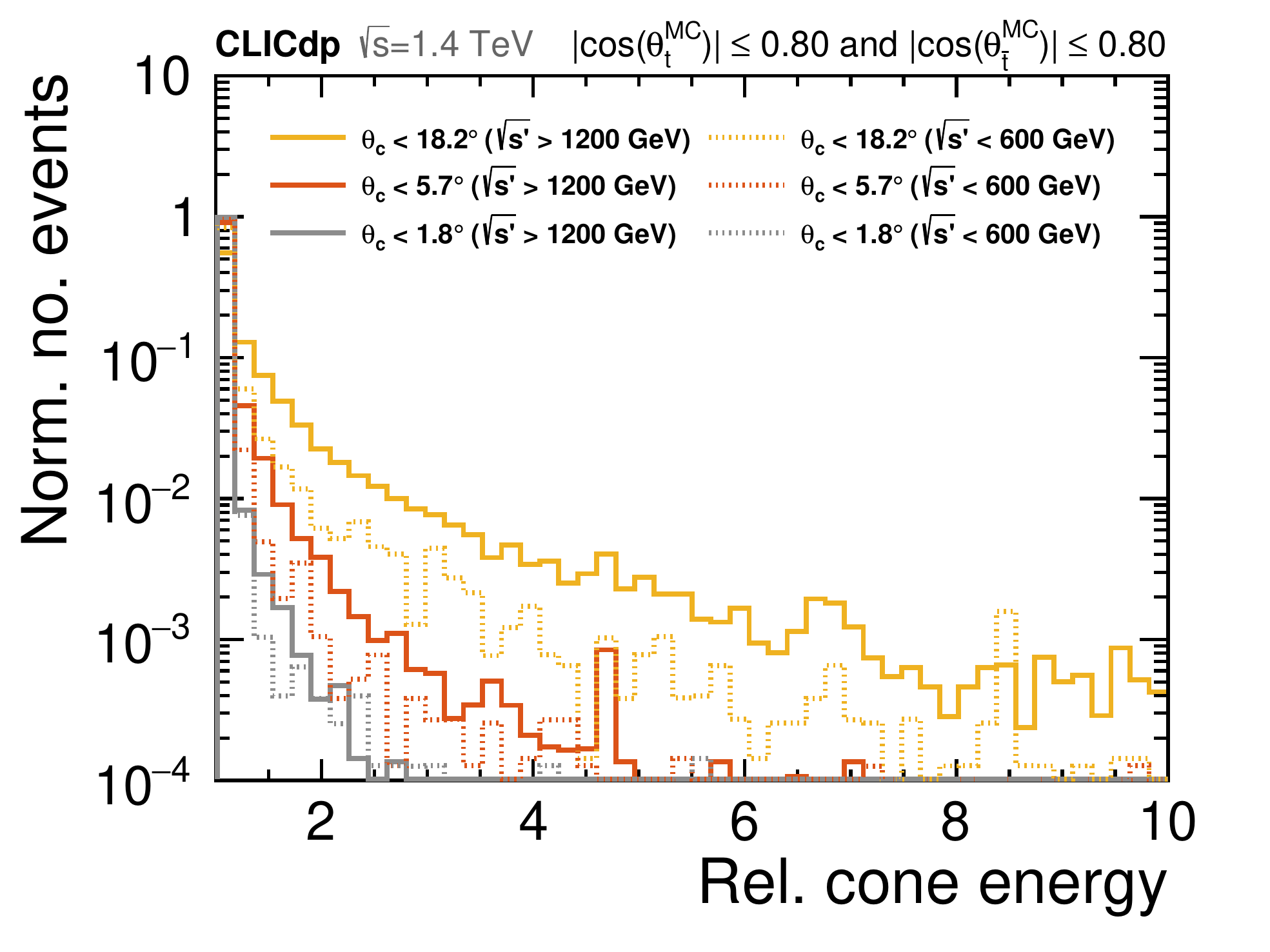}
\caption{The left panel shows $\pT$ distributions of photons for events with an effective collision energy above 1200 GeV (yellow) and below 600 GeV (red). Solid lines represent all reconstructed PFO photons while dotted lines represent the subset of PFOs matched to the ISR photons at parton-level. The right panel shows the relative cone energy (see text for details) of photons matched to the ISR photons at parton-level for events with an effective collision energy above 1200 GeV (solid) and below 600 GeV (dashed) and for three different cone sizes.\label{fig:analysis:photonreco:basicfeatures}}
\end{figure}

The left panel in \Cref{fig:analysis:photonreco:basicfeatures} shows the $\pT$ distribution for all reconstructed PFO photons (solid) and those matched to ISR photons at parton-level (dotted). The yellow lines represent events with an $\rootsprime$ close to the nominal collision energy ($>1200\,\gev$) and the red lines represent events with a significantly lower $\rootsprime$ ($<600\,\gev$). While the distribution drops sharply for events close to the nominal collision energy, it instead reaches a plateau around 100\,\gev for radiative events, above which the presence of high-energy ISR photons are clearly seen. For practical reasons we only include photons with an energy in excess of 5\,\gev.

Isolation of the candidate ISR photons is studied by looking at the total energy in a cone around the particle as a function of energy. In particular, we study the so-called relative cone energy, defined as the the sum of energies from all reconstructed PFOs located inside a cone of $\cos(\theta_{c})=0.995\,(\approx5.7^\circ)$ around the particle, divided by the energy of the particle itself. The right panel of \Cref{fig:analysis:photonreco:basicfeatures} shows the relative cone energy distribution for three different cone sizes, $\theta$.

Isolated high-energy photons are identified as photon PFOs with a $\pT$ in excess of $75\,\mathrm{GeV}$ and with a relative cone energy below 1.2. In addition, we require that the polar angle of the candidate photons are in the range $10^\circ\leq\theta\leq170^\circ$. Events with one of more identified high-energy isolated photons are vetoed and excluded from further analysis.
\clearpage

\section{Reconstruction of the effective collision energy}
\label{sec:effcom}

To reconstruct the effective centre-of-mass energy $\rootsprime$ we first assume that the missing transverse momentum, estimated by adding up the 4-vectors of the two large-$R$ jets and the isolated charged lepton, can be used as an estimator for the neutrino transverse momentum components. Here we neglect the effect from unidentified ISR and beamstrahlung photons. The $z$-component of the neutrino momentum, $p_{\PGn,z}$, is retrieved by solving 
\begin{equation}\label{eq:boosted:sprimereco}
M_{\PW}^2 = m_{\Pl}^2 + 2(E_{\Pl}E_{\PGn} - \vec{p}_{\Pl}\cdot\vec{p}_{\PGn}),
\end{equation}
given a constraint on the mass of the leptonically decaying $\PW$ boson, $M_{\PW}$. Here, the indices $\Pl$ and $\PGn$ denote the lepton and neutrino candidate quantities, respectively. 

\Cref{eq:boosted:sprimereco} is quadratic in $p_{\PGn,z}$ and its two solutions are shown in \Cref{eq:boosted:sprimerecosol1}. Note that no real-valued solutions can be obtained if the observed missing transverse energy, $\HepParticle{E}\!\!\!\!\!\!\!\slash_{T}$, fluctuates such that the invariant mass of the combined neutrino-lepton system is above $M_{\PW}$, i.e. $X$ must be greater or equal to zero. If not, the missing transverse energy is scaled to provide a real solution ($X=0$).

\begin{equation}\label{eq:boosted:sprimerecosol1}
p_z^{\PGn} = \frac{1}{2((p_z^{\Pl})^2-E_{\Pl}^2)}\big( p_z^{\Pl}m_{\Pl} - p_z^{\Pl}M_{\PW} - 2p_x^{\Pl}p_z^{\Pl}p_x^{\PGn} - 2p_y^{\Pl}p_z^{\Pl}p_y^{\PGn} \pm X \big)
\end{equation}
where
\begin{equation*}
X = \sqrt{ E_{\Pl}^2\big[( M_{\PW}^2 - m_{\Pl}^2 + 2(p_x^{\Pl}p_x^{\PGn} + p_y^{\Pl}p_y^{\PGn}) )^2 + 4\,\HepParticle{E}\!\!\!\!\!\!\!\slash_{T}^2(-E_{\Pl}^2+(p_z^{\Pl})^2) \big] }
\end{equation*}
and
\begin{equation*}
\HepParticle{E}\!\!\!\!\!\!\!\slash_{T} = \sqrt{ (p_x^{\PGn})^2 + (p_y^{\PGn})^2}.
\end{equation*}

The resulting neutrino-lepton system solutions are combined with each of the large-$R$ jets and the final candidate is chosen as the one that yields a mass closest to the generated top-quark mass.

\begin{figure}
\centering
\includegraphics[width=0.48\columnwidth]{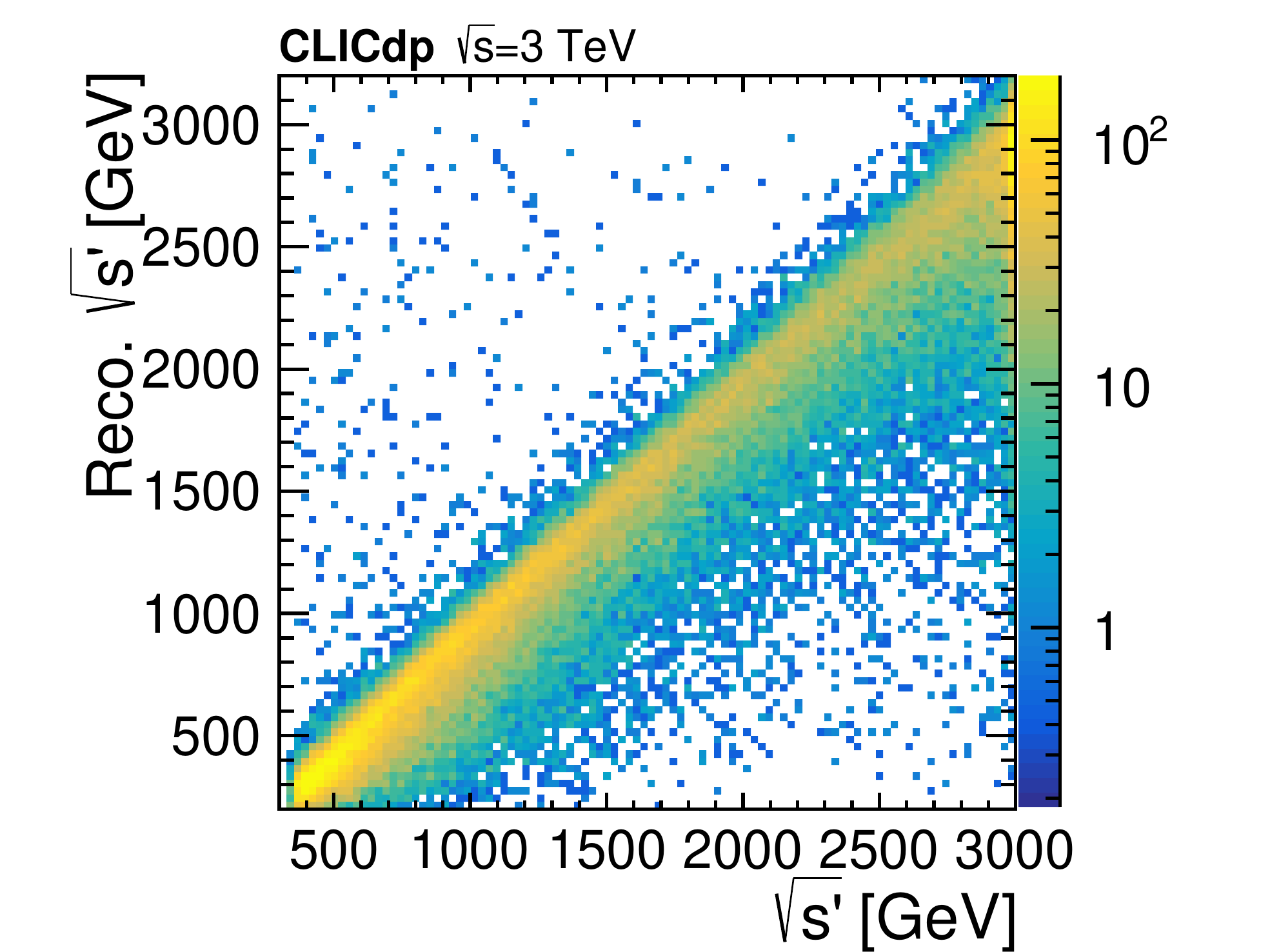}
~~~
\includegraphics[width=0.48\columnwidth]{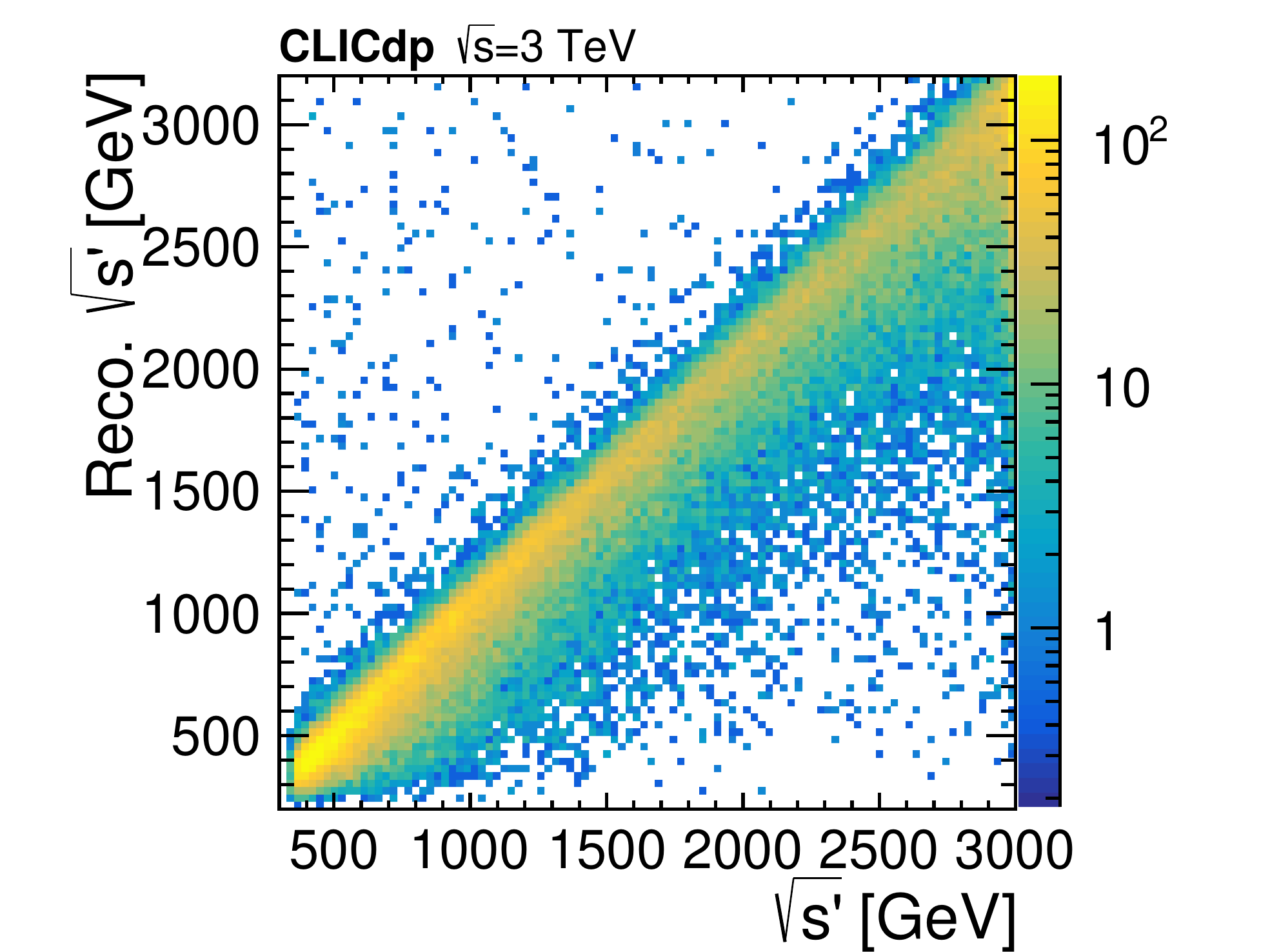}
\caption{Reconstructed centre-of-mass energy vs the generated collision energy, including the effects of the luminosity spectrum and ISR (left). The corresponding distribution after applying a bias correction based on the median pull (right). \label{fig:analysis:effcom:comVScomRaw}}
\end{figure}

\begin{figure}
\centering
\includegraphics[width=0.48\columnwidth]{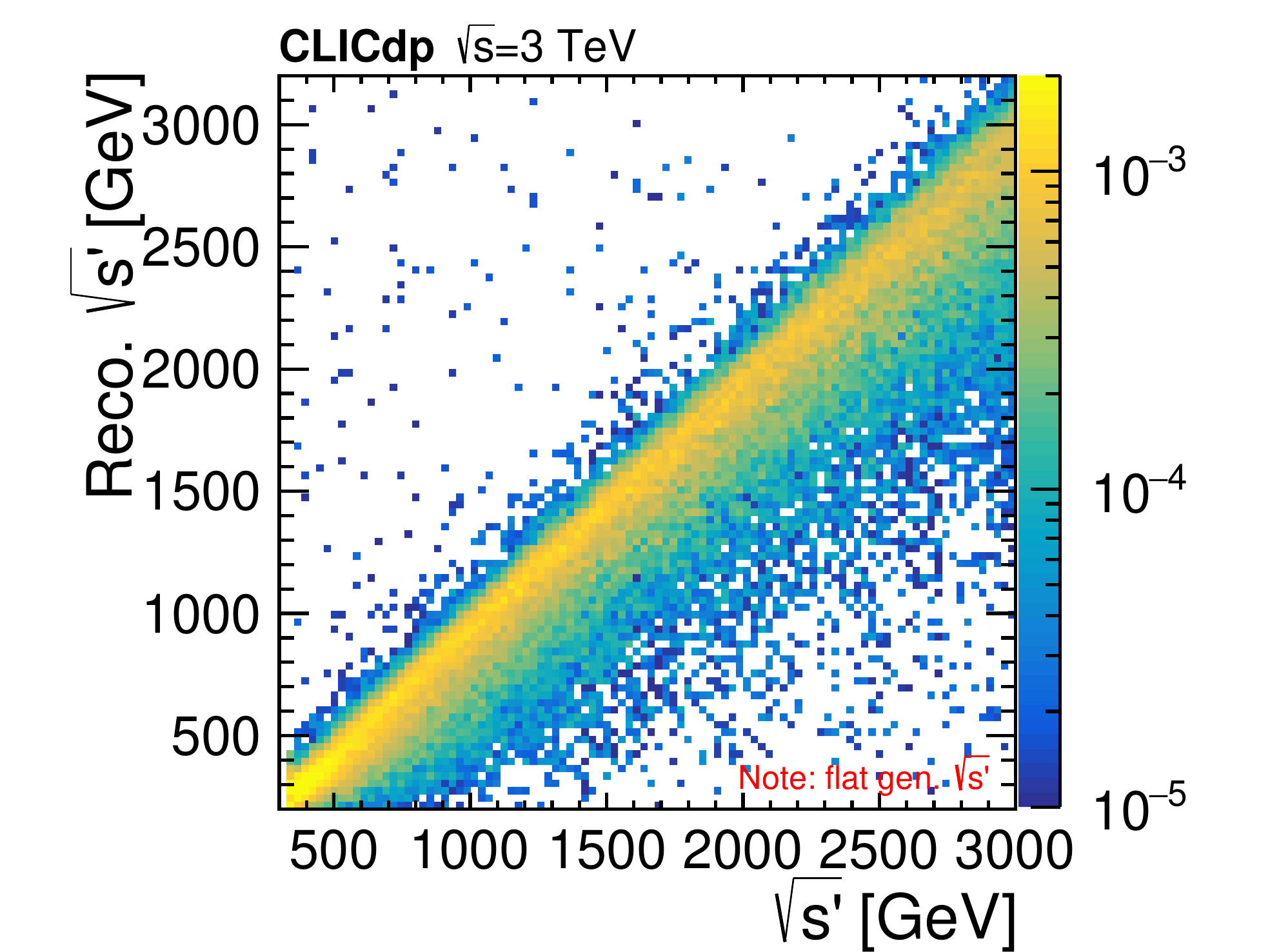}
~~~
\includegraphics[width=0.48\columnwidth]{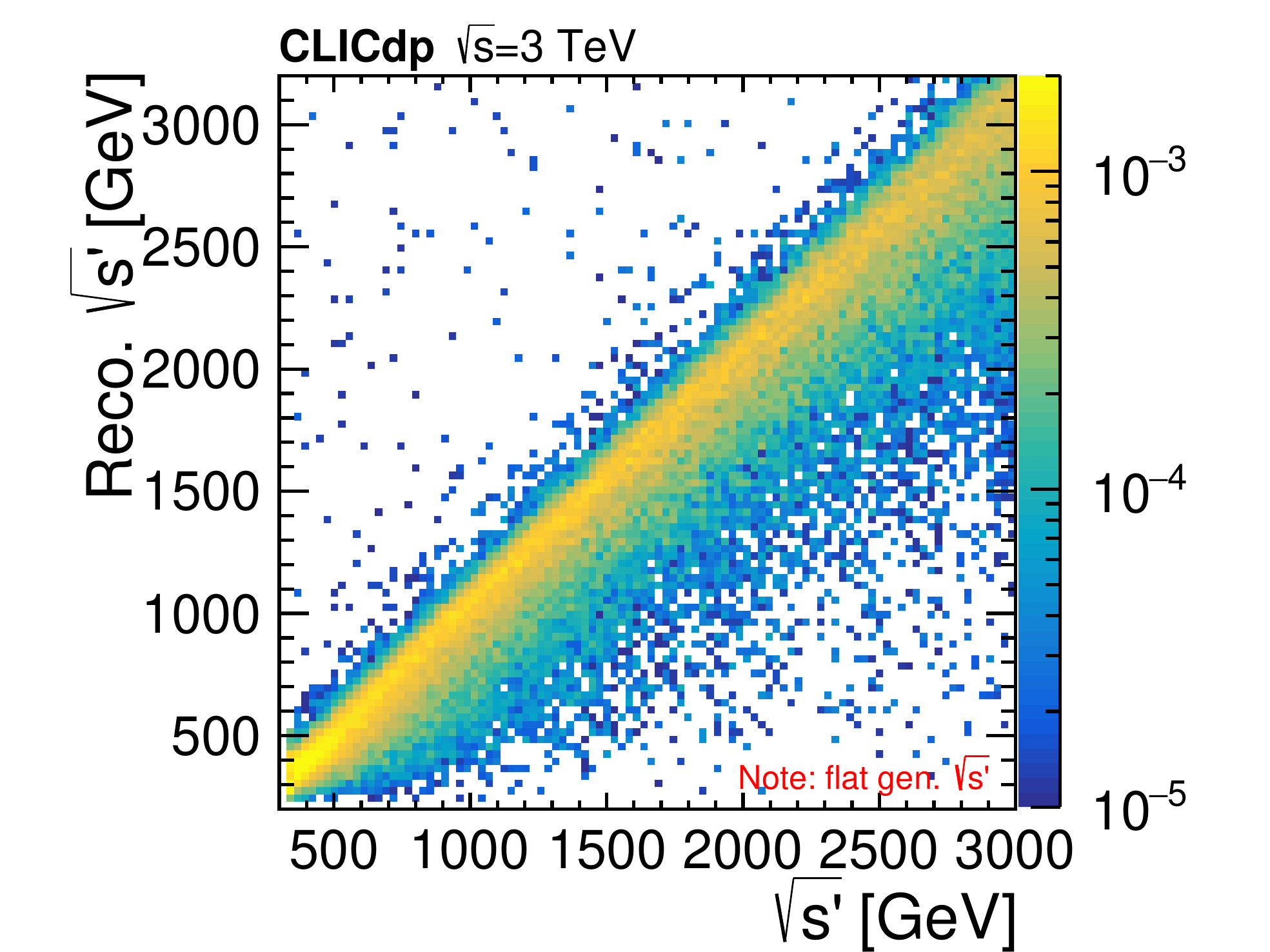}
\caption{Reconstructed centre-of-mass energy vs the normalised generated collision energy, including the effects of the luminosity spectrum and ISR. See caption of \Cref{fig:analysis:effcom:comVScomRaw} and text for details. \label{fig:analysis:effcom:comVScom}}
\end{figure}

The reconstructed effective centre-of-mass energy, denoted $\rootsprimereco$, is shown as function of the corresponding parton-level value $\rootsprime$ in the left panel of \Cref{fig:analysis:effcom:comVScomRaw}. The right panel shows the corresponding distribution after applying a bias correction based on the median pull. All following results and figures refer to the bias-corrected distribution. To illustrate the correlation down to lower values of $\rootsprime$, the same distributions are re-drawn in \Cref{fig:analysis:effcom:comVScom}, normalised so that each $\rootsprime$ bin contains the same number of entries, leading to a flat distribution in $\rootsprime$. 

\begin{figure}
\centering
\includegraphics[width=0.48\columnwidth]{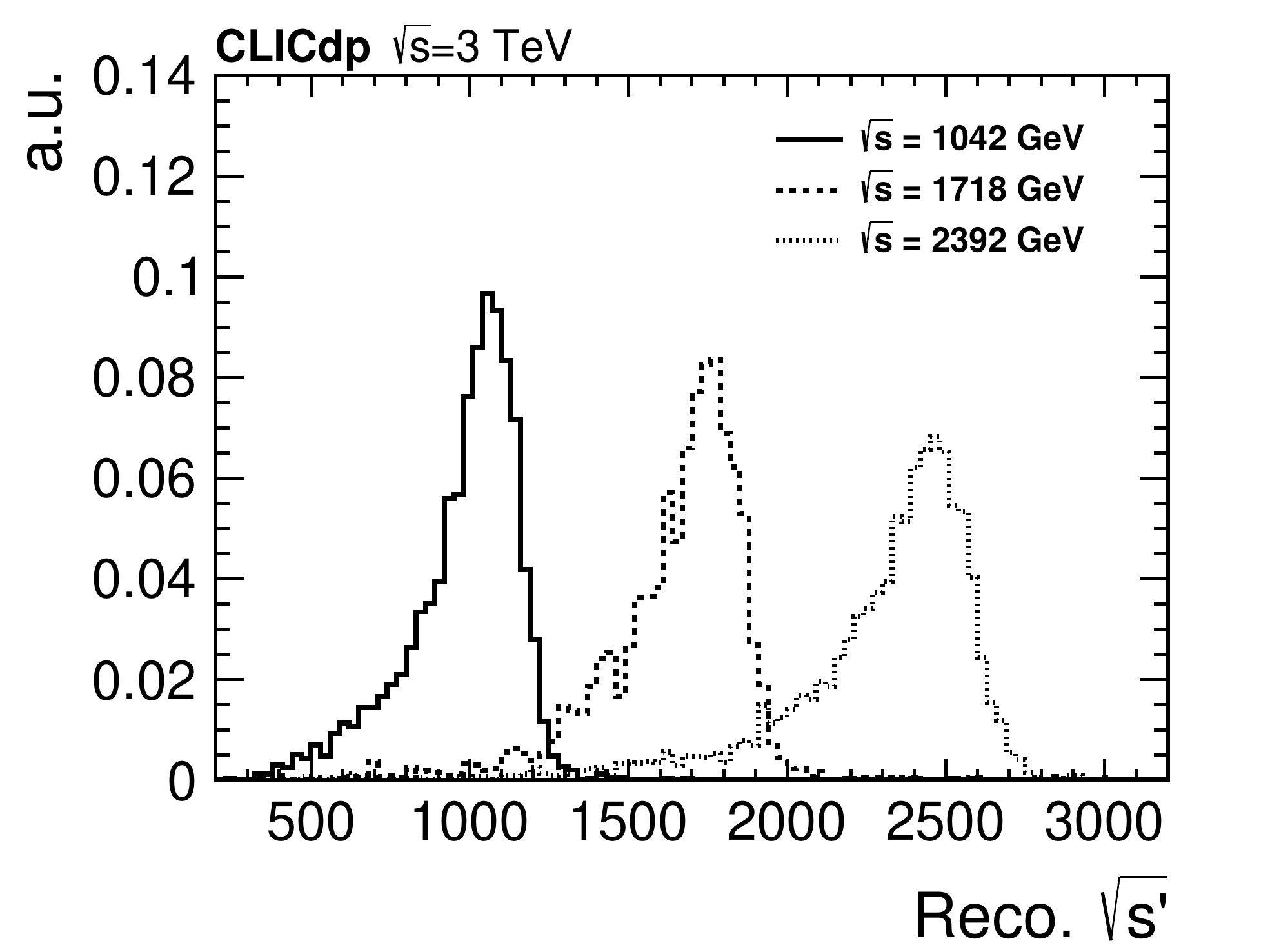}
~~~
\includegraphics[width=0.48\columnwidth]{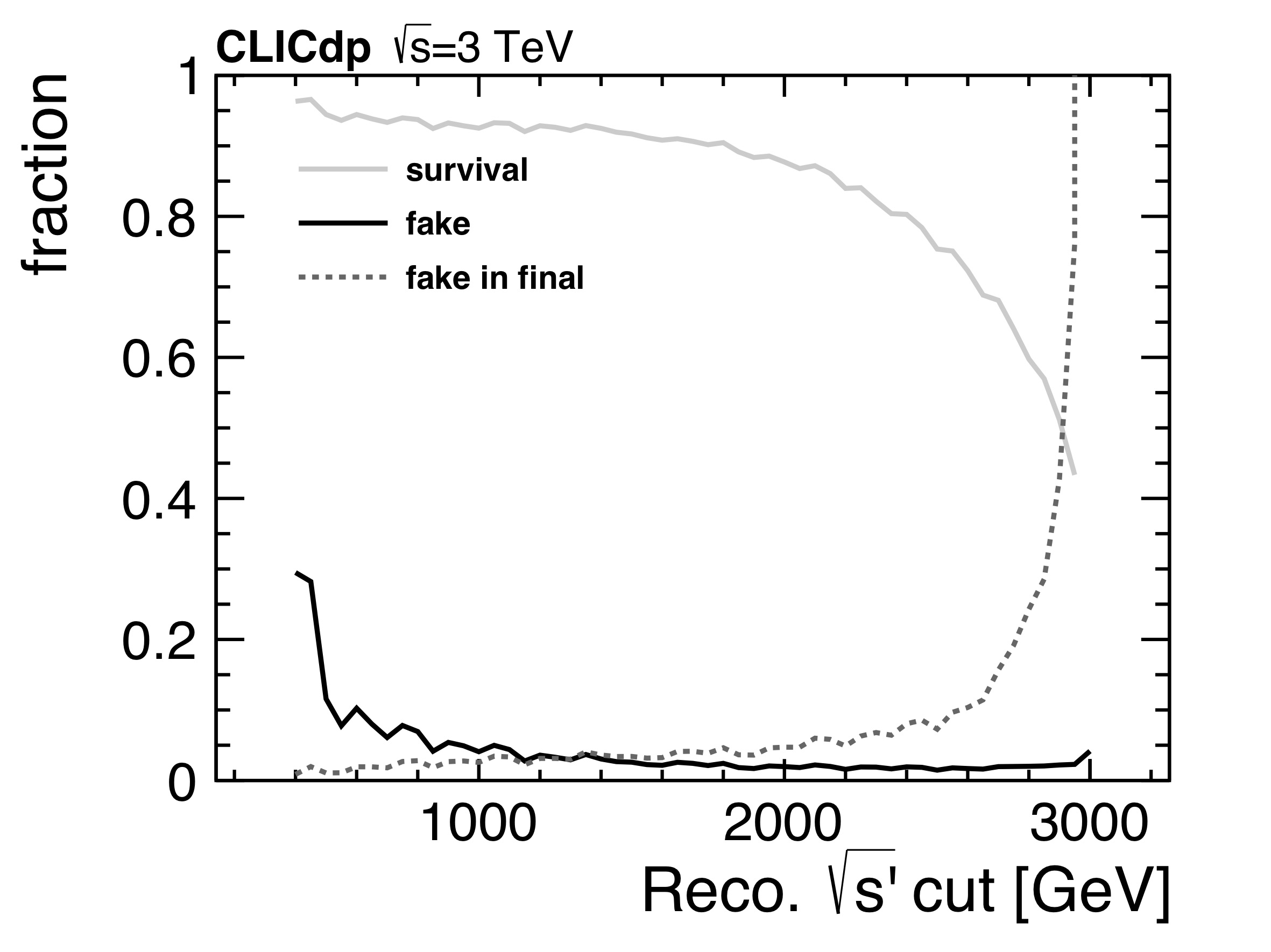}
\caption{Reconstructed centre-of-mass energy for three example values of $\rootsprime$ (left). Survival and fake classification fractions as a function of a cut on the reconstructed $\rootsprimereco$ (right). \label{fig:analysis:effcom:extra} }
\end{figure}

The left panel of \Cref{fig:analysis:effcom:extra} shows the projection of $\rootsprimereco$ at 3\,\tev for three example values of $\rootsprime$. The reconstruction yields RMS values between 100 and 160\,\gev for operation at 1.4\,\tev and between 140 and 360\,\gev at $\roots=3\,\tev$. The right panel shows the signal efficiency and fake classification fraction as a function of a cut on $\rootsprimereco$. Here 'survival' denote the fraction of events generated above the cut that are also reconstructed above. 'Fake' refers to the number of events generated below the cut that are reconstructed above and `fake in total' refers to the overall fraction of `fake' events in the final sample after applying the cut.

\begin{figure}
\centering
\includegraphics[width=0.48\columnwidth]{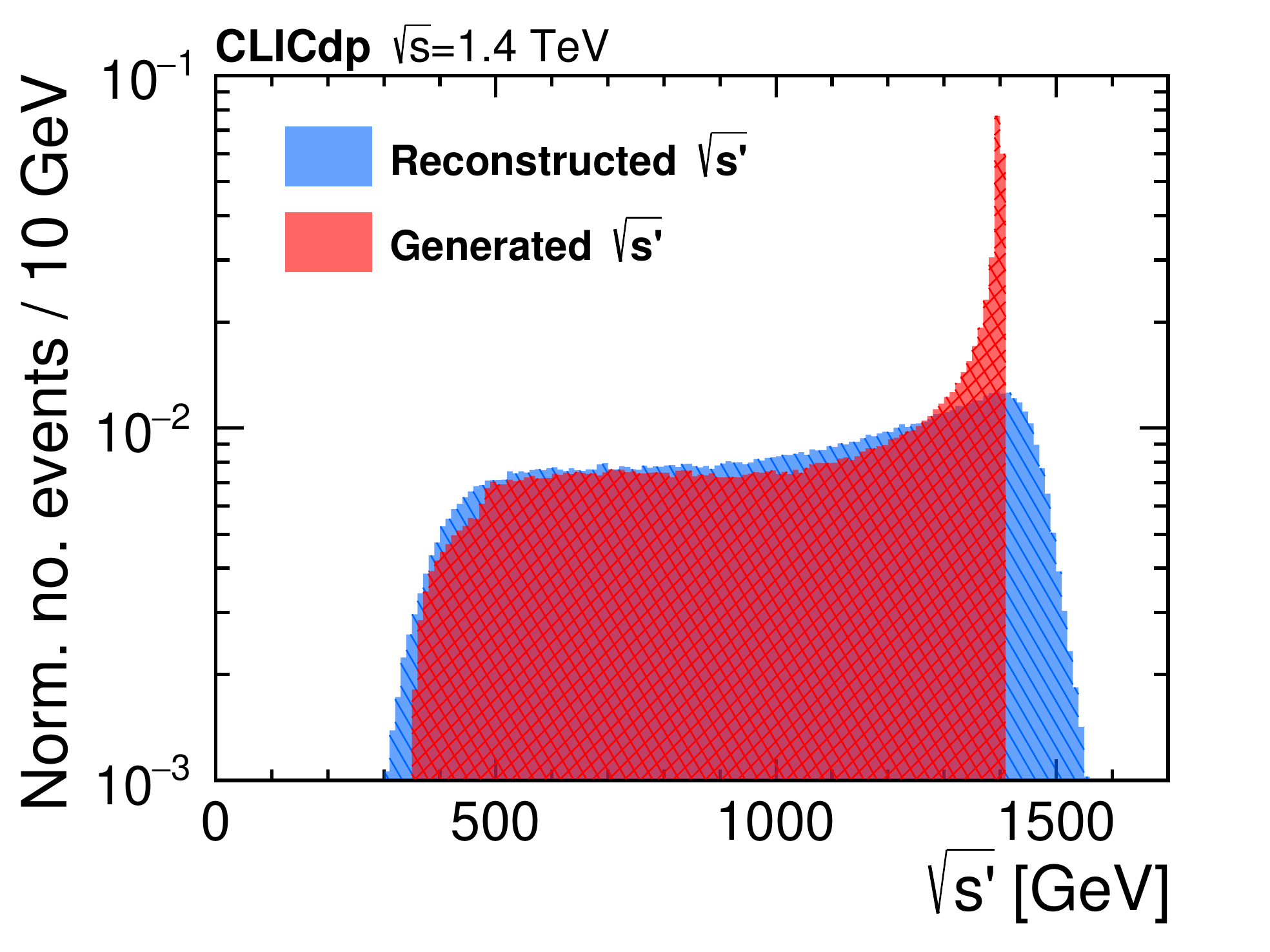}
~~~
\includegraphics[width=0.48\columnwidth]{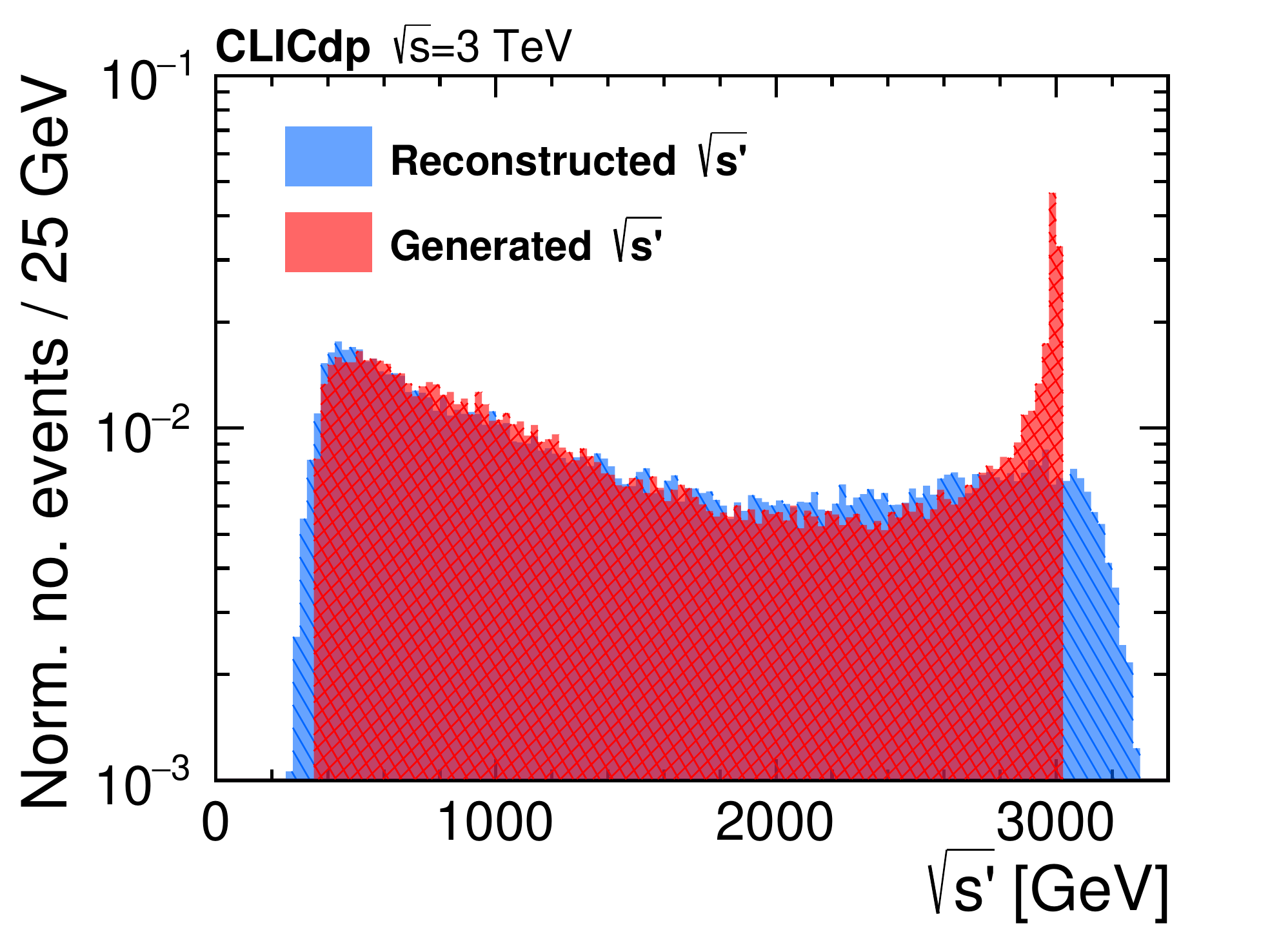}
\caption{Comparison of reconstructed and generated centre-of-mass energies of signal $\ttbar$ events for operation at $\roots=1.4\,\tev$ (left) and $\roots=3\,\tev$ (right).\label{fig:analysis:effcom:distcomp}}
\end{figure}

\begin{figure}
\centering
\includegraphics[width=0.48\columnwidth]{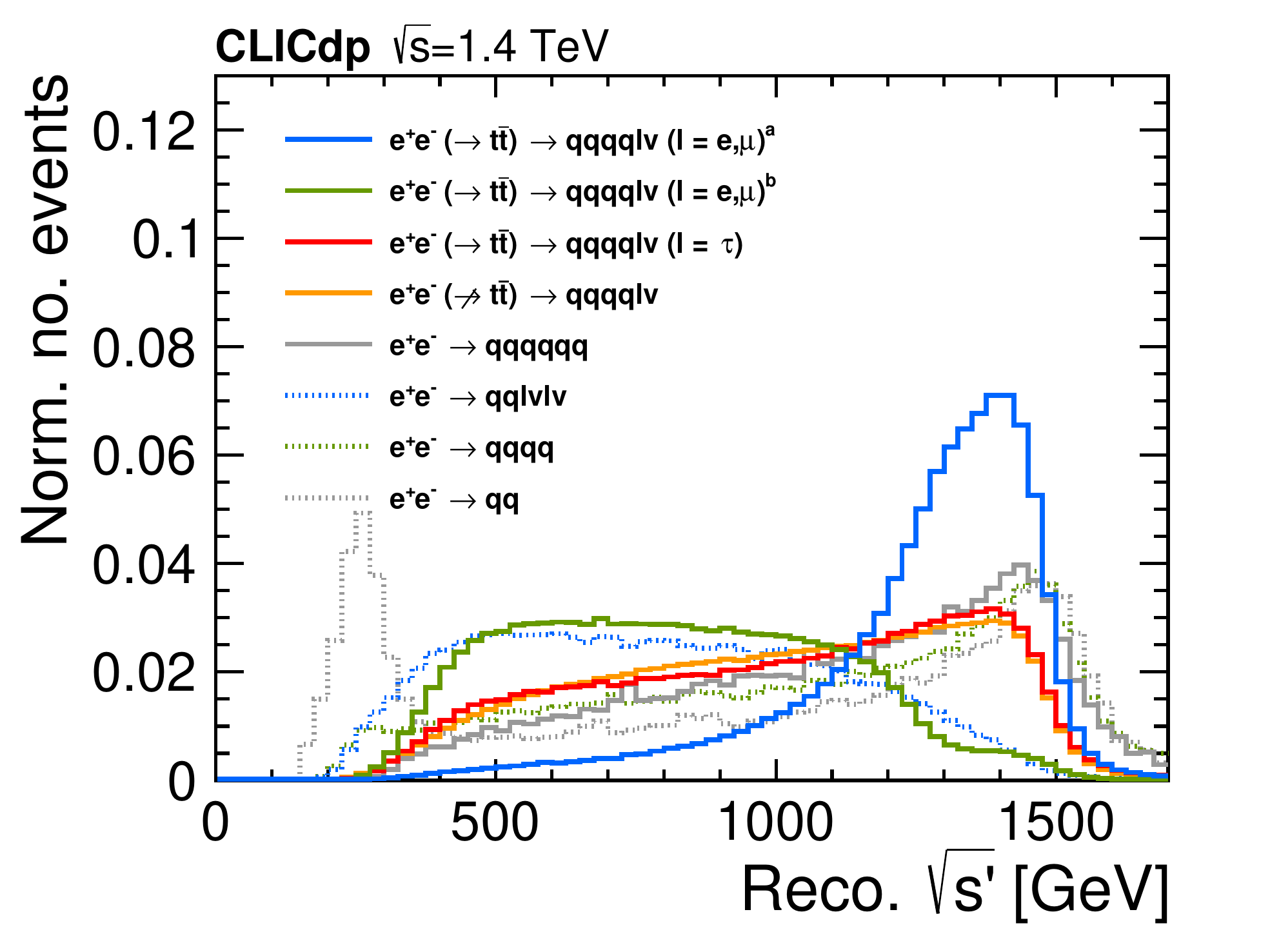}
~~~
\includegraphics[width=0.48\columnwidth]{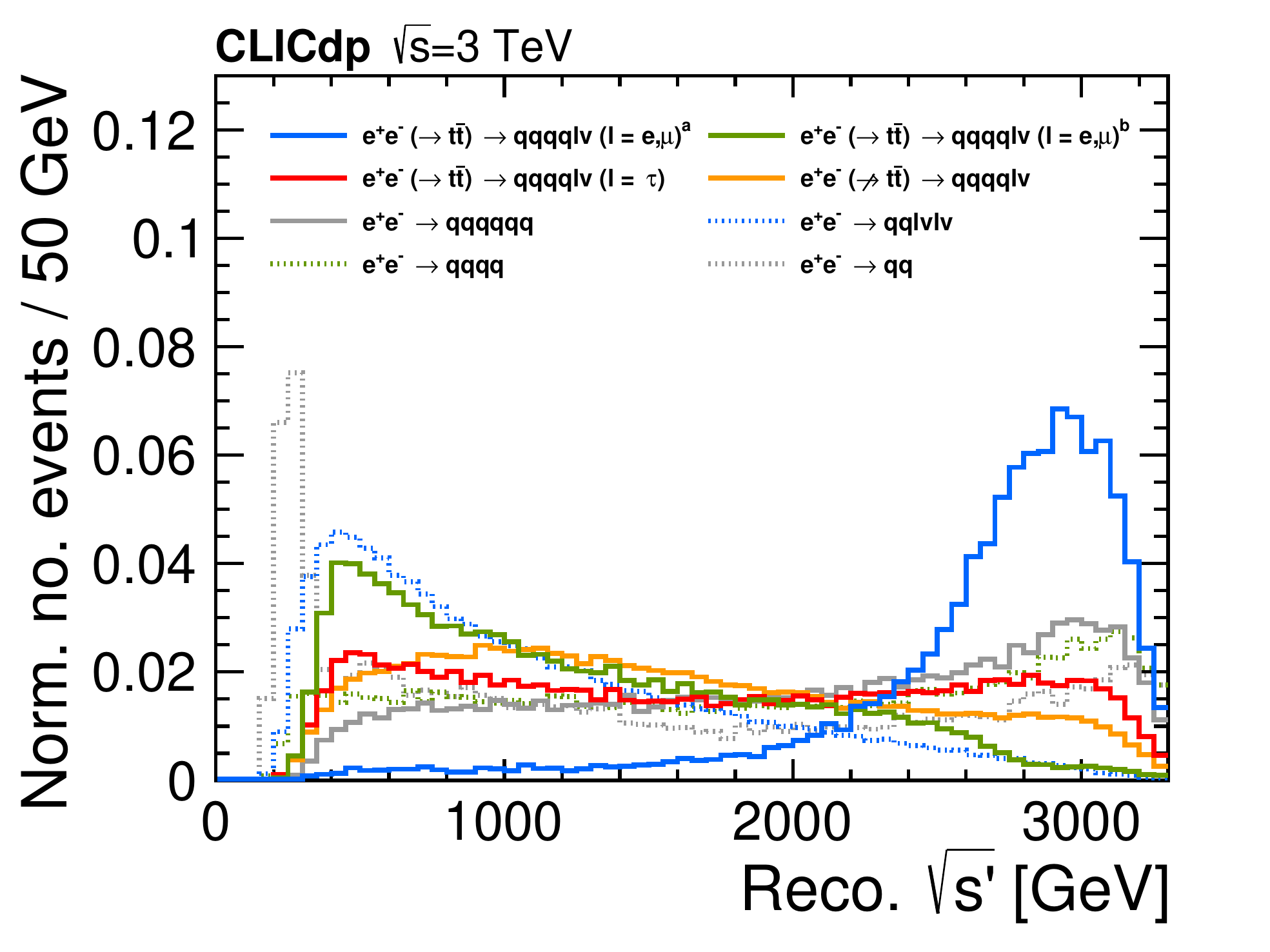}
\caption{Reconstructed centre-of-mass energy for the signal and backgrounds considered in the analysis, for operation at $\roots=1.4\,\tev$ (left) and $\roots=3\,\tev$ (right). The superscript `a' (`b') refers to the kinematic region $\rootsprime\geq1.2\,\tev$ ($\rootsprime<1.2\,\tev$). \label{fig:analysis:effcom:recodist}}
\end{figure}

A comparison of the generated and reconstructed distributions for signal $\ttbar$ events is shown in \Cref{fig:analysis:effcom:distcomp}. \Cref{fig:analysis:effcom:recodist} shows the reconstructed centre-of-mass energy individually for all the signal and background samples considered in the analysis. A cut is applied at 1.2\,\tev (2.6\,\tev) for operation at 1.4\,\tev (3\,\tev) ($\sim$85\% of $\roots$), corresponding to the kinematic region of the signal. For $\roots=1.4\,\tev$ this corresponds to a retention of about 36\% of the signal events while for the highest energy stage the corresponding number is about 18\%.

\clearpage
\section{Multivariate analysis}
\label{sec:mva}

Events with one isolated charged lepton in association with one large-R top-tagged jet, no isolated high-energy photons and with a reconstructed centre-of-mass energy $\geq1.2\,\tev$ ($\geq2.6\,\tev$) for operation at 1.4 TeV (3 TeV) are analysed using multivariate classification algorithms based on BDTs that simultaneously analyse all input variables in a multi-dimensional space, producing a final score for each event indicating whether it is signal- or background-like. The MVA training is performed using Scikit-learn~\cite{scikit-learn} with the AdaBoost-SAMME algorithm~\cite{adaboost}.

Due to the large variety of the background processes considered two initial MVAs are trained each focussing on distinguishing the signal events from a certain class of background events: the first MVA is trained to distinguish the signal from backgrounds with two quarks and either 0, 1, or 2 charged leptons, while the second MVA is trained to distinguish the signal from the fully-hadronic four-quark and six-quark jet backgrounds. The final MVA considers all backgrounds and takes the resulting classification score from the two initial MVAs as input. Separate algorithms are trained and applied for the 1.4\,\tev and 3\,\tev samples and for the two different polarisations considered. The following sections describe the input variables, training, and results of the MVA algorithms. 

\subsection{Input variables}
The MVAs initially consider a large number (66) of variables as input, whereas the final training for each MVA is performed only on the 20 variables with the largest separation between signal and background. The variables considered describe the kinematics of both the hadronically and leptonically decaying top quarks, total event $E_\mathrm{T}$, event missing $\pT$, visible energy, event shape, kinematics of the identified isolated lepton, flavour tagging information, jet splitting scales, and the jet substructure. The variables are described in more detail in the list below. The variables with the strongest separation power are displayed in the figures below and in the Appendix. In general, we show the variables for the sample at a nominal collision energy of $1.4\,\tev$. The corresponding figures for the sample at $3\,\tev$ are only included for cases where a significantly different behaviour is observed compared to the $1.4\,\tev$ sample. In general, we observe a slightly worse separation for the $3\,\tev$ sample. This is expected since the separation of the individual jet constituents decreases for a higher boost.

\paragraph{List of MVA variables}

\begin{itemize}

\item Large-R jet variables:
\begin{itemize}
\item Energy of the leading jet, see \Cref{fig:analysis:mva:variables:jetE},
\item \pT of the leading jet,
\item Energy of the next-to-leading jet, see \Cref{fig:analysis:mva:variables:jetE},
\item \pT of the next-to-leading jet,
\item Invariant mass of the total large-R jet system ($\mathrm{m}_{\mathrm{j1,j2}}\,(\mathrm{N}_{\mathrm{j}}=2)$), see \Cref{fig:analysis:mva:variables:invmass},
\end{itemize}
\begin{figure}
	\centering
	\includegraphics[width=0.48\columnwidth]{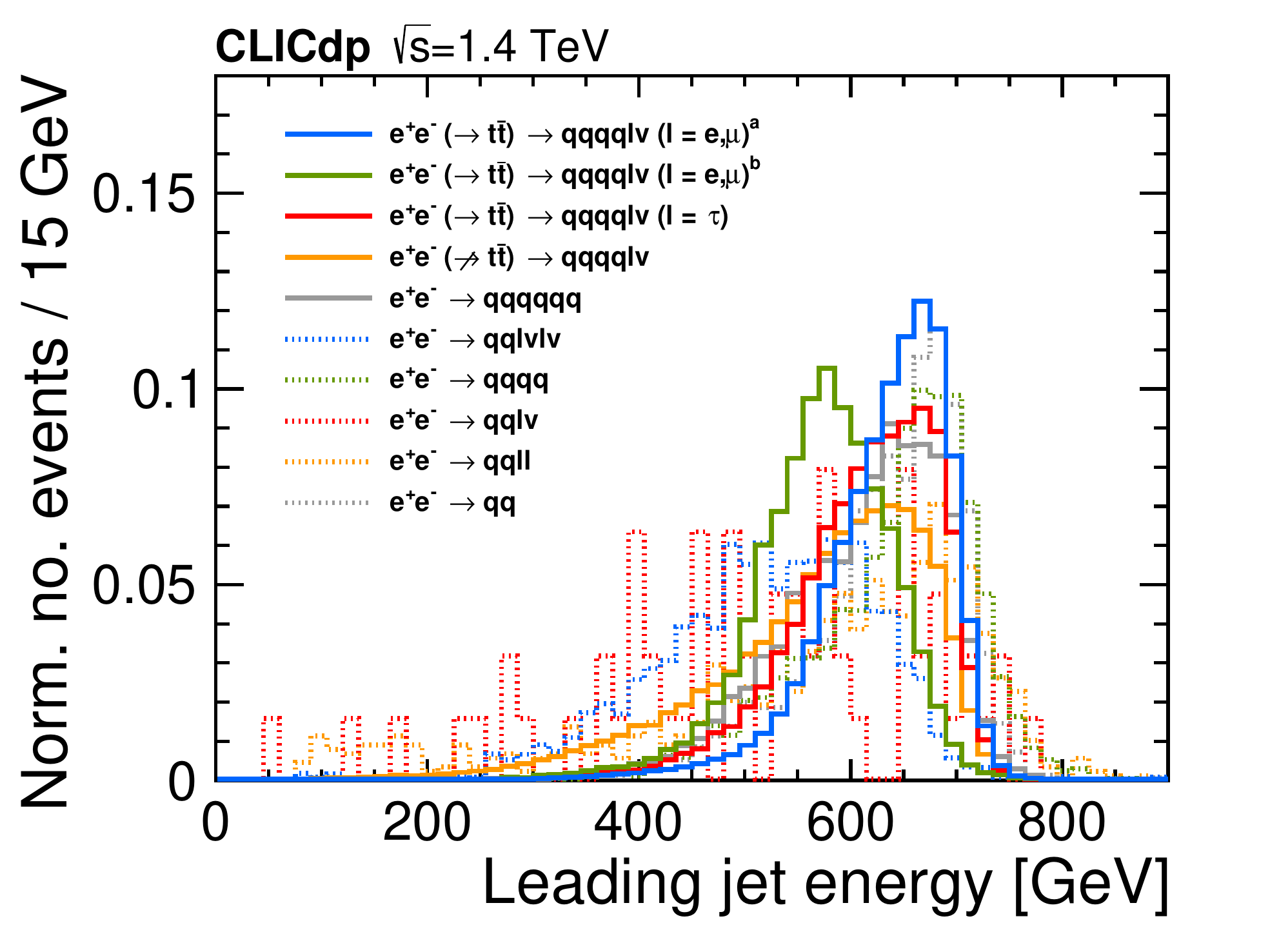}
	~~~
	\includegraphics[width=0.48\columnwidth]{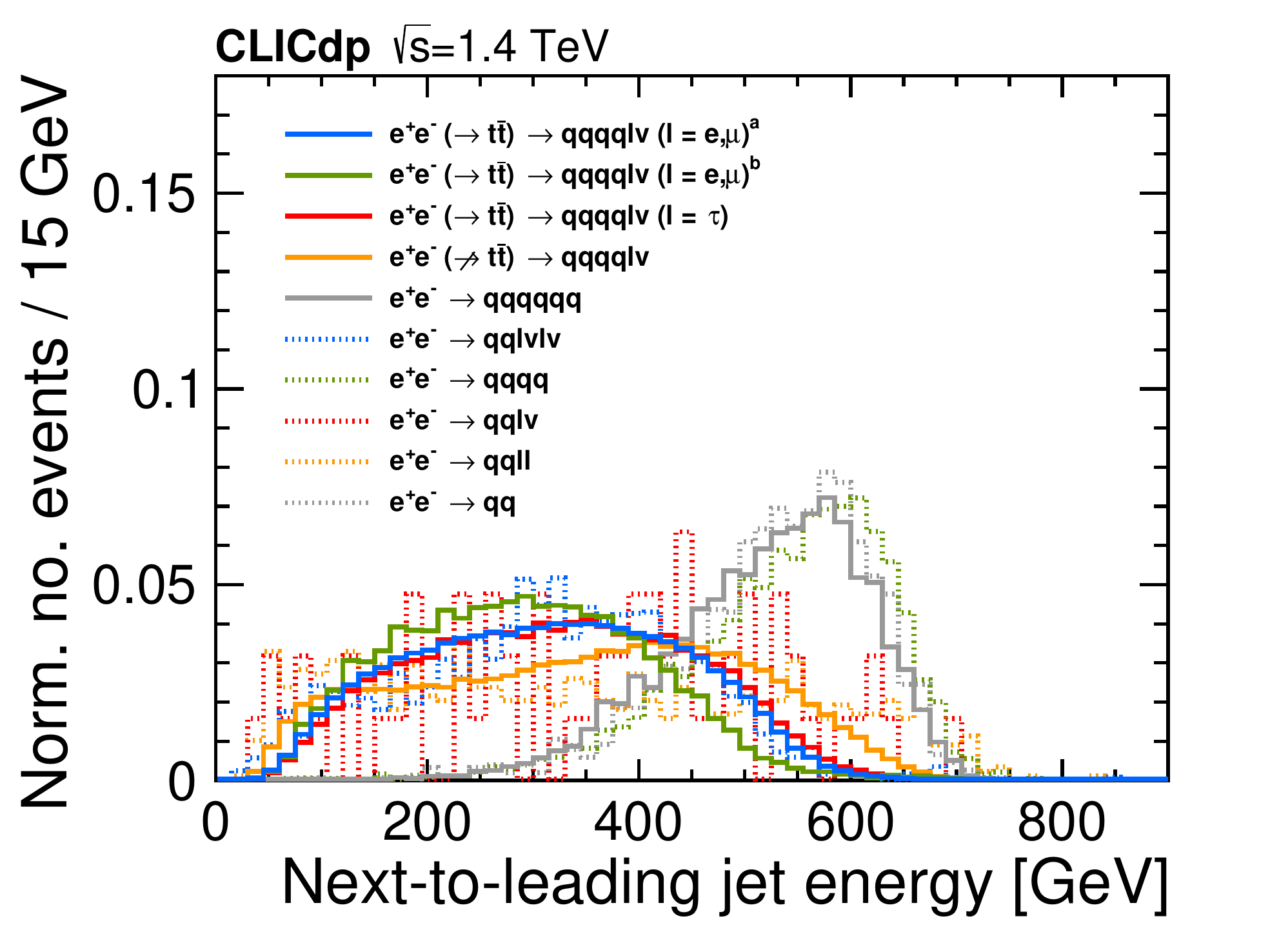}
\caption{Energy of the leading (left) and next-to-leading (right) large-R jets for operation at $\roots=\,1.4\tev$. The distribution is shown after the application of pre-cuts. The superscript `a' (`b') refers to the kinematic region $\rootsprime\geq1.2\,\tev$ ($\rootsprime<1.2\,\tev$). \label{fig:analysis:mva:variables:jetE}}
\end{figure}

\begin{figure}
	\centering
	\includegraphics[width=0.48\columnwidth]{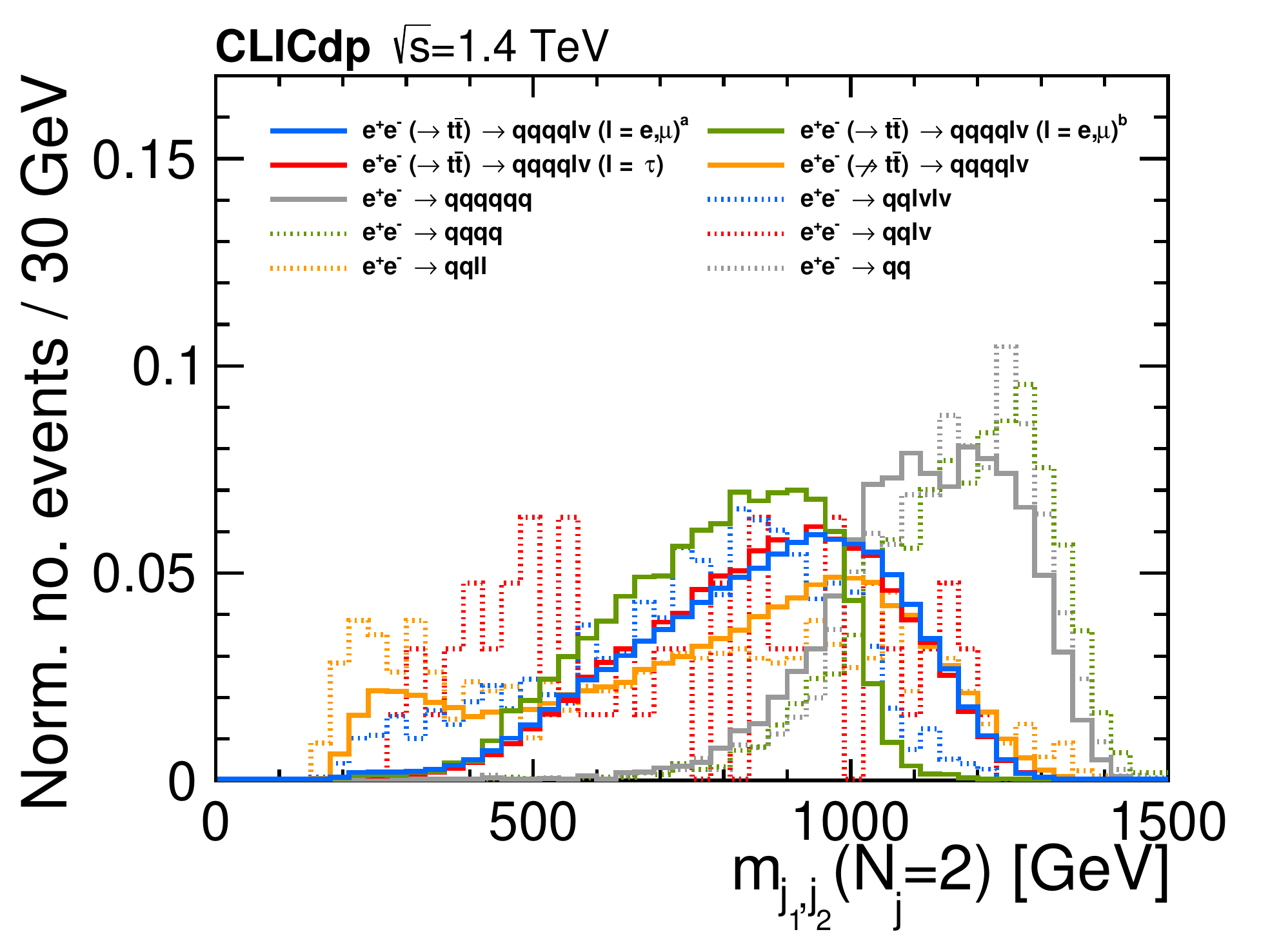}
	~~~
	\includegraphics[width=0.48\columnwidth]{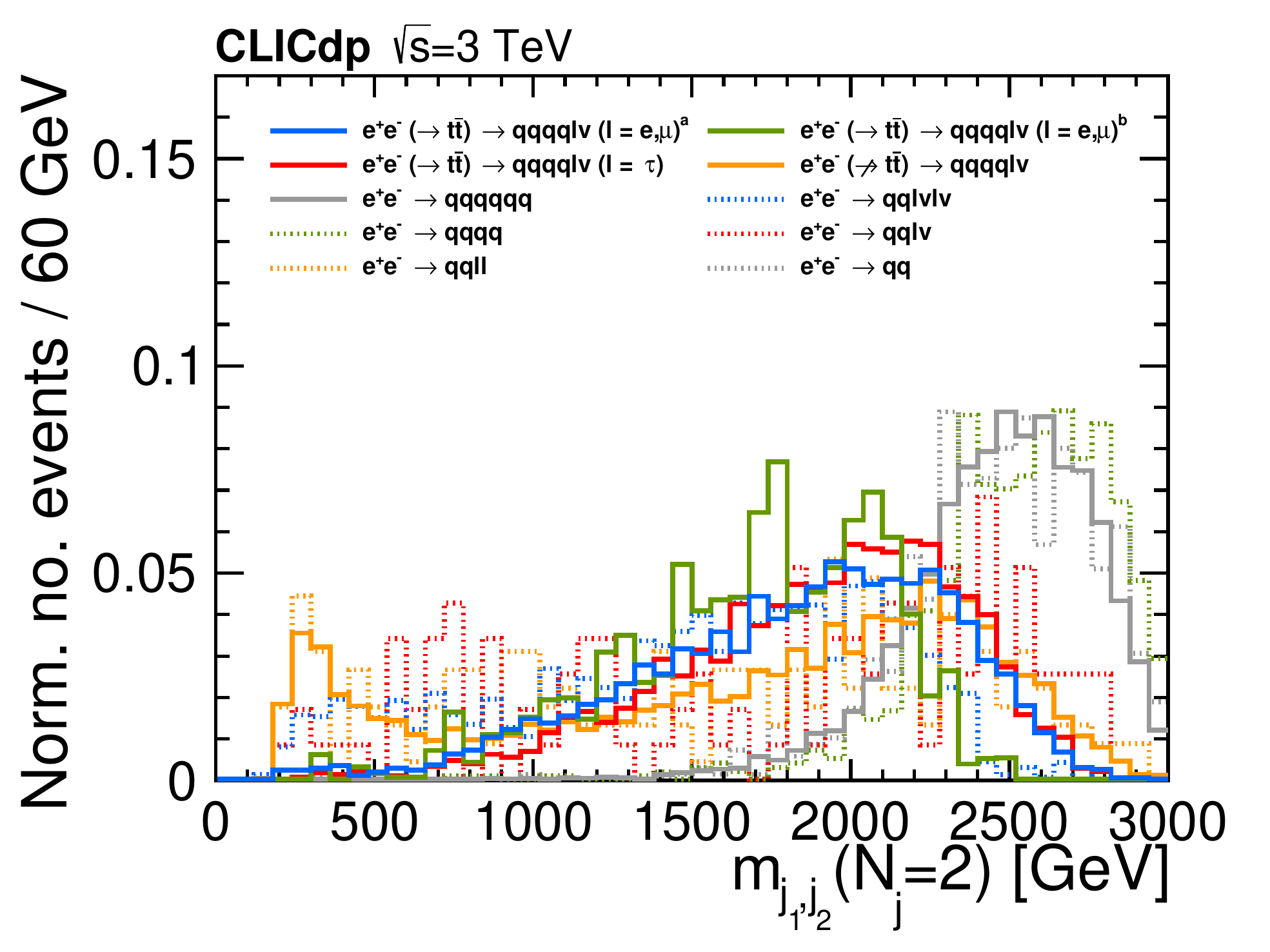}
\caption{Invariant mass of the total large-R jet system, for operation at $\roots=\,1.4\tev$ (left) and $\roots=\,3\tev$ (right). The distributions are shown after the application of pre-cuts. The superscript `a' (`b') refers to the kinematic region $\rootsprime\geq1.2\,\tev$ ($\rootsprime<1.2\,\tev$). \label{fig:analysis:mva:variables:invmass}}
\end{figure}

\item Kinematics of the hadronically decaying top-quark jet as identified by the top-quark tagger:
\begin{itemize}
\item Invariant mass of the top-quark candidate jet ($\mathrm{m}_{\PQt}$), see \Cref{fig:analysis:mva:variables:topMass},
\item Energy of the top-quark candidate jet ($E_{\PQt}$),
\item \pT of the top-quark candidate jet ($\pT^{\PQt}$),
\item Invariant mass of the W boson candidate sub-jet ($m_{\PW}$), see \Cref{fig:analysis:mva:variables:WMass},
\item Energy of the W boson candidate sub-jet ($E_{\PW}$),
\item \pT of the W boson candidate sub-jet ($\pT^{\PW}$),
\item Helicity angle $\theta_{\mathrm{W}}$, measured in the rest frame of the reconstructed $\PW$ boson and defined as the opening angle of the top quark to the softer of the two $\PW$ boson decay subjets. Note that too shallow an angle would be an indication of a false splitting, where one of the pairs of subjets produces a small mass compatible with QCD-like emission. See figure \Cref{fig:analysis:mva:variables:helicityangle}.
\end{itemize}
\begin{figure}
	\centering
	\includegraphics[width=0.48\columnwidth]{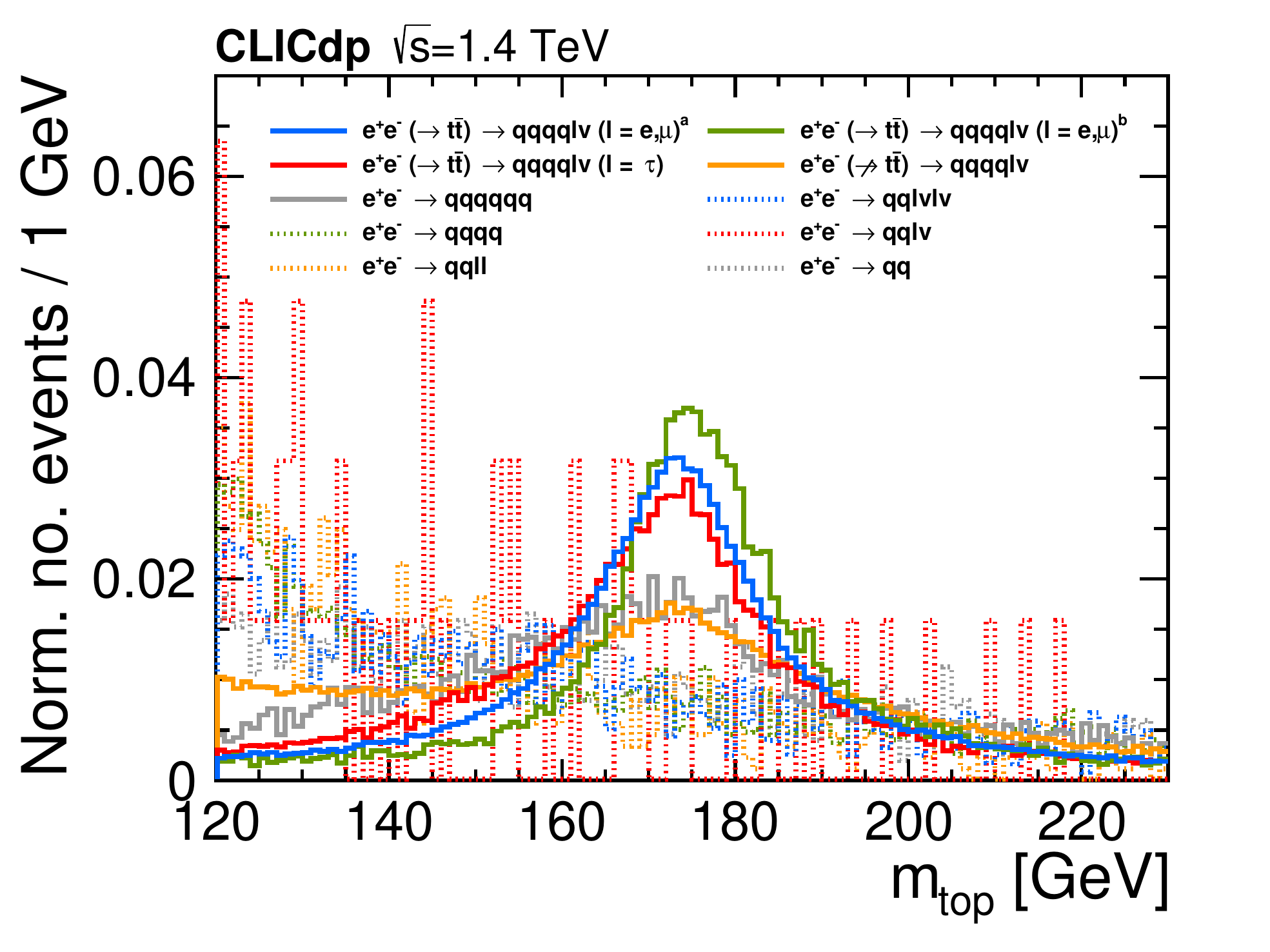}
	~~~
	\includegraphics[width=0.48\columnwidth]{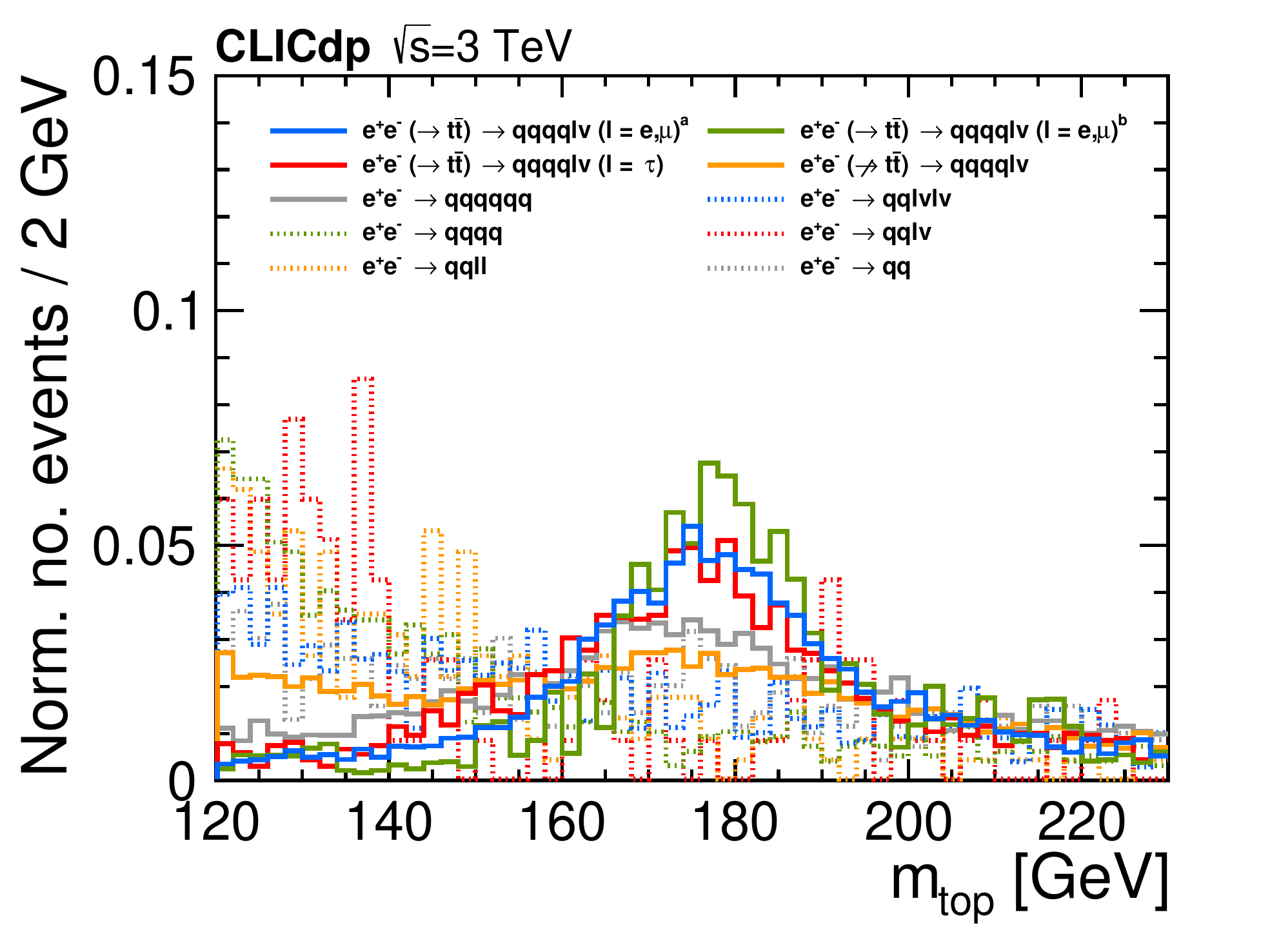}
\caption{Invariant mass of the top-quark candidate jet as identified by the top-quark tagger, for operation at $\roots=\,1.4\tev$ (left) and $\roots=\,3\tev$ (right). The distributions are shown after the application of pre-cuts. The superscript `a' (`b') refers to the kinematic region $\rootsprime\geq1.2\,\tev$ ($\rootsprime<1.2\,\tev$). \label{fig:analysis:mva:variables:topMass}}
\end{figure}
\begin{figure}
	\centering
	\includegraphics[width=0.48\columnwidth]{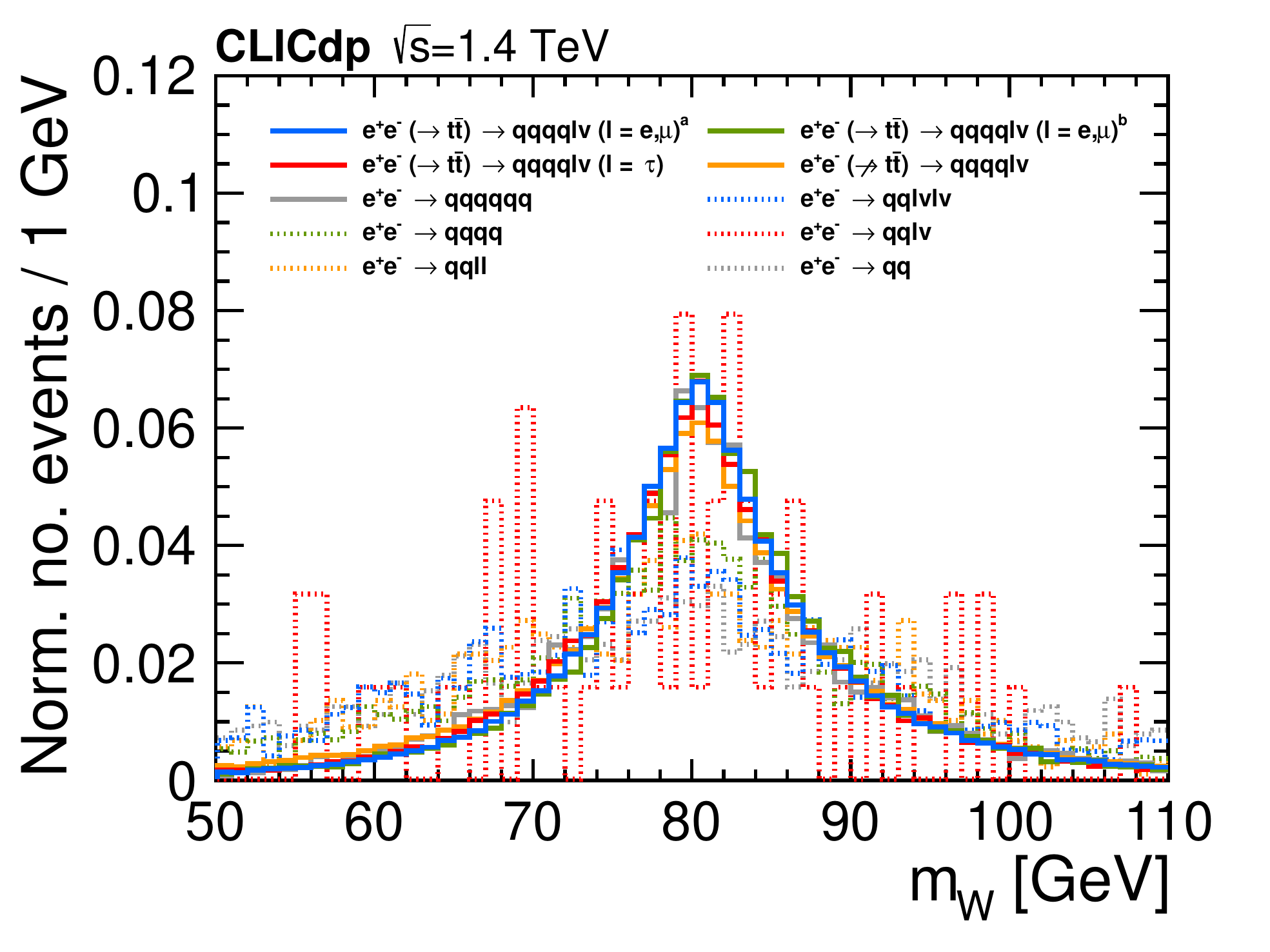}
	~~~
	\includegraphics[width=0.48\columnwidth]{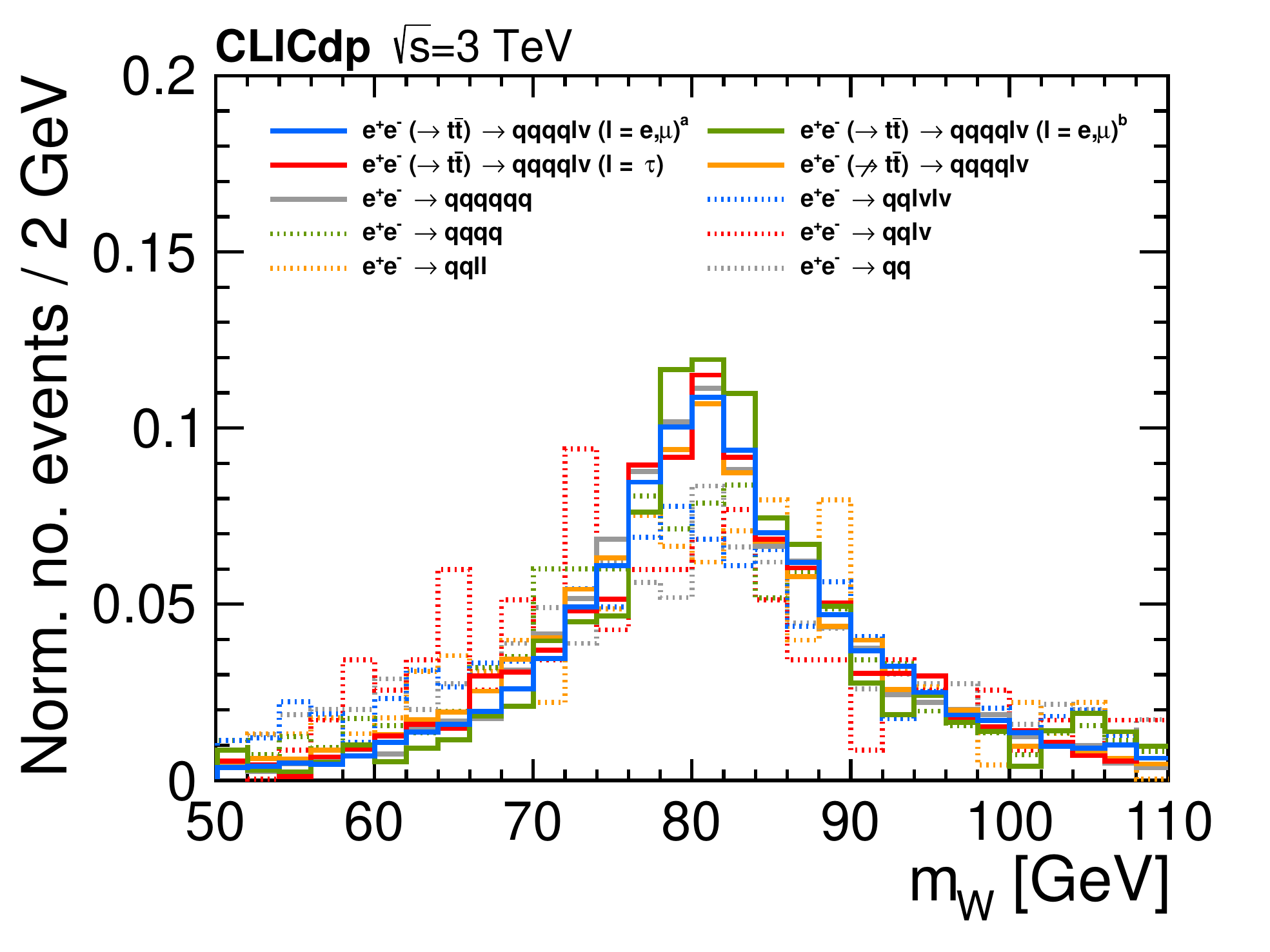}
\caption{Invariant mass of the W-boson candidate jet (hadronically decaying) as identified by the top-quark tagger, for operation at $\roots=\,1.4\tev$ (left) and $\roots=\,3\tev$ (right). The distributions are shown after the application of pre-cuts. The superscript `a' (`b') refers to the kinematic region $\rootsprime\geq1.2\,\tev$ ($\rootsprime<1.2\,\tev$). \label{fig:analysis:mva:variables:WMass}}
\end{figure}
\begin{figure}
	\centering
	\includegraphics[width=0.48\columnwidth]{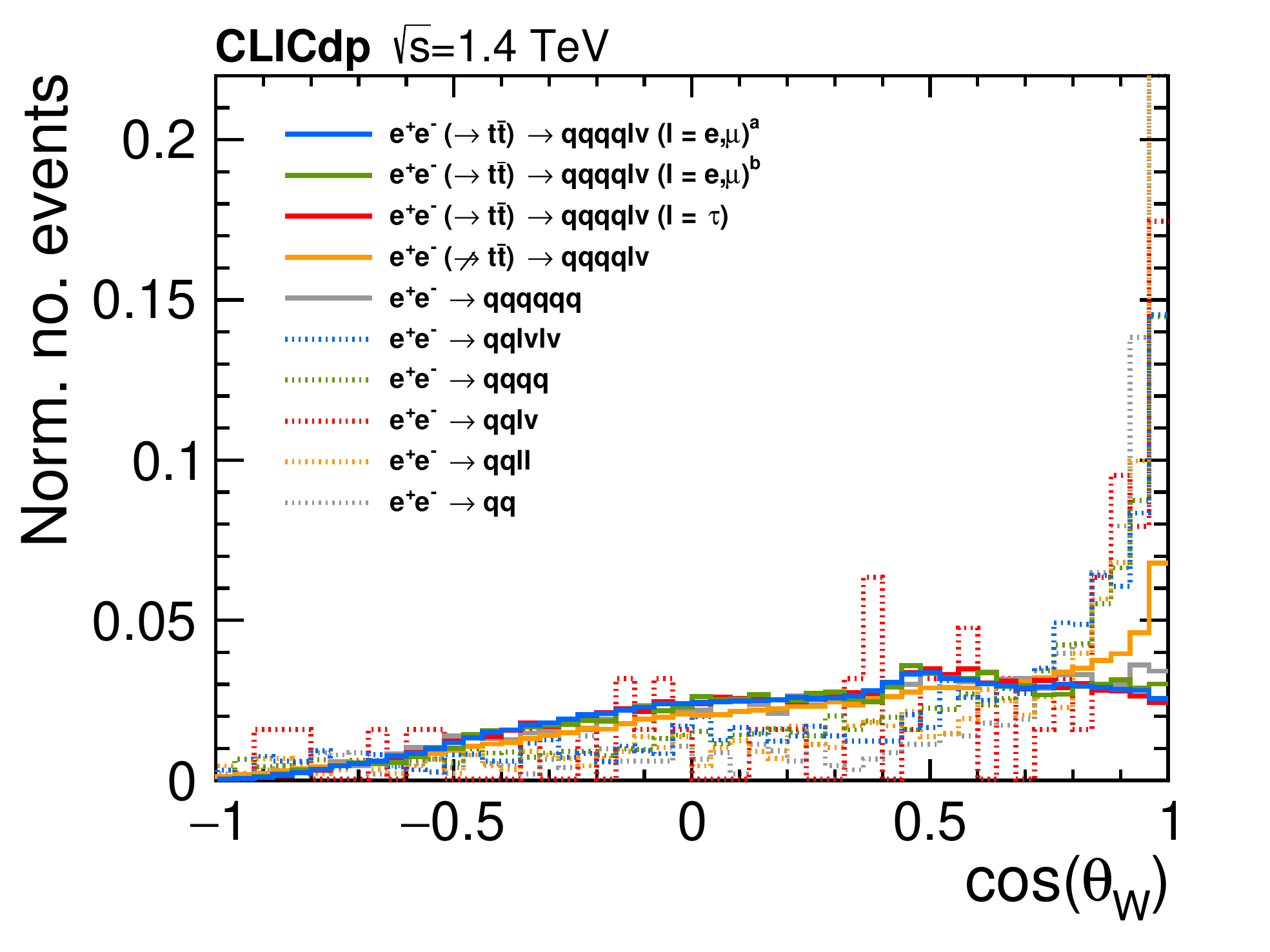}
	~~~
	\includegraphics[width=0.48\columnwidth]{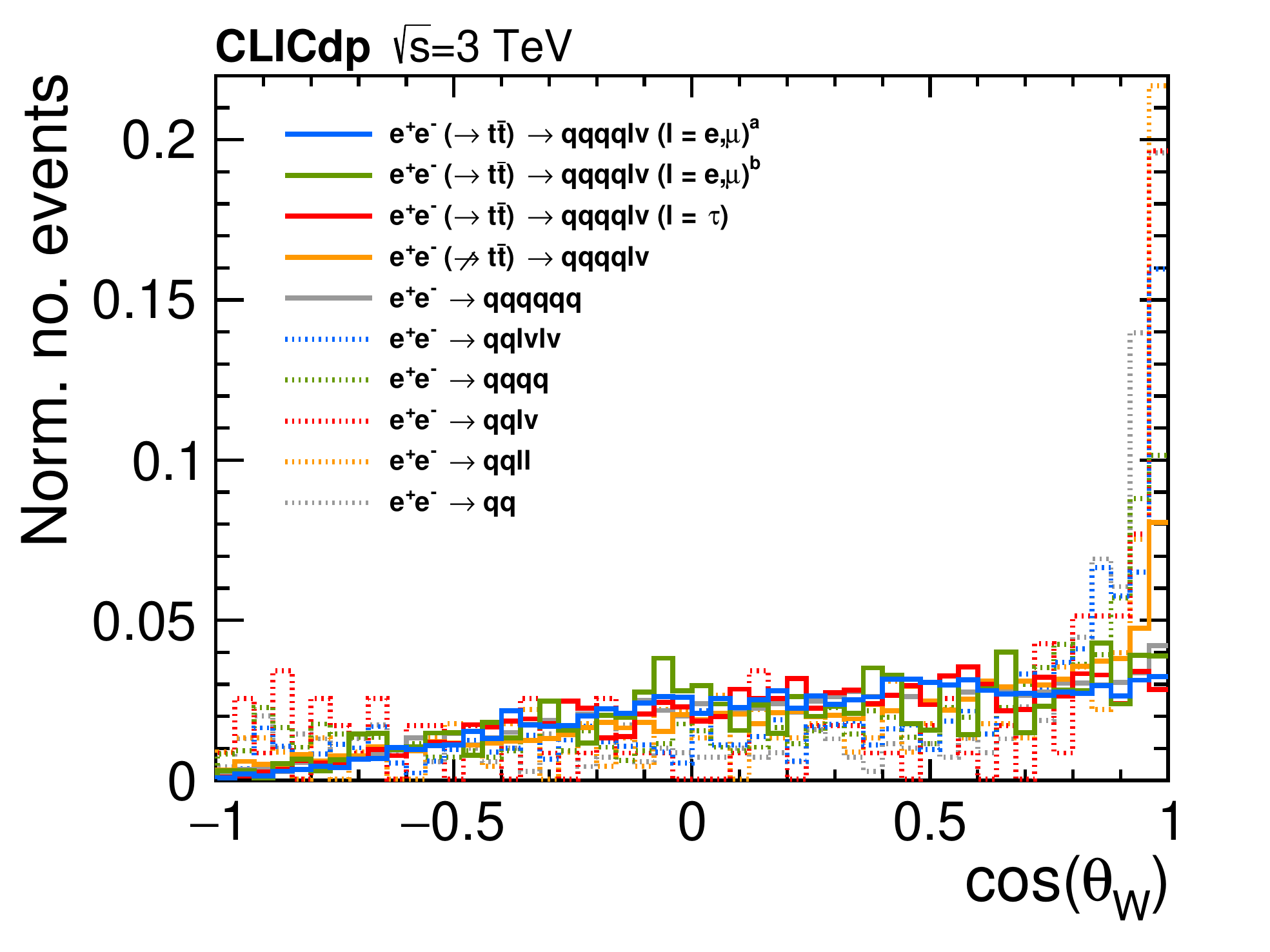}	
\caption{Top-quark decay helicity angle, $\theta_{\mathrm{W}}$, for operation at $\roots=\,1.4\tev$ (left) and $\roots=\,3\tev$ (right). The distributions are shown after the application of pre-cuts. The superscript `a' (`b') refers to the kinematic region $\rootsprime\geq1.2\,\tev$ ($\rootsprime<1.2\,\tev$). \label{fig:analysis:mva:variables:helicityangle}}
\end{figure}

\item Kinematics of the leptonically decaying top-quark (reconstructed by the method described in \Cref{sec:effcom}):
\begin{itemize}
\item Reconstructed effective centre-of-mass energy, see \Cref{fig:analysis:effcom:recodist},
\item Invariant mass of the leptonically decaying top-quark, see \Cref{fig:analysis:mva:variables:semilepmass},
\item Energy of the leptonically decaying top-quark,
\item $\pT$ of the leptonically decaying top-quark,
\item $p_{z}$ (z-component of the momentum) of the leptonically decaying top-quark,
\end{itemize}
\begin{figure}
	\centering
	\includegraphics[width=0.48\columnwidth]{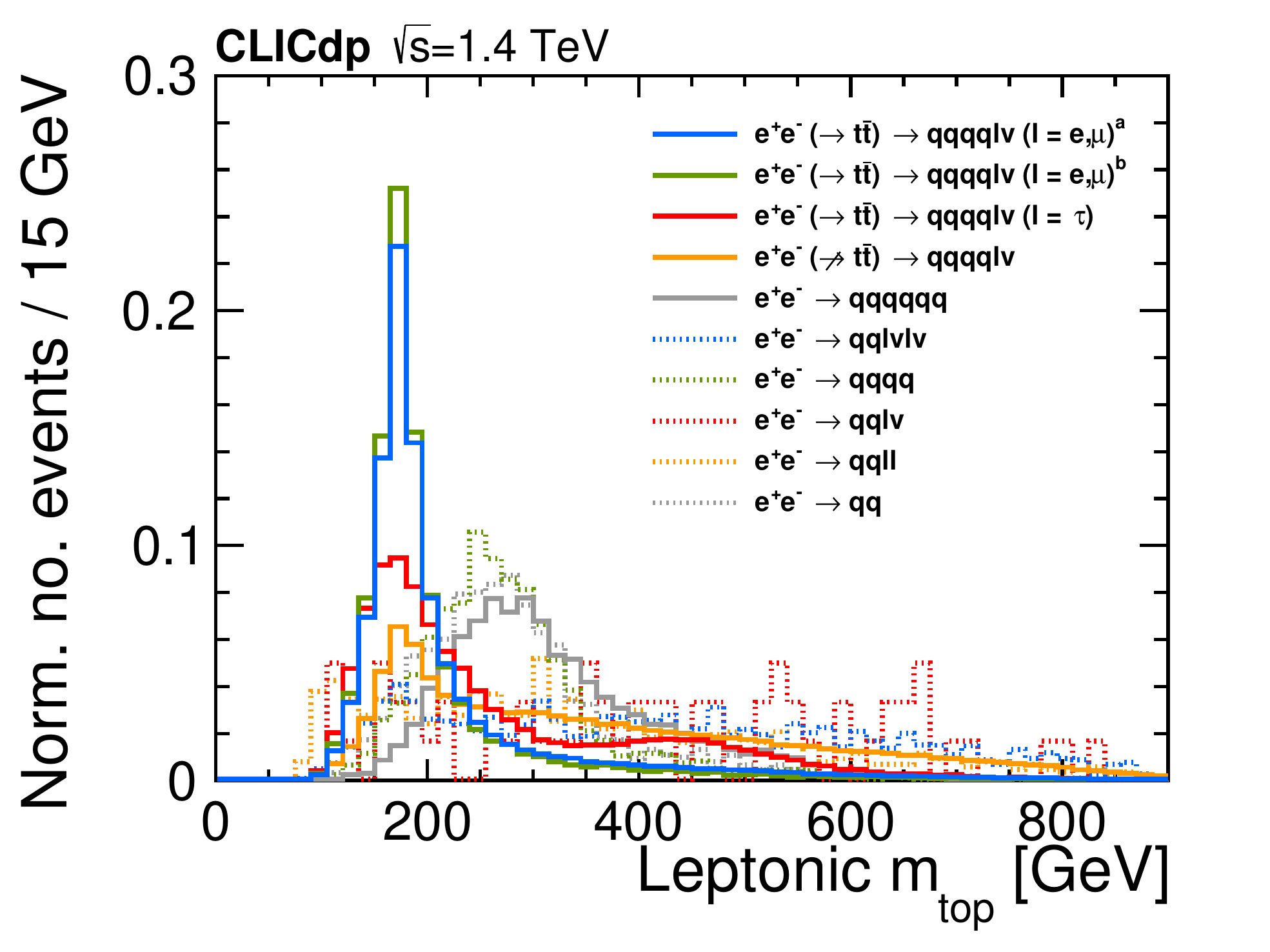}
	~~~
	\includegraphics[width=0.48\columnwidth]{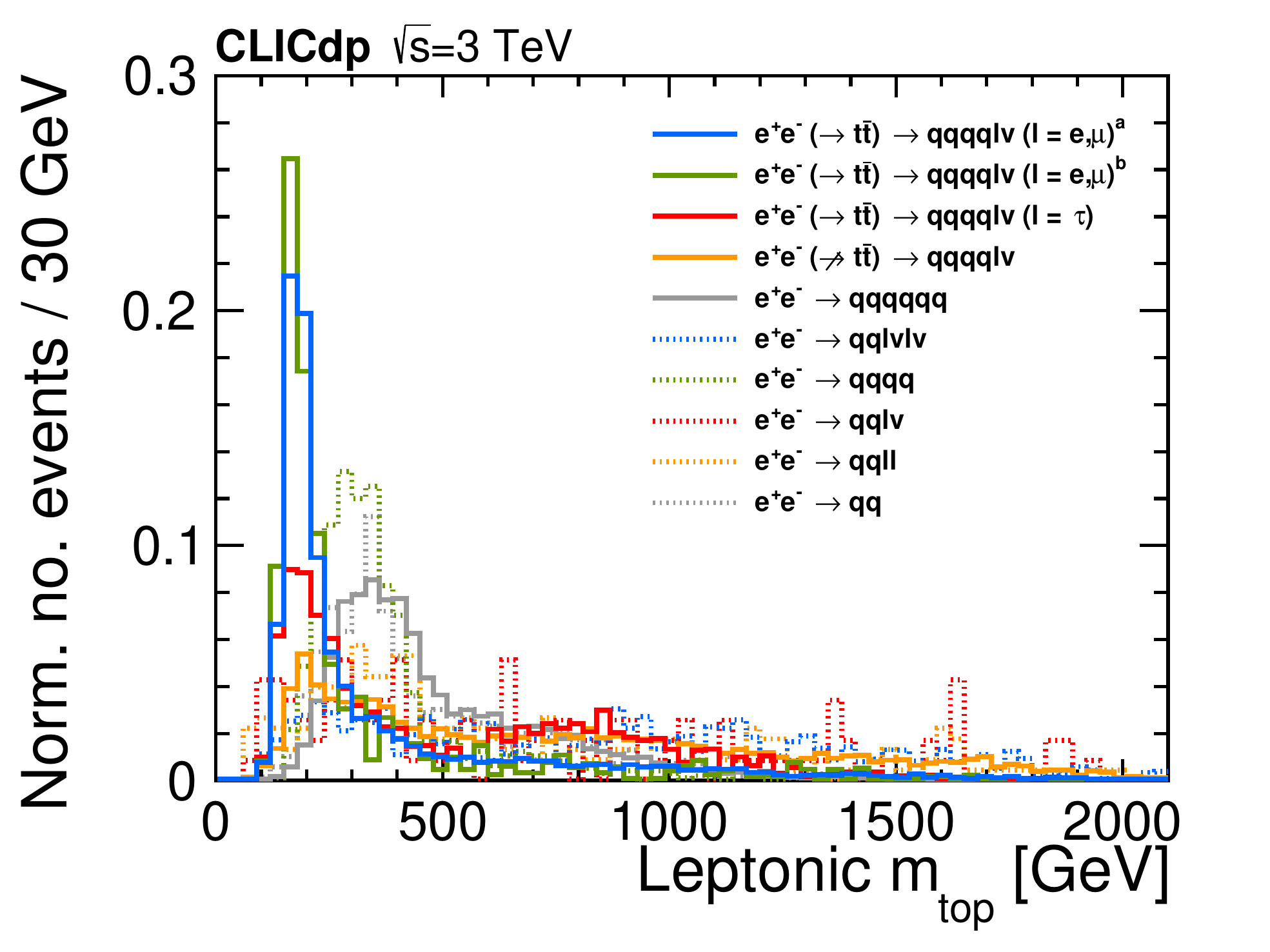}
\caption{Invariant mass of the leptonically decaying top-quark reconstructed by the method described in \Cref{sec:effcom}, for operation at $\roots=\,1.4\tev$ (left) and $\roots=\,3\tev$ (right). The distributions are shown after the application of pre-cuts. The superscript `a' (`b') refers to the kinematic region $\rootsprime\geq1.2\,\tev$ ($\rootsprime<1.2\,\tev$).  \label{fig:analysis:mva:variables:semilepmass}}
\end{figure}

\item Sub-structure variables for the individual large-R jets (see definitions in \Cref{sec:substructure}):
\begin{itemize}
\item N-subjettiness $\tau_{32}$, $\tau_{21}$, and $\tau_{31}$ for the leading large-R jet, see \Cref{fig:analysis:mva:variables:NsubjettinessJ1} ($\roots=1.4\,\tev$) and \Cref{fig:analysis:mva:variables:NsubjettinessJ1:3tev} ($\roots=3\,\tev$),
\item N-subjettiness, $\tau_{32}$, $\tau_{21}$, and $\tau_{31}$ for the next-to-leading large-R jet, see \Cref{fig:analysis:mva:variables:NsubjettinessJ2} ($\roots=1.4\,\tev$) and \Cref{fig:analysis:mva:variables:NsubjettinessJ2:3tev} ($\roots=3\,\tev$),
\item Energy correlation functions $\mathrm{C}_{2}$ and $\mathrm{C}_{3}$ for the leading large-R jet, see \Cref{fig:analysis:mva:variables:energycorrCJ1} ($\roots=1.4\,\tev$) and \Cref{fig:analysis:mva:variables:energycorrCJ1:3tev} ($\roots=3\,\tev$),
\item Energy correlation functions $\mathrm{C}_{2}$ and $\mathrm{C}_{3}$ for the next-to-leading large-R jet, see \Cref{fig:analysis:mva:variables:energycorrCJ2} ($\roots=1.4\,\tev$) and \Cref{fig:analysis:mva:variables:energycorrCJ2:3tev} ($\roots=3\,\tev$),
\item Energy correlation functions $\mathrm{D}_{2}$ and $\mathrm{D}_{3}$ for the leading large-R jet, see \Cref{fig:analysis:mva:variables:energycorrDJ1} ($\roots=1.4\,\tev$) and \Cref{fig:analysis:mva:variables:energycorrDJ1:3tev} ($\roots=3\,\tev$),
\item Energy correlation functions $\mathrm{D}_{2}$ and $\mathrm{D}_{3}$ for the next-to-leading large-R jet, see \Cref{fig:analysis:mva:variables:energycorrDJ2} ($\roots=1.4\,\tev$) and \Cref{fig:analysis:mva:variables:energycorrDJ2:3tev} ($\roots=3\,\tev$).
\end{itemize}

\item Flavour tagging of the individual large-R jets:
\begin{itemize}
\item Sum of b-tags, see \Cref{fig:analysis:mva:variables:flavourtagging},
\item Sum of c-tags, see \Cref{fig:analysis:mva:variables:flavourtagging},
\item Sum of c-tag/(b-tag+c-tag) ratio,
\item Highest b-tag,
\item Lowest b-tag,
\item Leading jet b-tag,
\item Leading jet c-tag,
\item Leading jet c-tag/(b-tag+c-tag) ratio,
\item Next-to-leading jet b-tag,
\item Next-to-leading jet c-tag,
\item Next-to-leading jet c-tag/(b-tag+c-tag) ratio,
\item Invariant mass of jets with a b-tag above 0.9.
\end{itemize}

\item Jet splitting scales, $d_{ij}$, defined as the jet clustering distance parameters for the last merging steps (going from j to i jets) in the exclusive large-$R$ jet clustering (see \Cref{fig:analysis:mva:variables:jetsplitting}):
\begin{itemize}
\item $d_{23}$, 
\item $d_{34}$,
\item $d_{45}$,
\item $d_{56}$,
\end{itemize}

\item Kinematics of the identified isolated lepton:
\begin{itemize}
\item Energy of the isolated lepton, see \Cref{fig:analysis:mva:variables:lepE},
\item \pT of the the isolated lepton,
\end{itemize}
\begin{figure}
	\centering
	\includegraphics[width=0.48\columnwidth]{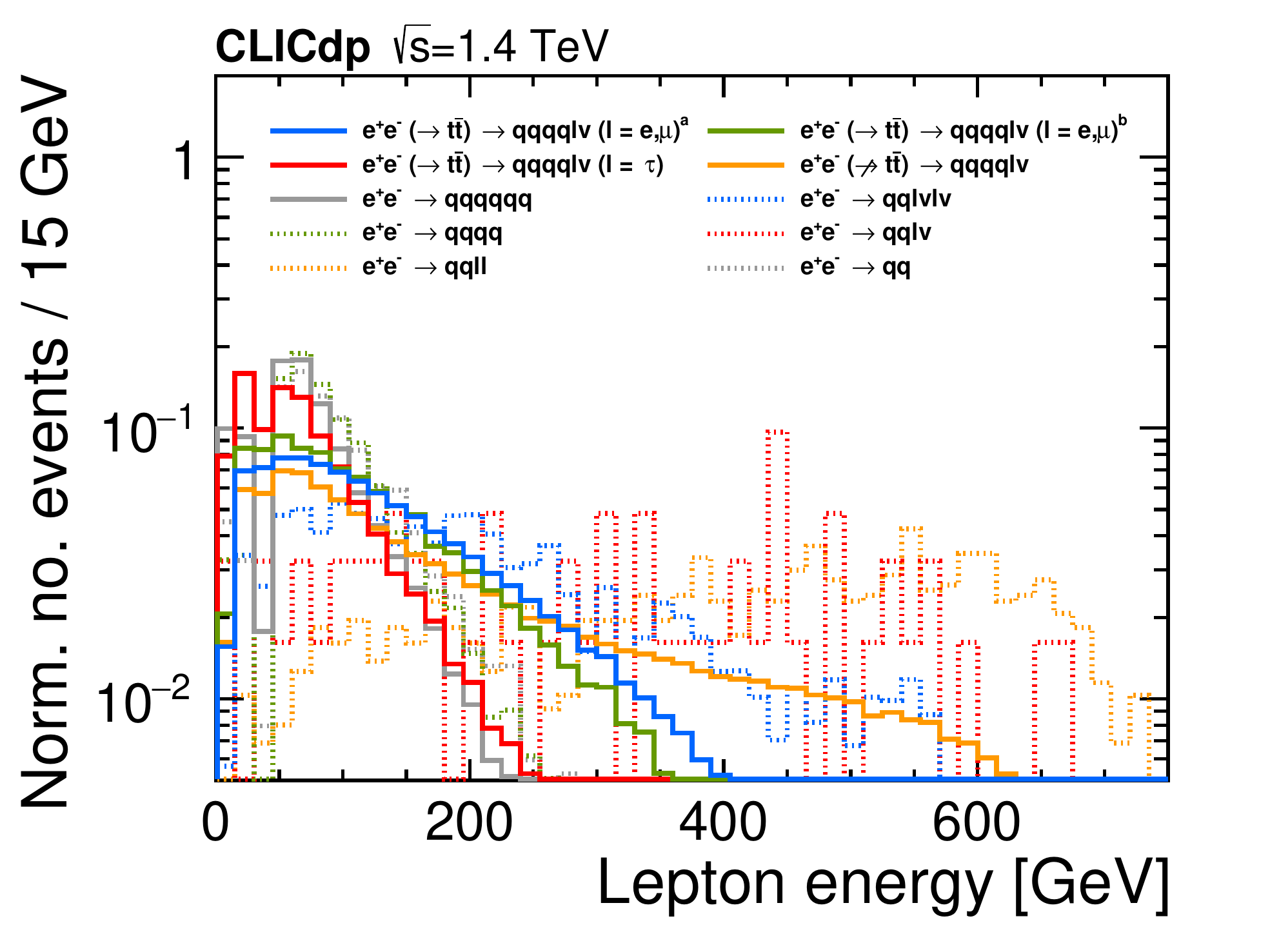}
	\caption{Energy of the identified isolated lepton, for operation at $\roots=\,1.4\tev$. The distribution is shown after the application of pre-cuts. The superscript `a' (`b') refers to the kinematic region $\rootsprime\geq1.2\,\tev$ ($\rootsprime<1.2\,\tev$). \label{fig:analysis:mva:variables:lepE}}
\end{figure}

\item Event/jet shape variables:
\begin{itemize}
\item Thrust of all PFOs (see \Cref{fig:analysis:mva:variables:eventshape}), leading jet, and next-to-leading jet,
\item Oblateness of all PFOs (see \Cref{fig:analysis:mva:variables:eventshape}), leading jet, and next-to-leading jet,
\item TSphericity of all PFOs (see \Cref{fig:analysis:mva:variables:eventshape}), leading jet, and next-to-leading jet,
\item Aplanarity of all PFOs (see \Cref{fig:analysis:mva:variables:eventshape}), leading jet, and next-to-leading jet,
\end{itemize}

\item Miscellaneous:
\begin{itemize}
\item Event $E_{\mathrm{T}}$,
\item Event visible energy $E_{\mathrm{visible}}$,
\item Event missing $\pT$,
\item Polar angle of the PFO with highest $\pT$ ($\cos(\theta_{\mathrm{max},\pT})$), see \Cref{fig:analysis:mva:variables:misc2},
\item Invariant mass of the two jets with highest $\pT$ (The event is clustered into four exclusive jets, using the VLC algorithm with a radius of 0.6).
\end{itemize}
\begin{figure}
	\centering
	\includegraphics[width=0.48\columnwidth]{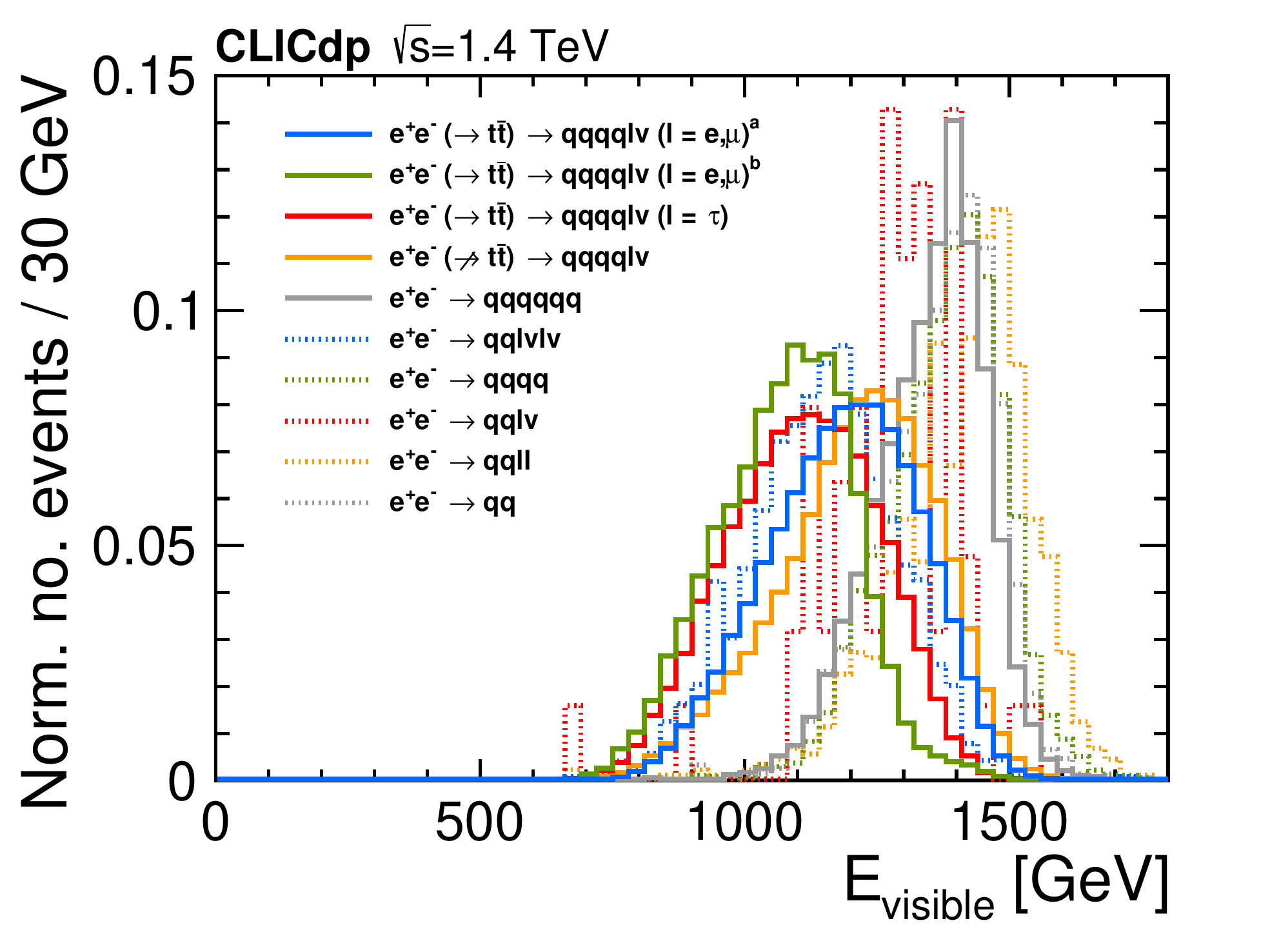}
	~~~
	\includegraphics[width=0.48\columnwidth]{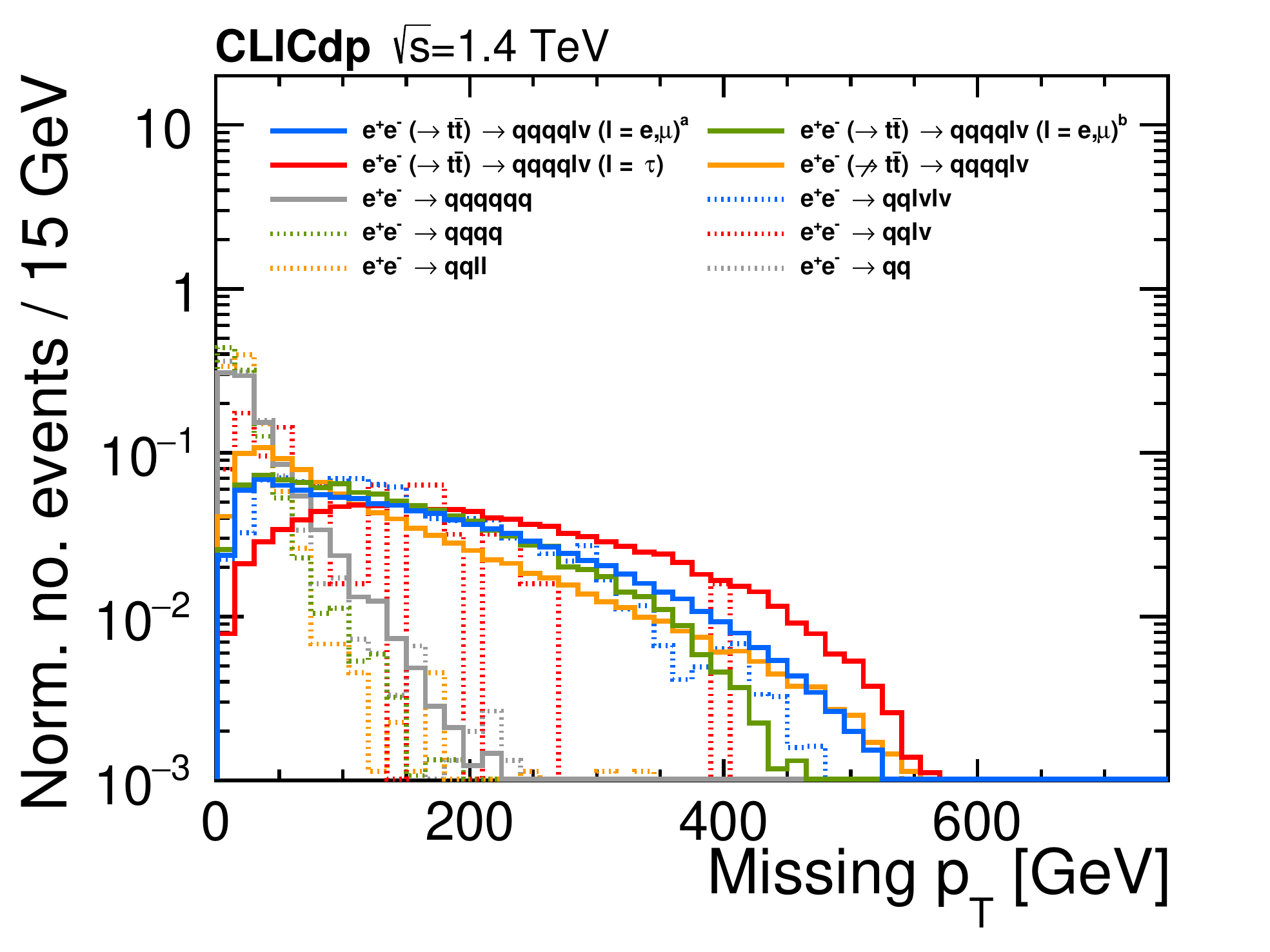}
\caption{Event visible energy (left) and missing $\pT$ (right), for $\roots=1.4\,\tev$. The distributions are shown after the application of pre-cuts. The superscript `a' (`b') refers to the kinematic region $\rootsprime\geq1.2\,\tev$ ($\rootsprime<1.2\,\tev$). \label{fig:analysis:mva:variables:misc1}}
\end{figure}
\begin{figure}
	\centering
	\includegraphics[width=0.48\columnwidth]{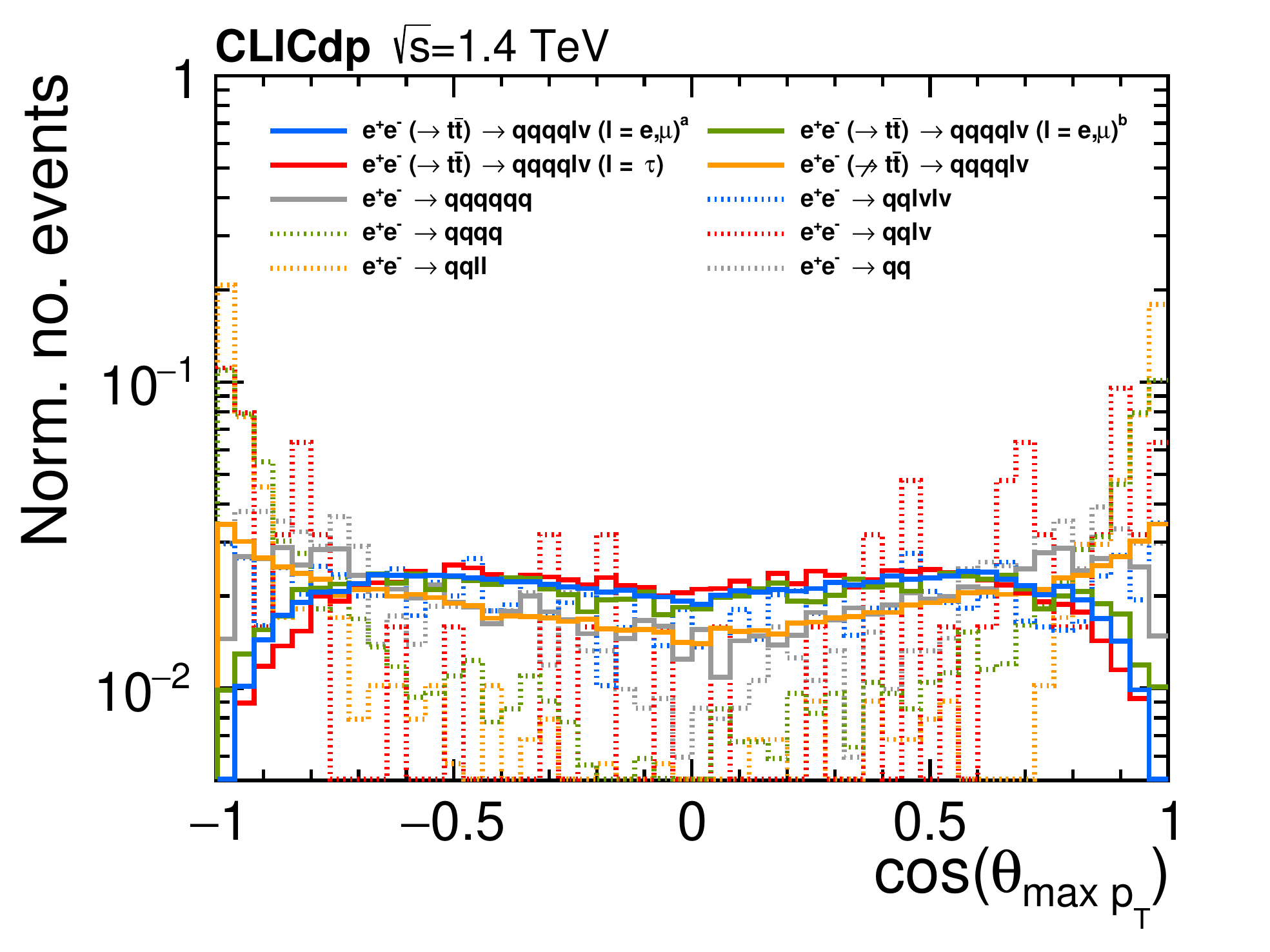}
\caption{Polar angle, $\theta$, of the PFO with highest $\pT$, for $\roots=1.4\,\tev$. The distribution is shown after the application of pre-cuts. The superscript `a' (`b') refers to the kinematic region $\rootsprime\geq1.2\,\tev$ ($\rootsprime<1.2\,\tev$). \label{fig:analysis:mva:variables:misc2}}
\end{figure}

\end{itemize}

\subsection{Training}

While the initial training considers a large number of variables, the final classifier is re-trained on the 20 variables with the largest separation between signal and background. This is done to reduce the dimensionality of the phase space as a means to lower the risk for overtraining. The MVAs are trained and tested using dedicated signal and background samples that are excluded from the final analysis. Further, to optimise the parameters of the individual algorithms separate validation samples are derived from the training samples. The final parameters are tuned to reduce overtraining and a 3-fold cross validation splitting strategy is applied in order to reduce bias in the choice of training, test, and validation sets further. The latter is also helpful to retain high statistics while partitioning the available data as outlined above. The variables used for each final MVA in the different event samples are presented in \Cref{tab:mva:training:variables1} and \Cref{tab:mva:training:variables2}, where the columns marked 1st, 2nd, and 3rd indicate the variables used in the two initial MVAs and the final (3rd) MVA, respectively.

\begin{table}[p!]
\centering
\begin{minipage}{\columnwidth}
\centering
\resizebox{1.0\textwidth}{!}{
\begin{tabular}{lcc|cc}
\toprule
\vspace{1.0mm}
{} & \multicolumn{2}{c}{$\roots=1.4\,\tev$} & \multicolumn{2}{c}{$\roots=3\,\tev$} \\
\vspace{1.0mm}
P(\Pem) & -80\% & +80\% & -80\% & +80\% \\
\midrule
{} & 1st / 2nd / 3rd & 1st / 2nd / 3rd & 1st / 2nd / 3rd & 1st / 2nd / 3rd \\
\textbf{MVA score} & & & & \\
Score of 1st initial MVA & - / - / \checkmark & - / - / \checkmark & - / - / \checkmark & - / - / \checkmark \\
Score of 2nd initial MVA & - / - / \checkmark & - / - / \checkmark & - / - / \checkmark & - / - / \checkmark \\

\textbf{Large-R jet variables} & & & & \\
Leading jet $E$ & - / \checkmark / - & \checkmark / \checkmark / - & - / - / \checkmark & - / - / - \\
Leading jet $\pT$ & - / - / \checkmark & - / - / - & - / - / \checkmark & - / - / \checkmark \\
Next-to-leading jet $E$ & \checkmark / \checkmark / \checkmark & \checkmark / \checkmark / \checkmark & \checkmark / \checkmark / \checkmark & \checkmark / - / - \\
Next-to-leading jet $\pT$ & - / - / \checkmark & - / - / \checkmark & - / - / \checkmark & - / \checkmark / - \\
$m_{\mathrm{j1,j2}}$ & \checkmark / - / \checkmark & - / \checkmark / - & - / \checkmark / - & \checkmark / \checkmark / \checkmark \\

\textbf{Hadronic top} & & & & \\
$m_{\PQt}$ & \checkmark / \checkmark / - & \checkmark / \checkmark / \checkmark & \checkmark / \checkmark / - & \checkmark / \checkmark / \checkmark \\
$E_{\PQt}$ & \checkmark / \checkmark / \checkmark & - / \checkmark / \checkmark & \checkmark / \checkmark / - & \checkmark / - / \checkmark \\
$\pT^{\PQt}$ & - / - / - & - / - / \checkmark & \checkmark / - / - & - / - / - \\
$m_{\PW}$ & - / - / - & - / - / - & - / - / \checkmark & - / - / - \\
$E_{\PW}$ & - / - / - & - / - / - & - / - / \checkmark & - / - / - \\
$\pT^{\PW}$ & - / - / - & - / - / - & - / - / - & - / - / - \\
$\theta_{\mathrm{W}}$ & \checkmark / - / - & - / - / - & - / \checkmark / \checkmark & - / \checkmark / \checkmark \\

\textbf{Leptonic top} & & & & \\
$m$ & \checkmark / \checkmark / \checkmark & \checkmark / \checkmark / \checkmark & \checkmark / \checkmark / \checkmark & \checkmark / - / \checkmark \\
$E$ & - / \checkmark / \checkmark & \checkmark / \checkmark / \checkmark & - / \checkmark / - & - / \checkmark / \checkmark \\
$\pT$ & - / - / - & - / - / - & - / - / - & - / \checkmark / - \\
$p_{z}$ & - / - / - & - / - / - & - / - / - & \checkmark / \checkmark / - \\

\textbf{Leading jet sub-structure} & & & & \\
$\tau_{32}$ & \checkmark / - / - & - / - / \checkmark & - / - / - & - / - / - \\
$\tau_{21}$ & - / - / - & - / - / - & - / - / - & - / - / - \\
$\tau_{31}$ & - / - / - & - / - / - & - / - / - & - / - / - \\
$\mathrm{C}_{2}$ & - / - / - & - / - / - & - / - / - & - / - / - \\
$\mathrm{C}_{3}$ & - / - / - & - / - / - & - / - / - & - / - / - \\
$\mathrm{D}_{2}$ & - / - / - & - / - / - & - / - / - & - / - / - \\
$\mathrm{D}_{3}$ & - / - / - & - / - / - & - / - / - & - / - / - \\

\textbf{Next-to-leading jet sub-structure} & & & & \\
$\tau_{32}$ & - / - / \checkmark & - / - / \checkmark & \checkmark / \checkmark / \checkmark & \checkmark / \checkmark / \checkmark \\
$\tau_{21}$ & - / - / - & - / - / - & - / - / - & - / - / - \\
$\tau_{31}$ & - / - / - & - / - / - & - / - / - & - / - / - \\
$\mathrm{C}_{2}$ & - / - / - & - / - / - & - / - / - & - / - / - \\
$\mathrm{C}_{3}$ & - / - / - & - / - / - & - / - / - & - / - / - \\
$\mathrm{D}_{2}$ & - / - / - & - / - / - & - / - / - & - / - / - \\
$\mathrm{D}_{3}$ & - / - / - & - / - / - & - / - / - & - / - / - \\

\bottomrule
\end{tabular}
}
\end{minipage}
\caption{List of variables used to train the multivariate discriminant for each sample. The 20 most powerful variables for each MVA are indicated with checkmarks. The columns marked 1st, 2nd, and 3rd indicate the variables used in the two initial MVAs and the final (3rd) MVA, respectively.
 \label{tab:mva:training:variables1}}
\end{table}

\begin{table}[p!]
\centering
\begin{minipage}{\columnwidth}
\centering
\resizebox{1.0\textwidth}{!}{
\begin{tabular}{lcc|cc}
\toprule
\vspace{1.0mm}
{} & \multicolumn{2}{c}{$\roots=1.4\,\tev$} & \multicolumn{2}{c}{$\roots=3\,\tev$} \\
\vspace{1.0mm}
P(\Pem) & -80\% & +80\% & -80\% & +80\% \\
\midrule
{} & 1st / 2nd / 3rd & 1st / 2nd / 3rd & 1st / 2nd / 3rd & 1st / 2nd / 3rd \\

\textbf{Flavour tagging variables} & & & & \\
$\sum$ b-tags & \checkmark / \checkmark / \checkmark & \checkmark / \checkmark / - & \checkmark / \checkmark / \checkmark & \checkmark / \checkmark / - \\
$\sum$ c-tags & - / - / - & - / - / - & - / - / - & - / - / - \\
$\sum$ c-tag/(b-tag+c-tag) & - / - / - & \checkmark / \checkmark / \checkmark & \checkmark / - / - & - / \checkmark / \checkmark \\
Highest b-tag & \checkmark / \checkmark / \checkmark & \checkmark / \checkmark / \checkmark & \checkmark / - / \checkmark & \checkmark / \checkmark / \checkmark \\
Lowest b-tag & \checkmark / \checkmark / - & \checkmark / \checkmark / - & \checkmark / \checkmark / \checkmark & \checkmark / \checkmark / - \\
Leading jet b-tag & \checkmark / \checkmark / - & \checkmark / - / - & \checkmark / - / - & \checkmark / - / - \\
Leading jet c-tag & \checkmark / - / - & - / - / - & - / - / \checkmark & - / - / - \\
Leading jet c-tag/(b-tag+c-tag) & - / \checkmark / - & - / - / - & \checkmark / - / - & - / - / - \\
Next-to-leading jet b-tag & \checkmark / \checkmark / - & - / - / - & \checkmark / - / \checkmark & \checkmark / \checkmark / - \\
Next-to-leading jet c-tag & - / - / - & - / - / - & - / - / - & - / - / - \\
Next-to-leading jet c-tag/(b-tag+c-tag) & - / \checkmark / - & \checkmark / - / - & - / - / - & \checkmark / \checkmark / - \\
Invariant mass of jets with a b-tag & - / - / - & \checkmark / - / - & - / \checkmark / - & \checkmark / - / - \\

\textbf{Jet splitting scales} & & & & \\
$d_{23}$ & - / - / \checkmark & - / - / \checkmark & - / \checkmark / \checkmark & - / - / \checkmark \\
$d_{34}$ & \checkmark / \checkmark / \checkmark & \checkmark / \checkmark / \checkmark & \checkmark / \checkmark / - & \checkmark / - / \checkmark \\
$d_{45}$ & - / \checkmark / - & \checkmark / \checkmark / \checkmark & \checkmark / \checkmark / - & - / \checkmark / - \\
$d_{56}$ & - / - / \checkmark & - / - / - & - / - / - & - / - / - \\

\textbf{Isolated lepton} & & & & \\
$E$ & - / - / \checkmark & \checkmark / \checkmark / \checkmark & - / \checkmark / - & \checkmark / - / \checkmark \\
$\pT$ & - / - / - & \checkmark / \checkmark / - & \checkmark / \checkmark / \checkmark & \checkmark / \checkmark / \checkmark \\

\textbf{Event/jet shape variables} & & & & \\
PFO Thrust & \checkmark / \checkmark / \checkmark & - / \checkmark / - & - / - / - & - / - / - \\
PFO Oblateness & - / - / - & - / - / - & - / - / \checkmark & - / - / \checkmark \\
PFO TSphericity & - / \checkmark / - & \checkmark / - / - & - / \checkmark / - & - / \checkmark / \checkmark \\
PFO Aplanarity & \checkmark / \checkmark / \checkmark & \checkmark / \checkmark / - & \checkmark / \checkmark / \checkmark & \checkmark / \checkmark / - \\
Leading jet Thrust & - / - / - & - / - / - & - / - / - & - / - / - \\
Leading jet Oblateness & - / - / - & - / - / - & - / - / - & - / - / - \\
Leading jet TSphericity & - / - / - & - / - / - & - / - / - & - / - / - \\
Leading jet Aplanarity & - / - / - & - / - / - & - / - / - & - / - / - \\
Next-to-leading jet Thrust & - / - / - & - / - / - & - / - / - & - / - / - \\
Next-to-leading jet Oblateness & - / - / - & - / - / - & - / - / - & - / - / - \\
Next-to-leading jet TSphericity & - / - / - & - / - / - & - / - / - & - / - / - \\
Next-to-leading jet Aplanarity & - / - / - & - / - / - & - / - / - & - / - / - \\

\textbf{Miscellaneous} & & & & \\
Reco. $\rootsprime$ & \checkmark / - / - & - / - / \checkmark & \checkmark / - / - & - / - / - \\
$E_{\mathrm{T}}$ & \checkmark / - / - & - / - / - & - / - / - & - / - / - \\
$E_{\mathrm{visible}}$ & \checkmark / \checkmark / \checkmark & \checkmark / \checkmark / \checkmark & \checkmark / \checkmark / - & \checkmark / - / - \\
Event missing $\pT$ & \checkmark / \checkmark / \checkmark & \checkmark / \checkmark / - & \checkmark / \checkmark / - & \checkmark / \checkmark / \checkmark \\
$\cos(\theta_{\mathrm{max},\pT})$ & - / - / - & - / - / \checkmark & - / - / - & - / - / - \\
Inv. mass of highest $\pT$ jets & - / - / - & - / \checkmark / - & - / - / - & - / \checkmark / \checkmark \\

\bottomrule
\end{tabular}
}
\end{minipage}
\caption{(Cont.) List of variables used to train the multivariate discriminant for each sample. The 20 most powerful variables for each MVA are indicated with checkmarks. The columns marked 1st, 2nd, and 3rd indicate the variables used in the two initial MVAs and the final (3rd) MVA, respectively.\label{tab:mva:training:variables2}}
\end{table}

\subsection{Results}

\begin{figure}
	\centering
	\includegraphics[width=0.48\columnwidth]{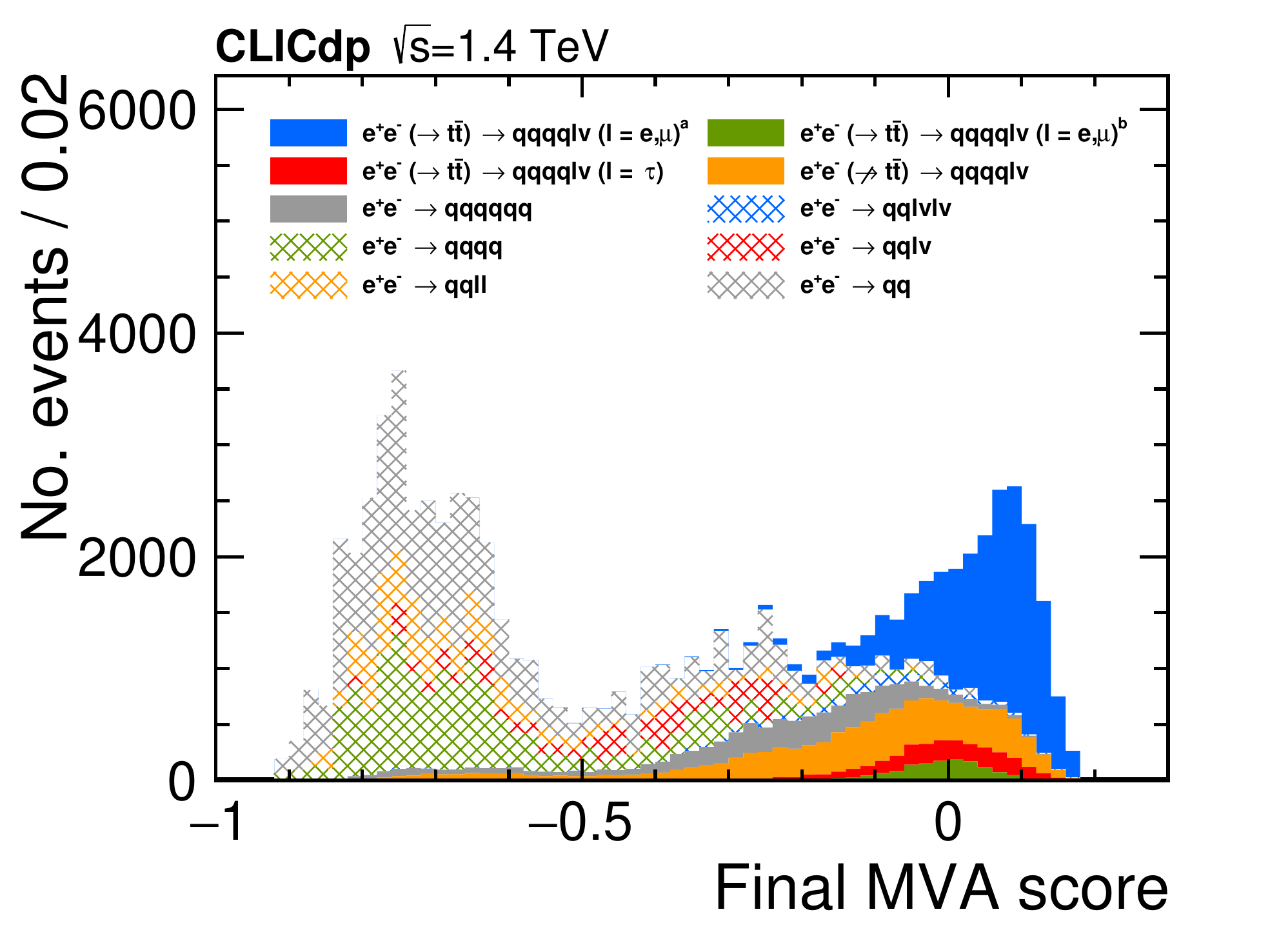}
	~~~
	\includegraphics[width=0.48\columnwidth]{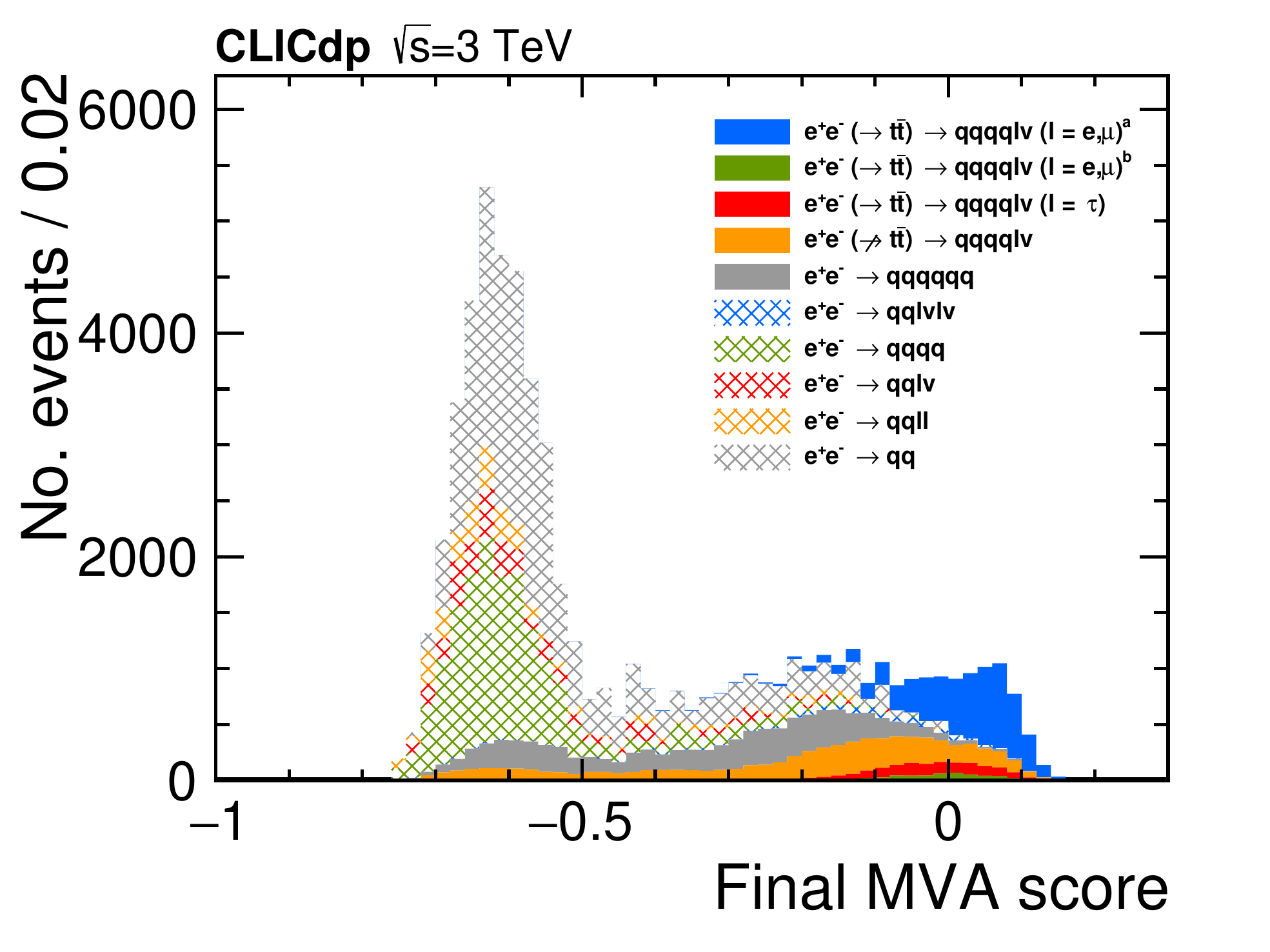}
\caption{Final MVA classification score for $\roots=1.4\,\tev$ (left) and $\roots=3\,\tev$ (right), at -80\% polarisation. The distributions are stacked and normalised to the corresponding yield assuming an integrated luminosity of $2.0\,\abinv$ and $4.0\,\abinv$, respectively. The superscript `a' (`b') refers to the kinematic region $\rootsprime\geq1.2\,\tev$ ($\rootsprime<1.2\,\tev$). \label{fig:analysis:mva:results:scores}}
\end{figure}

The resulting final classification scores for the signal and background samples for $\roots=1.4\,\tev$ and $\roots=3\,\tev$, are presented in \Cref{fig:analysis:mva:results:scores}. In general, strong suppression is achieved for the two quark samples and the fully-hadronic four-quark and six-quark samples.

\begin{figure}
	\centering
	\includegraphics[width=0.48\columnwidth]{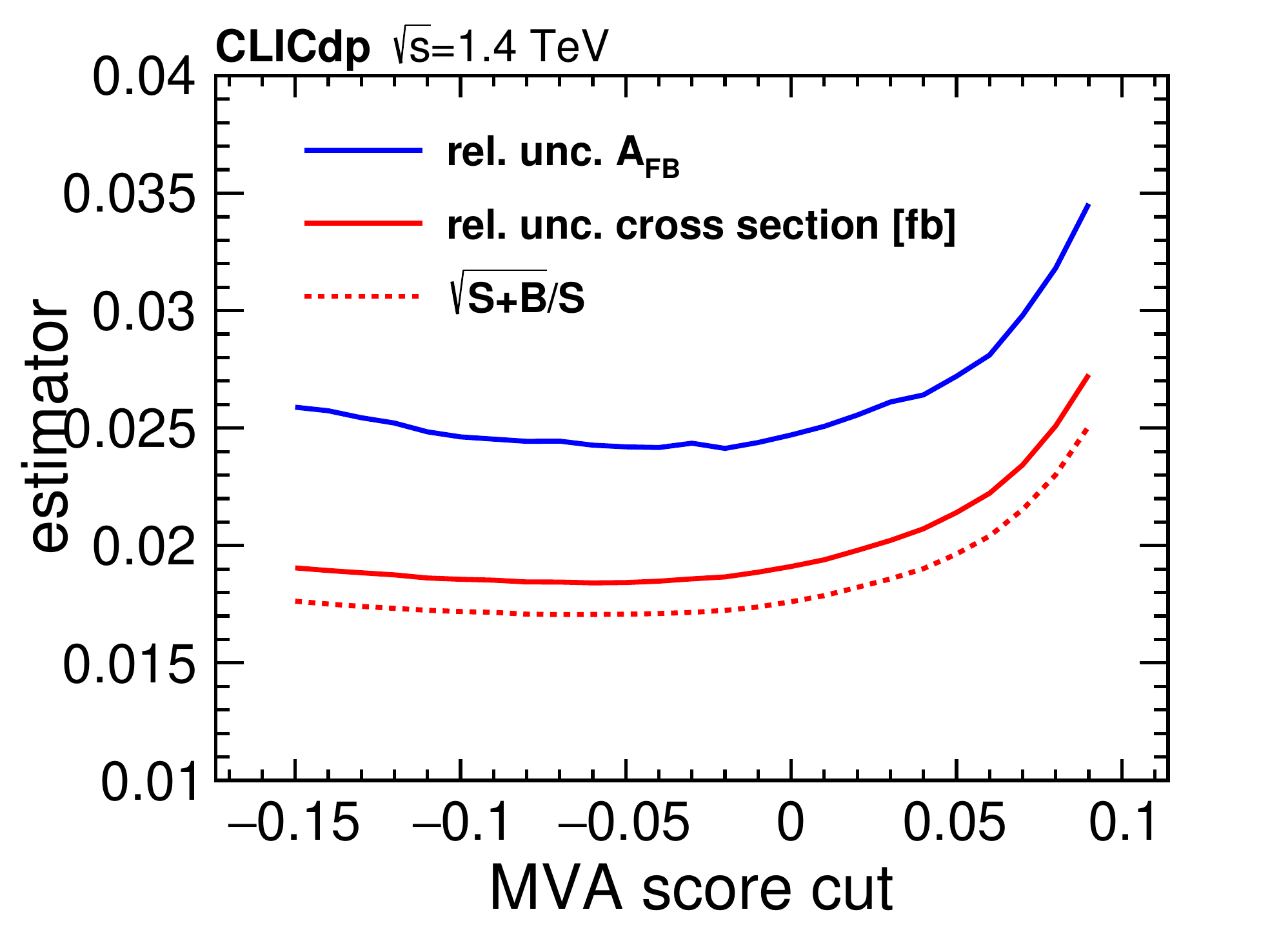}
	~~~
	\includegraphics[width=0.48\columnwidth]{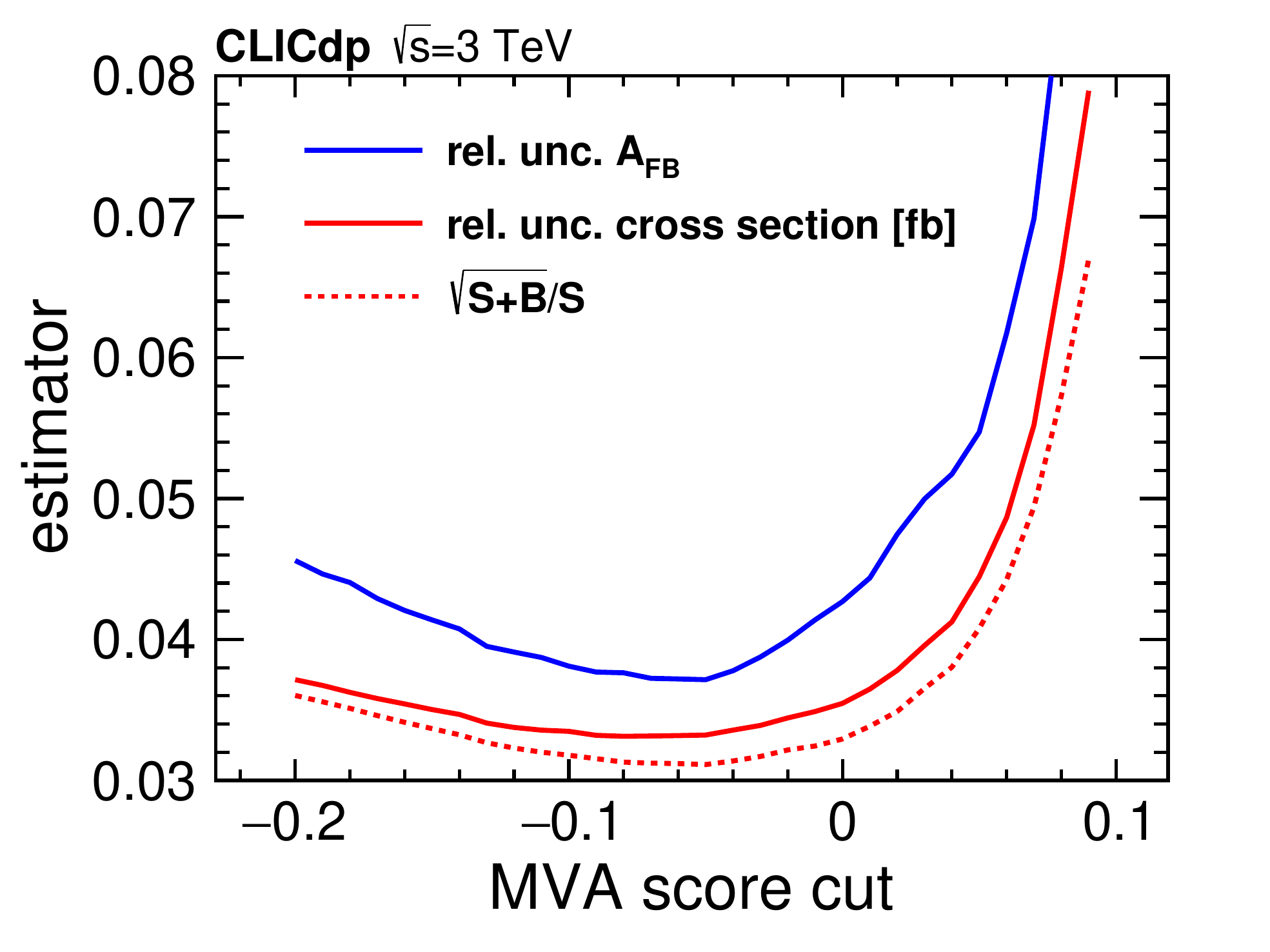}
\caption{Optimisation of the MVA classification score cut for $\roots=1.4\,\tev$ (left) and $\roots=3\,\tev$ (right), both at -80\% polarisation. The relative statistical uncertainty on $\afb$ ($\csttbar$) is shown in blue (red) solid lines. In addition, the red dashed line shows the result for a simple estimator of the statistical uncertainty of the cross section. The optimal cut is chosen as a trade-off between the two main estimators (solid lines).\label{fig:analysis:mva:results:opt}}
\end{figure}

Signal-like events are selected by applying a cut to the classification score, removing events with a value below the cut. The cut is chosen to minimise the statistical uncertainty on the two extracted observables $\afb$ and $\csttbar$, as defined in \Cref{eq:totcs} and \Cref{eq:afb}. These estimators are shown in \Cref{fig:analysis:mva:results:opt} as a function of the applied cut: the relative uncertainty on $\afb$ ($\csttbar$) is presented in blue (red) solid lines. In addition, we display the results for a simple estimator defined as $\sqrt{S+B}/S$, where $S$ and $B$ are the number of signal and background events after the cut. The final cuts are selected as a trade off between the relative uncertainties on the two main estimators: -0.04 at $\roots=1.4\,\tev$ and -0.05 at $\roots=3\,\tev$. The corresponding values for +80\% polarisation are -0.02 and -0.05, respectively.

\section{Event selection efficiencies}
\label{sec:eventselsum}

The event selection efficiencies for the signal and background samples along with the total number of events selected at final level are shown in \Cref{tab:ttbar:boosted:selection1400} and \Cref{tab:ttbar:boosted:selection3000}, for the samples at nominal collision energies of 1.4 and 3\,TeV, respectively. The cross sections quoted as defined in the kinematic regions $\rootsprime\geq1.2\,\tev$ and $\rootsprime\geq2.6\,\tev$, respectively. Contributions from other backgrounds such as $\epem\to\PQq\PQq\PGn\PGn$ and additional six-fermion processes are found to be negligibly small.

\begin{figure}
  \centering
  \includegraphics[width=0.48\columnwidth]{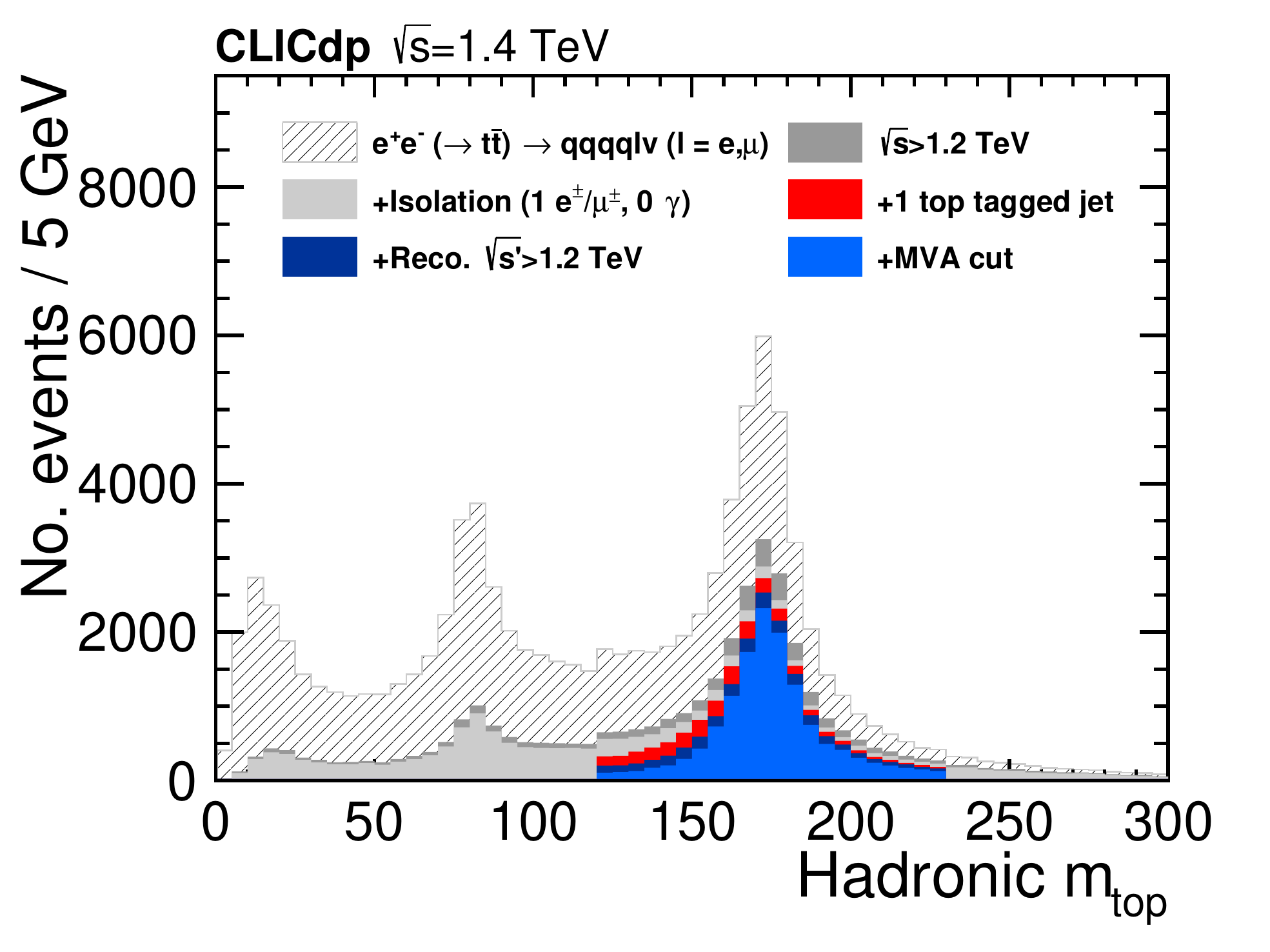}
  ~~~
  \includegraphics[width=0.48\columnwidth]{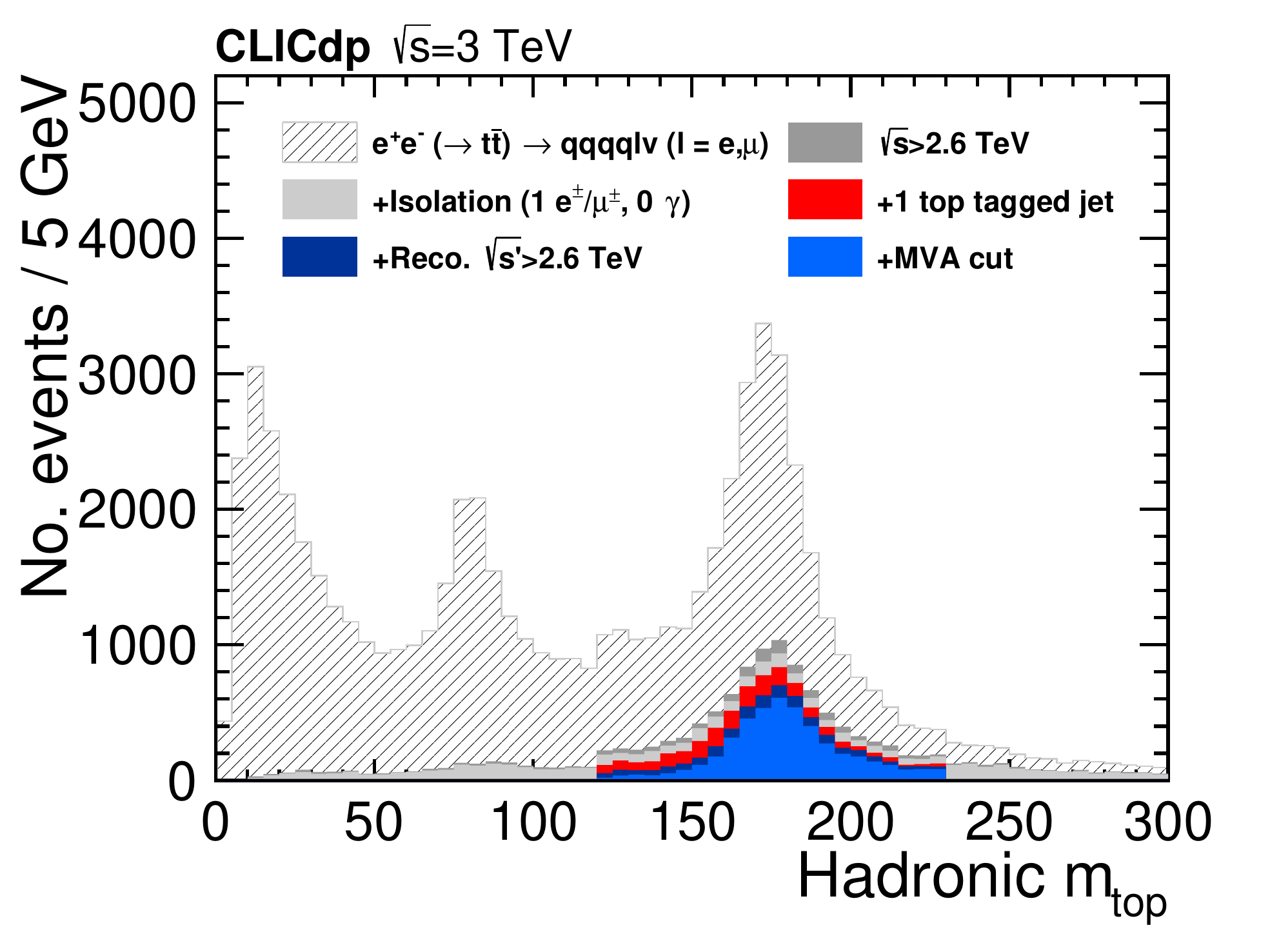}
\caption{Reconstructed invariant mass of the hadronically decaying top-quark candidates at different levels of the event selection, for operation at $P(\Pem)=\text{-}80\%$ and a nominal collision energy of $1.4\,\tev$ (left) and $3\,\tev$ (right), assuming an integrated luminosity of $2.0\,\abinv$ and $4.0\,\abinv$, respectively. \label{fig:ttbar:summary:invmasslevels}}
\end{figure}

\begin{figure}
  \centering
  \includegraphics[width=0.48\columnwidth]{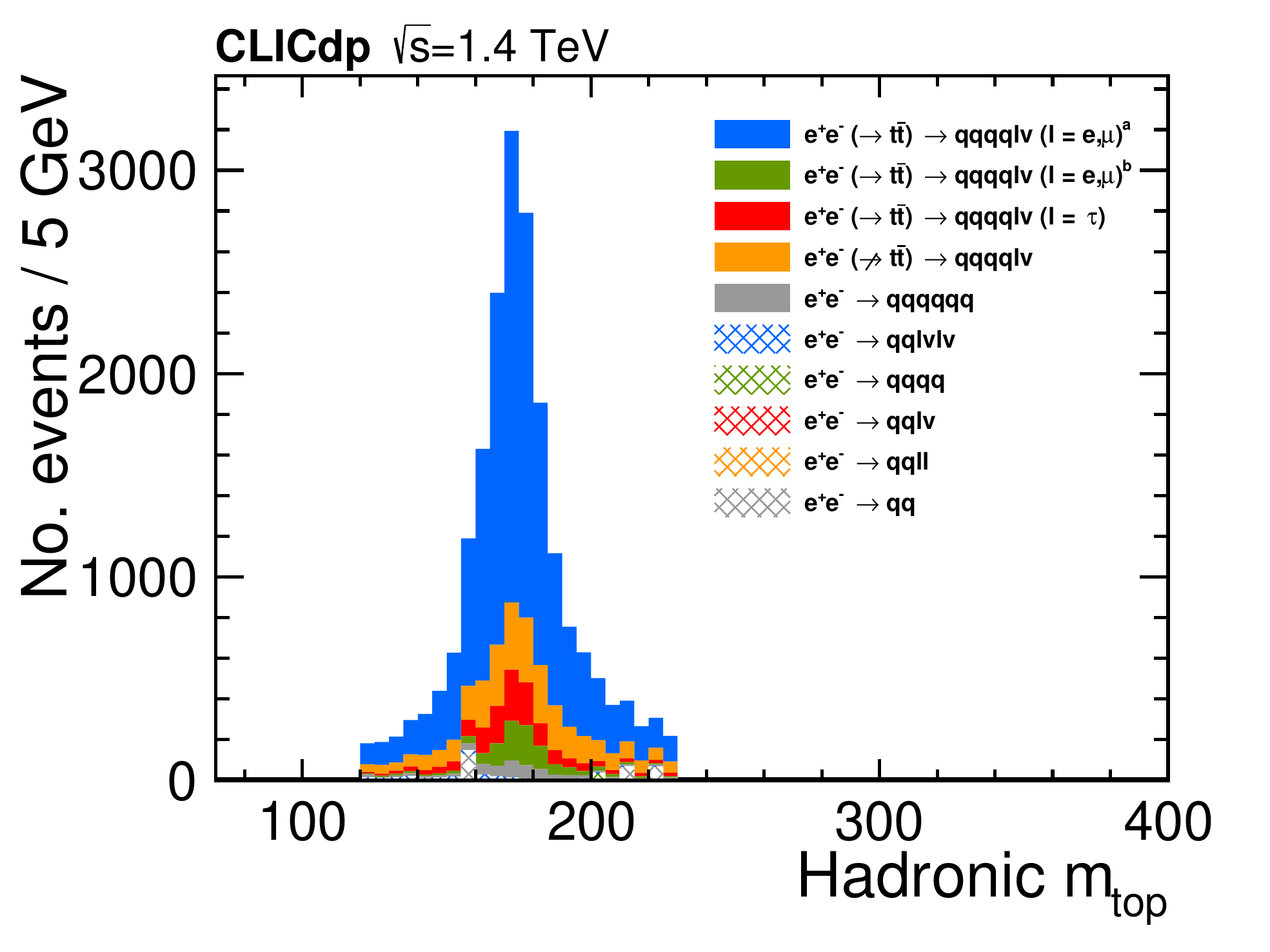}
  ~~~
  \includegraphics[width=0.48\columnwidth]{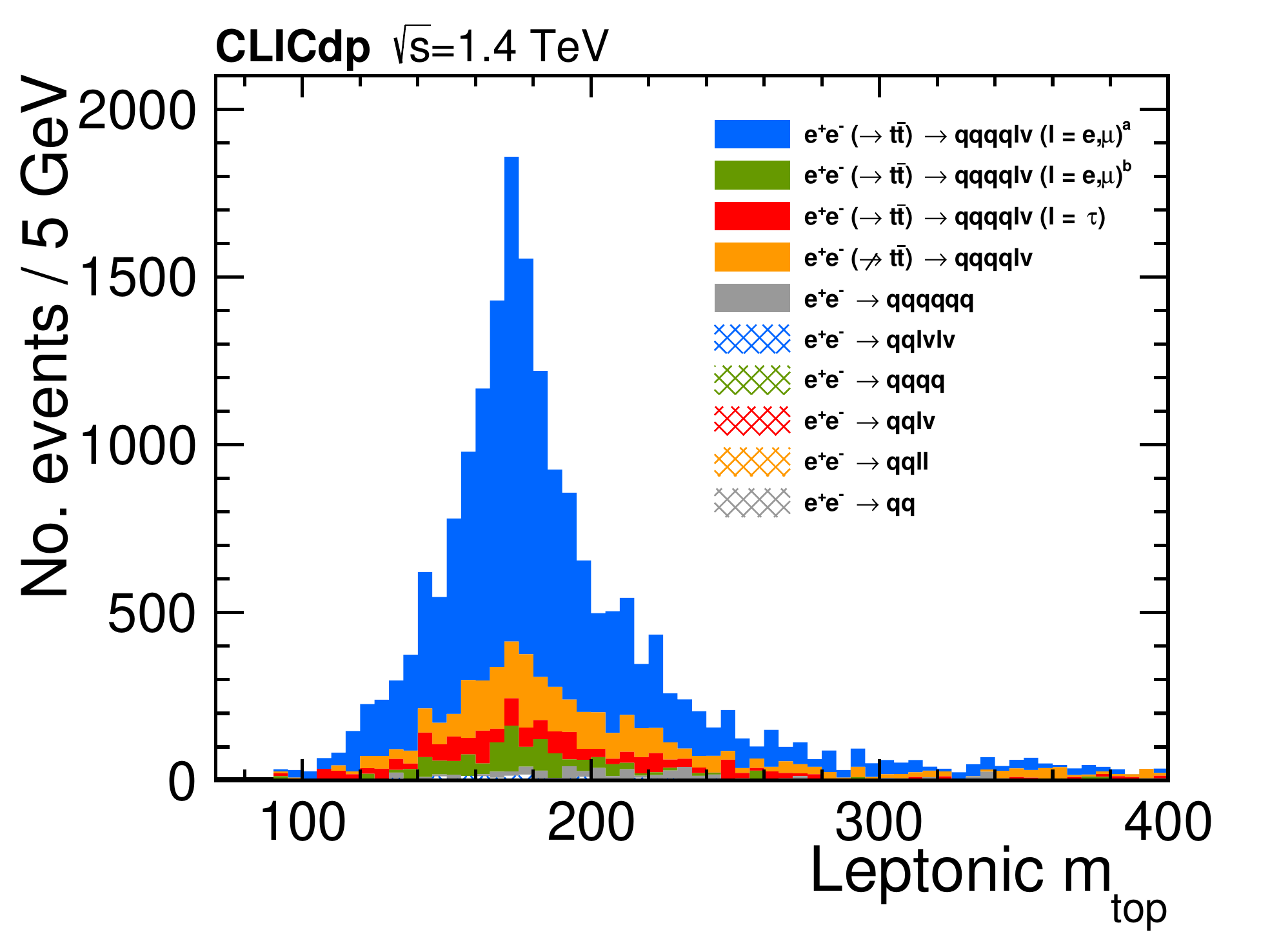}
  \caption{Reconstructed invariant mass of the hadronically (left) and leptonically (right) decaying top-quark candidates at final level of the event selection. The distributions are stacked and correspond to operation at $\roots=1.4\,\tev$ and $P(\Pem)=\text{-}80\%$ with an integrated luminosity of $2.0\,\abinv$. The superscript `a' (`b') refers to the kinematic region $\rootsprime\geq1.2\,\tev$ ($\rootsprime<1.2\,\tev$). \label{fig:ttbar:summary:invmass}}
\end{figure}

\begin{figure}
  \centering
  \includegraphics[width=0.48\columnwidth]{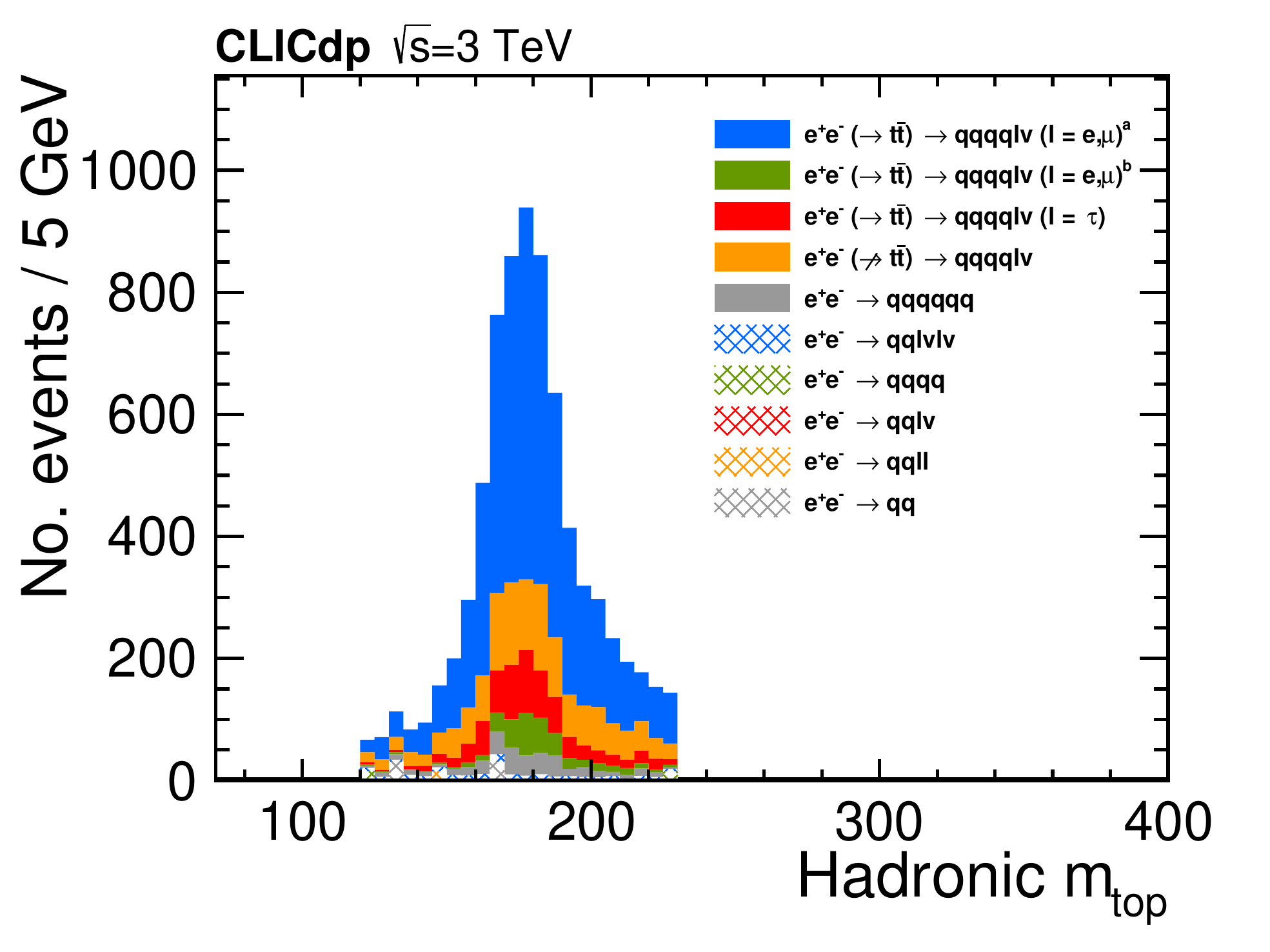}
  ~~~
  \includegraphics[width=0.48\columnwidth]{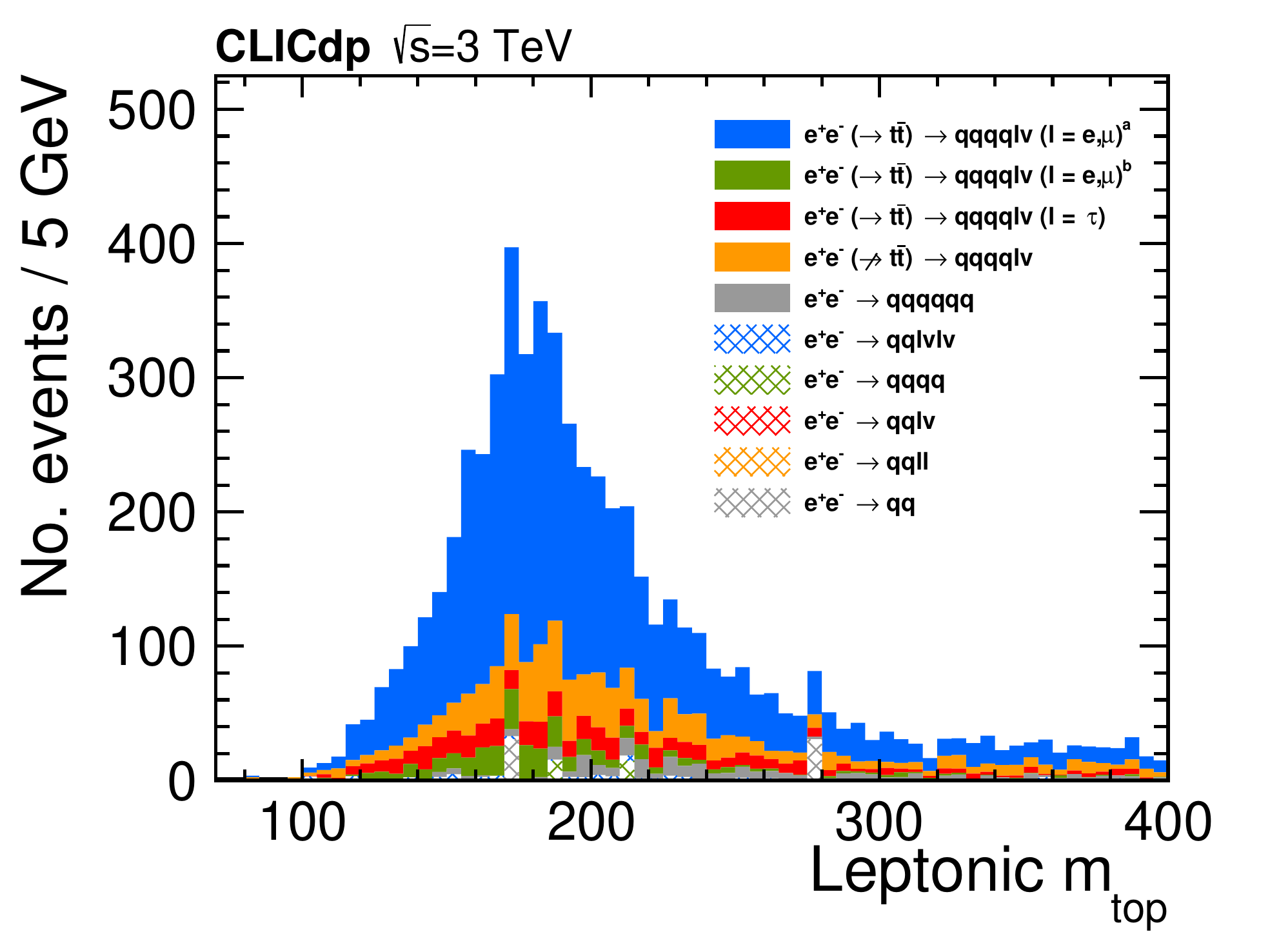}
\caption{Reconstructed invariant mass of the hadronically (left) and leptonically (right) decaying top-quark candidates at final level of the event selection. The distributions are stacked and correspond to operation at $\roots=3\,\tev$ and $P(\Pem)=\text{-}80\%$ with an integrated luminosity of $4.0\,\abinv$. The superscript `a' (`b') refers to the kinematic region $\rootsprime\geq1.2\,\tev$ ($\rootsprime<1.2\,\tev$). \label{fig:ttbar:summary:invmass3tev}}
\end{figure}

The reconstructed invariant mass of the hadronically decaying top-quarks for the signal sample is shown at different levels of the event selection in \Cref{fig:ttbar:summary:invmasslevels} (solid colours). The distributions show the mass of the tagged top-quark candidate or the corresponding paired large-R jet mass for tagged top-quarks (depending on level) and the large-R jet mass of the leading jet for non-tagged events. \Cref{fig:ttbar:summary:invmass} and \Cref{fig:ttbar:summary:invmass3tev} show the reconstructed invariant mass of the hadronically (left) and leptonically (right) decaying top-quarks (stacked distribution) at final level for the signal and different background samples. The mass of the leptonically decaying top-quark is reconstructed by using the neutrino hypothesis outlined in \autoref{sec:effcom} and used in the reconstructed of the effective centre-of-mass energy; the missing transverse momentum is used as an estimator for the neutrino transverse momentum components and $p_{\PGn,z}$ is retrieved by solving \autoref{eq:boosted:sprimereco}. 

The remaining background events are predominantly from processes with the same final state top-quark pair topology as the signal, while other background have been effectively eradicated. The purity of the event sample at final level, defined as the number of remaining signal events divided by the total number of events, is between around 70\% at $\roots=1.4\,\tev$ and between 60-66\% at $\roots=3.0\,\tev$.

\begin{figure}
  \centering
  \includegraphics[width=0.48\columnwidth]{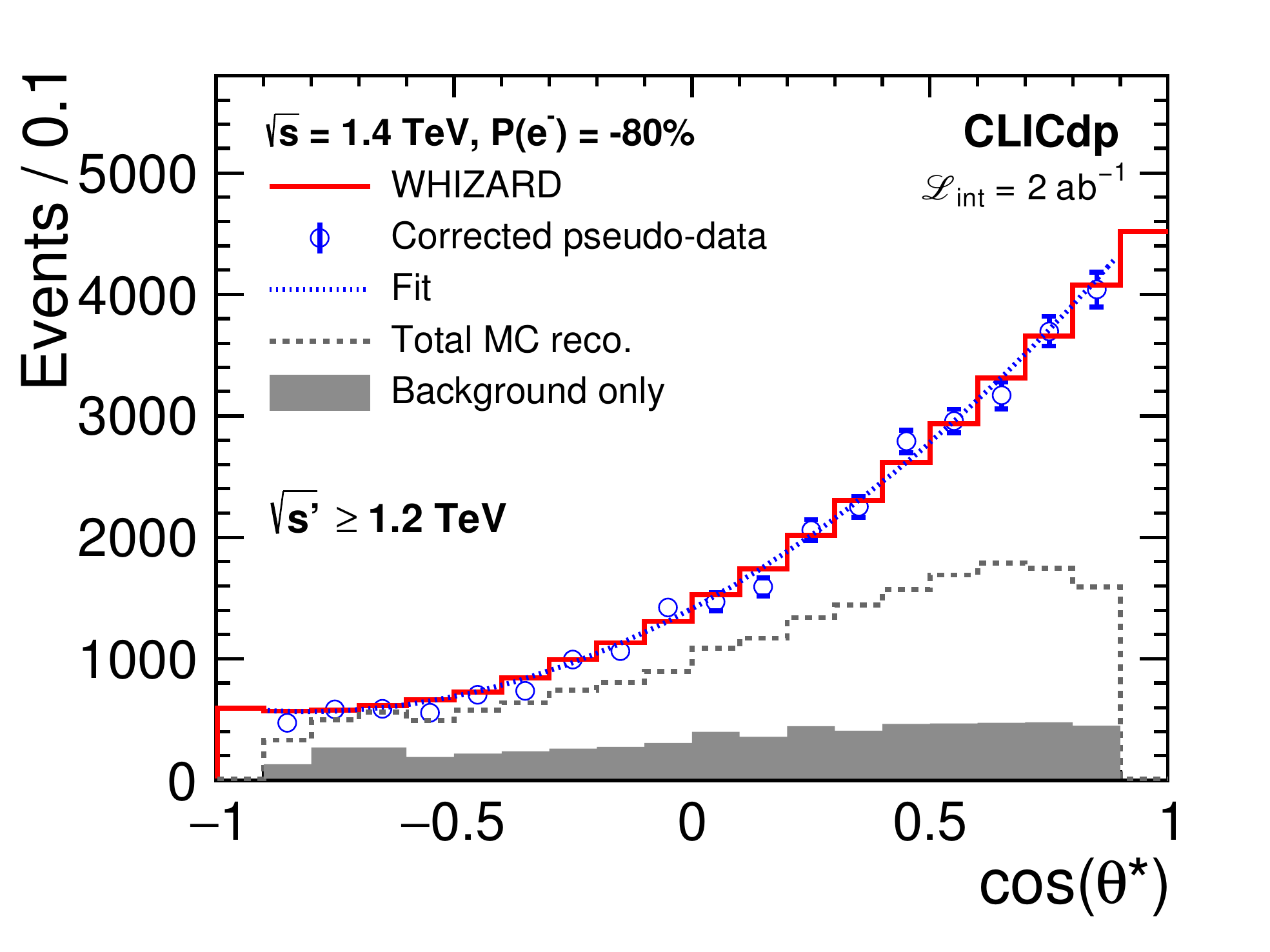}
  ~~~~
   \includegraphics[width=0.48\columnwidth]{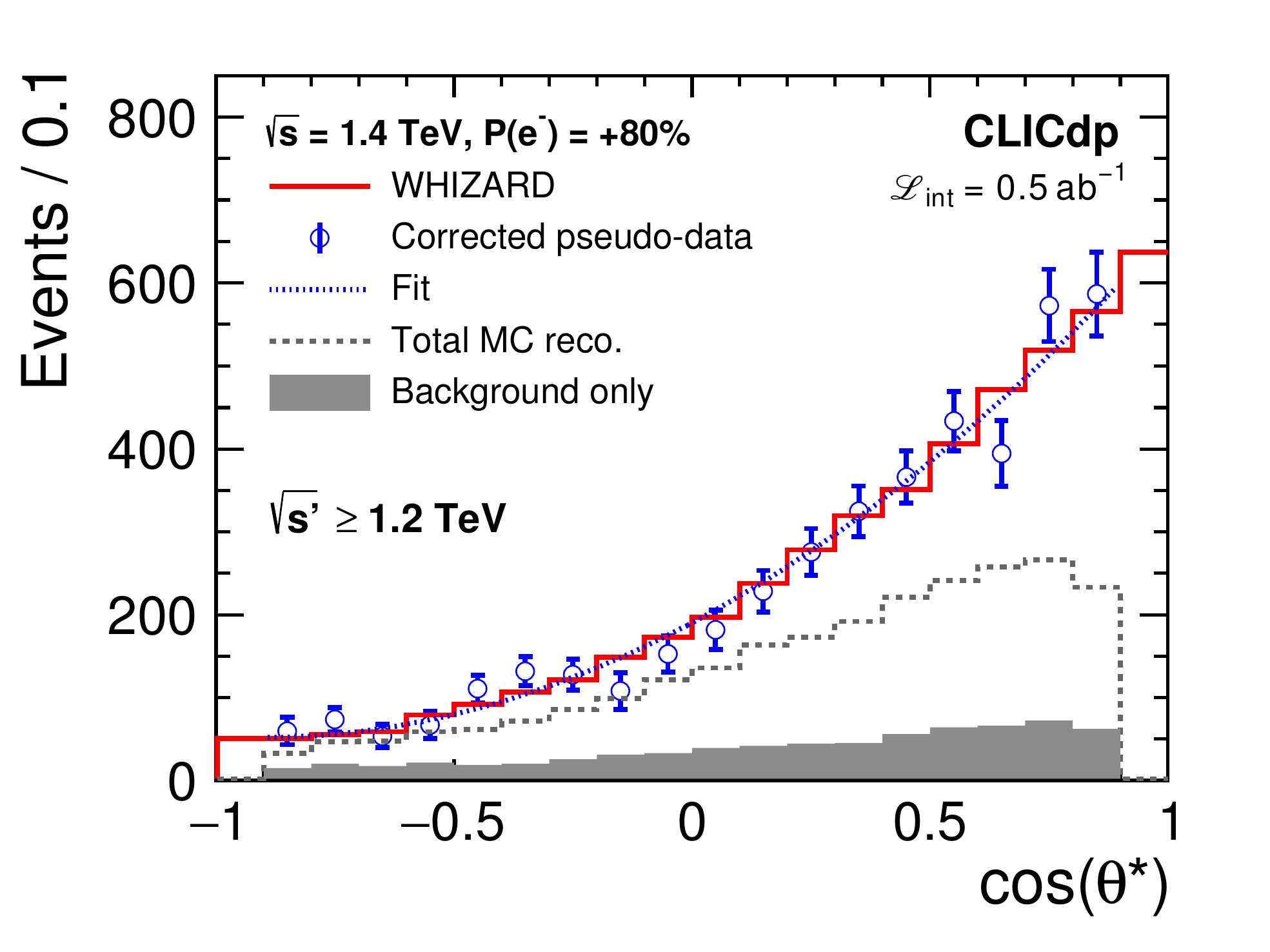}
  \caption{Polar angle distributions of the hadronically decaying top-quark candidates at final level of the event selection at a nominal collision energy of 1.4\,\tev for $P(\Pem)=\text{-}80\%$ (left) and $P(\Pem)=\text{+}80\%$ (right), and an integrated luminosity of $2.0\,\abinv$ and $0.5\,\abinv$, respectively. Figure taken from \cite{Abramowicz:2018rjq}. \label{fig:ttbar:boosted:costheta}}
\end{figure}

\begin{figure}
  \centering
    \includegraphics[width=0.48\columnwidth]{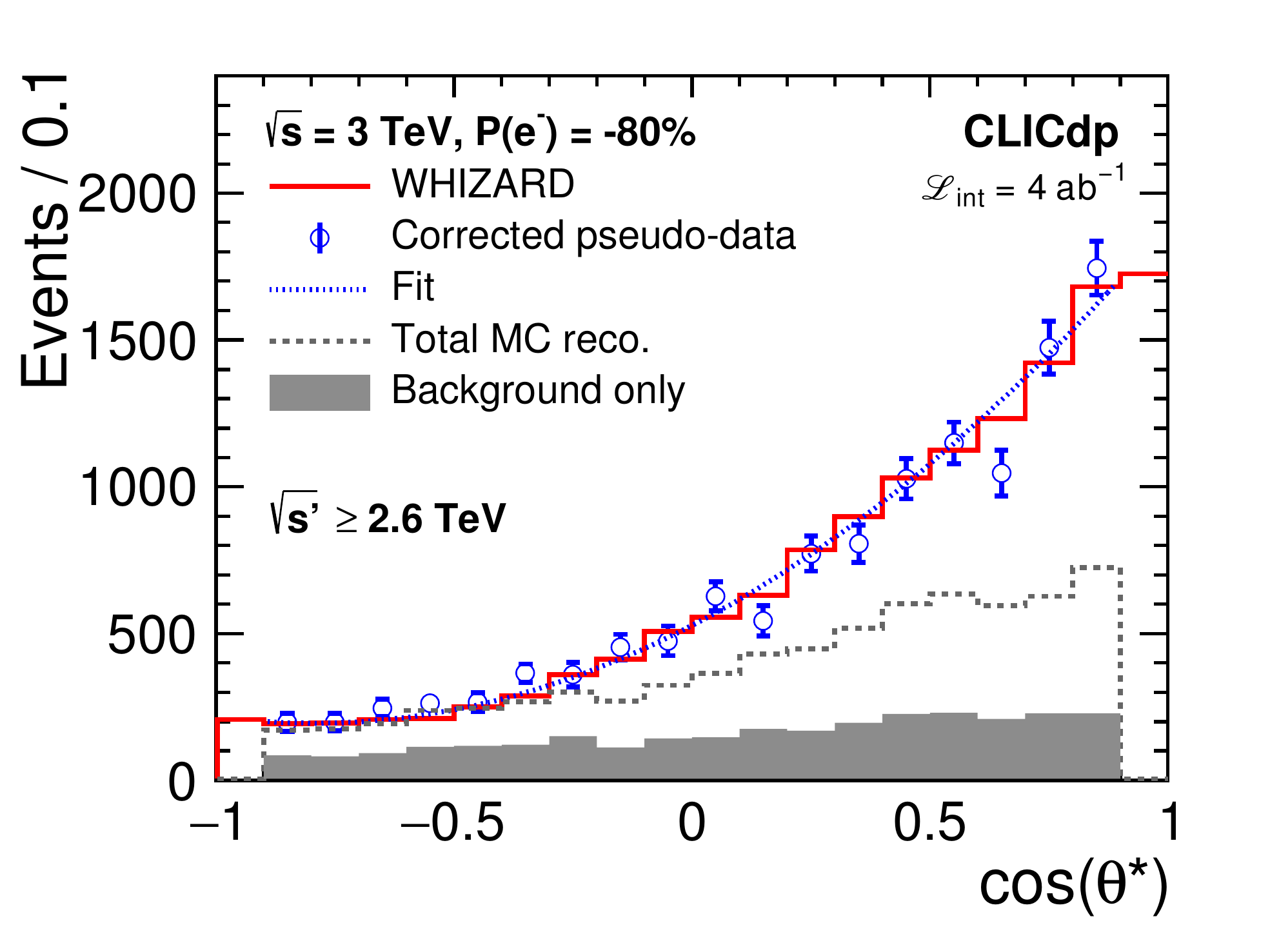}
   ~~~~
  \includegraphics[width=0.48\columnwidth]{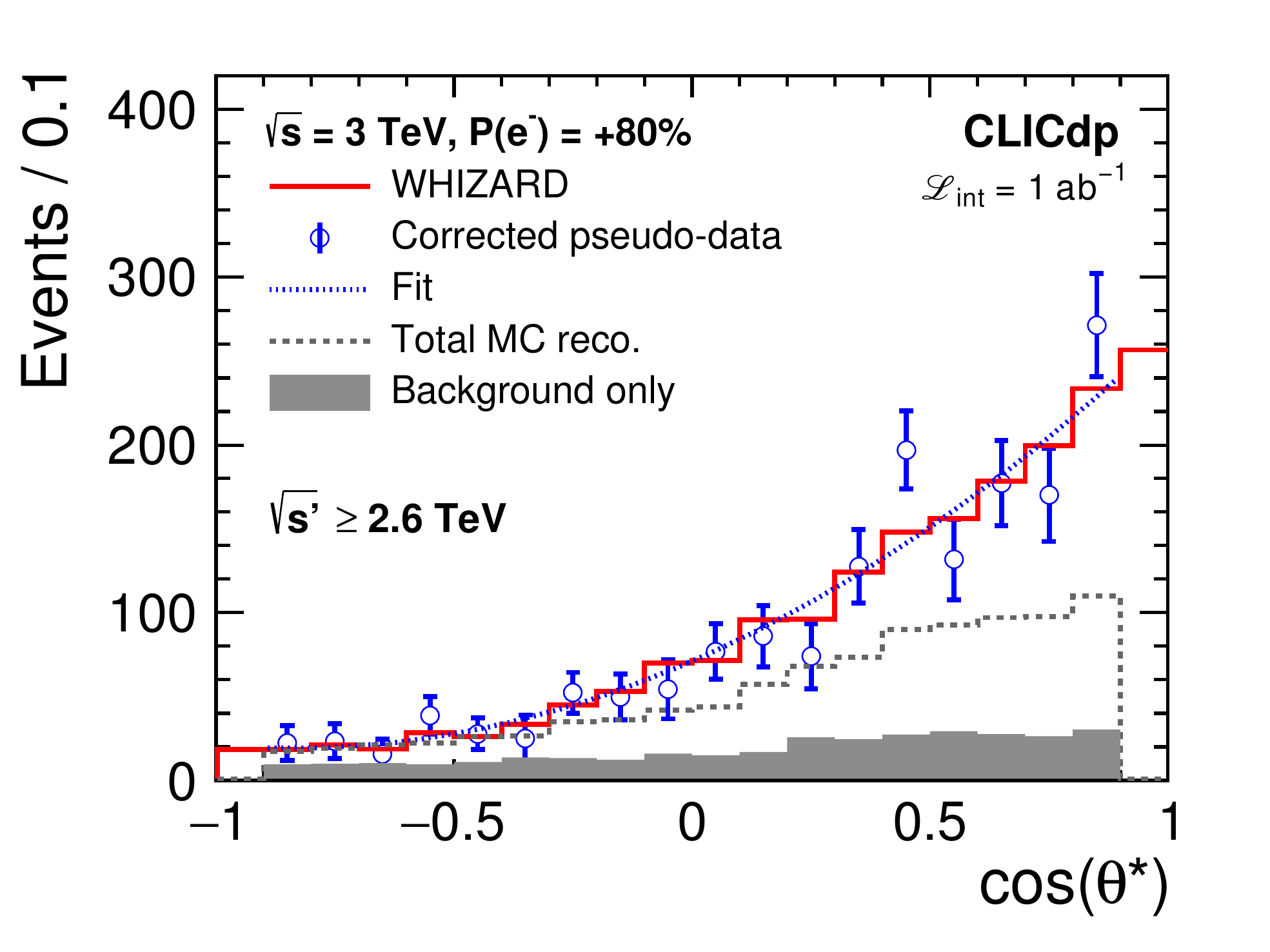}
  \caption{Polar angle distributions of the hadronically decaying top-quark candidates at final level of the event selection at a nominal collision energy of 3\,\tev for $P(\Pem)=\text{-}80\%$ (left) and $P(\Pem)=\text{+}80\%$ (right), and an integrated luminosity of $4.0\,\abinv$ and $1.0\,\abinv$, respectively. Figure taken from \cite{Abramowicz:2018rjq}.
\label{fig:ttbar:boosted:costheta3tev}}
\end{figure}

The polar-angle distributions of the hadronically decaying top-quark candidates are shown in \Cref{fig:ttbar:boosted:costheta} and \Cref{fig:ttbar:boosted:costheta3tev}. The dashed grey curve represents the reconstructed distribution at final level and the grey area indicates the level of the background events only. These both include the effects of detector modelling, event reconstruction, and candidate selection. The blue data points shows the outcome of one pseudo-experiment performed for the given luminosity, after subtraction of background and correction for finite selection efficiencies. The blue dotted line shows the fit performed to the pseudo-experiment data and is used to extract the two variables $\csttbar$ and $\afb$ as defined in \Cref{eq:totcs} and \Cref{eq:afb}. Finally, the red solid line shows the simulated distribution at parton-level (\whizard). The distributions are shown for the fiducial region $-0.9\leq\cos\theta^{*}\leq0.9$. The selection efficiency in the region $-0.7\leq\cos\theta^{*}\leq0.7$ is generally flat with a central value of about 50\% for at both $\roots=1.4\,\tev$ and $\roots=3.0\,\tev$. In the forward regions the efficiency drops to 30\%.

\begin{table}
\centering
\begin{minipage}{\columnwidth}
\resizebox{1.0\textwidth}{!}{
\begin{tabular}{lcccccccc}
\toprule
\vspace{1.0mm}
{} & \multicolumn{2}{c}{$\sigma\,[\fb]$} & \multicolumn{2}{c}{$\epsilon_{\,\mathrm{Pre}}\,[\%]$} & \multicolumn{2}{c}{$\epsilon_{\,\mathrm{MVA}}\,[\%]$} & \multicolumn{2}{c}{$N$} \rule{0pt}{3ex} \\
\vspace{1.0mm}
P(\Pem) & -80\% & +80\% & -80\% & +80\% & -80\% & +80\% & -80\% & +80\% \\
Process \\
\midrule
$\epem(\to\ttbar)\to\PQq\PQq\PQq\PQq\Pl\PGn\,(\Pl=\Pe,\PGm)$\footnote{Kinematic region defined as $\rootsprime\geq1.2\,\tev$} & 18.4 & 9.83 & 43 & 44 & 85 & 87 & 13,469 & 1,902 \\
\midrule
$\epem(\to\ttbar)\to\PQq\PQq\PQq\PQq\Pl\PGn\,(\Pl=\Pe,\PGm)$\footnote{$\rootsprime<1.2\,\tev$} & 28.5 & 14.9 & 2.5 & 2.7 & 68 & 56 & 952 & 111 \\
$\epem(\to\ttbar)\to\PQq\PQq\PQq\PQq\Pl\PGn\,(\Pl=\PGt)$ &23.2 & 12.3 & 4.7 & 4.8 & 63 & 57 & 1,379 & 167 \\
$\epem(\not\to\ttbar)\to\PQq\PQq\PQq\PQq\Pl\PGn$ & 72.2 & 16.5 & 6.0 & 7.2 & 35 & 59 & 3,032 & 348 \\
$\epem\to\PQq\PQq\PQq\PQq\PQq\PQq$ & 116 & 44.9 & 2.3 & 2.4 & 9.2 & 9.5 & 499 & 51 \\
$\epem\to\PQq\PQq\Pl\PGn\Pl\PGn$ & 44.1 & 15.3 & 1.2 & 1.5 & 27 & 40 & 285 & 45 \\
$\epem\to\PQq\PQq\PQq\PQq$ & 2,300 & 347 & 0.31 & 0.47 & 0.22 & 0.56 & 32 & 5 \\
$\epem\to\PQq\PQq\Pl\PGn$ & 6,980 & 1,640 & 0.02 & 0.01 & 0.00 & 0.00 & - & - \\
$\epem\to\PQq\PQq\Pl\Pl$ & 2,680 & 2,530 & 0.01 & 0.08 & 0.00 & 0.00 & - & - \\
$\epem\to\PQq\PQq$ & 4,840 & 3,170 & 0.21 & 0.16 & 1.3 & 0.00 & 259 & - \\
\bottomrule
\end{tabular}
}
\end{minipage}
\caption{Event selection summary for the analysis at $\roots=1.4\,\tev$, assuming $2.0\,\abinv$ and $0.5\,\abinv$ for $P(\Pem)=\text{-}80\%$ and $P(\Pem)=\text{+}80\%$, respectively. The cross section quoted for the signal sample in the uppermost row is defined in the kinematic region $\rootsprime\geq1.2\,\tev$. The fractional pre-selection and MVA selection efficiencies are shown in the subsequent columns along with the number of events in the final sample. Table taken from \cite{Abramowicz:2018rjq}. \label{tab:ttbar:boosted:selection1400}}
\end{table}

\begin{table}
\centering
\begin{minipage}{\columnwidth}
\resizebox{1.0\textwidth}{!}{
\begin{tabular}{lcccccccc}
\toprule
\vspace{1.0mm}
{} & \multicolumn{2}{c}{$\sigma\,[\fb]$} & \multicolumn{2}{c}{$\epsilon_{\,\mathrm{Pre}}\,[\%]$} & \multicolumn{2}{c}{$\epsilon_{\,\mathrm{MVA}}\,[\%]$} & \multicolumn{2}{c}{$N$} \rule{0pt}{3ex} \\
\vspace{1.0mm}
P(\Pem) & -80\% & +80\% & -80\% & +80\% & -80\% & +80\% & -80\% & +80\% \\
Process \\
\midrule
$\epem(\to\ttbar)\to\PQq\PQq\PQq\PQq\Pl\PGn\,(\Pl=\Pe,\PGm)$\footnote{Kinematic region defined as $\rootsprime\geq2.6\,\tev$} & 3.48 & 1.89 & 41 & 43 & 80 & 85 & 4,563 & 692 \\
\midrule
$\epem(\to\ttbar)\to\PQq\PQq\PQq\PQq\Pl\PGn\,(\Pl=\Pe,\PGm)$\footnote{$\rootsprime<2.6\,\tev$} & 13.7 & 7.26 & 0.98 & 0.86 & 65 & 76 & 352 & 48 \\
$\epem(\to\ttbar)\to\PQq\PQq\PQq\PQq\Pl\PGn\,(\Pl=\PGt)$ & 8.45 & 4.51 & 3.6 & 3.8 & 58 & 47 & 699 & 81 \\
$\epem(\not\to\ttbar)\to\PQq\PQq\PQq\PQq\Pl\PGn$ & 99.6 & 22.6 & 1.4 & 1.4 & 23 & 51 & 1,344 & 155 \\
$\epem\to\PQq\PQq\PQq\PQq\PQq\PQq$ & 54.0 & 18.0 & 3.4 & 3.8 & 4.7 & 6.1 & 344 & 41 \\
$\epem\to\PQq\PQq\Pl\PGn\Pl\PGn$ & 59.7 & 14.9 & 0.28 & 0.37 & 23 & 40 & 155 & 22 \\
$\epem\to\PQq\PQq\PQq\PQq$ & 963 & 130 & 0.36 & 0.38 & 0.21 & 0.39 & 29 & 2 \\
$\epem\to\PQq\PQq\Pl\PGn$ & 8,810 & 2,310 & 0.01 & 0.01 & 0.00 & 0.00 & - & - \\
$\epem\to\PQq\PQq\Pl\Pl$ & 3,230 & 3,060 & 0.02 & 0.02 & 0.44 & 0.00 & 13 & - \\
$\epem\to\PQq\PQq$ & 3,510 & 2,390 & 0.15 & 0.11 & 0.29 & 0.00 & 61 & - \\
\bottomrule
\end{tabular}
}
\end{minipage}
\caption{Event selection summary for the analysis of $\ttbar$ events at $\roots=3\,\tev$, assuming $4.0\,\abinv$ and $1.0\,\abinv$ for $P(\Pem)=\text{-}80\%$ and $P(\Pem)=\text{+}80\%$, respectively. The cross section quoted for the signal sample in the uppermost row is defined in the kinematic region $\rootsprime\geq2.6\,\tev$. The fractional pre-selection and MVA selection efficiencies are shown in the subsequent columns along with the number of events in the final sample. Table taken from \cite{Abramowicz:2018rjq}. \label{tab:ttbar:boosted:selection3000}}
\end{table}

\clearpage
\section{Cross section and asymmetry measurements}
\label{sec:results}

In this section we present the prospects for measurements of the top-quark pair total production cross section $\csttbar$ and forward-backward asymmetry $\afb$ at the higher energy stages of CLIC. Results are presented for both beam polarisations considered for a baseline with shared running time for -80\% and +80\% electron polarisation in the ratio 80:20. The observables are extracted from the polar-angle distributions presented in \Cref{fig:ttbar:boosted:costheta} and \Cref{fig:ttbar:boosted:costheta3tev}, after background subtraction and correction for finite selection efficiencies. 

\Cref{eq:AFBFit} is assumed to correctly describe the shape of the distributions in the full range, $-1.0\leq\cos(\theta^{*})\leq1.0$, and is fitted in the fiducial region $-0.9\leq\cos(\theta^{*})\leq0.9$, motivated by the limited reconstruction and event selection acceptance in the very forward region. An efficiency correction is applied, estimated bin-by-bin using half of the available sample and applied to the other half, and vice versa. Note that this procedure assumes that the MC correctly describes the selection efficiency in the polar-angle distribution. By construction, the procedure thus reproduces the corresponding parton-level results up to statistical fluctuations introduced by the procedure itself. The resulting parameters $\sigma_{1,2,3}$ are used to extract the observables, $\csttbar$ and $\afb$, in the full range, through \Cref{eq:totcs} and \Cref{eq:afb}. 

The results are presented in Table 11 of \cite{Abramowicz:2018rjq} and repeated here in \Cref{tab:results}. The expected precision on the top-quark pair production cross section and the forward-backward asymmetry are 1.1\%\,(2.0\%) and 1.4\%\,(2.3\%), respectively, for operation at 1.4\,\tev\,(3\,\tev) with -80\% electron polarisation~\cite{Abramowicz:2018rjq}. Operation at +80\% electron polarisation, leads to values that are about a factor 2.5 higher~\cite{Abramowicz:2018rjq}.

\subsection{Effective field theory interpretation}
The event selection presented here has also been studied in the context of Effective Field Theory (EFT) top-philic operators in \cite{Abramowicz:2018rjq}. These operators are used to parametrise new physics effects induced in the top-quark electroweak interactions and benefit from operation at different electron beam polarisation, since it allows to efficiently disentangle the photon and Z-boson contributions to the $\ttbar$ final state, as well as from operation at high centre-of-mass energies, since it allows to probe energy scales well beyond the nominal collision energy of CLIC. By including the results for the high-energy stages of CLIC presented in this paper, a clear improvement is seen with respect to operation at $\roots=380\,\gev$. In particular, the sensitivity to so-called four-fermion operators improves by more than one order of magnitude, allowing CLIC to probe new physics scales of the order of tens of TeV~\cite{Abramowicz:2018rjq}.

\begin{table}
\begin{minipage}{\columnwidth}
\begin{center}
\begin{tabular}{l|cccc}
\toprule
$\roots$ & \multicolumn{2}{c}{1.4\,\tev} & \multicolumn{2}{c}{3\,\tev} \rule{0pt}{3ex} \\
P(\Pem) & -80\% & +80\% & -80\% & +80\% \rule{0pt}{3ex} \\
\midrule
$\csttbar$~[fb] & 18.44 & 9.84 & 3.52 & 1.91 \rule{0pt}{3ex} \\
stat. unc.~[fb] & 0.21 & 0.29 & 0.07 & 0.09 \rule{0pt}{3ex} \\
\midrule
\afb & 0.567 & 0.620 & 0.596 & 0.645 \rule{0pt}{3ex}\\
stat. unc. & 0.008 & 0.020 & 0.014 & 0.034 \rule{0pt}{3ex}\\
\bottomrule
\end{tabular}
\end{center}
\caption{Results from the analysis of semi-leptonically decaying top quarks at the three stages of CLIC. The values are obtained from full simulation studies using the \clicild detector concept. Note that the cross section, $\csttbar$, and \afb are defined in the kinematic region of $\rootsprime\geq1.2\,(2.6)\,\tev$ for operation at $\roots=1.4\,\tev\,(3\,\tev)$. In addition, the cross sections represent a convolution of the total $\ttbar$ production cross section with the CLIC luminosity spectrum in the kinematic region studied. Table taken from \cite{Abramowicz:2018rjq}. \label{tab:results}}
\end{minipage}
\end{table}

\subsection{Systematic uncertainties}
\label{ssec:systematics}

The expected uncertainties given in \Cref{tab:results} are purely statistical and do not include the effect of potential sources of systematic uncertainty. However, the results illustrate the level of precision that would be desirable for the control of systematic effects. A full investigation of systematic uncertainties is beyond the scope of this paper, but the impact of some ad-hoc variations are discussed in a larger context of $\ttbar$ studies at CLIC in \cite{Abramowicz:2018rjq}. In conclusion, the effects on $\csttbar$ and $\afb$ indicate that this analysis would not be limited by systematics effects.


\section{Summary and conclusions}
\label{sec:summary}

The analysis and results presented in this paper focus on ``single lepton+jets'' $\ttbar$ final states and represent the first studies of $\ttbar$ production in full detector simulation for a multi-TeV $\epem$ collider. The main results from this paper and a summary of the analysis strategy were previously summarised in \cite{Abramowicz:2018rjq}. Here we outline the underlying analysis in greater detail and include for the first time the optimisation procedure for the boosted top-tagger.

We present results for the top-quark pair production cross section and the forward-backward asymmetry, constituting powerful tools for discovery and a deeper understanding of the nature of the electro-weak symmetry breaking. The main results from this paper were also summarised in \cite{Abramowicz:2018rjq} where a comprehensive view of the prospects of for the foreseen top-quark programme at CLIC was presented.

The highly collimated jet environments present for operation above 1\,\tev leads to highly boosted topologies where dedicated techniques based on the analysis of the sub-structure of large-R jets show a clear advantage. Top-tagging efficiencies of 70\% are achieved thanks to the low background levels expected at CLIC and the overall high granularity and excellent jet energy resolution of the CLIC detector concept. The use of boosted techniques enables the sensitivity to BSM physics to be extended to the highest collision energies and beyond; detailed studies of effective field theory top-philic operators illustrate that new physics scales of order of tens of TeV~\cite{Abramowicz:2018rjq} is within reach.

Further improvements can be made by including for example fully-hadronic final states, or semi-leptonic tau events where the tau decays leptonically. However, for the former, the jet charge reconstruction, needed for the reconstruction of observables such as the \afb, is challenging and needs to be studied in more detail. Further improvements of the jet reconstruction algorithms and boosted top-tagging strategies would also be beneficial since the performance of more complex final states in general is often limited by reconstruction issues like the confusion in the jet clustering.

\section*{Acknowledgements}
This work benefited from services provided by the ILC Virtual Organisation, supported by the national resource providers of the EGI Federation. This research was done using resources provided by the Open Science Grid, which is supported by the National Science Foundation and the U.S. Department of Energy's Office of Science. This project has received funding from the European Union's Horizon 2020 Research and Innovation programme under Grant Agreement no. 654168.

\printbibliography[title=References]

\clearpage
\appendix
\clearpage

\section{Additional jet clustering results}
\label{sec:appendix}


\begin{figure}[htpb]
\centering
\includegraphics[width=0.48\columnwidth]{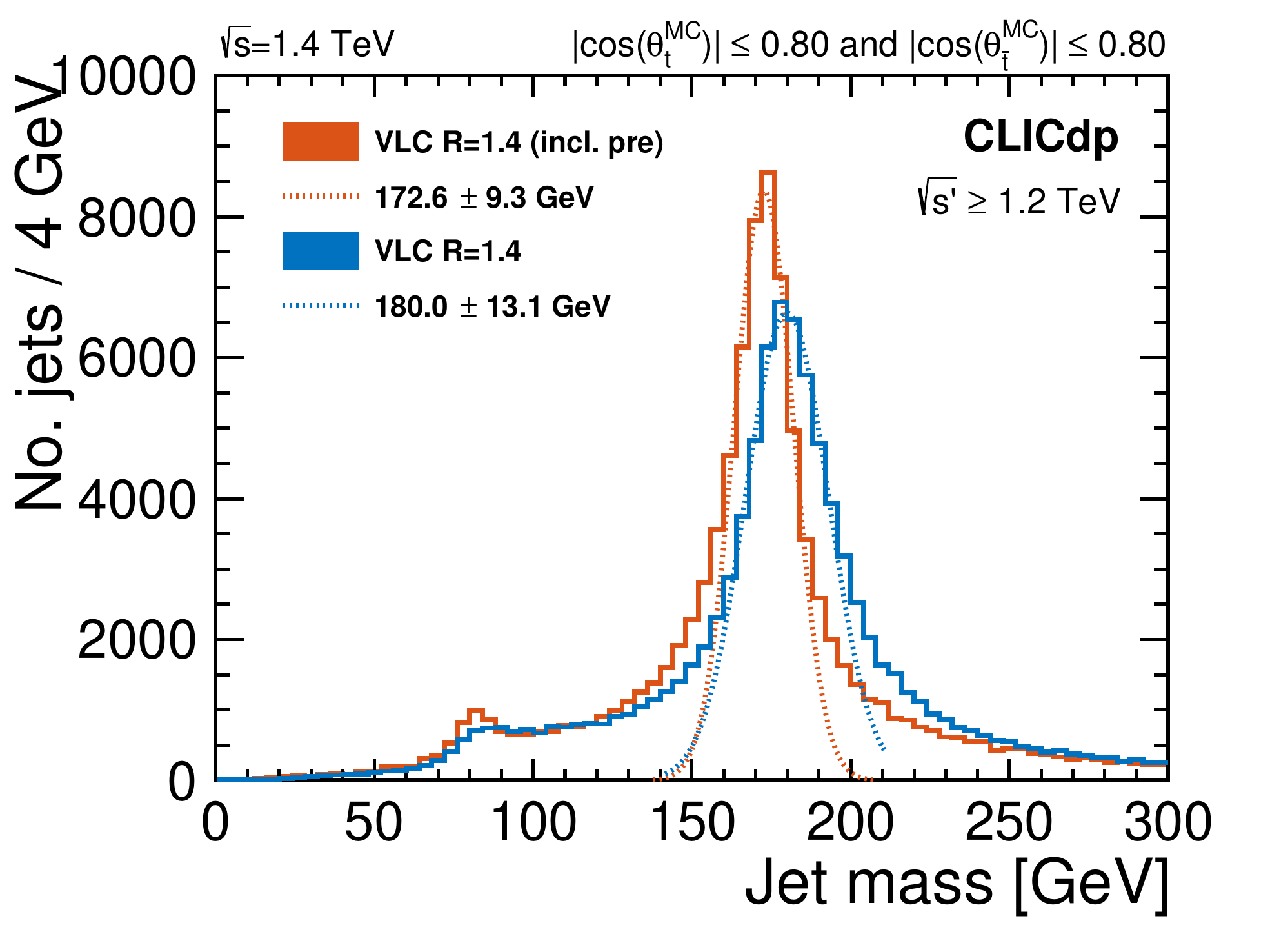}
~~~
\includegraphics[width=0.48\columnwidth]{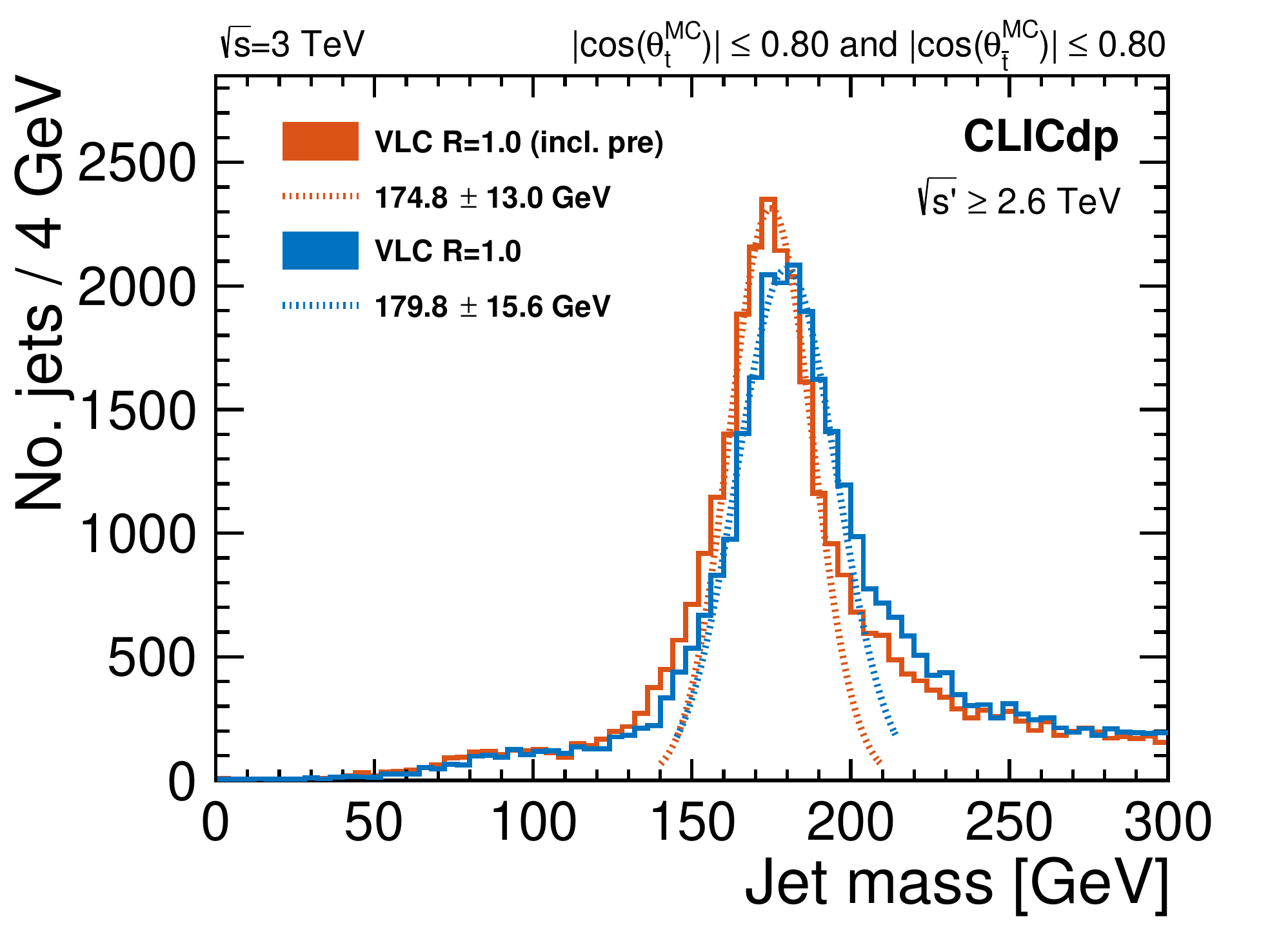}
\caption{Number of jets as a function of the reconstructed jet mass for fully-hadronic $\ttbar$ events at $\roots=1.4\,\tev$ (left) and $\roots=3\,\tev$ (right) and for $\rootsprime$ above 1.2\,\tev and 2.6\,\tev, respectively. The dashed lines represent the corresponding gaussian fit to the distributions. The distributions in red (blue) show the jet mass including (excluding) pre-clustering. \label{fig:jetmassopt}}
\end{figure}

\begin{figure}[htpb]
\centering
\includegraphics[width=0.48\columnwidth]{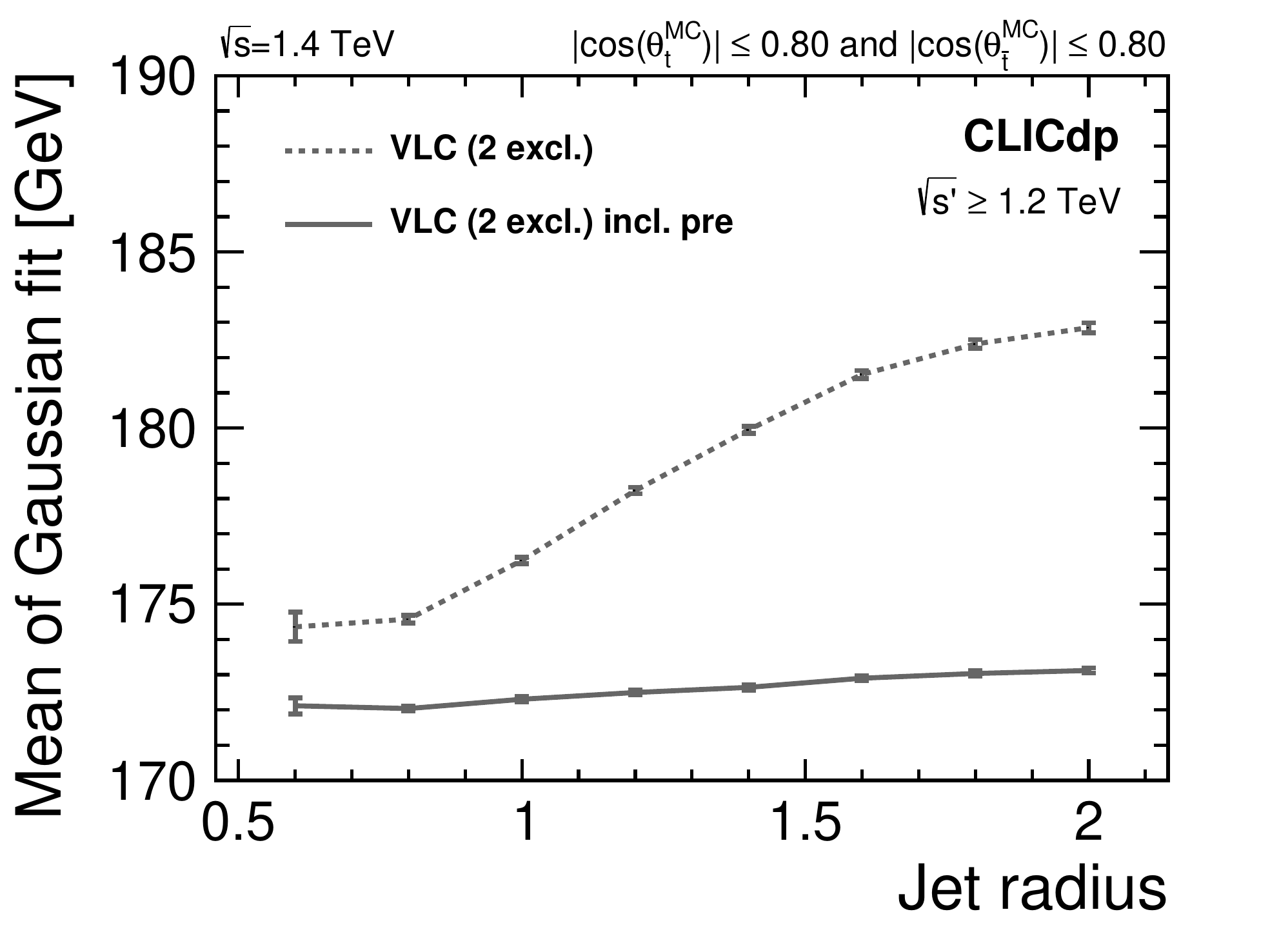}
~~~
\includegraphics[width=0.48\columnwidth]{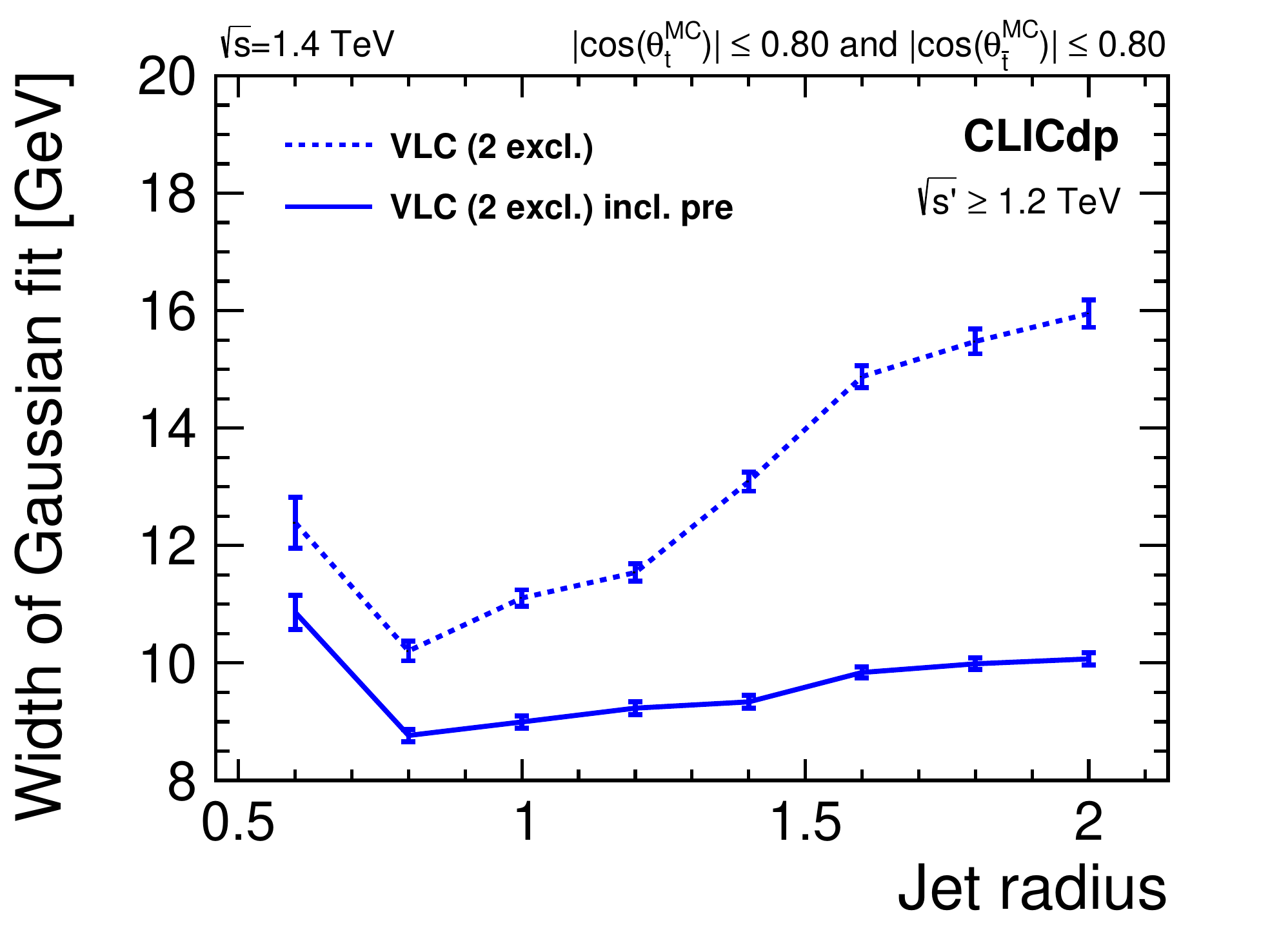}
\caption{Mean (left) and width (right) of the gaussian fit to the fully-hadronic $\ttbar$ large-R jet mass distribution at $\roots=1.4\,\tev$ as function of jet clustering radius, shown for the case of including pre-clustering (solid) and not (dashed). \label{fig:jetmassopt}}
\end{figure}

\begin{figure}[htpb]
\centering
\includegraphics[width=0.48\columnwidth]{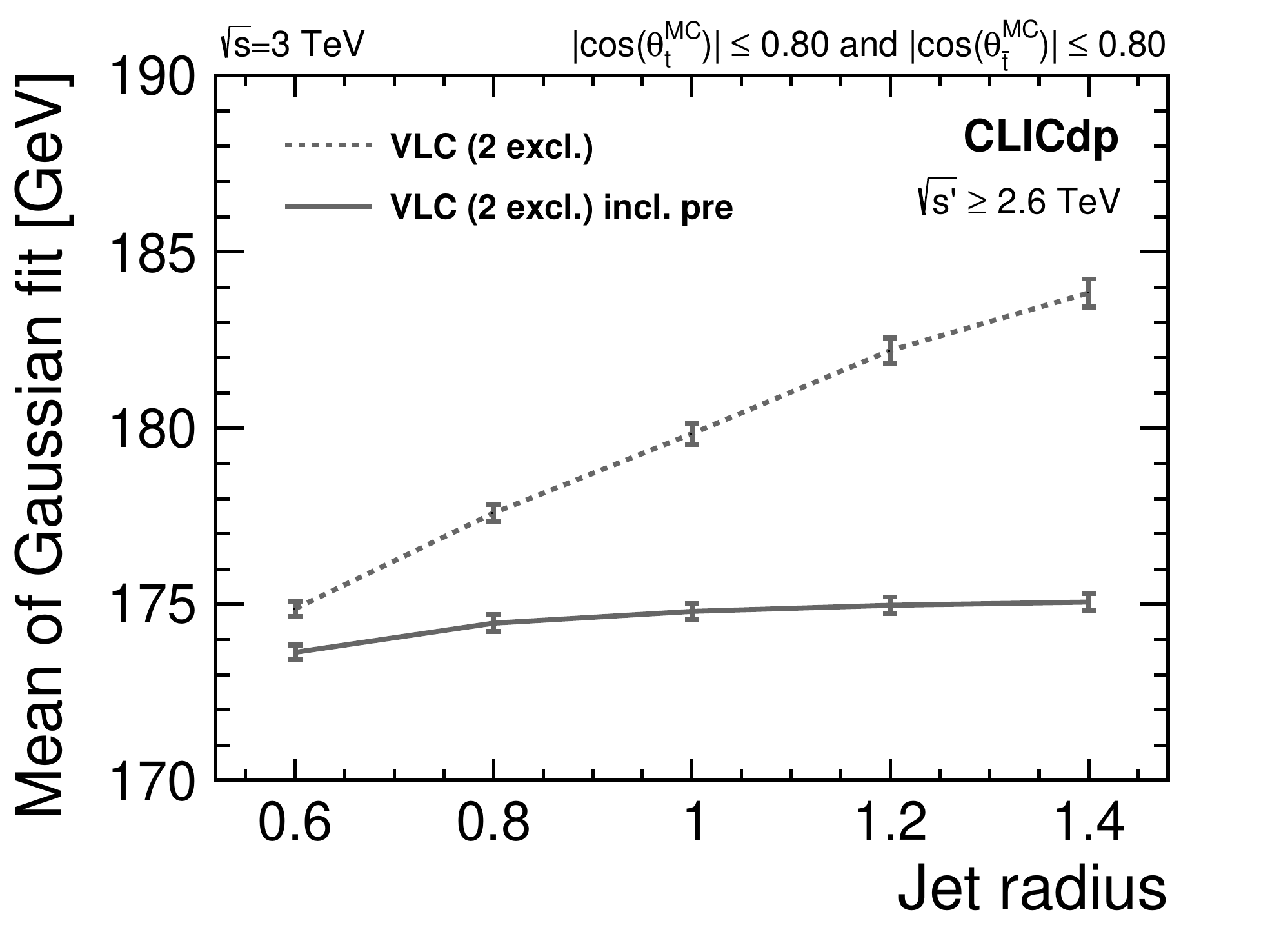}
~~~
\includegraphics[width=0.48\columnwidth]{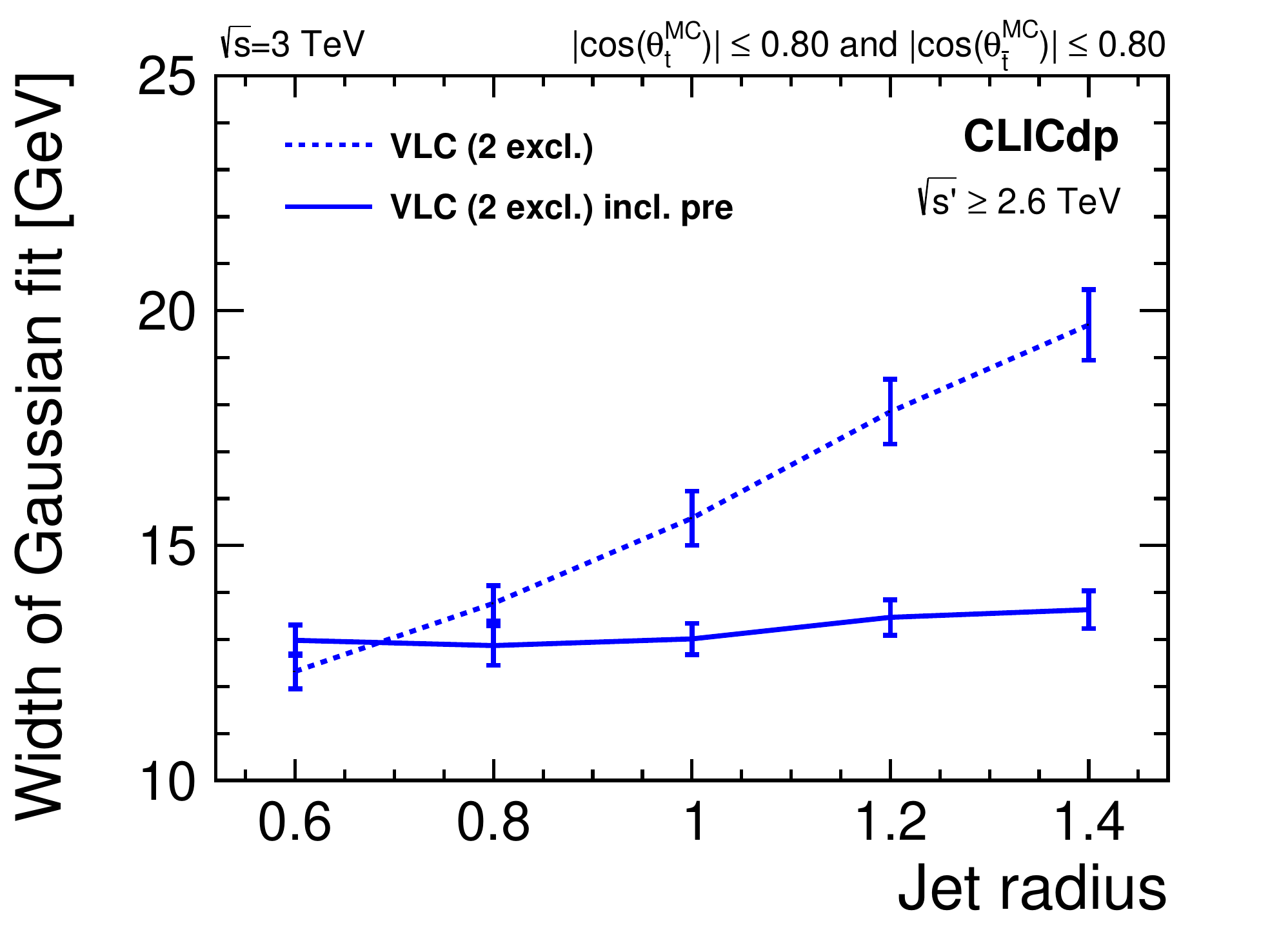}
\caption{Mean (left) and width (right) of the gaussian fit to the fully-hadronic $\ttbar$ large-R jet mass distribution at $\roots=3\,\tev$ as function of jet clustering radius, shown for the case with (solid) and without (dashed) pre-clustering. \label{fig:jetmassopt}}
\end{figure}

\begin{figure}[htpb]
\centering
\includegraphics[width=0.48\columnwidth]{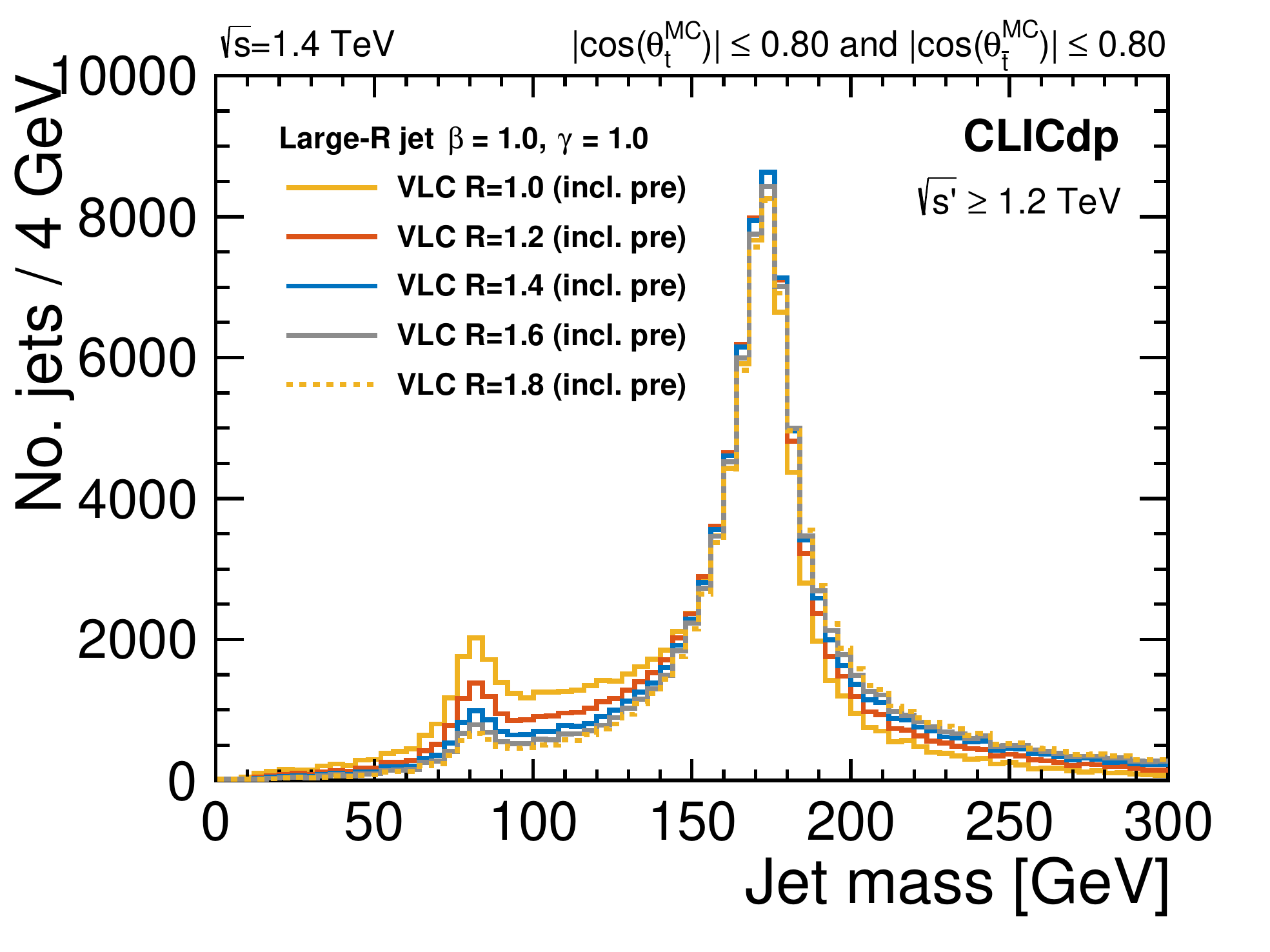}
~~~
\includegraphics[width=0.48\columnwidth]{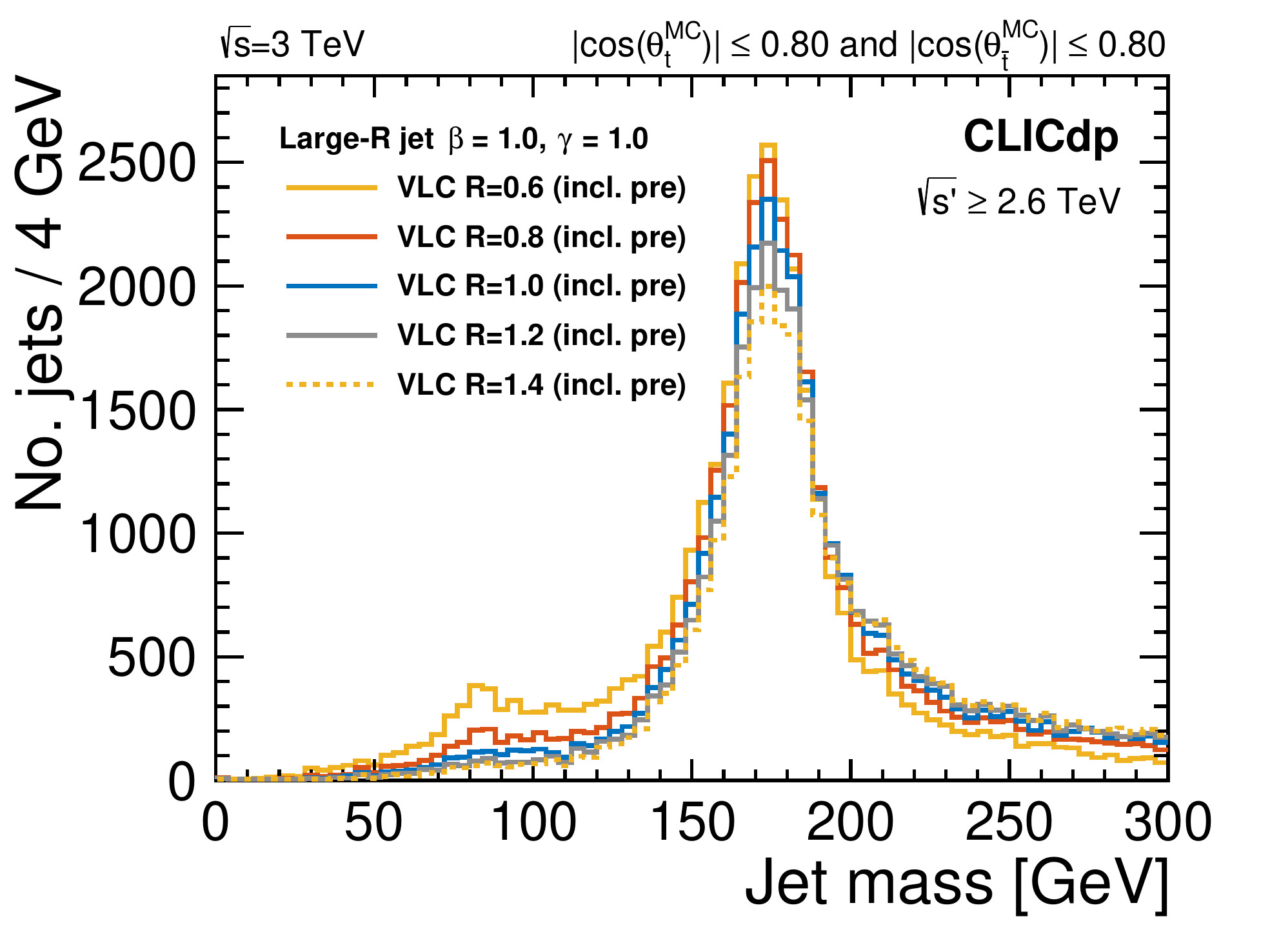}
\caption{Number of jets as a function of the reconstructed jet mass for fully-hadronic $\ttbar$ events at $\roots=1.4\,\tev$ (left) and $\roots=3\,\tev$ (right) and for $\rootsprime$ above 1.2\,\tev and 2.6\,\tev, respectively.\label{fig:analysis:jetreco:largeRradius:boosted}}
\end{figure}

\begin{figure}[htpb]
\centering
\includegraphics[width=0.48\columnwidth]{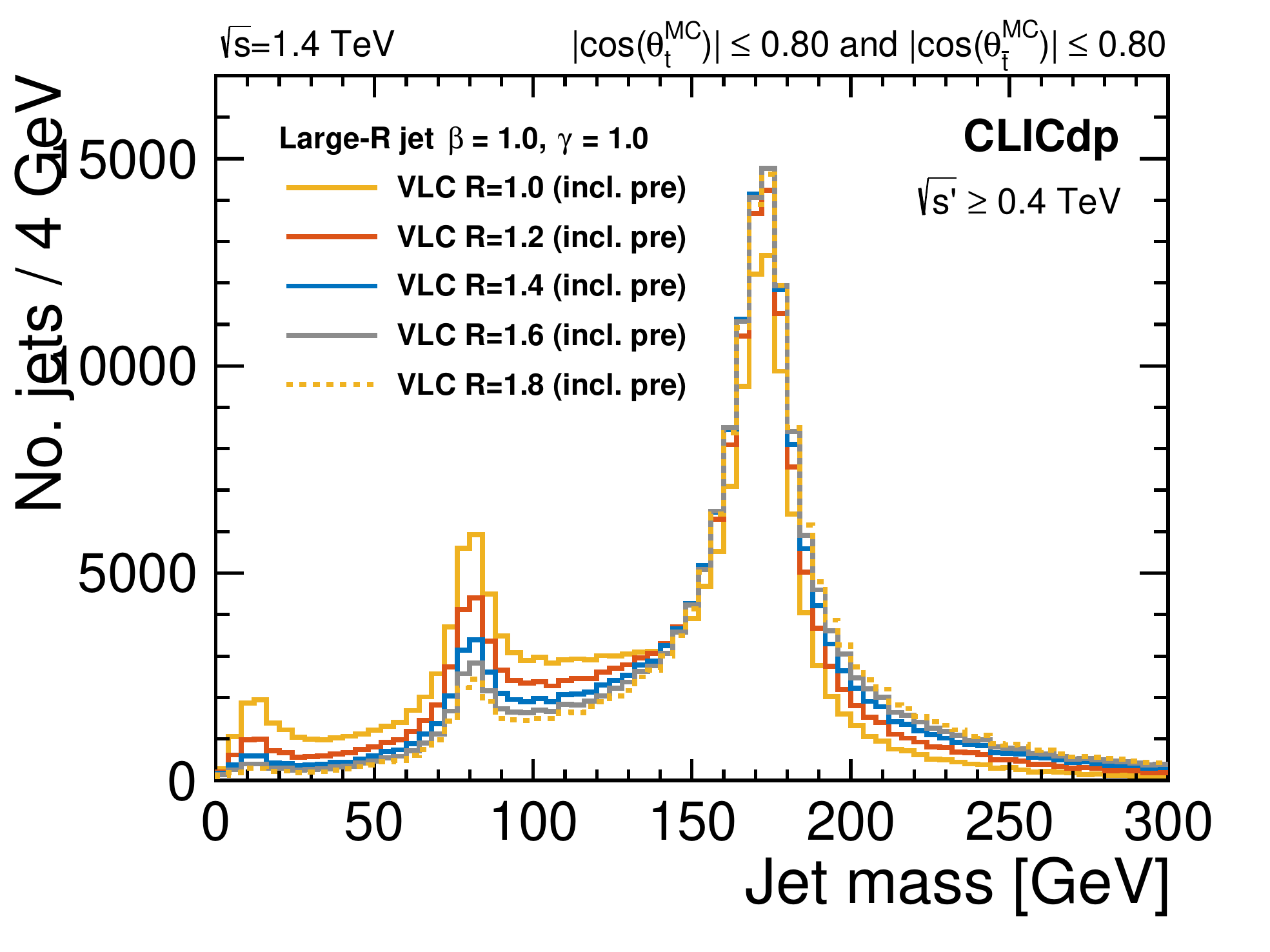}
~~~
\includegraphics[width=0.48\columnwidth]{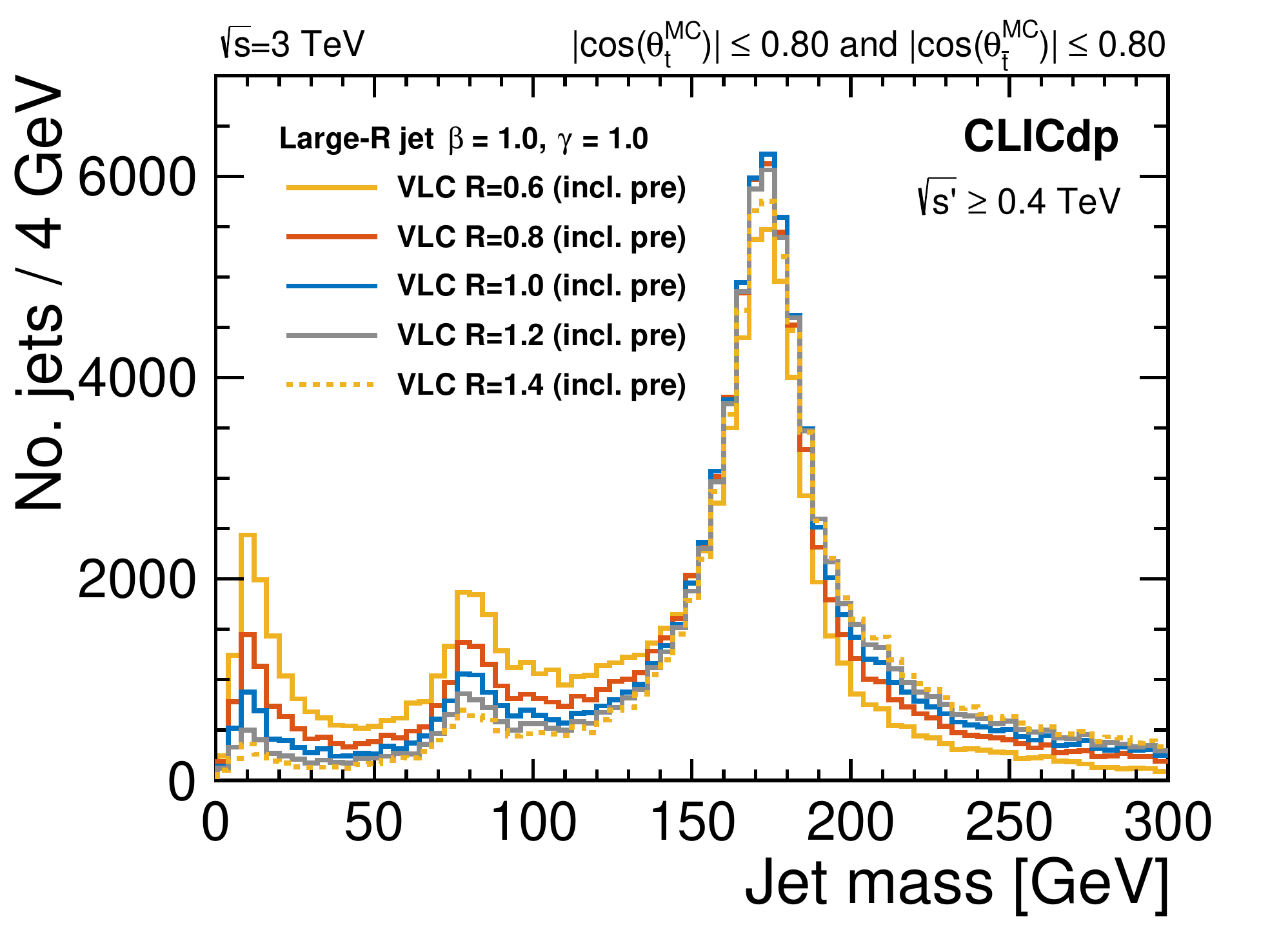}
\caption{Number of jets as a function of the reconstructed jet mass for fully-hadronic $\ttbar$ events at $\roots=1.4\,\tev$ (left) and $\roots=3\,\tev$ (right) and for $\rootsprime$ above 0.4\,\tev.\label{fig:analysis:jetreco:largeRradius:nonboosted}}
\end{figure}

\begin{figure}[htpb]
\centering
\includegraphics[width=0.48\columnwidth]{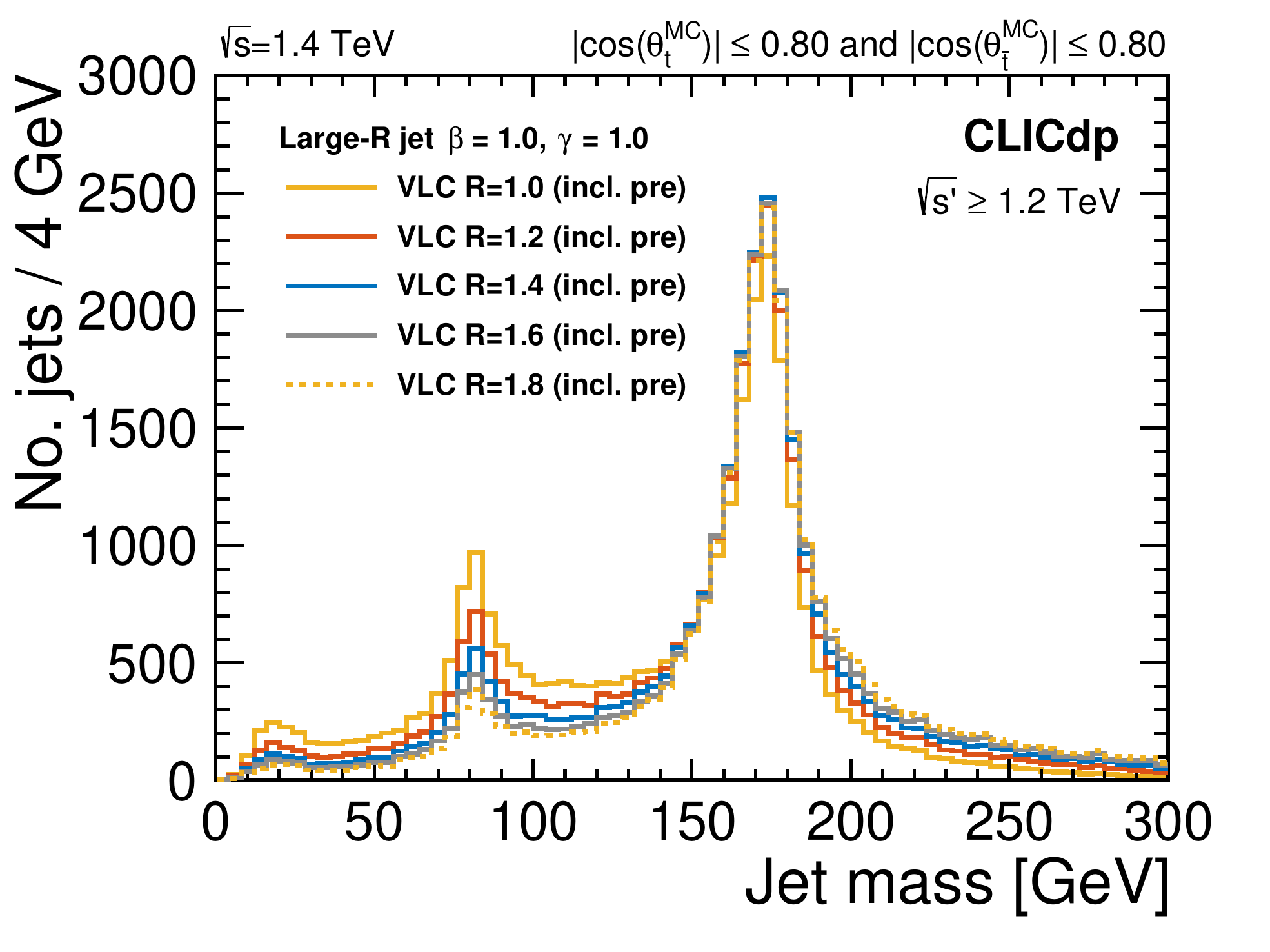}
~~~
\includegraphics[width=0.48\columnwidth]{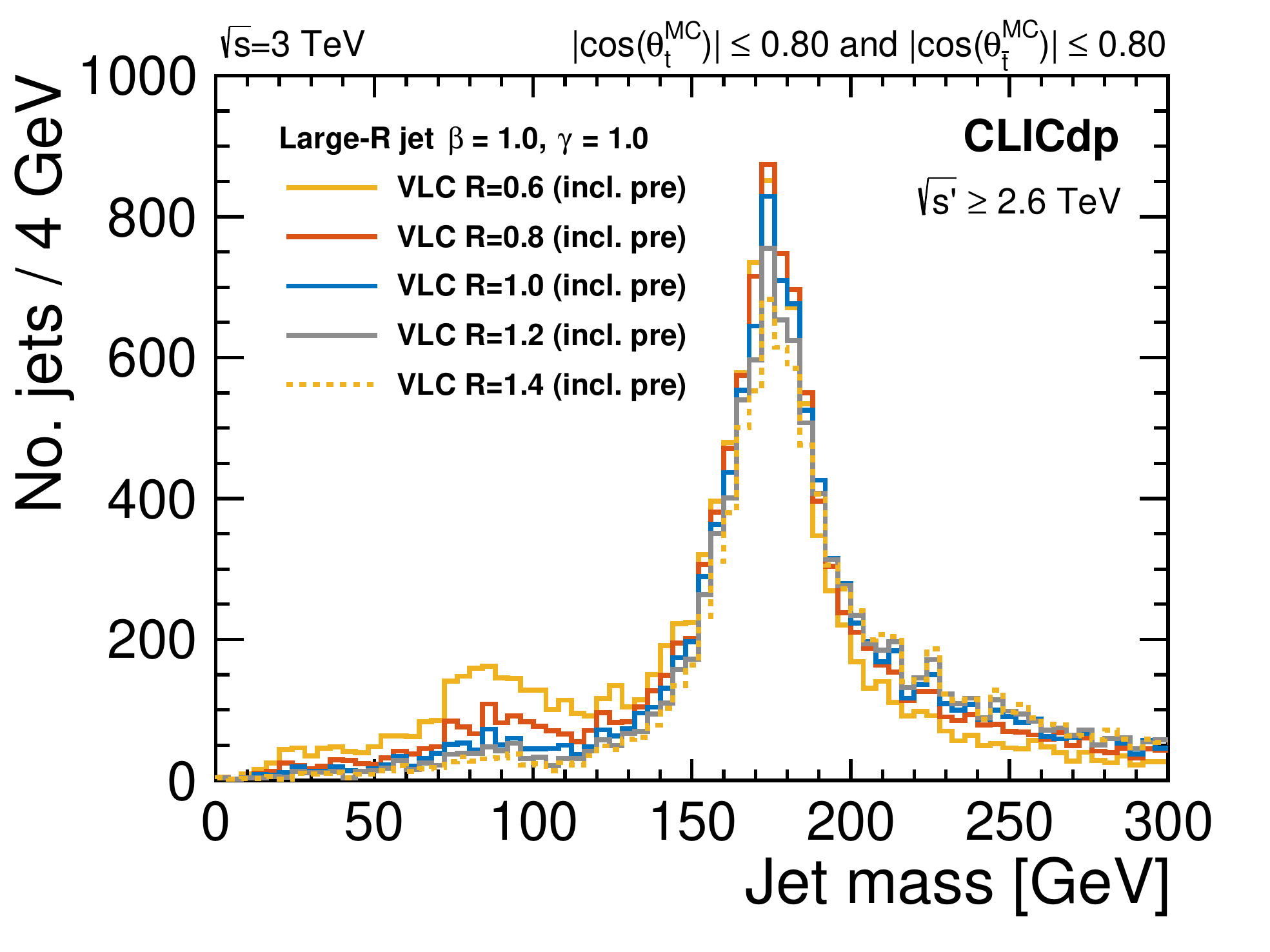}
\caption{Number of jets as a function of the reconstructed jet mass for semi-leptonic $\ttbar$ events at $\roots=1.4\,\tev$ (left) and $\roots=3\,\tev$ (right) and for $\rootsprime$ above 1.2\,\tev and 2.6\,\tev, respectively. \label{fig:analysis:jetreco:largeRradius:semilep:boosted}}
\end{figure}

\begin{figure}[htpb]
\centering
\includegraphics[width=0.48\columnwidth]{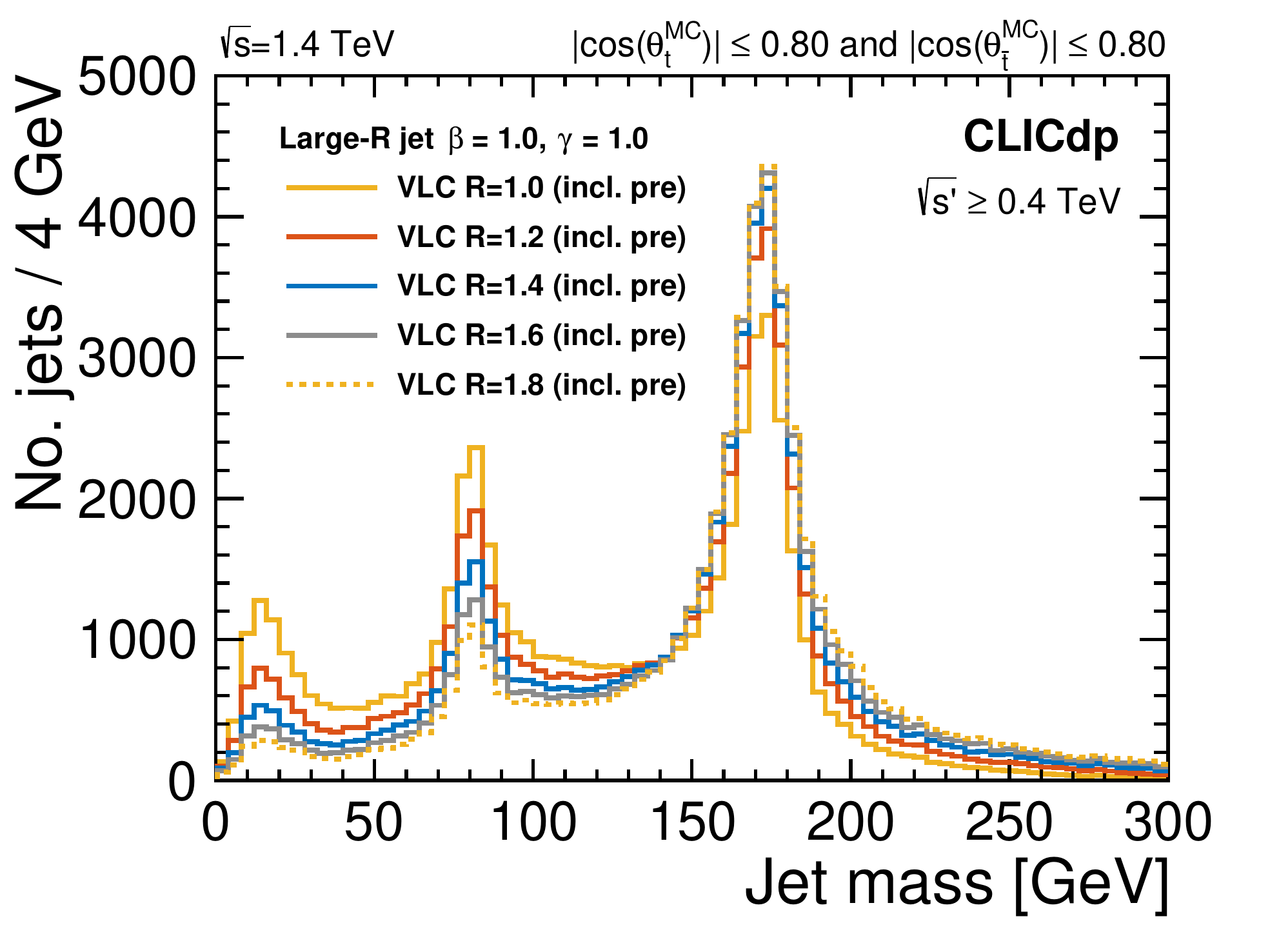}
~~~
\includegraphics[width=0.48\columnwidth]{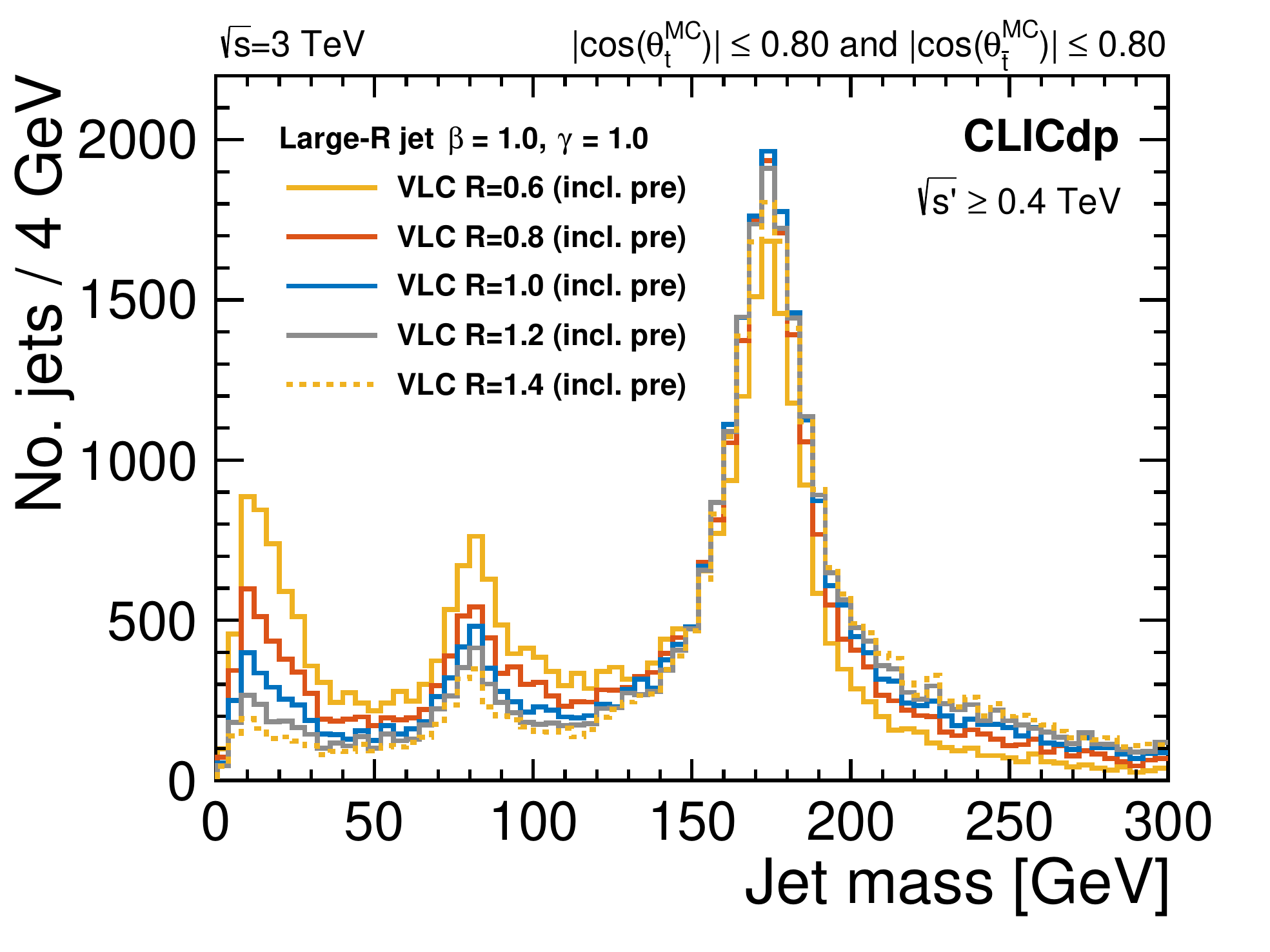}
\caption{Number of jets as a function of the reconstructed jet mass for semi-leptonic $\ttbar$ events at $\roots=1.4\,\tev$ (left) and $\roots=3\,\tev$ (right) and for $\rootsprime$ above 0.4\,\tev.\label{fig:analysis:jetreco:largeRradius:semilep:nonboosted}}
\end{figure}

\FloatBarrier

\clearpage

\section{Additional jet sub-structure variables}

\begin{figure}[htpb]
	\centering
	\begin{subfigure}{0.48\columnwidth}
	\includegraphics[width=\textwidth]{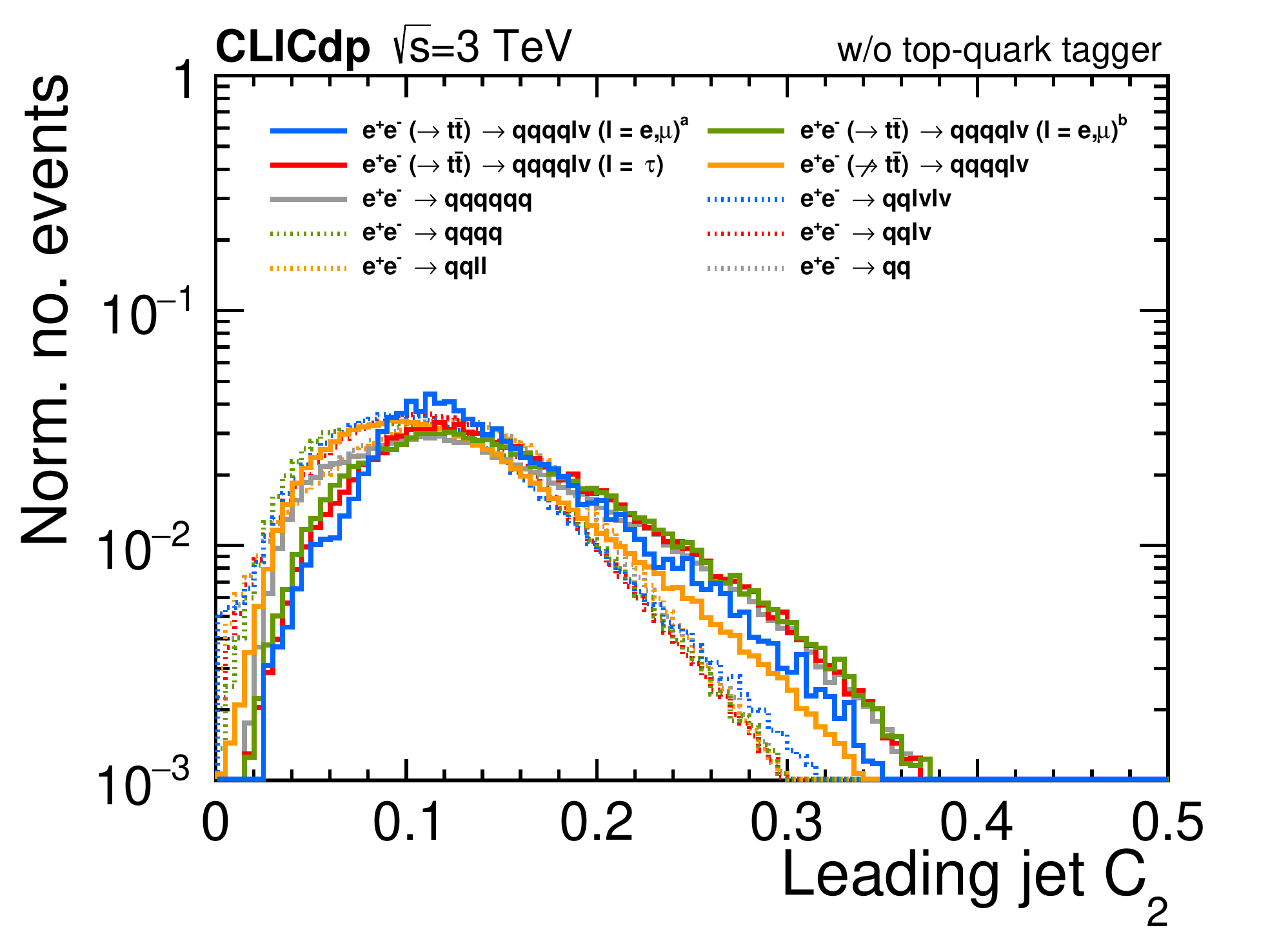}
	\caption{$C_{2}$ without applying the top-quark tagger.}
	\end{subfigure}
	~~~
	\begin{subfigure}{0.48\columnwidth}
	\includegraphics[width=\textwidth]{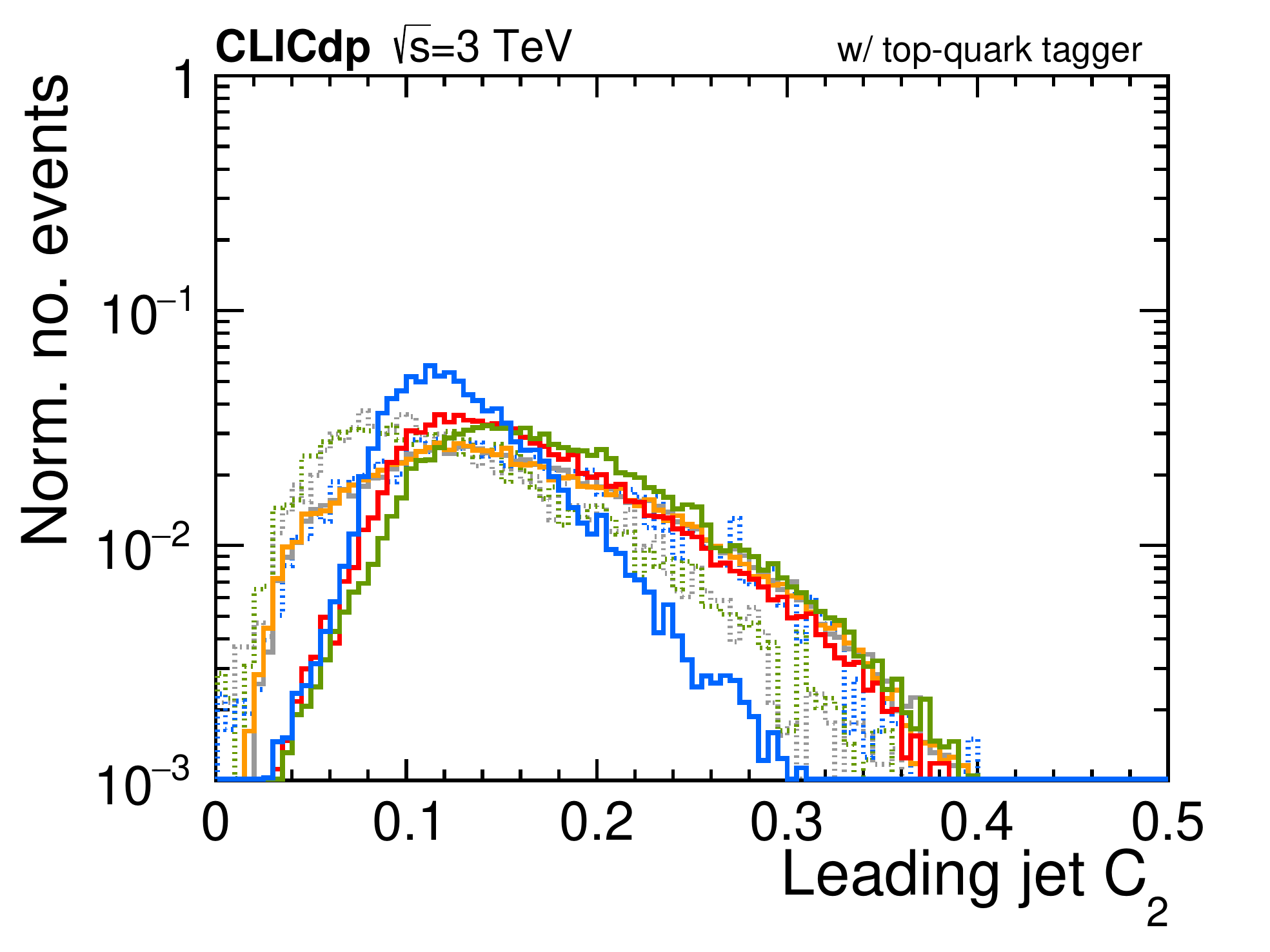}
	\caption{$C_{2}$ after applying the top-quark tagger.}
	\end{subfigure}\\
	\vspace{5mm}
	\begin{subfigure}{0.48\columnwidth}
	\includegraphics[width=\textwidth]{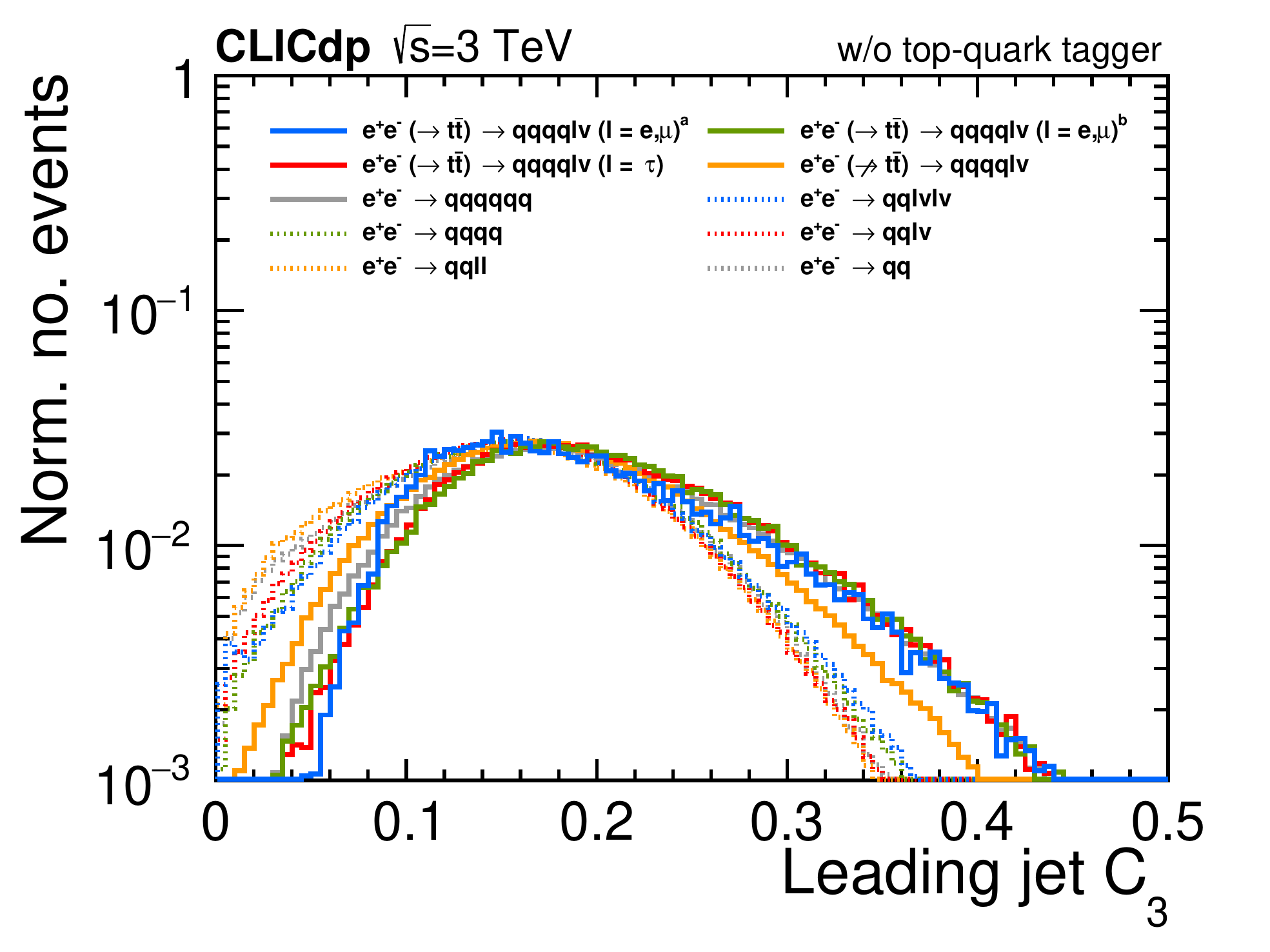}
	\caption{$C_{3}$ without applying the top-quark tagger.}
	\end{subfigure}
	~~~
	\begin{subfigure}{0.48\columnwidth}
	\includegraphics[width=\textwidth]{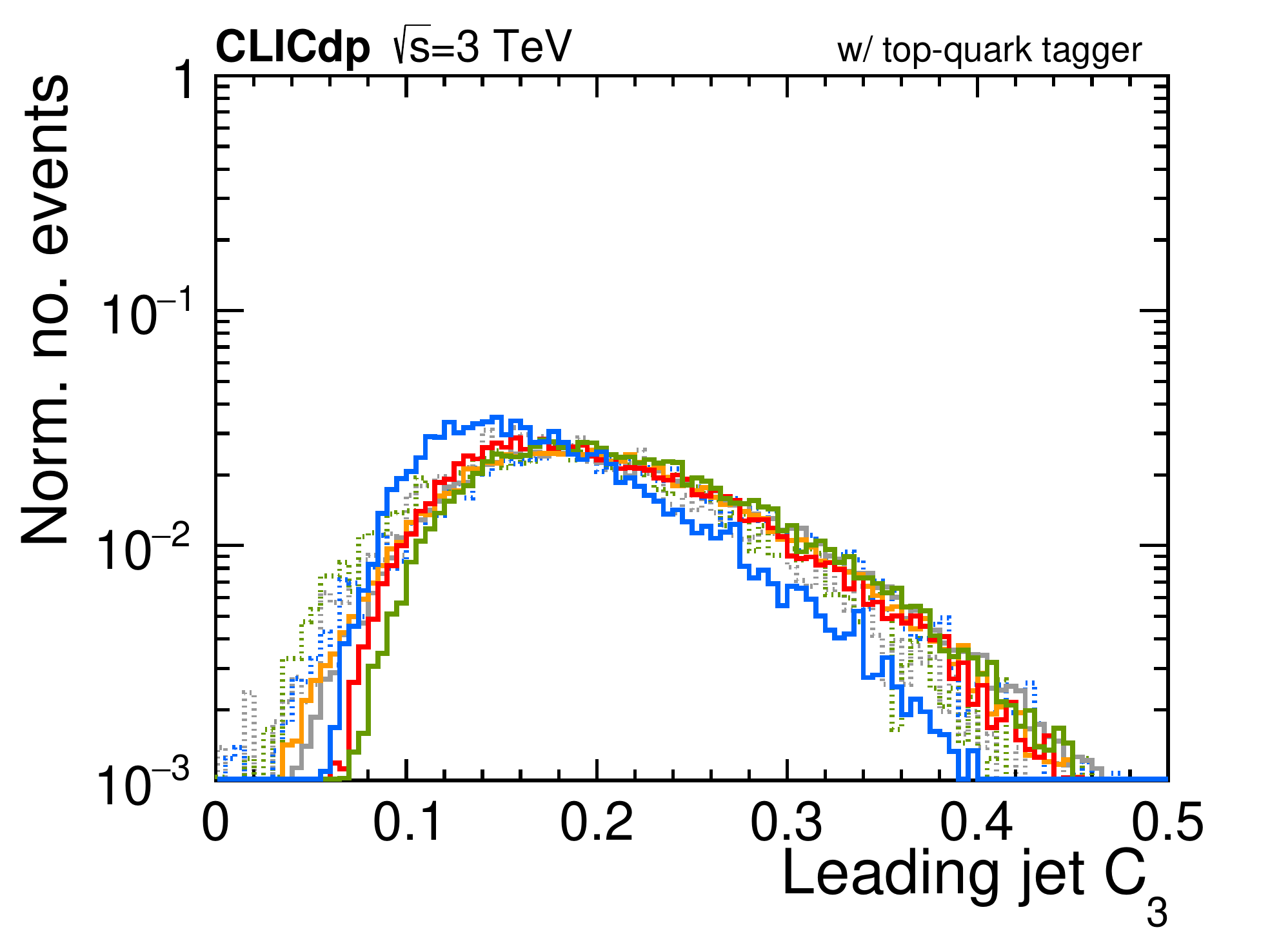}
	\caption{$C_{3}$ after applying the top-quark tagger.}
	\end{subfigure}\\
\caption{Substructure variables for the leading (highest energy) large-R jet. Note that the qqlv and qqll backgrounds have been omitted for the figures in the right column. The retention of these backgrounds is already very low after the pre-cuts. The superscript `a' (`b') refers to the kinematic region $\rootsprime\geq1.2\,\tev$ ($\rootsprime<1.2\,\tev$). \label{fig:analysis:mva:variables:energycorrCJ1:3tev}}
\end{figure}

\begin{figure}[htpb]
	\centering
	\begin{subfigure}{0.48\columnwidth}
	\includegraphics[width=\textwidth]{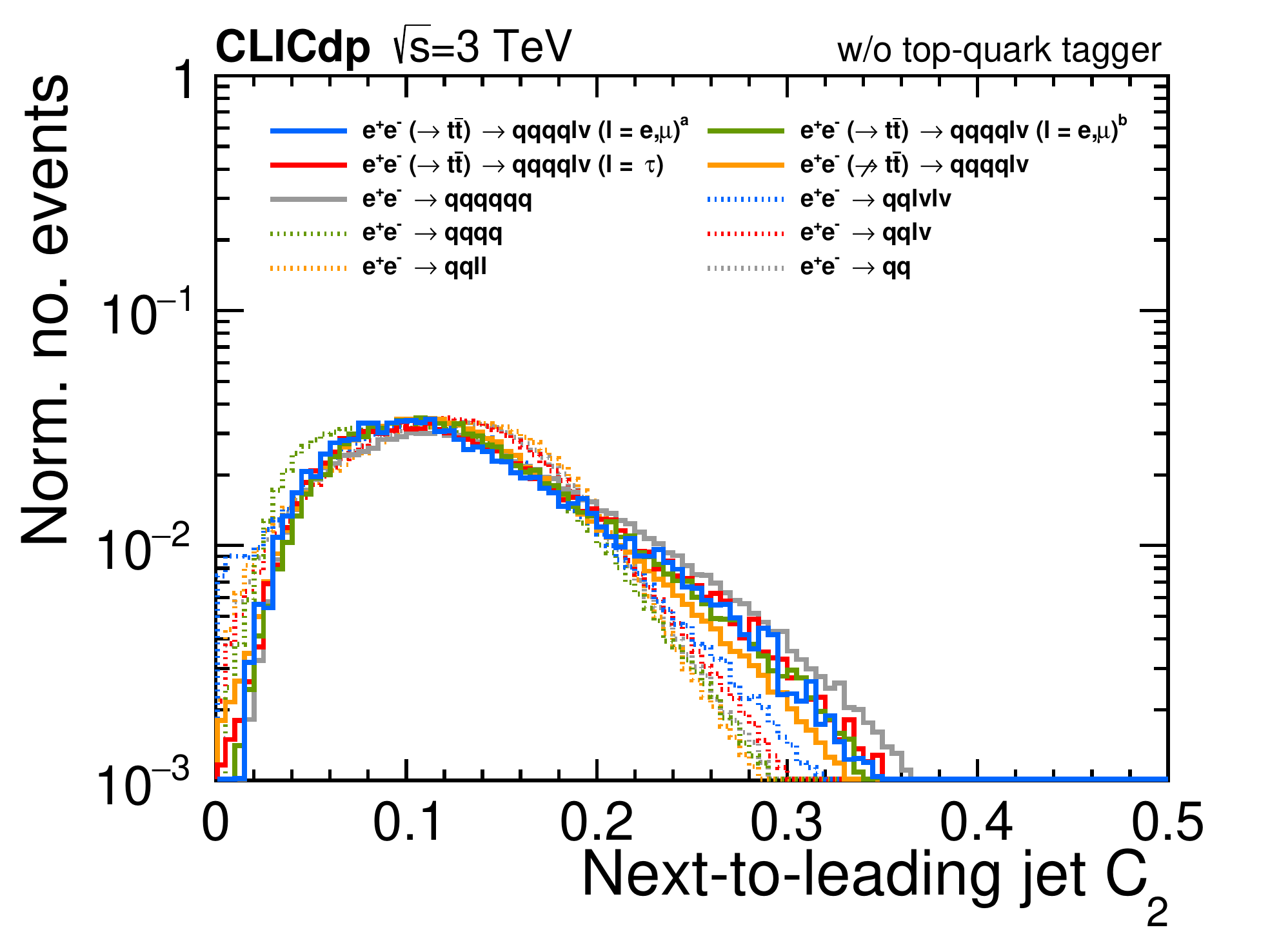}
	\caption{$C_{2}$ without applying the top-quark tagger.}
	\end{subfigure}
	~~~
	\begin{subfigure}{0.48\columnwidth}
	\includegraphics[width=\textwidth]{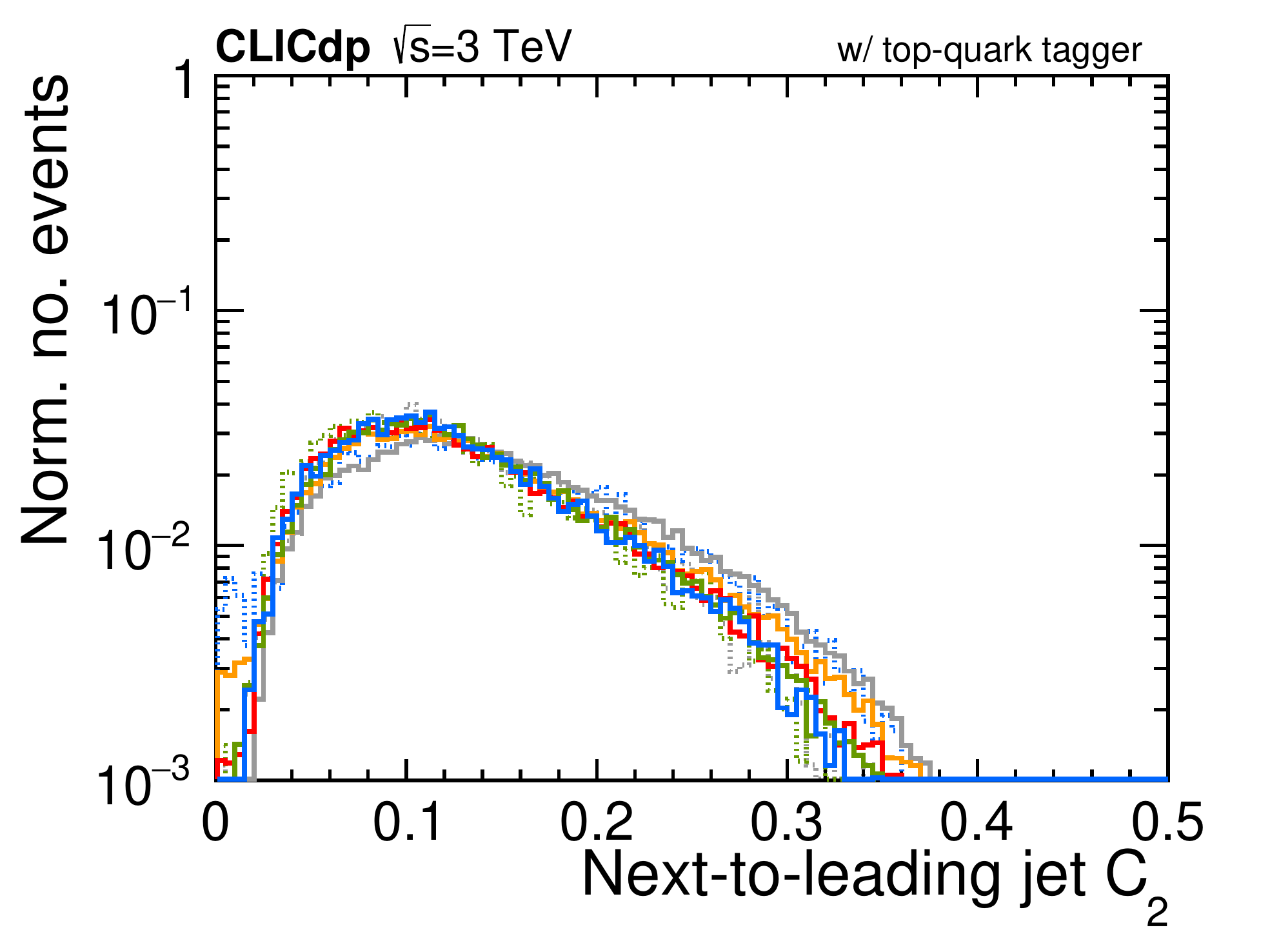}
	\caption{$C_{2}$ after applying the top-quark tagger.}
	\end{subfigure}\\
	\vspace{5mm}
	\begin{subfigure}{0.48\columnwidth}
	\includegraphics[width=\textwidth]{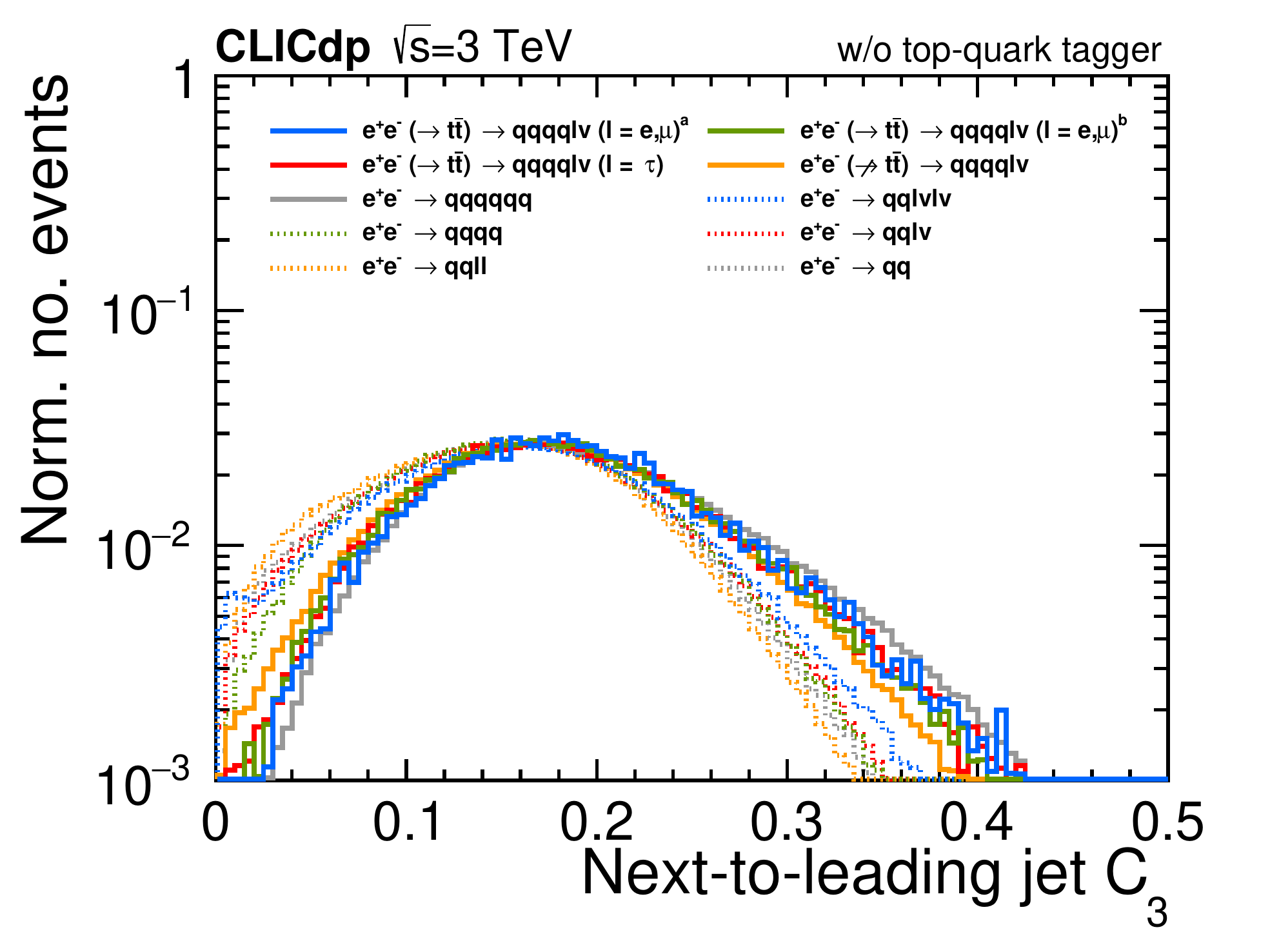}
	\caption{$C_{3}$ without applying the top-quark tagger.}
	\end{subfigure}
	~~~
	\begin{subfigure}{0.48\columnwidth}
	\includegraphics[width=\textwidth]{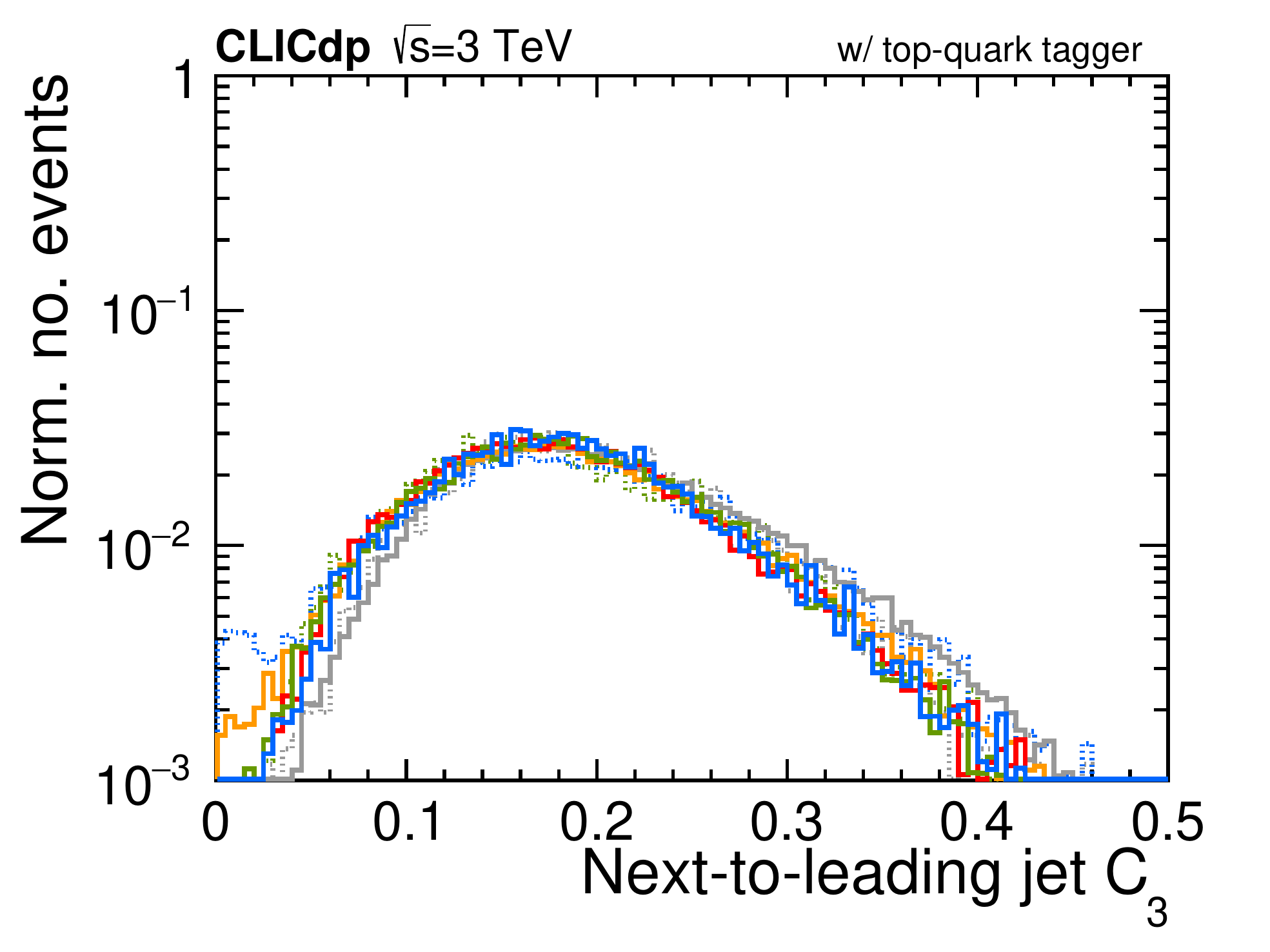}
	\caption{$C_{3}$ after applying the top-quark tagger.}
	\end{subfigure}\\
\caption{Substructure variables for the next-to-leading (lowest energy) large-R jet. Note that the qqlv and qqll backgrounds have been omitted for the figures in the right column. The retention of these backgrounds is already very low after the pre-cuts. The superscript `a' (`b') refers to the kinematic region $\rootsprime\geq1.2\,\tev$ ($\rootsprime<1.2\,\tev$). \label{fig:analysis:mva:variables:energycorrCJ2:3tev}}
\end{figure}

\begin{figure}[htpb]
	\centering
	\begin{subfigure}{0.48\columnwidth}
	\includegraphics[width=\textwidth]{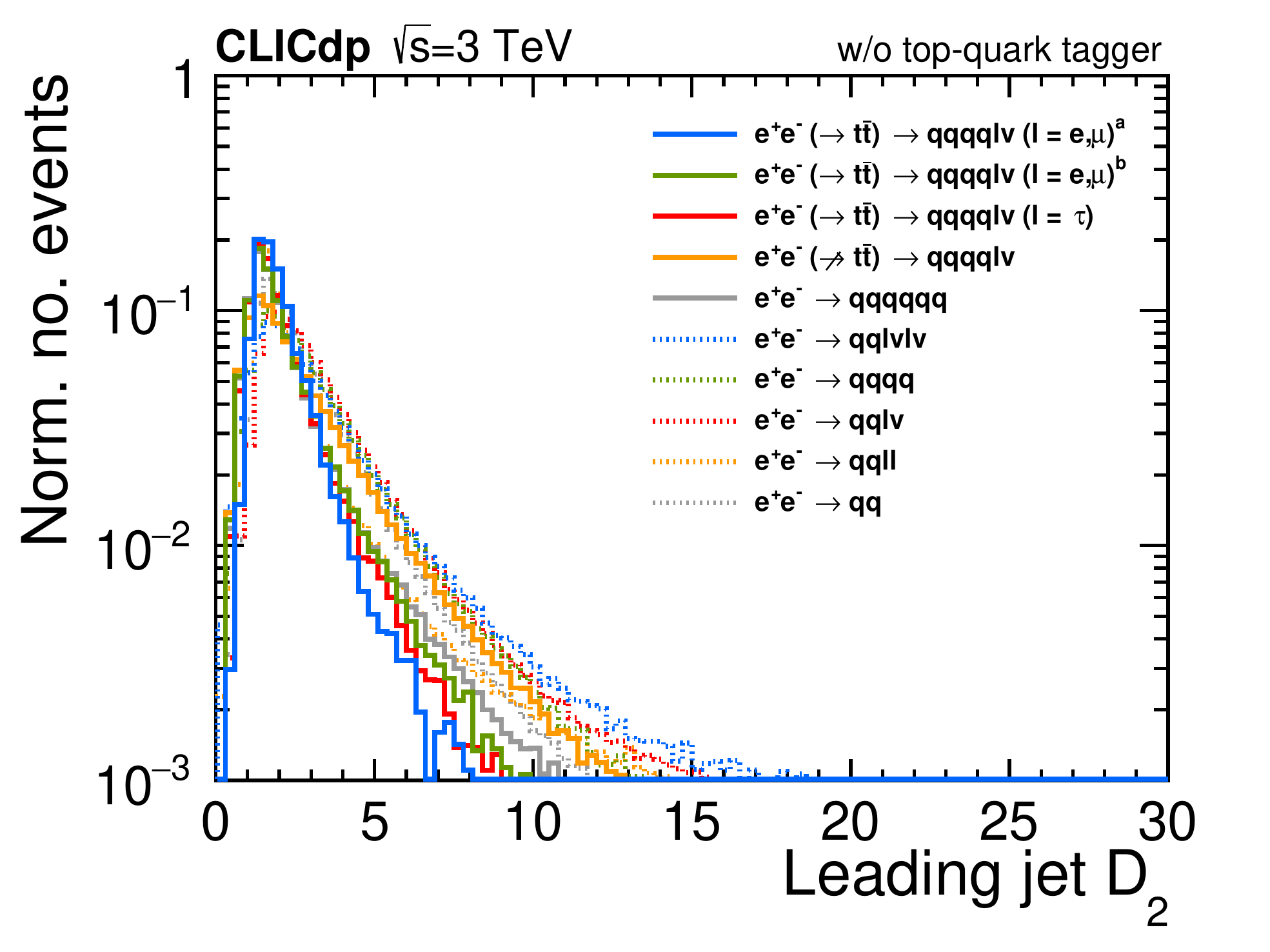}
	\caption{$D_{2}$ without applying the top-quark tagger.}
	\end{subfigure}
	~~~
	\begin{subfigure}{0.48\columnwidth}
	\includegraphics[width=\textwidth]{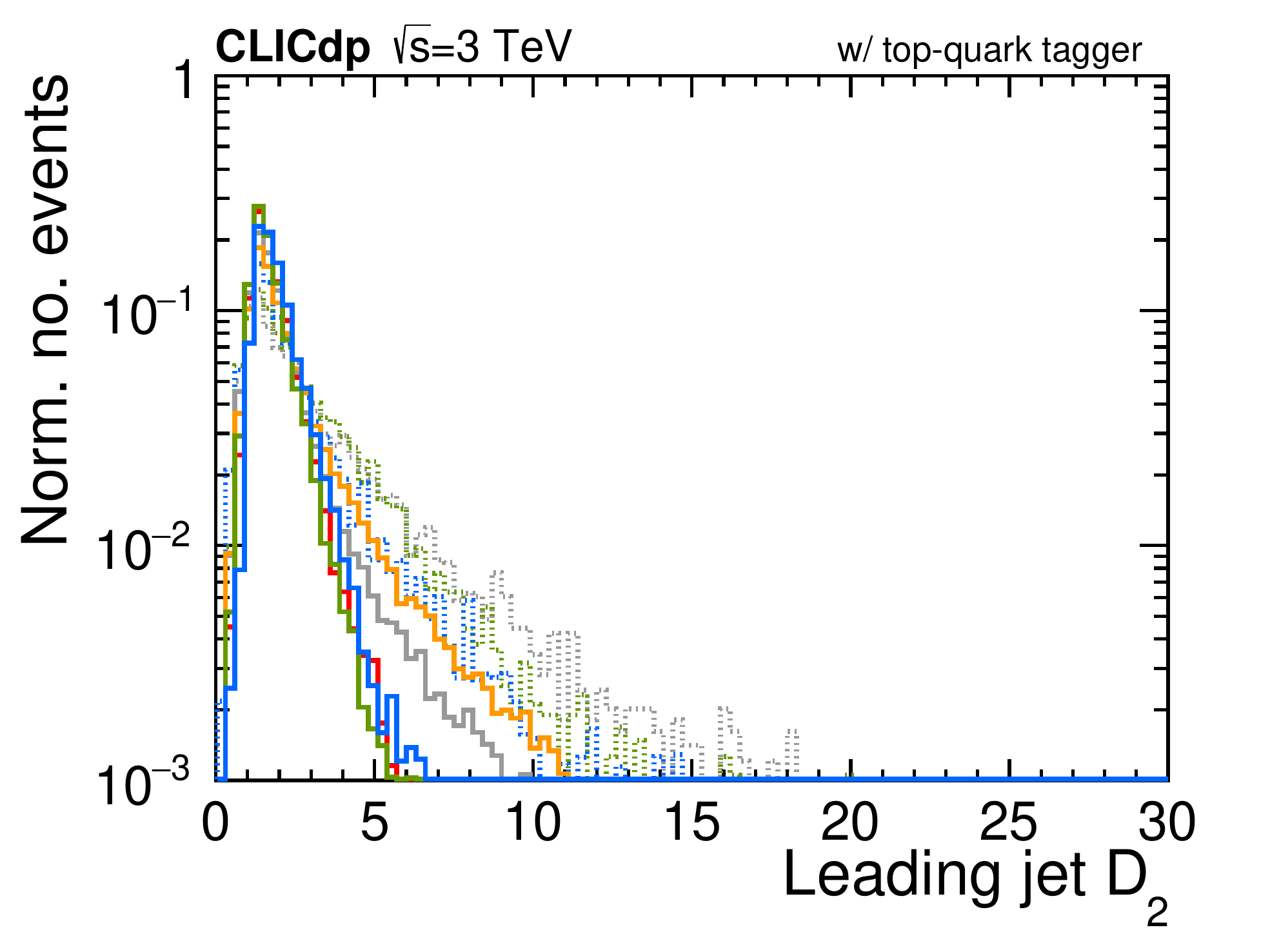}
	\caption{$D_{2}$ after applying the top-quark tagger.}
	\end{subfigure}\\
	\vspace{5mm}
	\begin{subfigure}{0.48\columnwidth}
	\includegraphics[width=\textwidth]{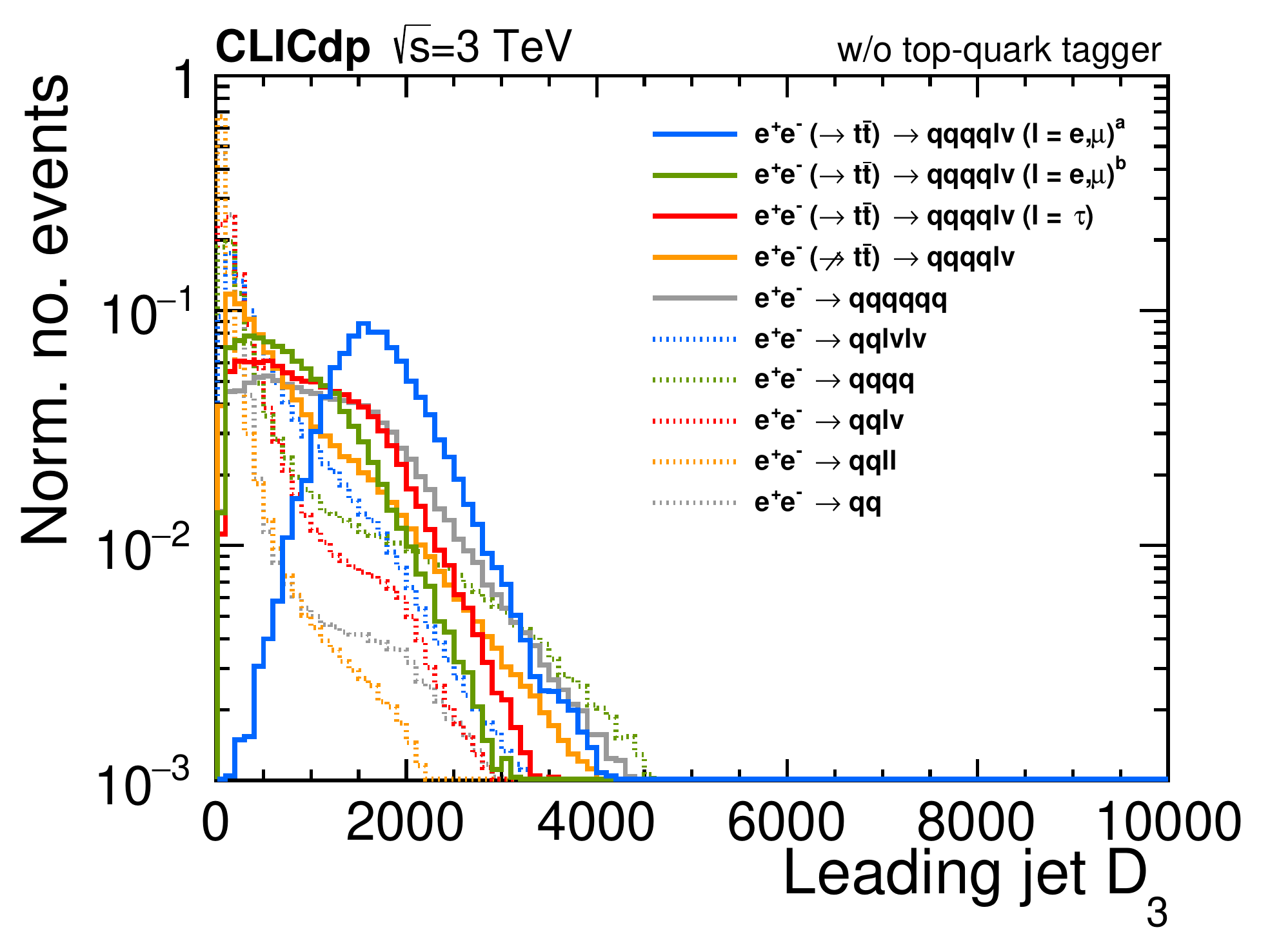}
	\caption{$D_{3}$ without applying the top-quark tagger.}
	\end{subfigure}
	~~~
	\begin{subfigure}{0.48\columnwidth}
	\includegraphics[width=\textwidth]{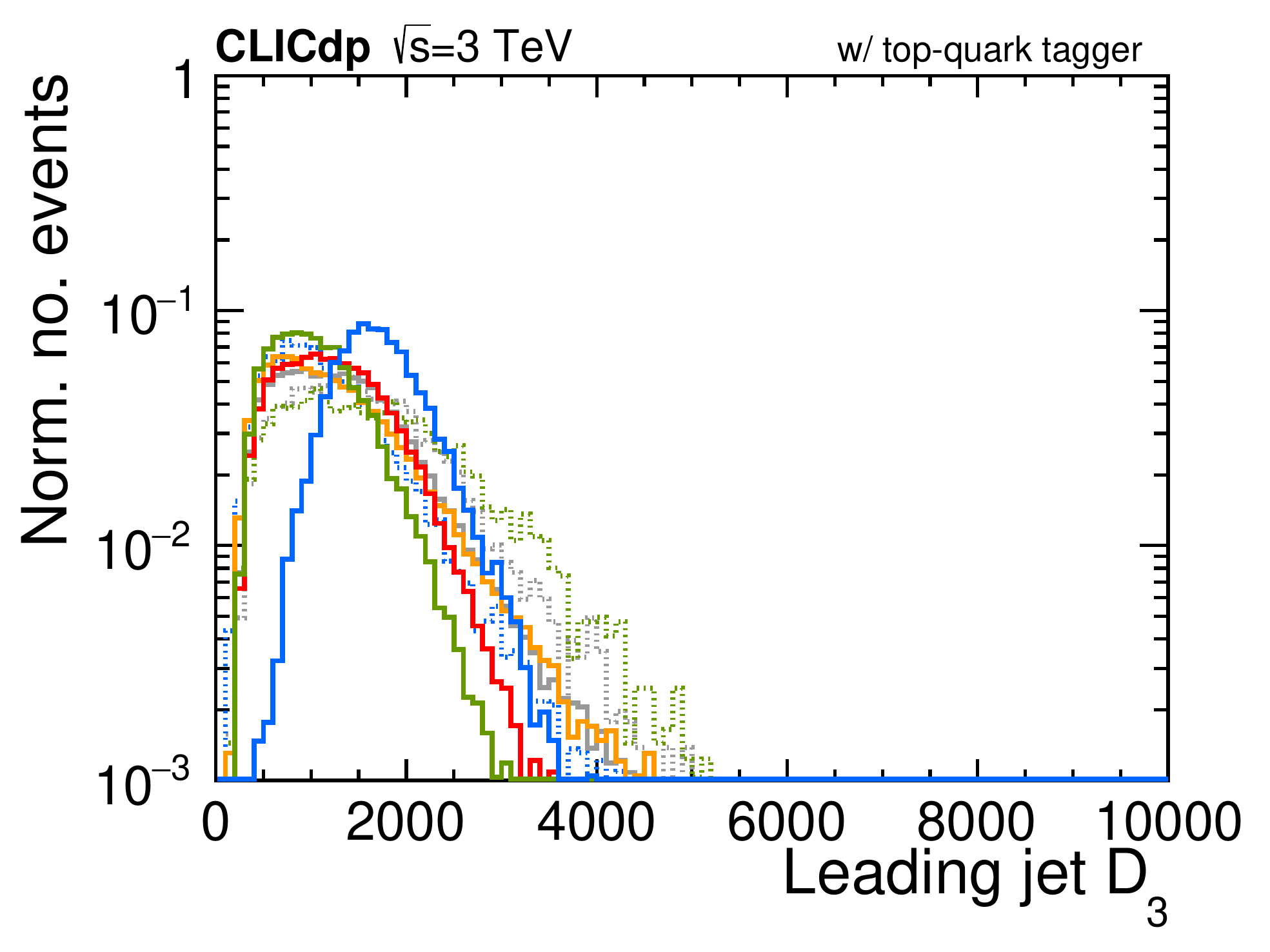}
	\caption{$D_{3}$ after applying the top-quark tagger.}
	\end{subfigure}\\
\caption{Substructure variables for the leading (highest energy) large-R jet. Note that the qqlv and qqll backgrounds have been omitted for the figures in the right column. The retention of these backgrounds is already very low after the pre-cuts. The superscript `a' (`b') refers to the kinematic region $\rootsprime\geq1.2\,\tev$ ($\rootsprime<1.2\,\tev$). \label{fig:analysis:mva:variables:energycorrDJ1:3tev}}
\end{figure}

\begin{figure}[htpb]
	\centering
	\begin{subfigure}{0.48\columnwidth}
	\includegraphics[width=\textwidth]{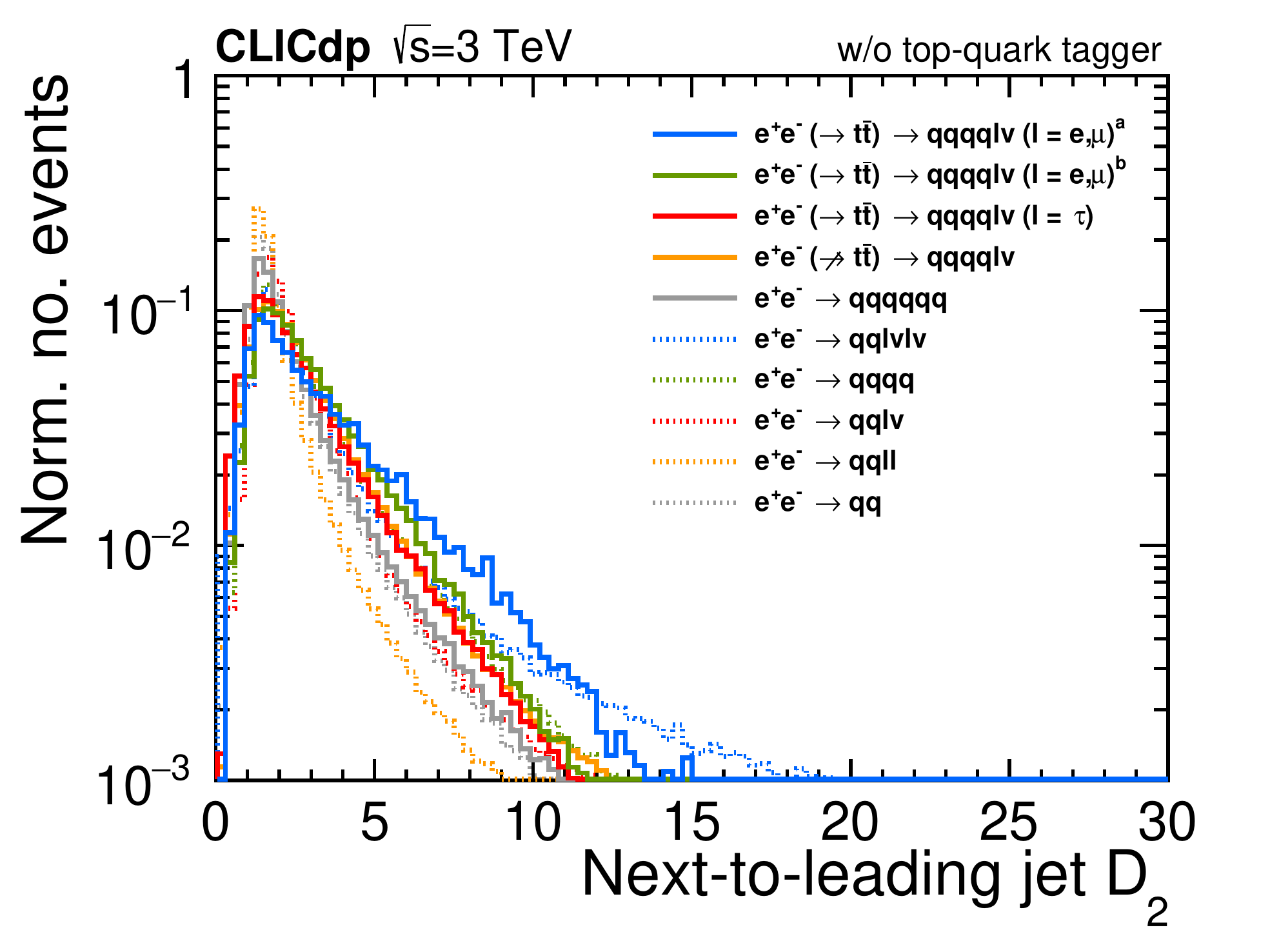}
	\caption{$D_{2}$ without applying the top-quark tagger.}
	\end{subfigure}
	~~~
	\begin{subfigure}{0.48\columnwidth}
	\includegraphics[width=\textwidth]{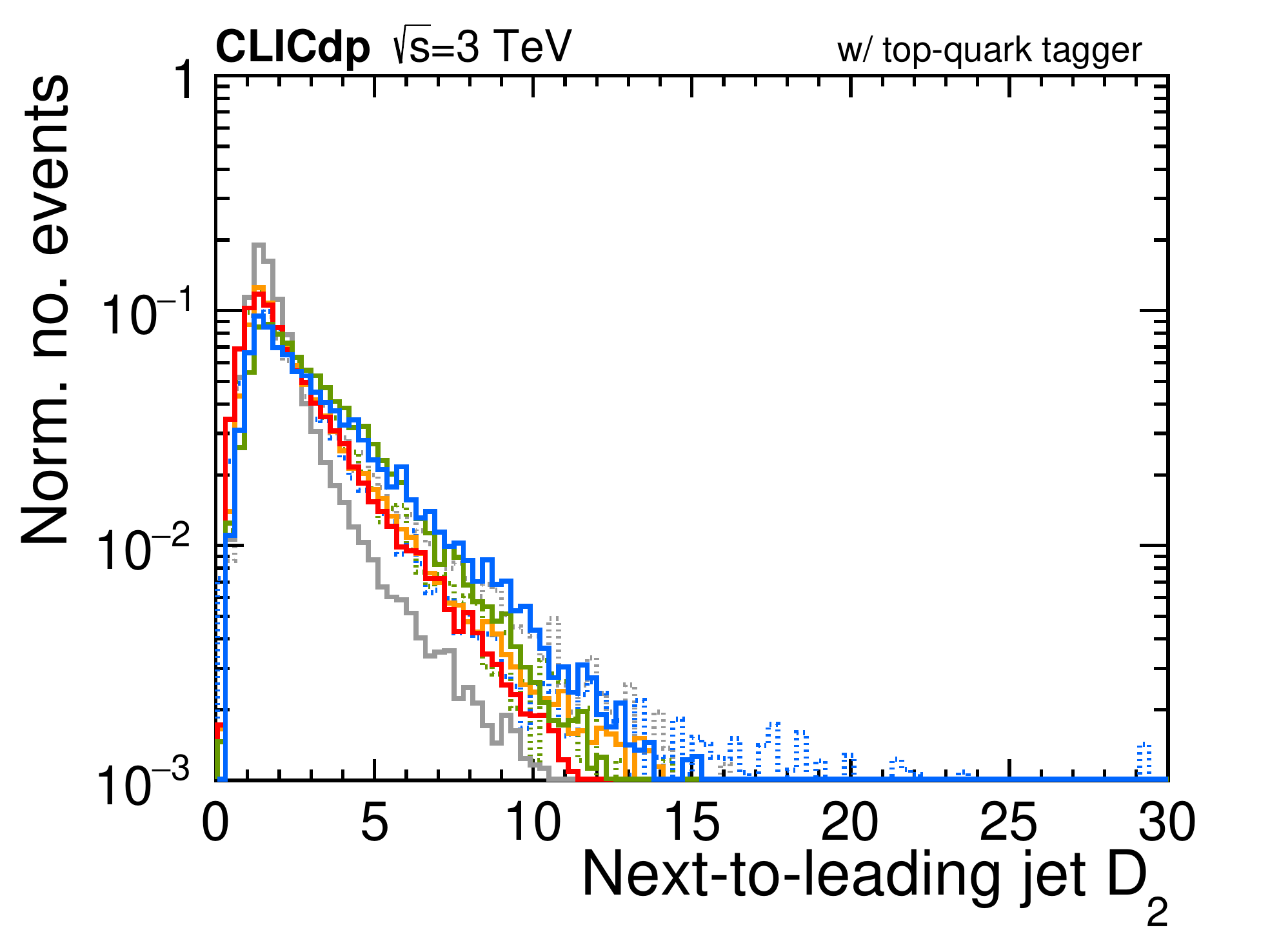}
	\caption{$D_{2}$ after applying the top-quark tagger.}
	\end{subfigure}\\
	\vspace{5mm}
	\begin{subfigure}{0.48\columnwidth}
	\includegraphics[width=\textwidth]{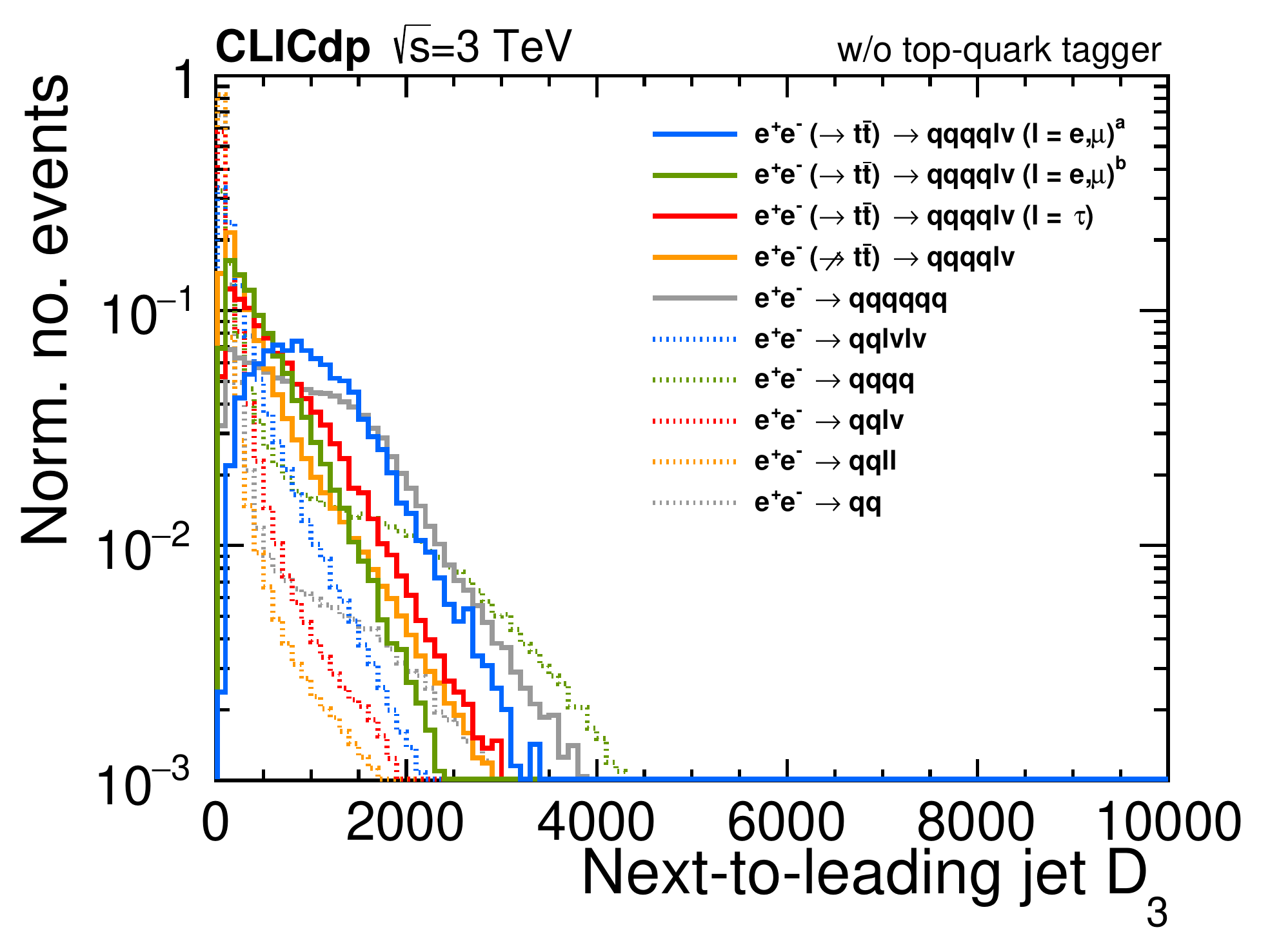}
	\caption{$D_{3}$ without applying the top-quark tagger.}
	\end{subfigure}
	~~~
	\begin{subfigure}{0.48\columnwidth}
	\includegraphics[width=\textwidth]{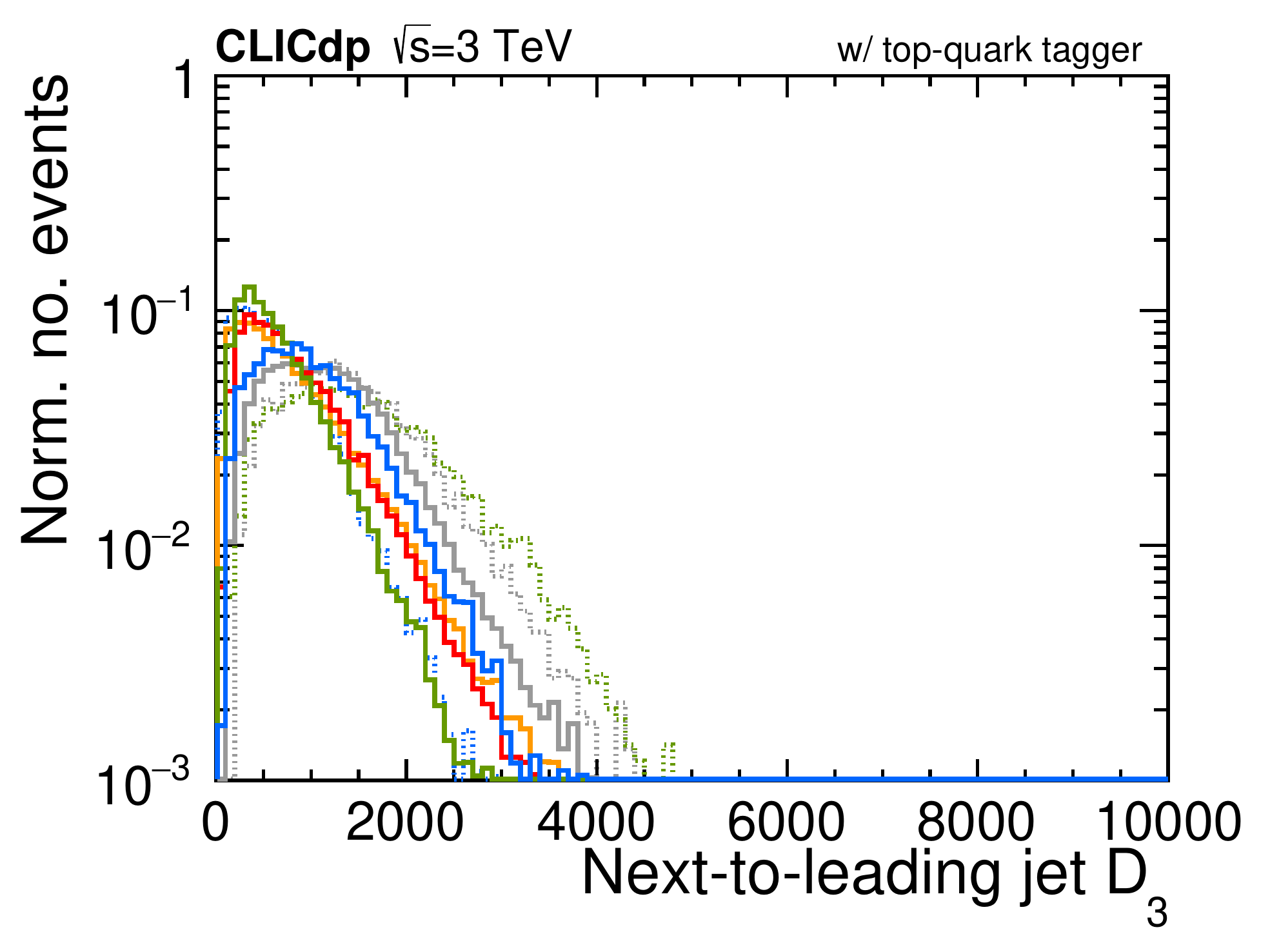}
	\caption{$D_{3}$ after applying the top-quark tagger.}
	\end{subfigure}\\
\caption{Substructure variables for the next-to-leading (lowest energy) large-R jet. Note that the qqlv and qqll backgrounds have been omitted for the figures in the right column. The retention of these backgrounds is already very low after the pre-cuts. The superscript `a' (`b') refers to the kinematic region $\rootsprime\geq1.2\,\tev$ ($\rootsprime<1.2\,\tev$). \label{fig:analysis:mva:variables:energycorrDJ2:3tev}}
\end{figure}

\begin{figure}[htpb]
	\centering
	\begin{subfigure}{0.48\columnwidth}
	\includegraphics[width=\textwidth]{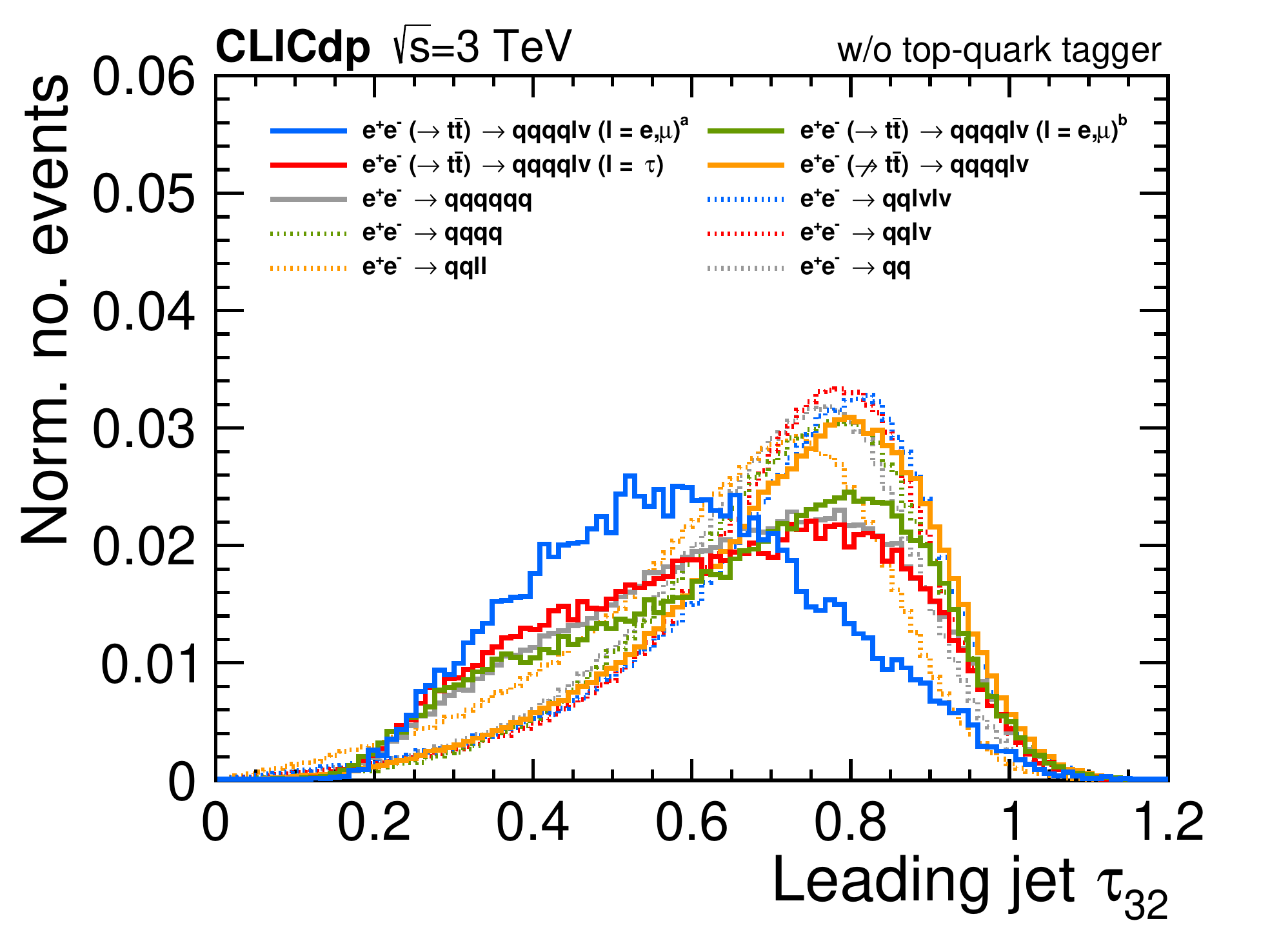}
	\caption{$\tau_{32}$ without applying the top-quark tagger.}
	\end{subfigure}
	~~~
	\begin{subfigure}{0.48\columnwidth}
	\includegraphics[width=\textwidth]{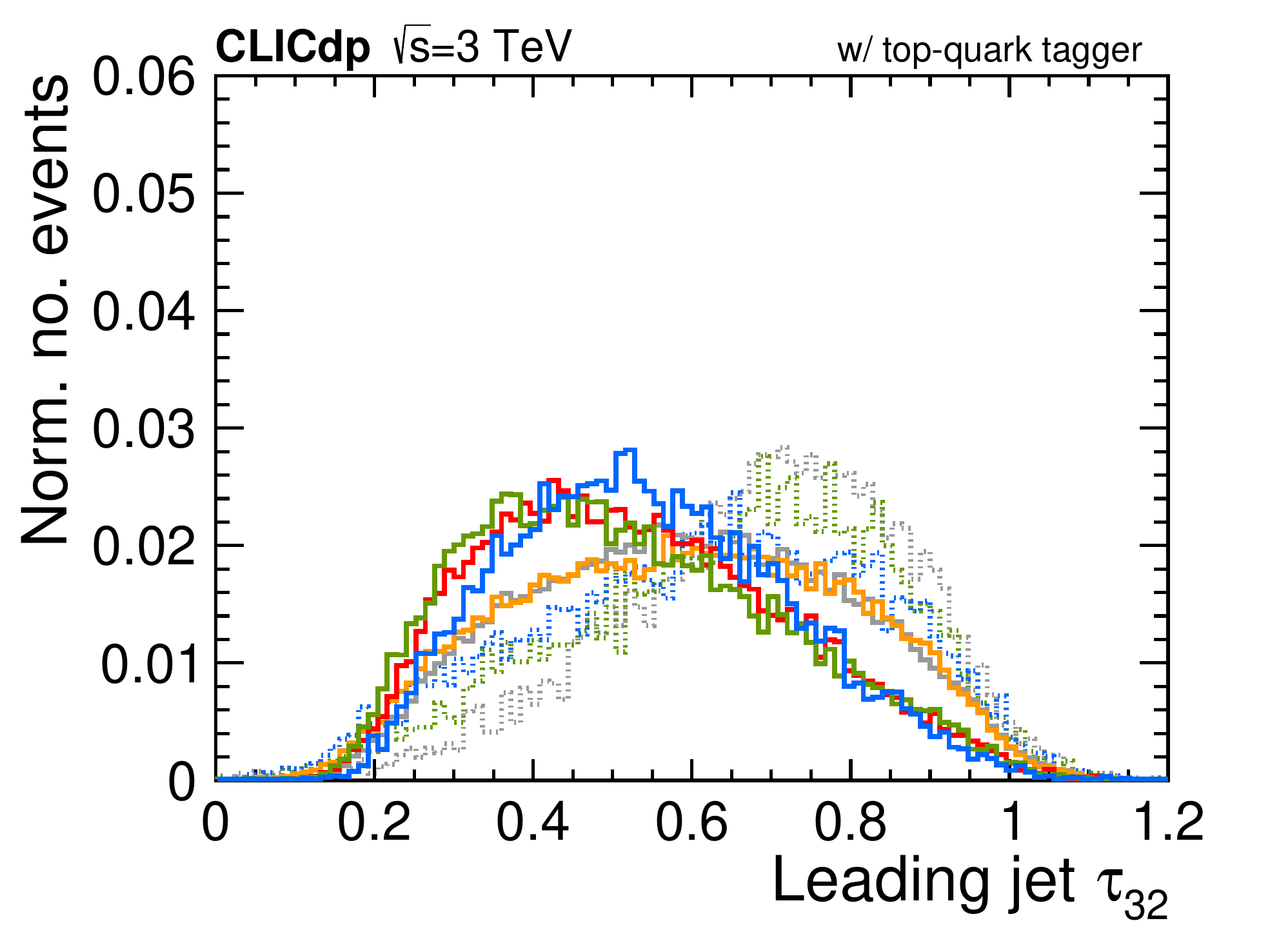}
	\caption{$\tau_{32}$ after applying the top-quark tagger.}
	\end{subfigure}\\
	\vspace{5mm}
	\begin{subfigure}{0.48\columnwidth}
	\includegraphics[width=\textwidth]{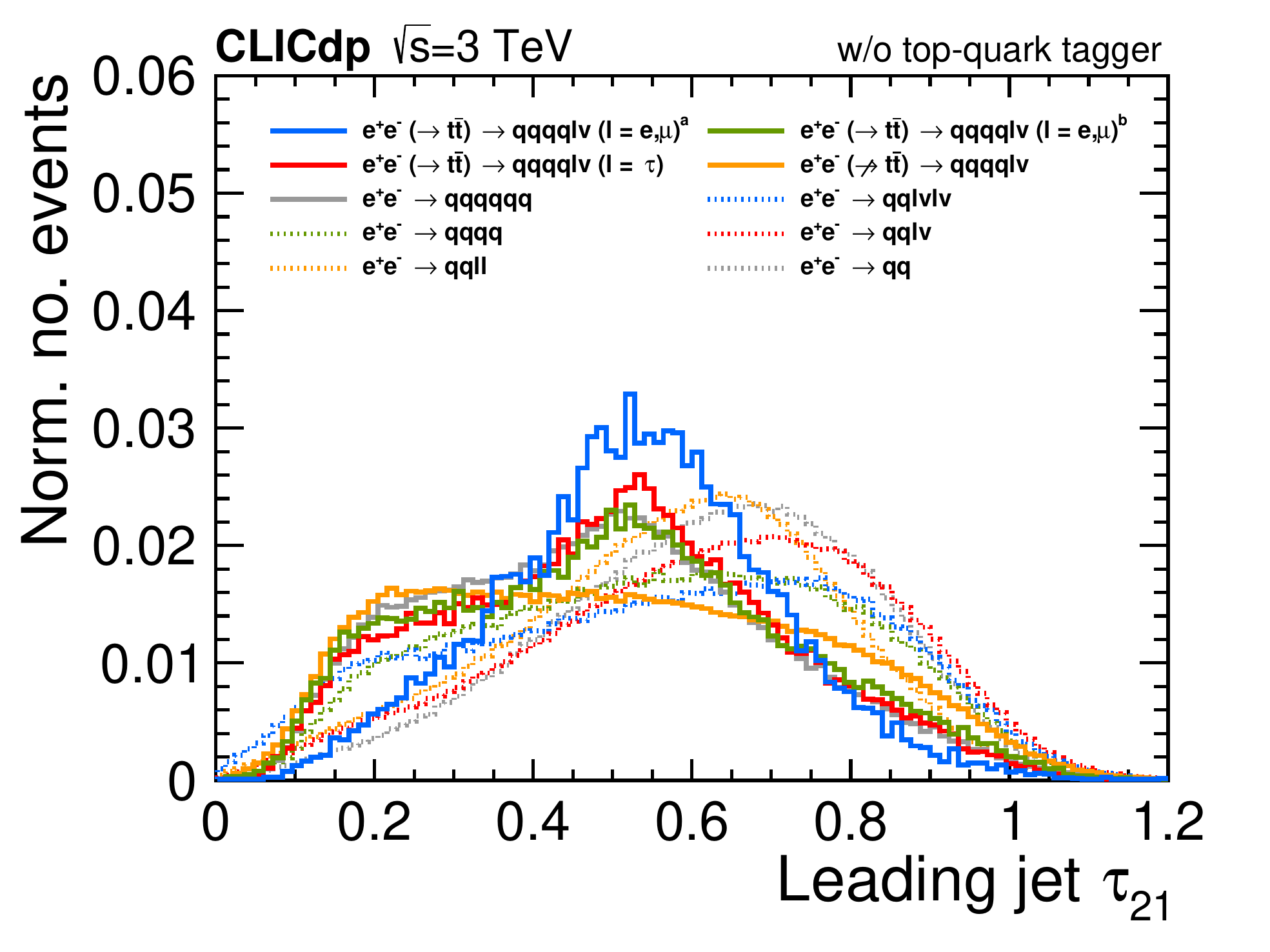}
	\caption{$\tau_{21}$ without applying the top-quark tagger.}
	\end{subfigure}
	~~~
	\begin{subfigure}{0.48\columnwidth}
	\includegraphics[width=\textwidth]{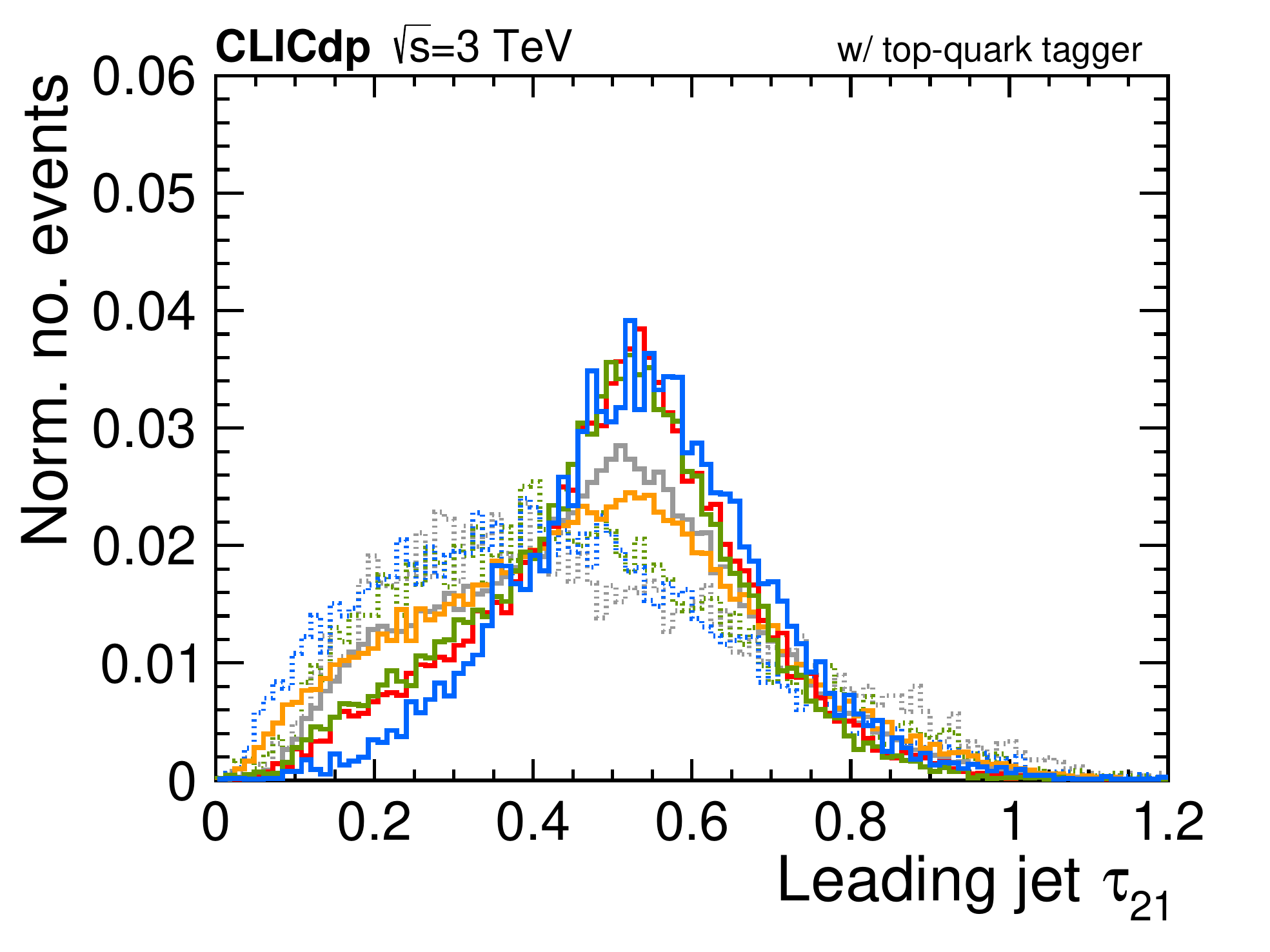}
	\caption{$\tau_{21}$ after applying the top-quark tagger.}
	\end{subfigure}\\
	\vspace{5mm}
	\begin{subfigure}{0.48\columnwidth}
	\includegraphics[width=\textwidth]{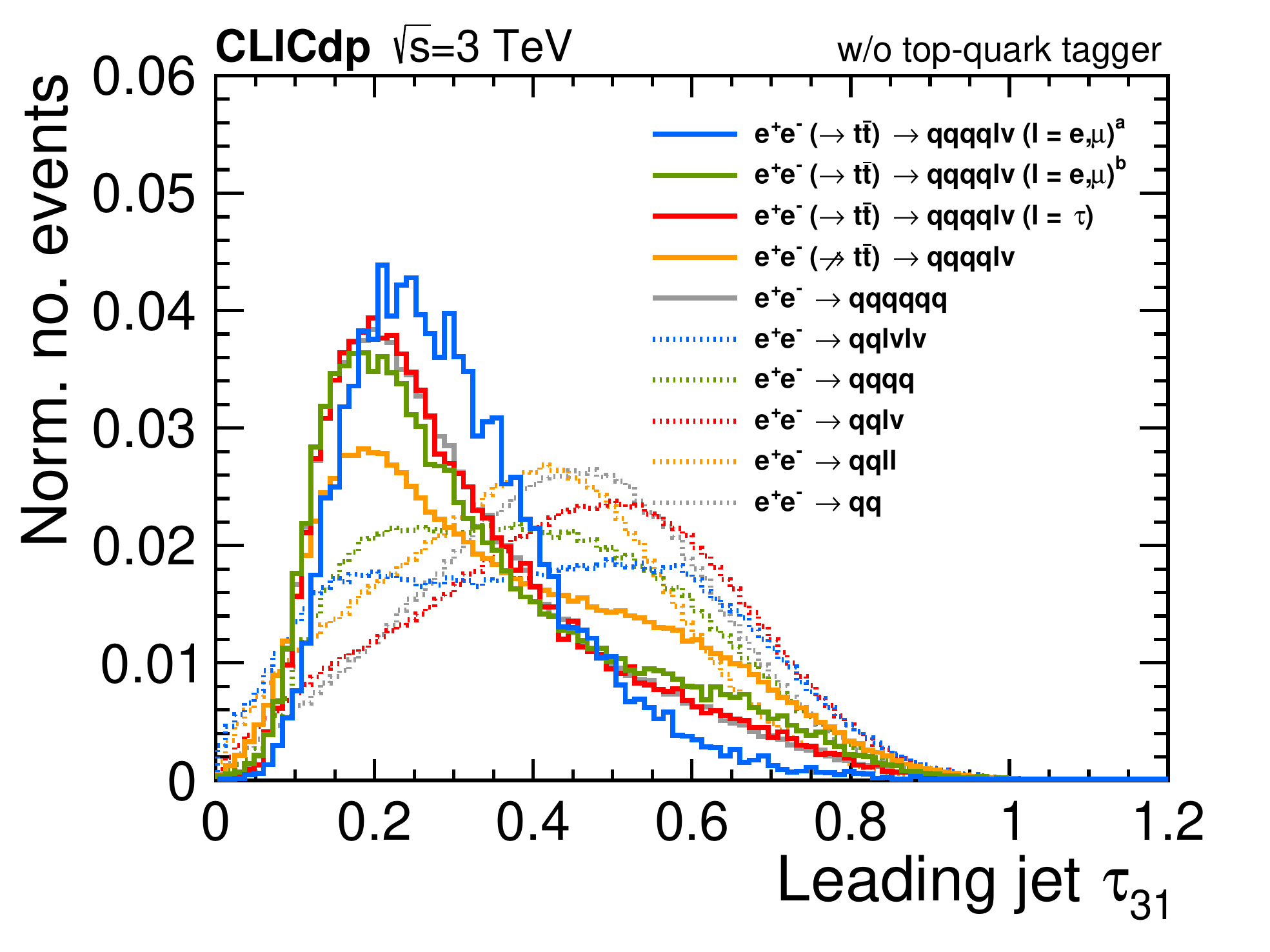}
	\caption{$\tau_{31}$ without applying the top-quark tagger.}
	\end{subfigure}
	~~~
	\begin{subfigure}{0.48\columnwidth}
	\includegraphics[width=\textwidth]{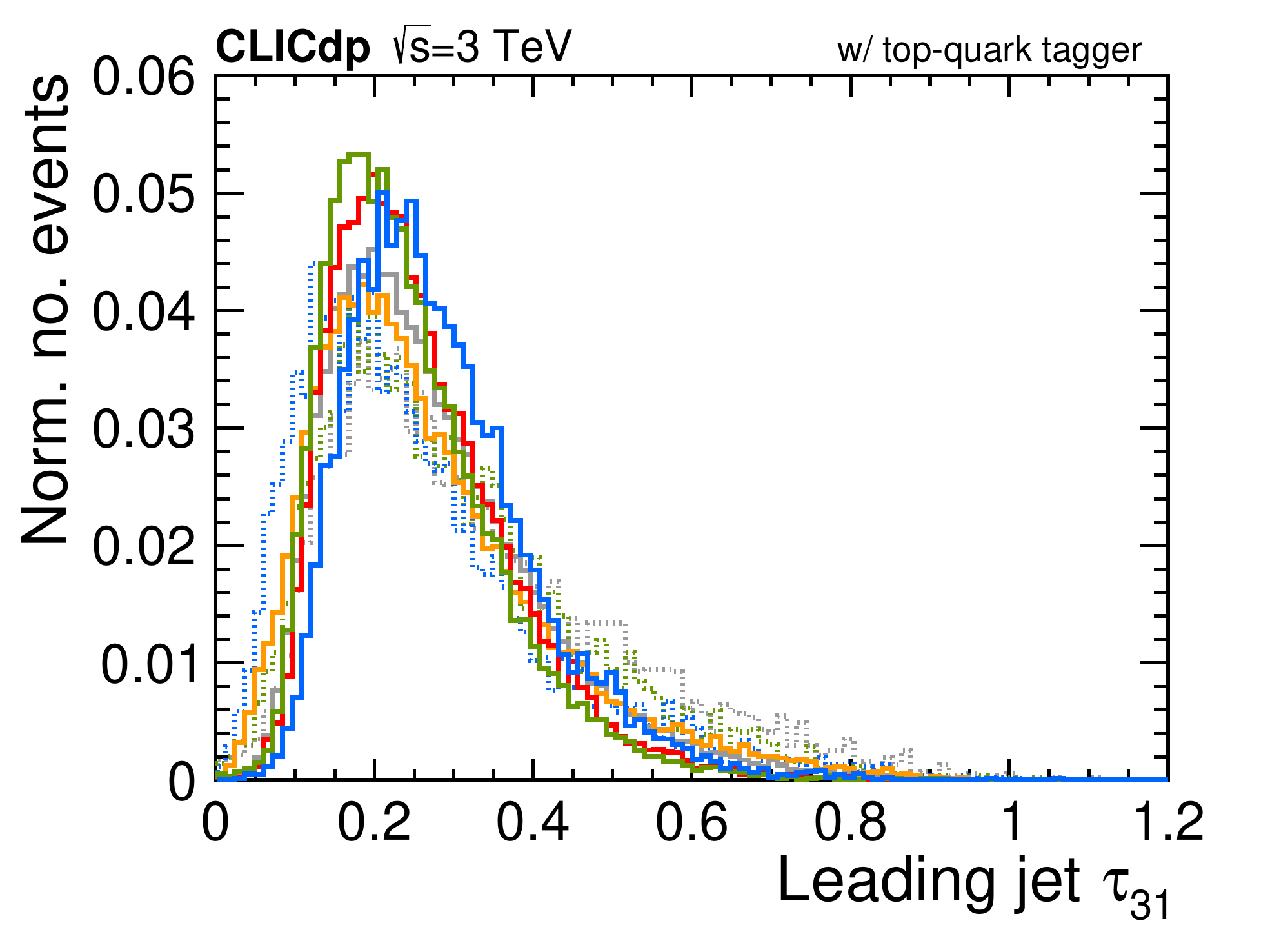}
	\caption{$\tau_{31}$ after applying the top-quark tagger.}
	\end{subfigure}
\caption{Substructure variables for the leading (highest energy) large-R jet. Note that the qqlv and qqll backgrounds have been omitted for the figures in the right column. The retention of these backgrounds is already very low after the pre-cuts. The superscript `a' (`b') refers to the kinematic region $\rootsprime\geq1.2\,\tev$ ($\rootsprime<1.2\,\tev$). \label{fig:analysis:mva:variables:NsubjettinessJ1:3tev}}
\end{figure}

\begin{figure}[htpb]
	\centering
	\begin{subfigure}{0.48\columnwidth}
	\includegraphics[width=\textwidth]{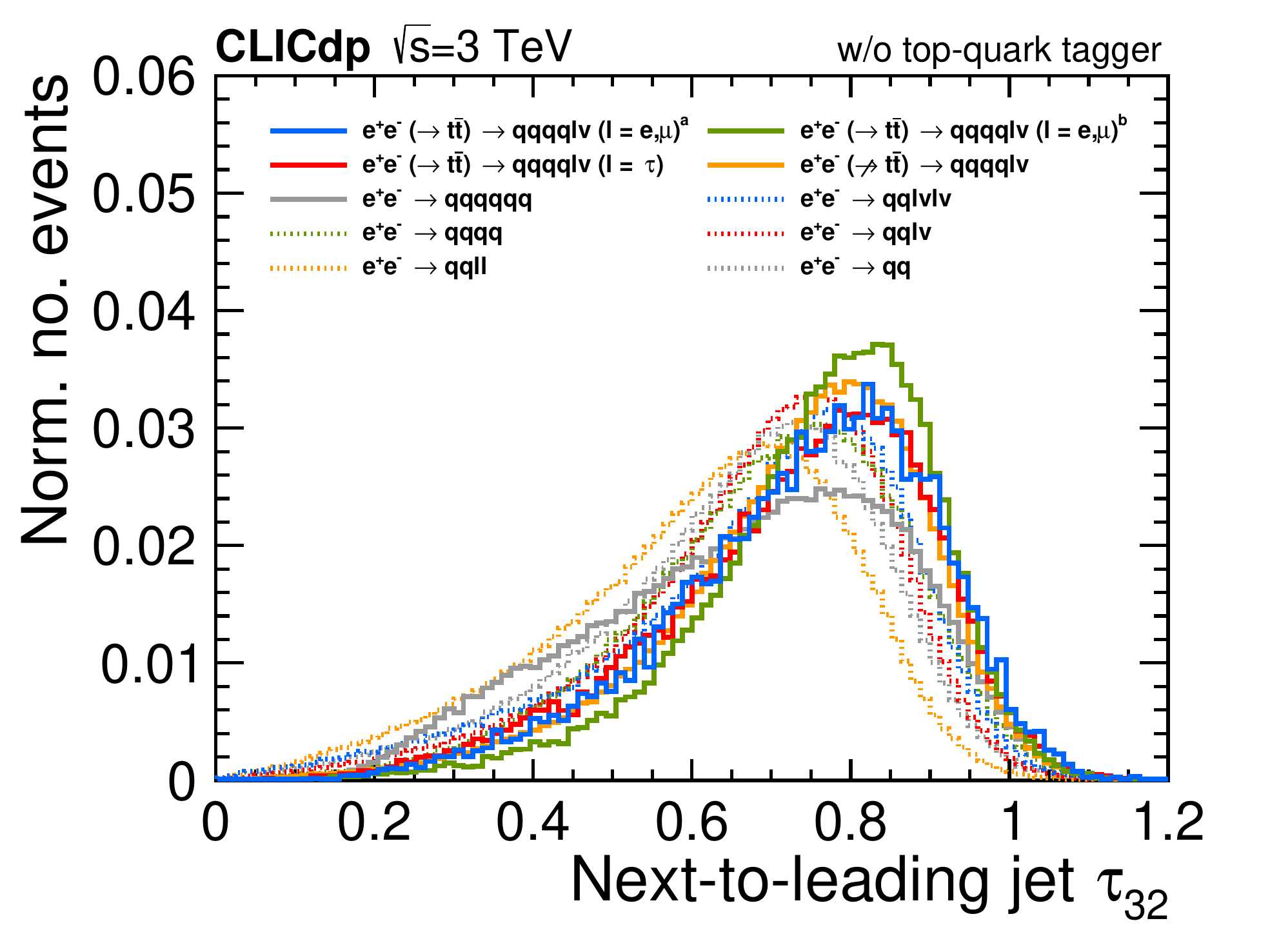}
	\caption{$\tau_{32}$ without applying the top-quark tagger.}
	\end{subfigure}
	~~~
	\begin{subfigure}{0.48\columnwidth}
	\includegraphics[width=\textwidth]{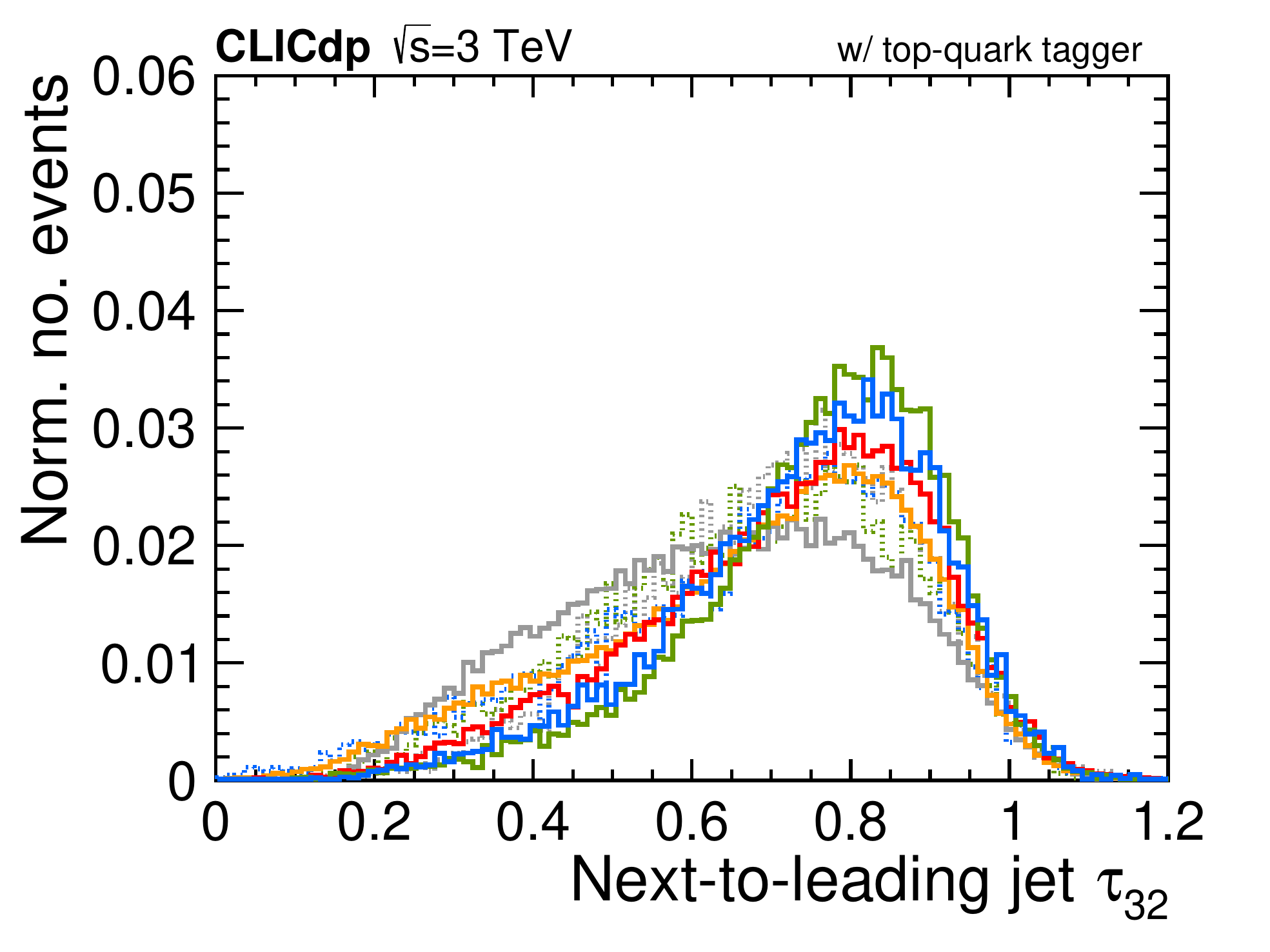}
	\caption{$\tau_{32}$ after applying the top-quark tagger.}
	\end{subfigure}\\
	\vspace{5mm}
	\begin{subfigure}{0.48\columnwidth}
	\includegraphics[width=\textwidth]{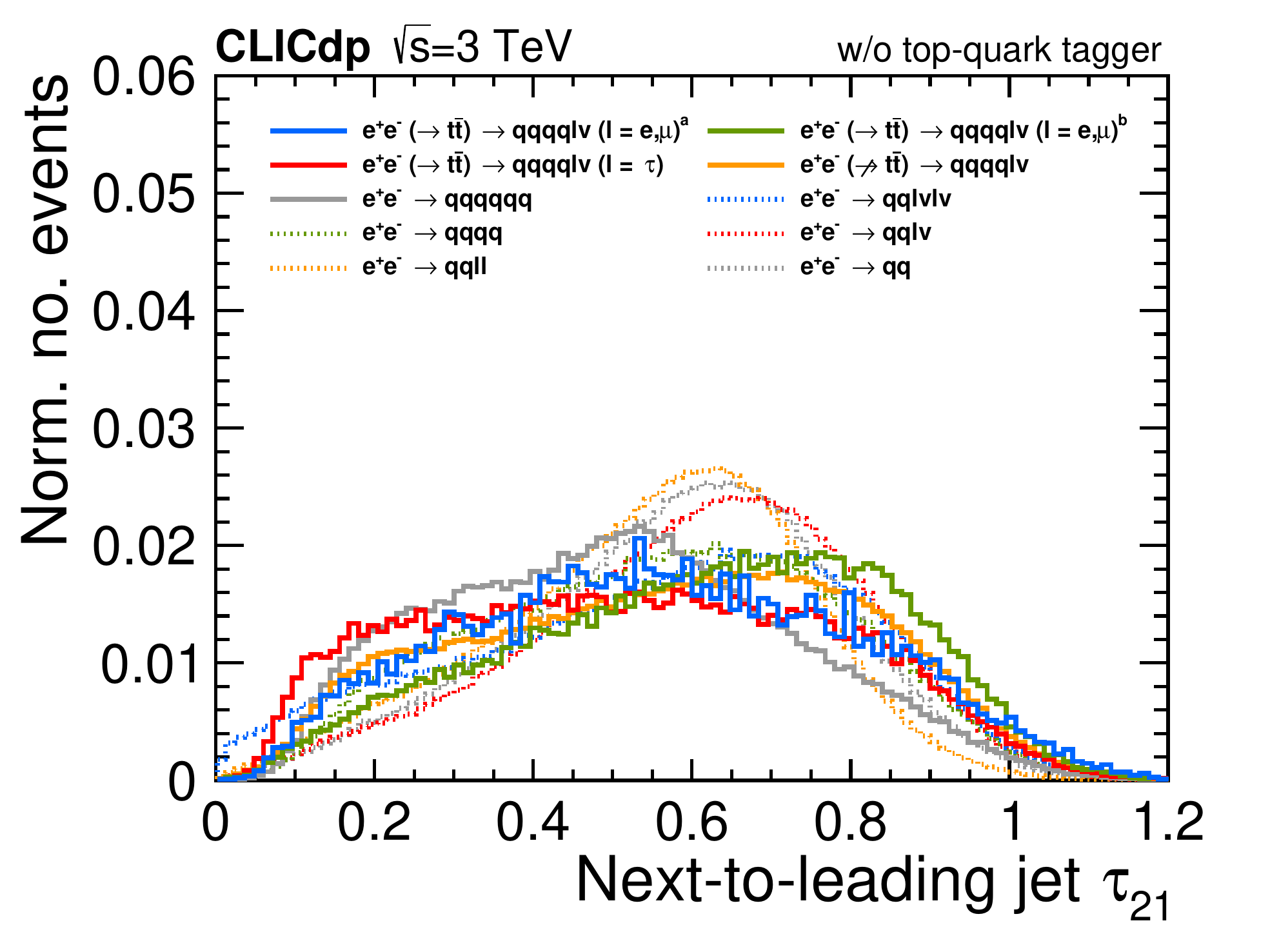}
	\caption{$\tau_{21}$ without applying the top-quark tagger.}
	\end{subfigure}
	~~~
	\begin{subfigure}{0.48\columnwidth}
	\includegraphics[width=\textwidth]{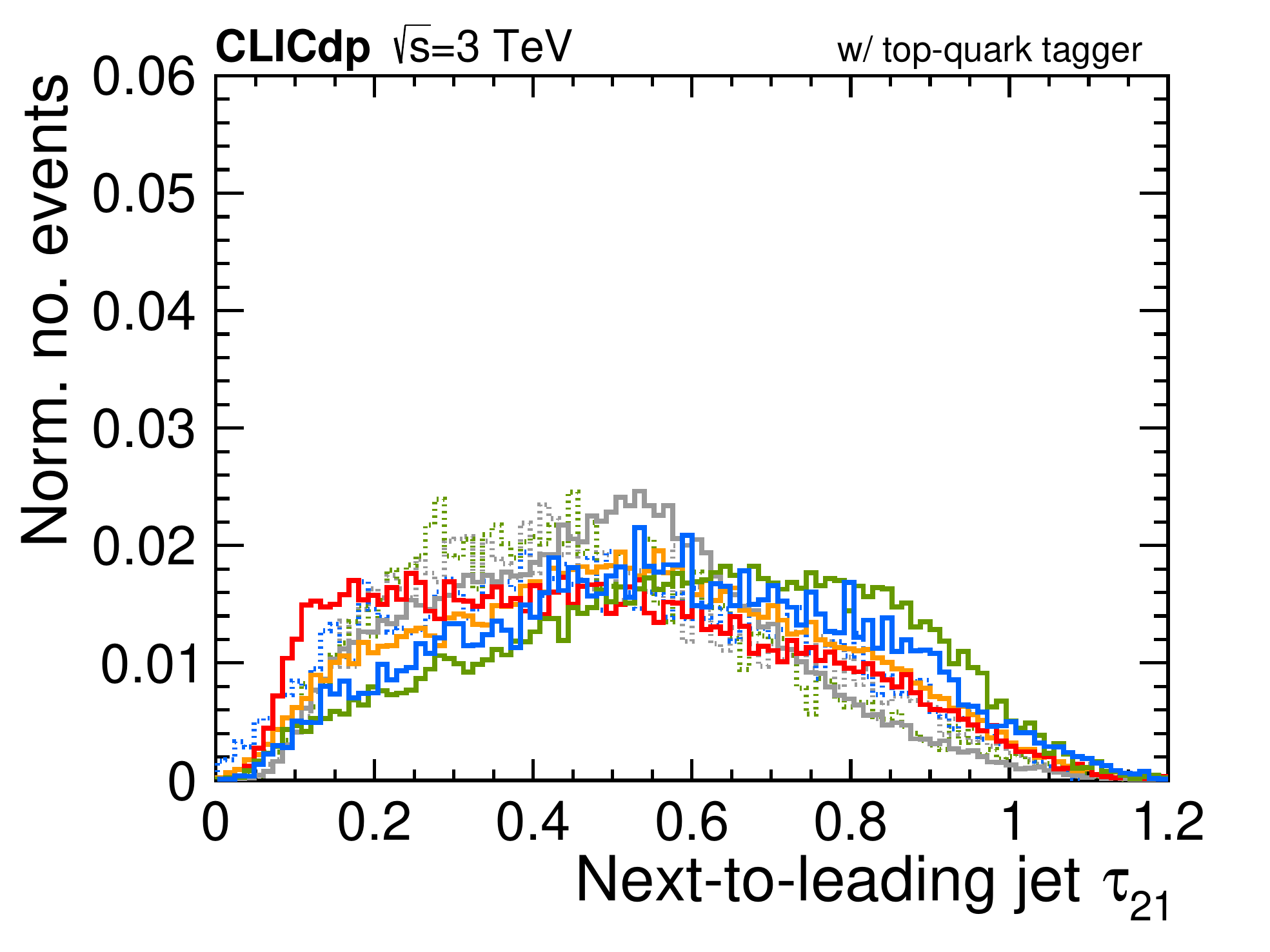}
	\caption{$\tau_{21}$ after applying the top-quark tagger.}
	\end{subfigure}\\
	\vspace{5mm}
	\begin{subfigure}{0.48\columnwidth}
	\includegraphics[width=\textwidth]{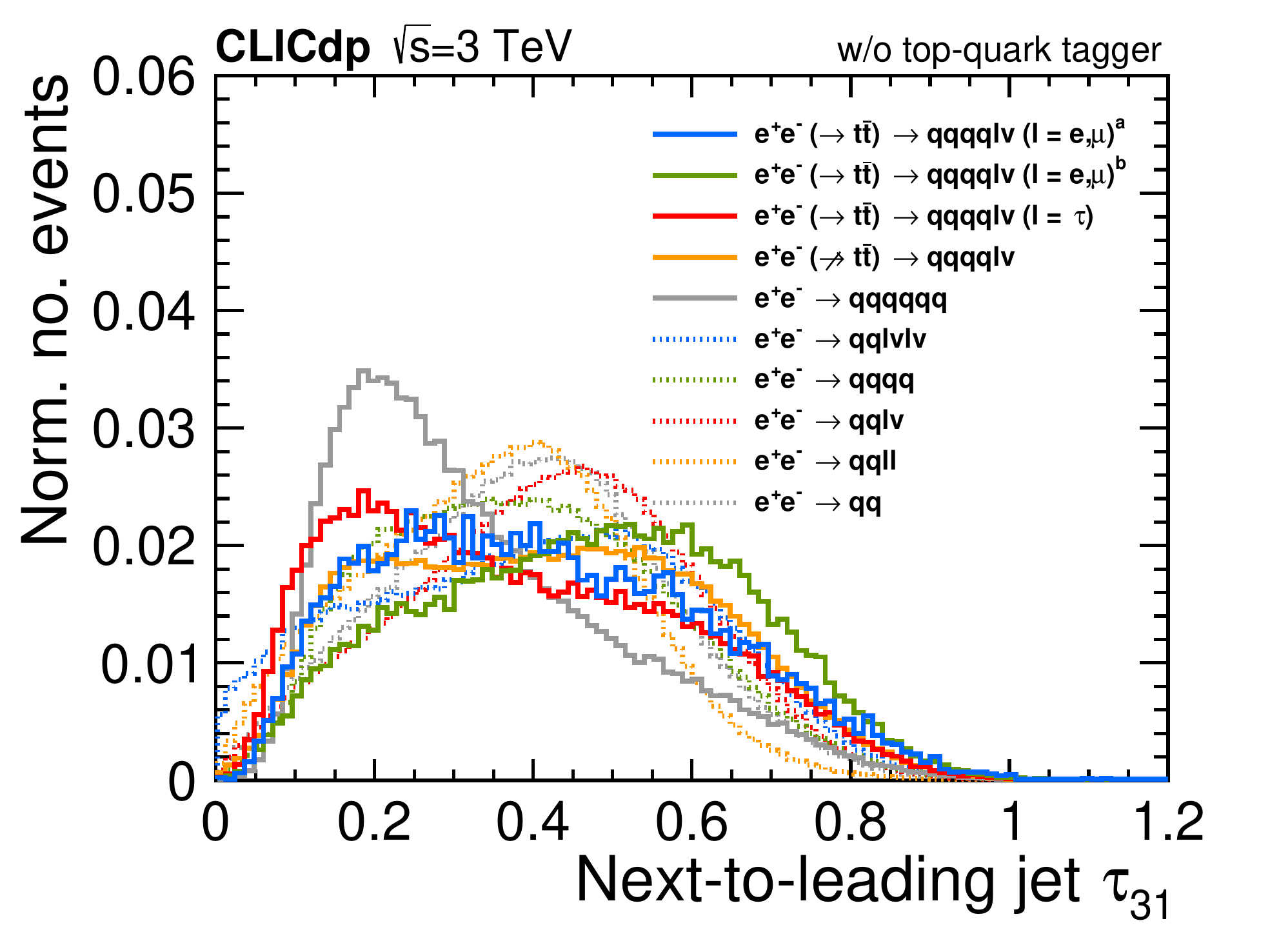}
	\caption{$\tau_{31}$ without applying the top-quark tagger.}
	\end{subfigure}
	~~~
	\begin{subfigure}{0.48\columnwidth}
	\includegraphics[width=\textwidth]{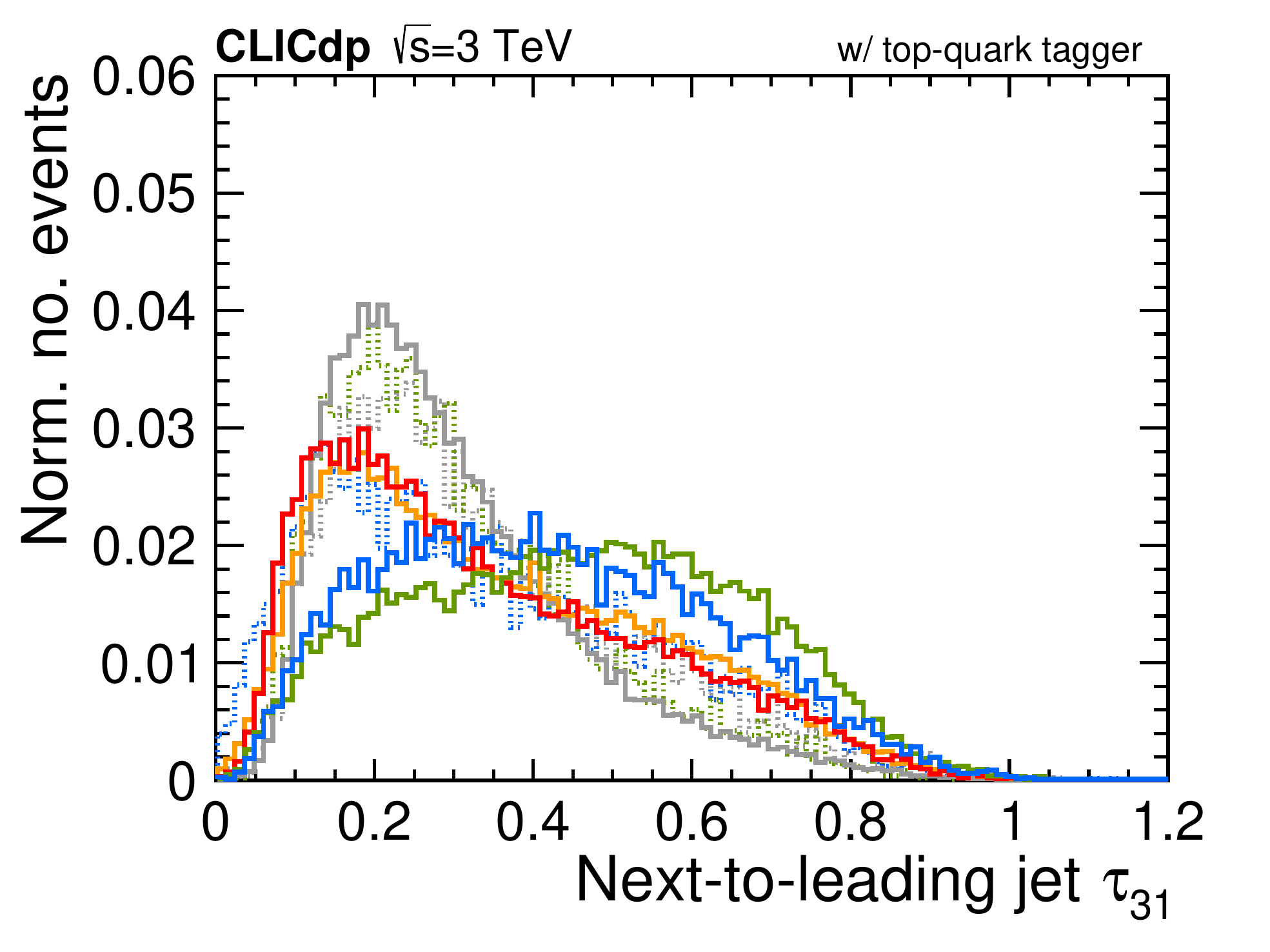}
	\caption{$\tau_{31}$ after applying the top-quark tagger.}
	\end{subfigure}
\caption{Substructure variables for the next-to-leading (lowest energy) large-R jet. Note that the qqlv and qqll backgrounds have been omitted for the figures in the right column. The retention of these backgrounds is already very low after the pre-cuts. The superscript `a' (`b') refers to the kinematic region $\rootsprime\geq1.2\,\tev$ ($\rootsprime<1.2\,\tev$). \label{fig:analysis:mva:variables:NsubjettinessJ2:3tev}}
\end{figure}

\FloatBarrier

\clearpage

\section{Additional MVA variables}

\begin{figure}[htpb]
	\centering
	\begin{subfigure}{0.48\columnwidth}
	\includegraphics[width=\textwidth]{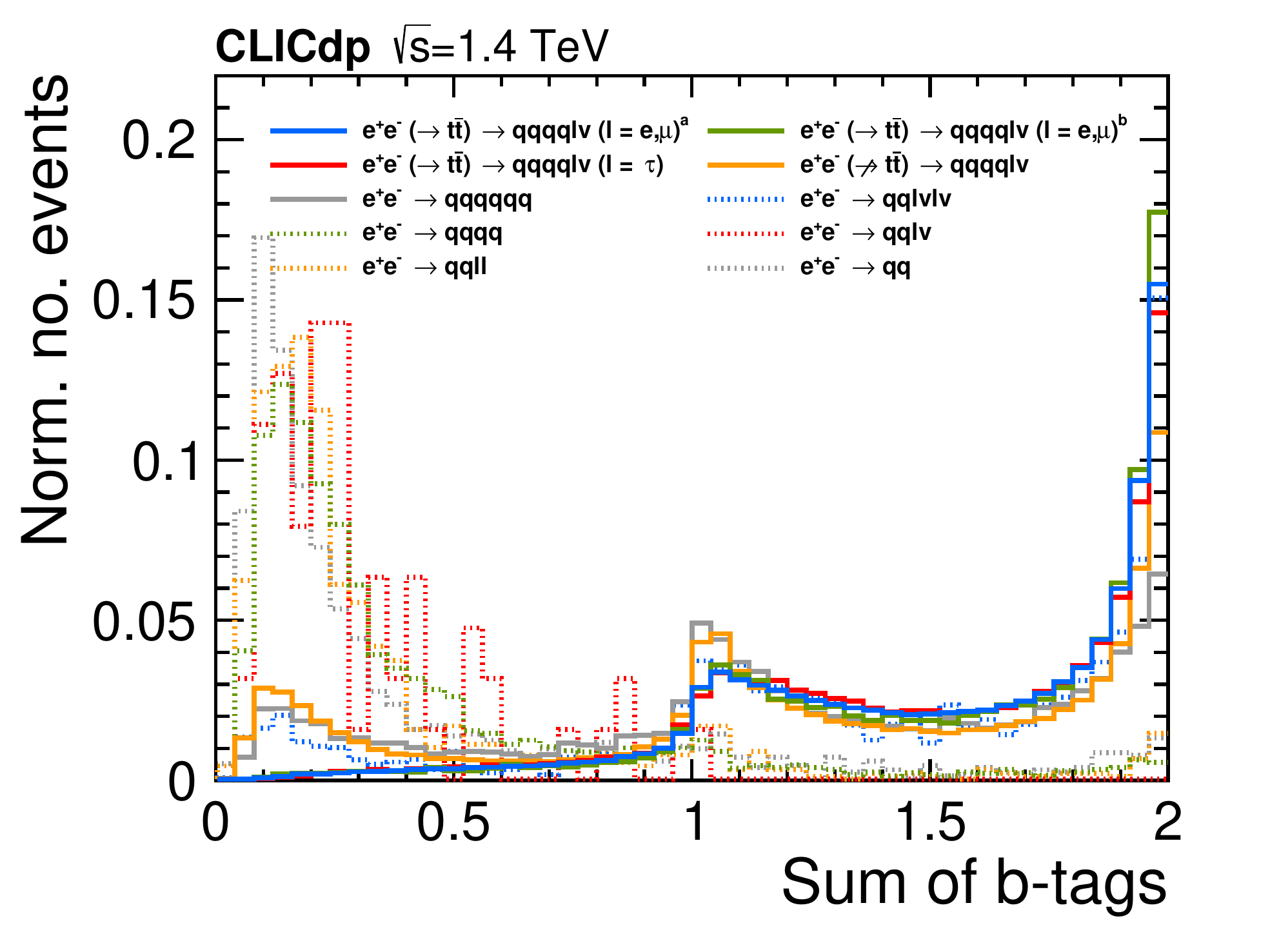}
	\caption{$\sum$ b-tags}
	\end{subfigure}
	~~~
	\begin{subfigure}{0.48\columnwidth}
	\includegraphics[width=\textwidth]{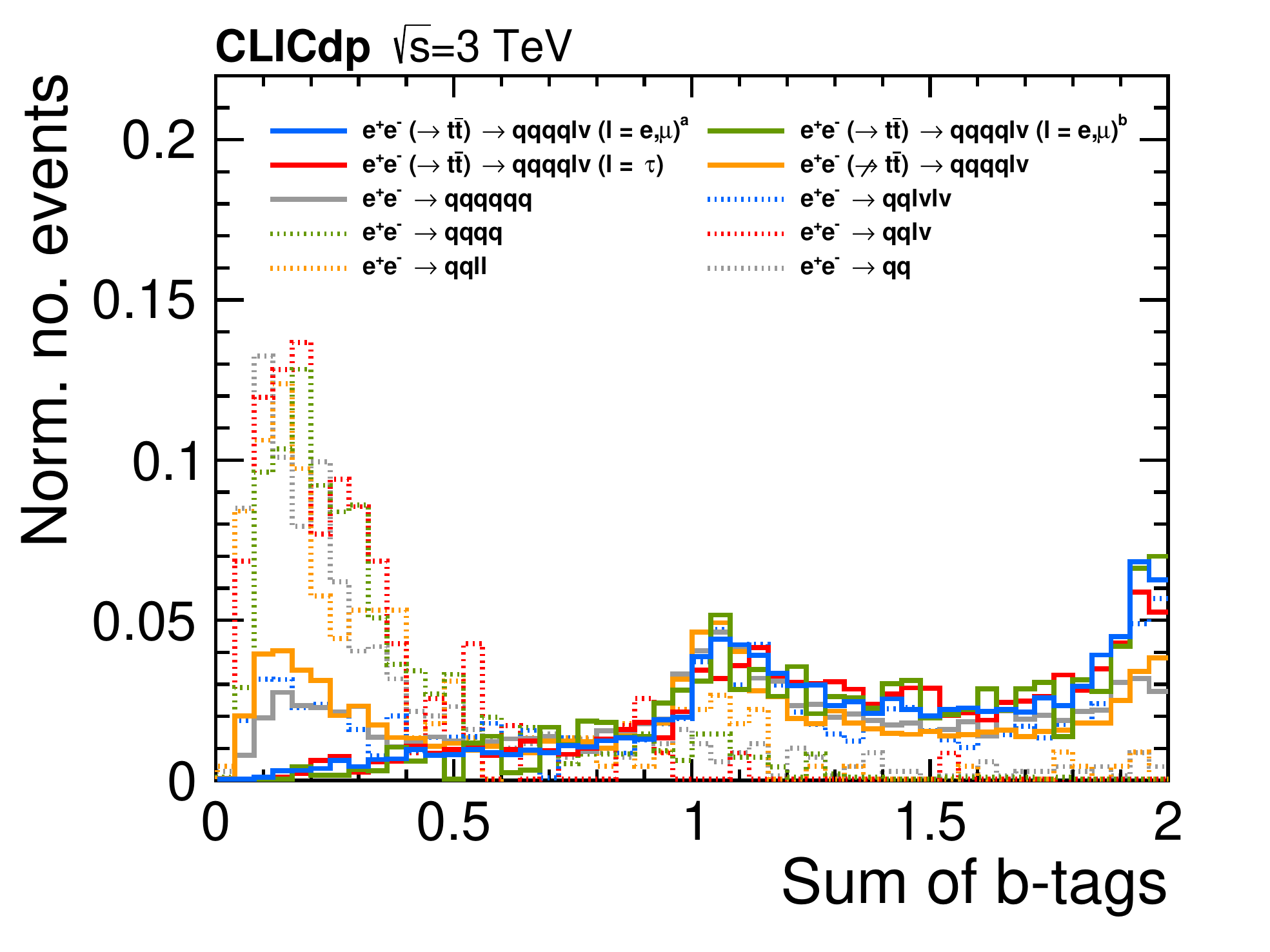}
	\caption{$\sum$ b-tags}
	\end{subfigure}\\
	\vspace{5mm}
	\begin{subfigure}{0.48\columnwidth}
	\includegraphics[width=\textwidth]{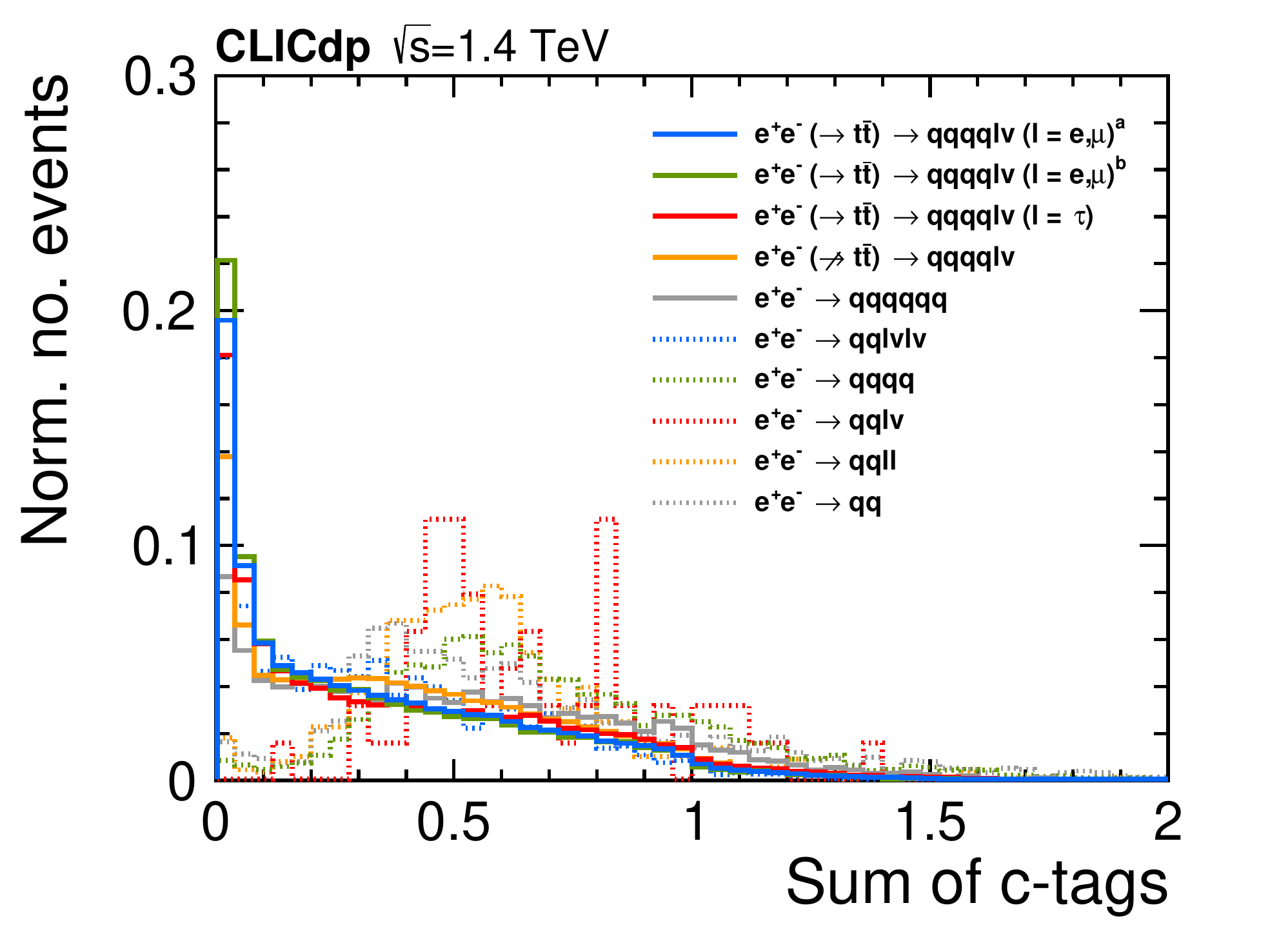}
	\caption{$\sum$ c-tags}
	\end{subfigure}
	~~~
	\begin{subfigure}{0.48\columnwidth}
	\includegraphics[width=\textwidth]{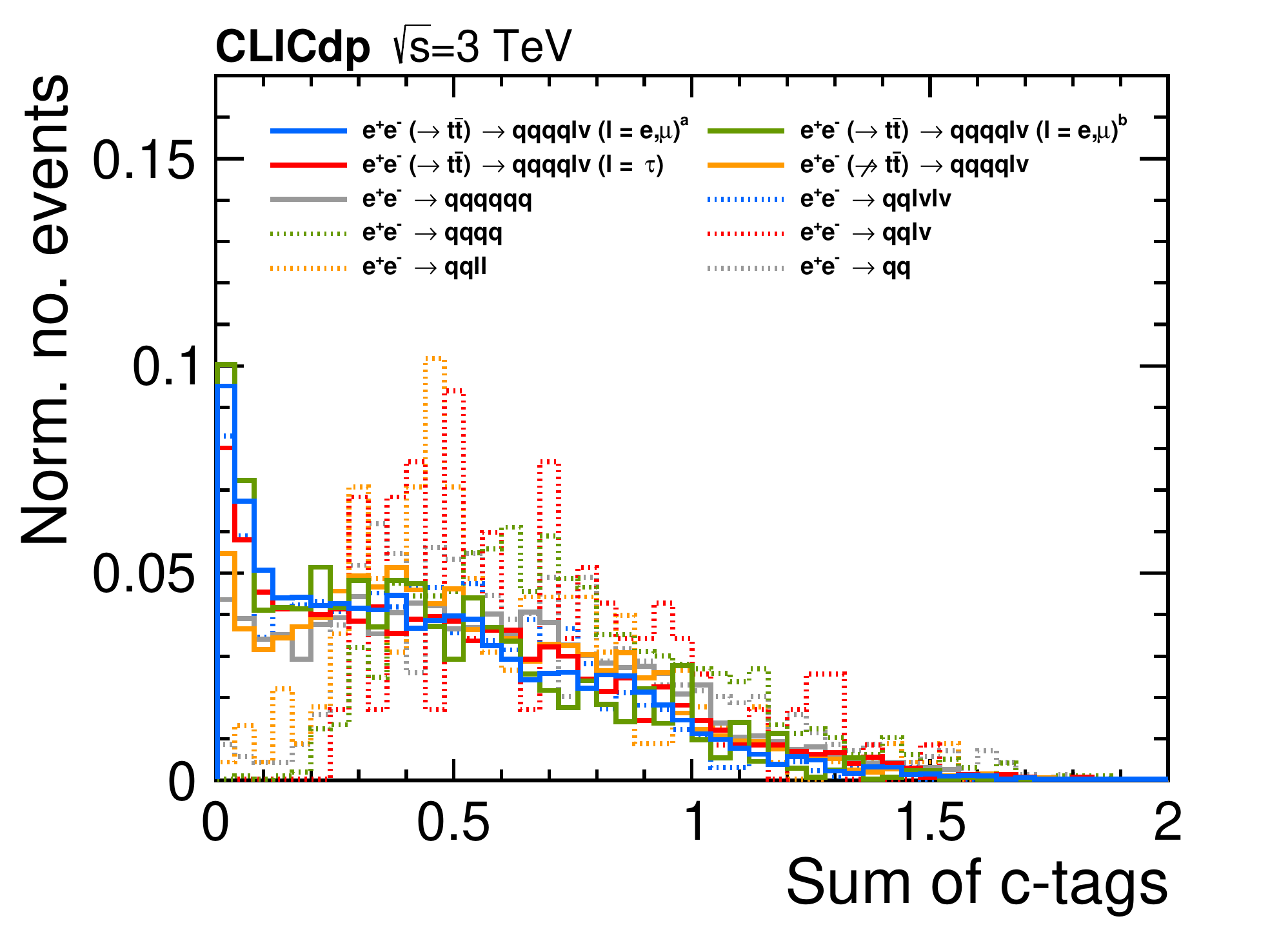}
	\caption{$\sum$ c-tags}
	\end{subfigure}
\caption{Flavour tagging variables for operation at $\roots=1.4\,\tev$ (left) and $\roots=3\,\tev$ (right). The distributions are shown after the application of pre-cuts. The superscript `a' (`b') refers to the kinematic region $\rootsprime\geq1.2\,\tev$ ($\rootsprime<1.2\,\tev$). \label{fig:analysis:mva:variables:flavourtagging}}
\end{figure}

\begin{figure}[htpb]
	\centering
	\begin{subfigure}{0.48\columnwidth}
	\includegraphics[width=\textwidth]{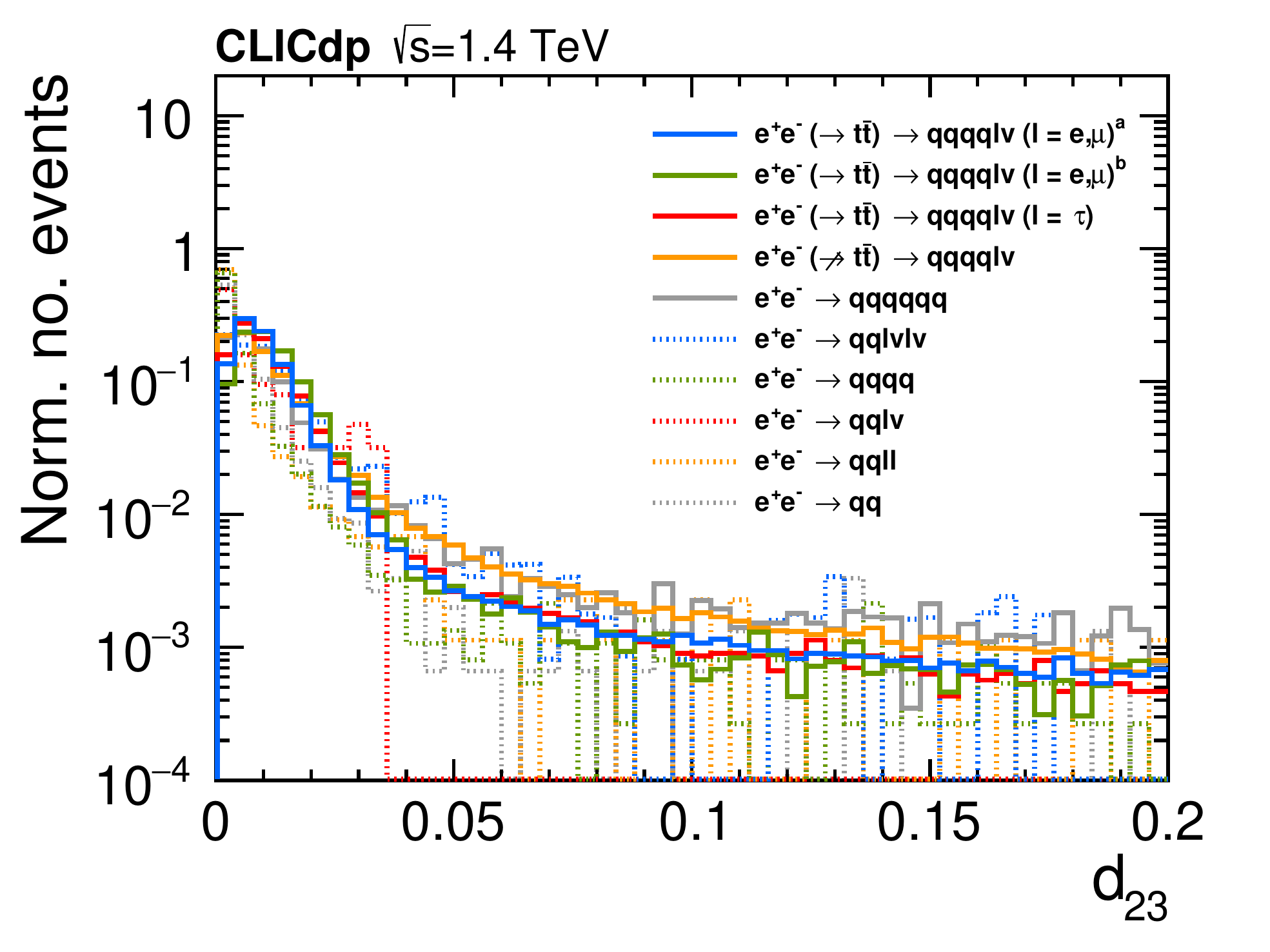}
	\caption{$y_{23}$}
	\end{subfigure}
	~~~
	\begin{subfigure}{0.48\columnwidth}
	\includegraphics[width=\textwidth]{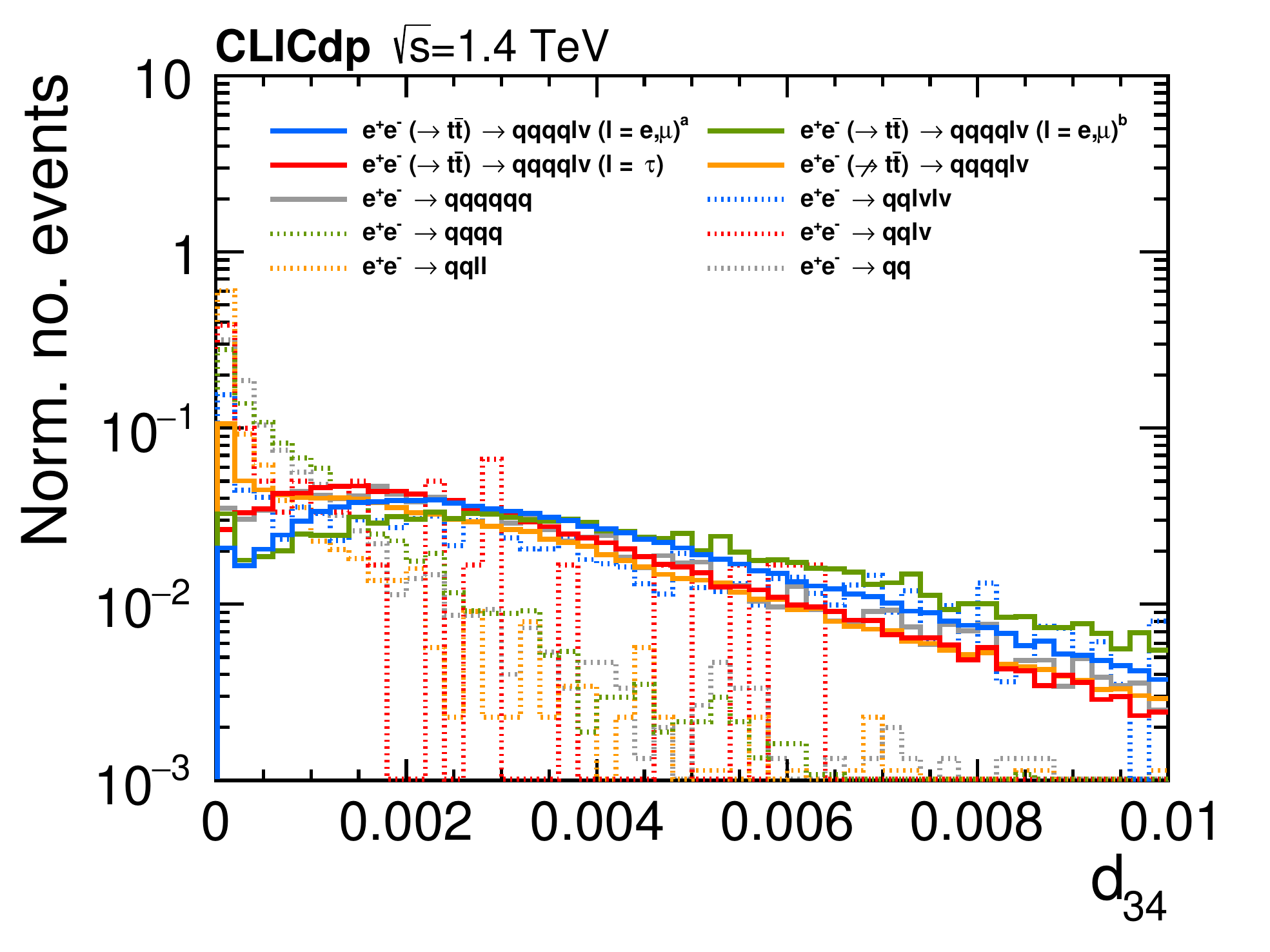}
	\caption{$y_{34}$}
	\end{subfigure}\\
	\vspace{5mm}
	\begin{subfigure}{0.48\columnwidth}
	\includegraphics[width=\textwidth]{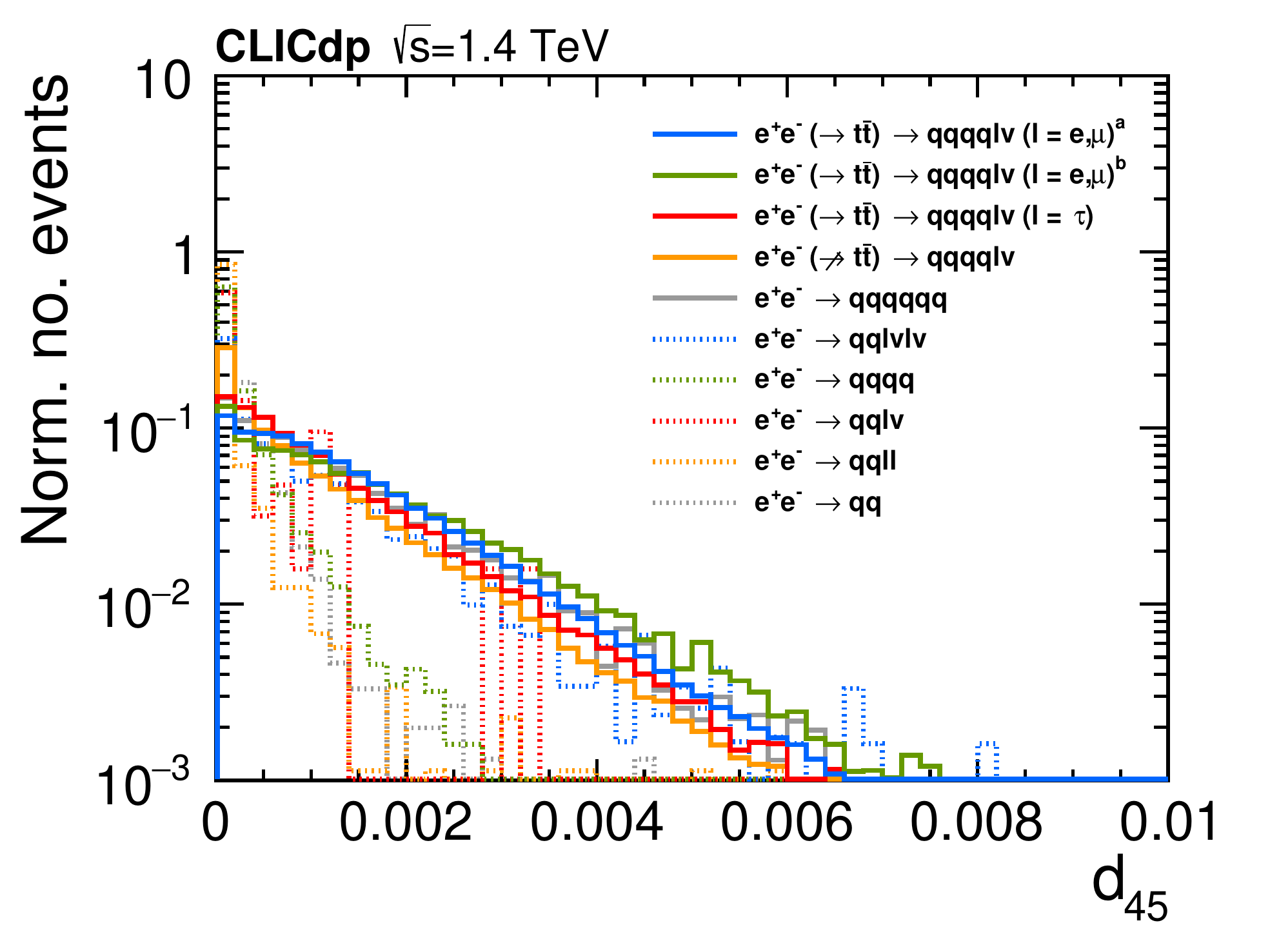}
	\caption{$y_{45}$}
	\end{subfigure}
	~~~
	\begin{subfigure}{0.48\columnwidth}
	\includegraphics[width=\textwidth]{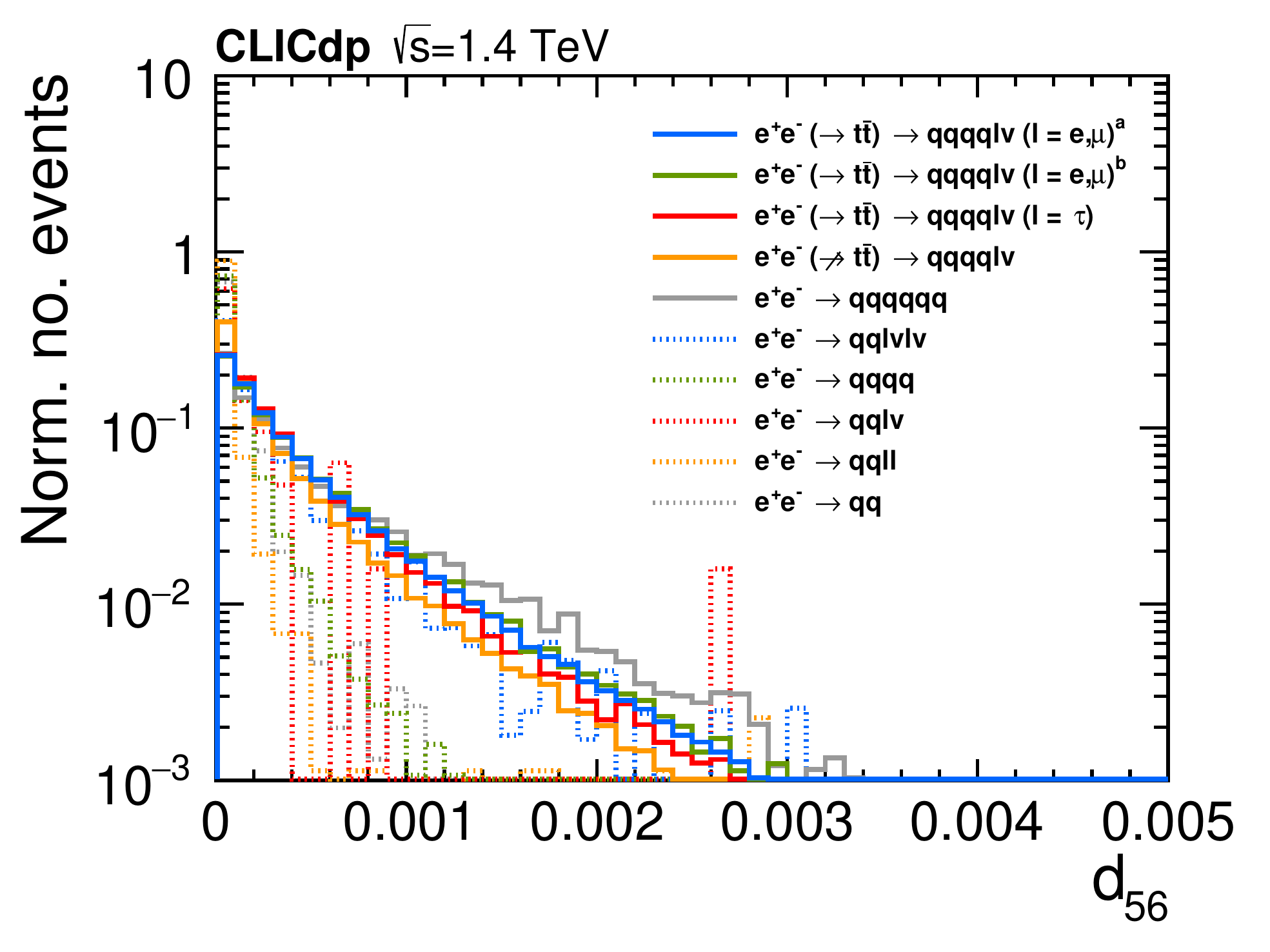}
	\caption{$y_{56}$}
	\end{subfigure}
\caption{Jet splitting scales for operation at $\roots=1.4\,\tev$ (left) and $\roots=3\,\tev$ (right). The distributions are shown after the application of pre-cuts. The superscript `a' (`b') refers to the kinematic region $\rootsprime\geq1.2\,\tev$ ($\rootsprime<1.2\,\tev$). \label{fig:analysis:mva:variables:jetsplitting}}
\end{figure}

\begin{figure}[htpb]
	\centering
	\begin{subfigure}{0.48\columnwidth}
	\includegraphics[width=\textwidth]{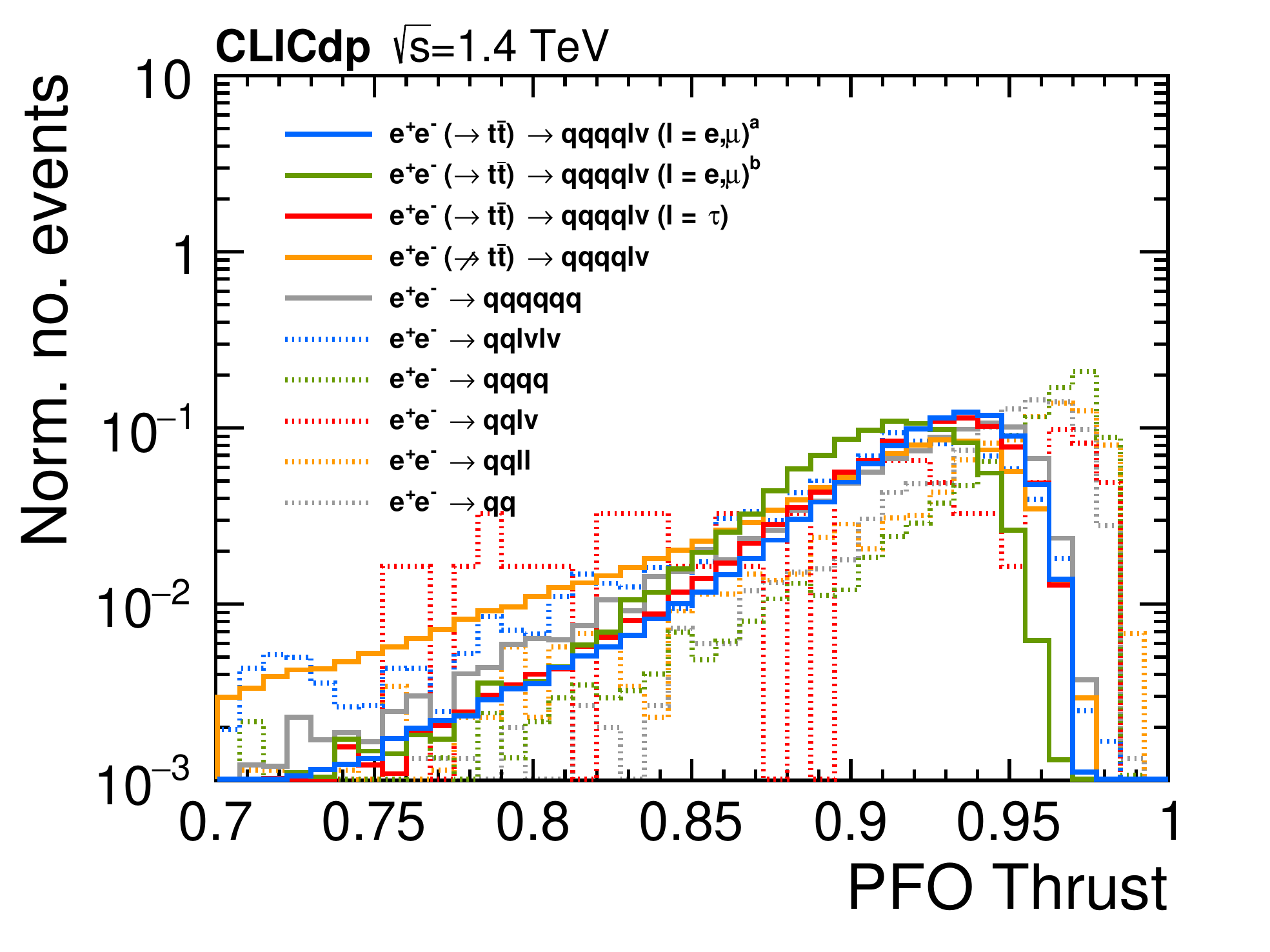}
	\caption{PFO Thrust}
	\end{subfigure}
	~~~
	\begin{subfigure}{0.48\columnwidth}
	\includegraphics[width=\textwidth]{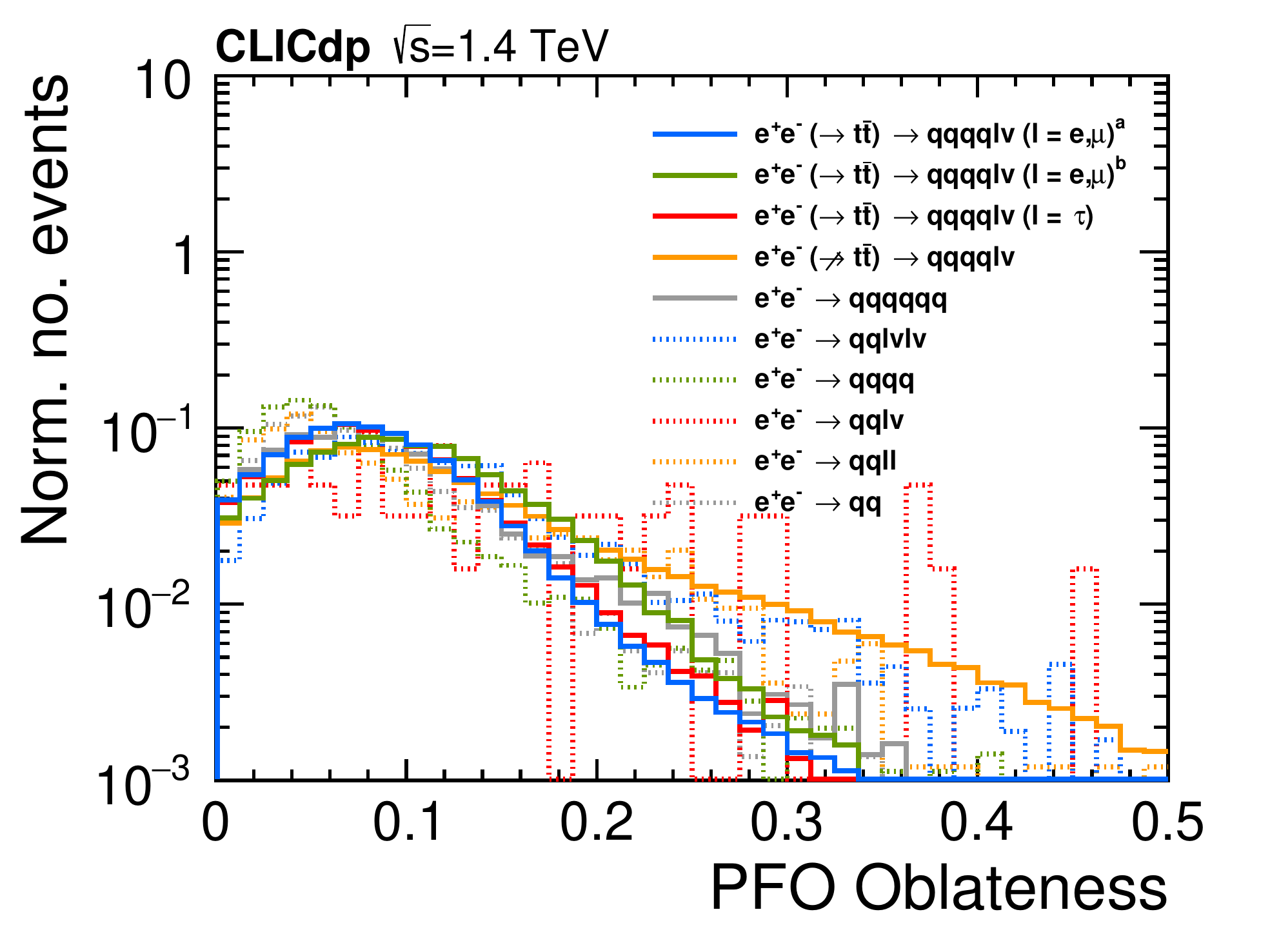}
	\caption{PFO Oblateness}
	\end{subfigure}\\
	\vspace{5mm}
	\begin{subfigure}{0.48\columnwidth}
	\includegraphics[width=\textwidth]{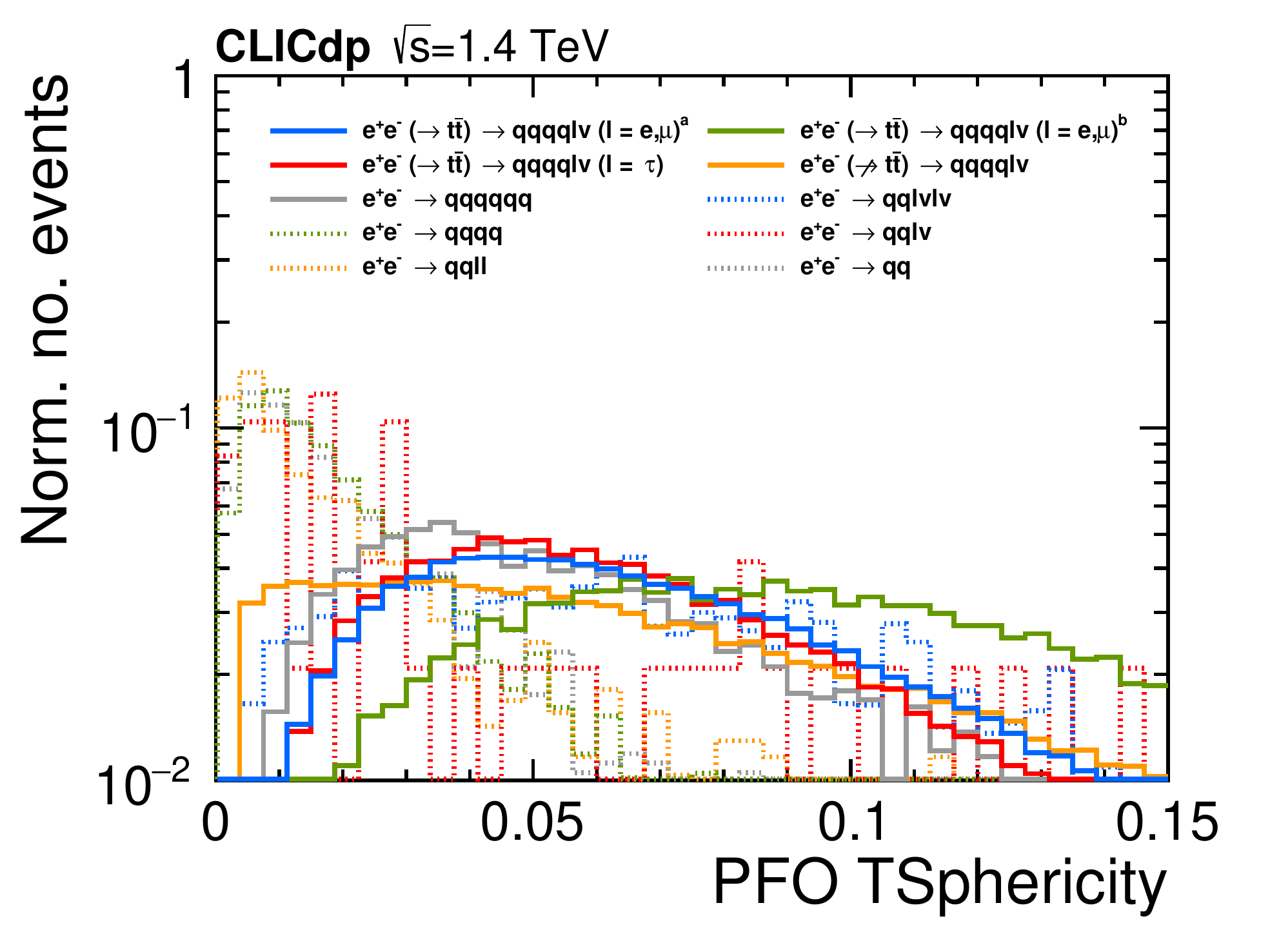}
	\caption{PFO TSphericity}
	\end{subfigure}
	~~~
	\begin{subfigure}{0.48\columnwidth}
	\includegraphics[width=\textwidth]{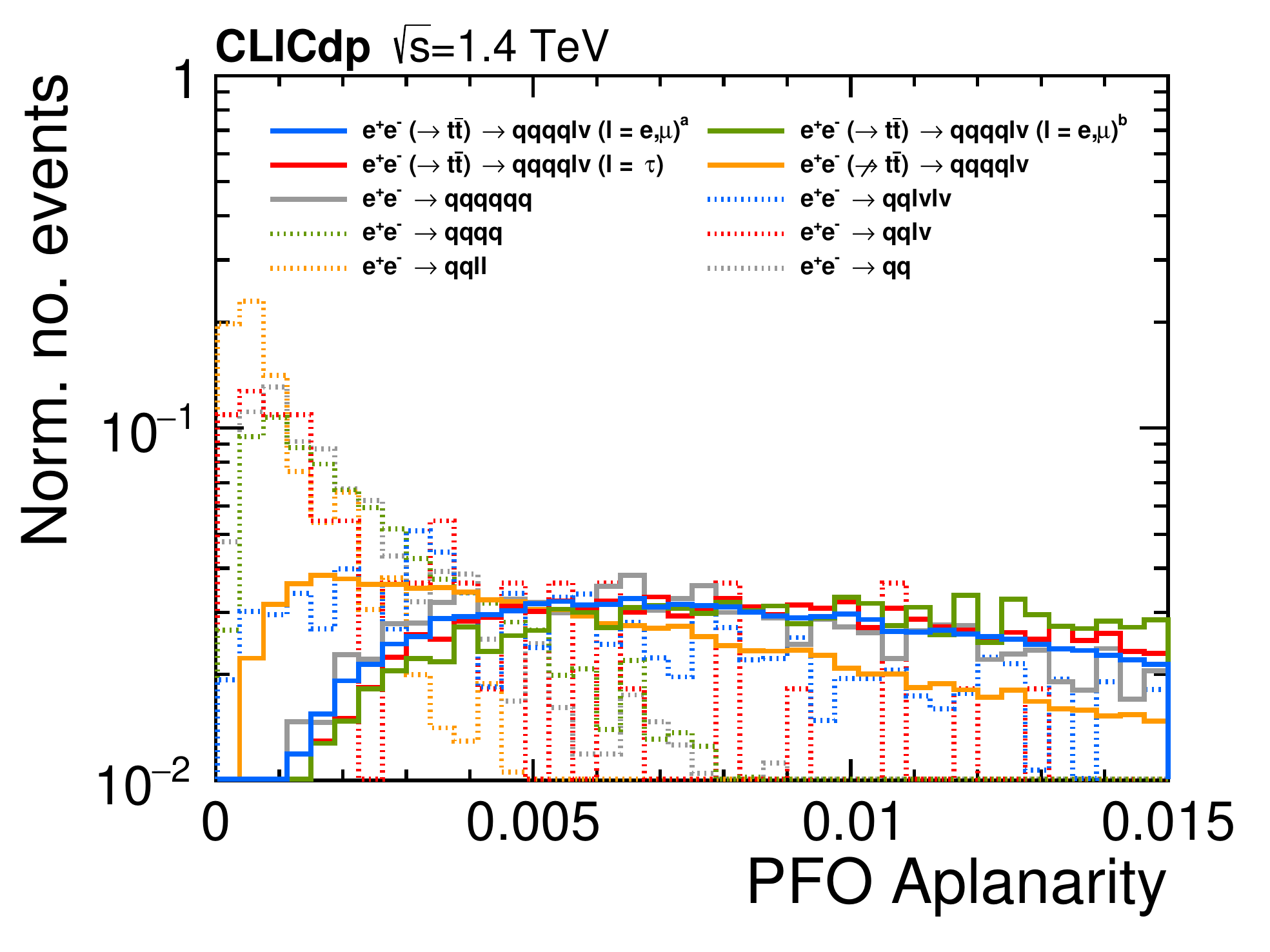}
	\caption{PFO Aplanarity}
	\end{subfigure}
\caption{Event shape variables for operation at $\roots=1.4\,\tev$ (left) and $\roots=3\,\tev$ (right). The distributions are shown after the application of pre-cuts. The superscript `a' (`b') refers to the kinematic region $\rootsprime\geq1.2\,\tev$ ($\rootsprime<1.2\,\tev$). \label{fig:analysis:mva:variables:eventshape}}
\end{figure}

\end{document}